\documentclass[11pt, reqno]{amsart}

   \usepackage{caption,float}
\usepackage[dcucite]{harvard}
\usepackage{fourier}

\usepackage{graphicx,amsmath,amssymb,epsfig,harvard}
\usepackage{graphicx,amsmath,amssymb,epsfig}
\usepackage{colortbl}

\usepackage{xr}
\usepackage{multirow}

\long\def\comment#1{} \oddsidemargin +0.2in \evensidemargin +0.2in
\usepackage{amssymb}
\usepackage{amsmath}
\usepackage{amsfonts}
\usepackage{graphicx}

\long\def\comment#1{}
\newtheorem{algorithm}{Algorithm}
\newtheorem{theorem}{Theorem}

\newtheorem{corollary}{Corollary}
\newtheorem{lemma}{Lemma}

\newtheorem{assumption}{Assumption}
\theoremstyle{definition}

\numberwithin{definition}{section}
\numberwithin{theorem}{section}
\numberwithin{lemma}{section}
\numberwithin{proposition}{section}
\numberwithin{corollary}{section}
\numberwithin{algorithm}{section}

\newtheorem{step}{Step}
\newtheorem{remark}{Remark}[section]
\numberwithin{remark}{section}
\newtheorem{example}{Example}
\numberwithin{example}{section}
\newcommand{\citen}{\citeasnoun}

\newcommand{\be}{\begin{eqnarray}}
\newcommand{\ee}{\end{eqnarray}}

\newcommand{\indep}{\perp\!\!\!\!\perp}

\newcommand{\ba}{\begin{array}}
\newcommand{\ea}{\end{array}}
\newcommand{\bs}{\begin{align}\begin{split}\nonumber}
\newcommand{\bsnumber}{\begin{align}\begin{split}}
\newcommand{\es}{\end{split}\end{align}}

\renewcommand{\(}{\left(}
\renewcommand{\)}{\right)}

\renewcommand{\hat}{\widehat}

\newcommand{\Ep}{{\mathrm{E}}}

\renewcommand{\Pr}{{\mathrm{P}}}

\renewcommand{\hat}{\widehat}
\renewcommand{\leq}{\leqslant}
\renewcommand{\geq}{\geqslant}
\newcommand{\Keywords}[1]{\par\noindent{\small{\em Keywords\/}: #1}}

\DeclareMathOperator*{\argmin}{arg\,min}
\DeclareMathOperator*{\argmax}{arg\,max}

\title[DR with Selection]{Distribution Regression with Sample Selection and UK Wage Decomposition }
\author[Chernozhukov, Fern\'andez-Val and Luo]{Victor Chernozhukov  \and Iv\'an Fern\'andez-Val \and Siyi Luo 
}

\date{\today}
\thanks{MIT, BU and BU. First Arxiv version: November 2018. We thank the editor James Heckman, three anonymous referees, Manuel Arellano, Marianne Bitler, Richard Blundell, Stephane Bonhomme, Shuowen Chen, Matt Hong,  Sylvia Klosin,  Dennis Kristensen, Victor Quintas Martinez, Enrique Sentana and the seminar participants at  Banco Central de Chile, Banco de Espa\~na, Bristol, BU, Cambridge, Cemfi, Chicago,  Erasmus, ESWC 2020, LSE, Michigan, Northwestern, NYU,  Oxford, PUC, Queen Mary, UCL and Virginia
for
helpful comments. We are extremely grateful to Richard Blundell and Barra Roantree at the IFS for providing us with the data. We gratefully acknowledge research support from the British Academy's visiting fellowships and National Science Foundation. Fern\'andez-Val and Chernozhukov were visiting UCL CEMMAP over different points in time, while working on this paper; they are both grateful for their hospitality. }
\begin{document}
\maketitle
\begin{abstract}
We develop a  distribution regression model under endogenous sample selection. This model
is a semi-parametric generalization of the Heckman selection model.  It accommodates much richer effects of the covariates on outcome distribution and patterns of heterogeneity in the selection process, and allows for drastic departures from the Gaussian error structure, while maintaining the same level tractability as the classical model. The model applies  to continuous, discrete and mixed outcomes.  We provide identification,   estimation, and inference methods, and apply them  to obtain wage decomposition for the UK. Here we  decompose the difference between the male and female wage distributions into composition, wage structure,  selection structure, and selection sorting effects.  After controlling for endogenous employment selection, we still find substantial gender wage gap -- ranging from 21\% to 40\% throughout the (latent) offered wage distribution that is not explained by composition. We also uncover positive sorting for single men and negative sorting  for married women that accounts for a substantive fraction of the gender wage gap at the top of the distribution.
\pagenumbering{gobble}
\bigskip

\end{abstract}

\newpage

\pagenumbering{arabic}
\section{introduction}\label{sec:intro}

Sample selection is ubiquitous in empirical economics. For example, it arises naturally in the estimation of wage equations because we do not observe wages of individuals who do not work \cite{gronau74,heckman74}.  Sample selection biases the estimation of causal and predictive effects when the reasons for not observing the data are related to the outcome of interest. In the wage example, there is sample selection bias whenever the employment status  and offered wage depend on common unobserved variables such as ability, motivation or skills. The most widely used solution to the sample selection bias is the Heckman selection model (HSM) introduced in \citen{heckman74}. The classical HSM provides a parsimonious and convenient way to account for sample selection by assuming parametric Gaussian distributions for the outcome and selection processes and imposing strong homogeneity assumptions on the impact of exogenous covariates. In this paper, we present a generalization of the HSM that allows for expressly non-Gaussian structures and eliminates the strong homogeneity restrictions, resulting in a semi-parametric model with some key parameters represented as nonparametric functions. This generalization enables a tractable econometric approach, similar in ease to the classical HSM.

To illustrate the central concepts, we take labor supply as an example. Let $Y^*$ denote the latent offered wage (outcome), and $D^*$ be the net disutility from working (selection), e.g., difference between reservation and offered wage.  We observe the employment status indicator equal to one when the offered wage is greater than the reservation wage, 
$D = 1(D^* \leq 0)$, and wage, $Y= Y^*$, only in the case of employment, $D=1$. Assume that the drivers of the outcome and selection obey the additive structure:\footnote{We can have other exogenous covariates $X$ affecting all model components.  We interpret the discussion here as conditional on $X$ having taken a fixed value.  We suppress the dependence on $X$ here for notational convenience.}
\begin{equation}\label{eq:hsm}
Y^* = \mu + U, \quad D^* = \nu(Z) + V,
\end{equation}
where $(\mu, \nu(Z))$ are the means of the outcome and selection, $Z$ is an instrumental variable that shifts the disutility of working but does not affect the offered wage, and $(U,V)$ are centered stochastic shocks independent of $Z$. The classical HSM results from restricting $(U,V)$ to follow a Gaussian distribution: 
\begin{equation}\label{heckman}
\Pr (U \leq u, V \leq v) = \Phi_2(u/\sigma_U,v/\sigma_V; \rho),
\end{equation}
where $\Phi_2(\cdot,\cdot;\rho)$ denotes the standard bivariate normal distribution function with correlation $\rho$, and $\sigma_U$ and $\sigma_V$ the standard deviations of $U$ and $V$. This model allows one to identify the distribution of the offered wage  from the distribution of observed wage and provides tractable inference.  The classical HSM has been challenged for its reliance on parametric  assumptions to obtain identification, and the data often reject these assumptions (e.g., due to clustering of offered and observed wages at the minimum wage and other levels). These challenges motivate the generalizations we consider in this study.


We generalize the classical HSM in several dimensions. We start by relaxing the Gaussian assumption on the marginal distributions of the stochastic shocks in (\ref{heckman}):
\begin{equation}\label{paranorormal}
\Pr(U \leq u, V \leq v) = \Phi_2 ( \Phi^{-1}\left(F_U(u)\right), \Phi^{-1}\left(F_V(v)\right); \rho),
\end{equation}
where $F_U$ and $F_V$ are non-parametric marginal distributions of the stochastic shocks $U$ and $V$, which can be continuous, discrete or mixed,  and $\Phi^{-1}$ is the quantile function of the standard normal. The distribution of the latent offered wage $Y^*$ is characterized by the parameters $(\mu, F_U)$. Despite the still partly Gaussian nature, the model can astutely generate drastically non-Gaussian distributional shapes, as demonstrated by \citen{wasserman-nonparanormal} and illustrated in Figure 1. Our contribution is to demonstrate that the distribution of the latent offered wage $Y^*$ is still identified from the observed distribution of wages and employment if the instrumental variable $Z$ takes on at least two values, and to provide tractable estimation and inference methods.  
Thus, by using (\ref{paranorormal}) we break away from parametric assumptions and Gaussian behaviors of the classical sample selection model but maintain the same level of tractability.

\begin{figure}
    \includegraphics[height=.32\textwidth, width=0.32\textwidth]{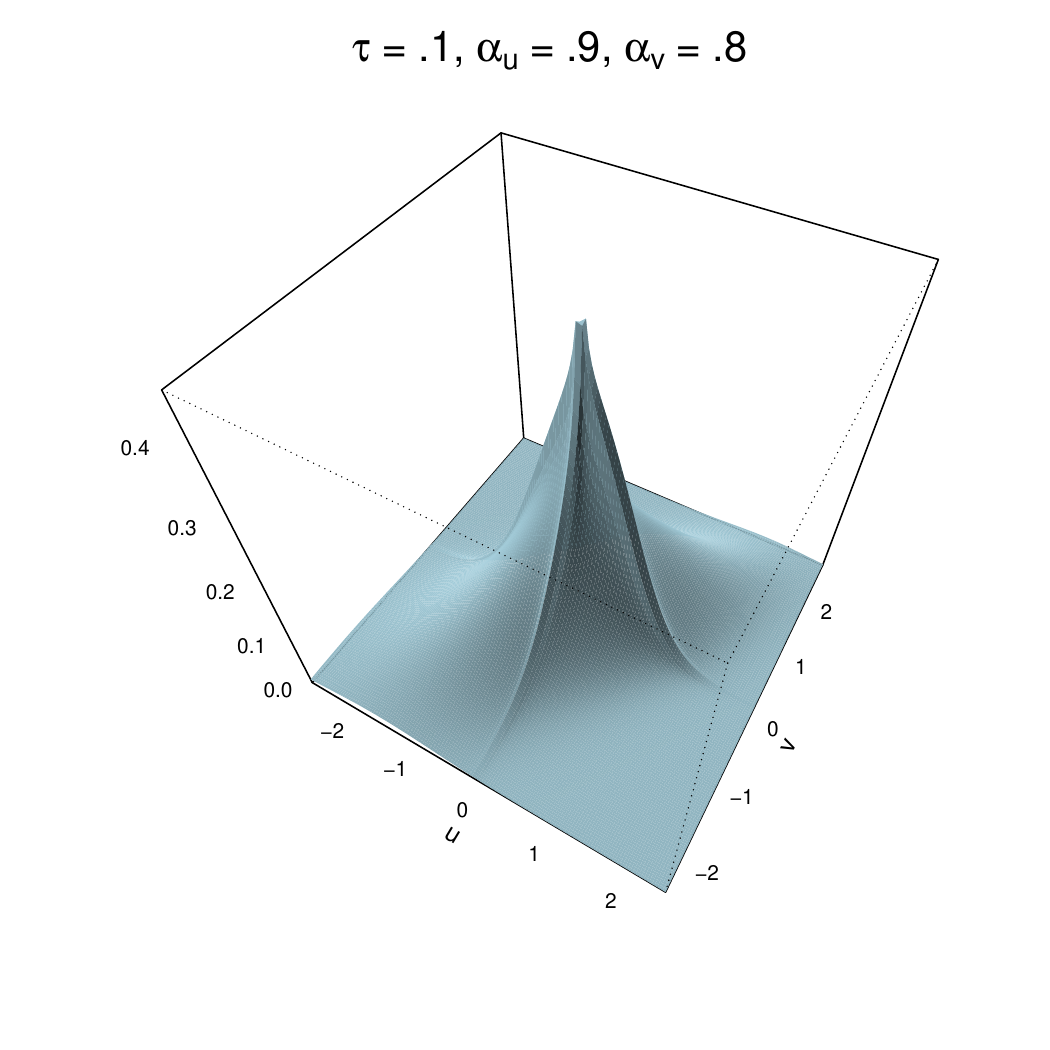}
    \includegraphics[height=.32\textwidth, width=0.32\textwidth]{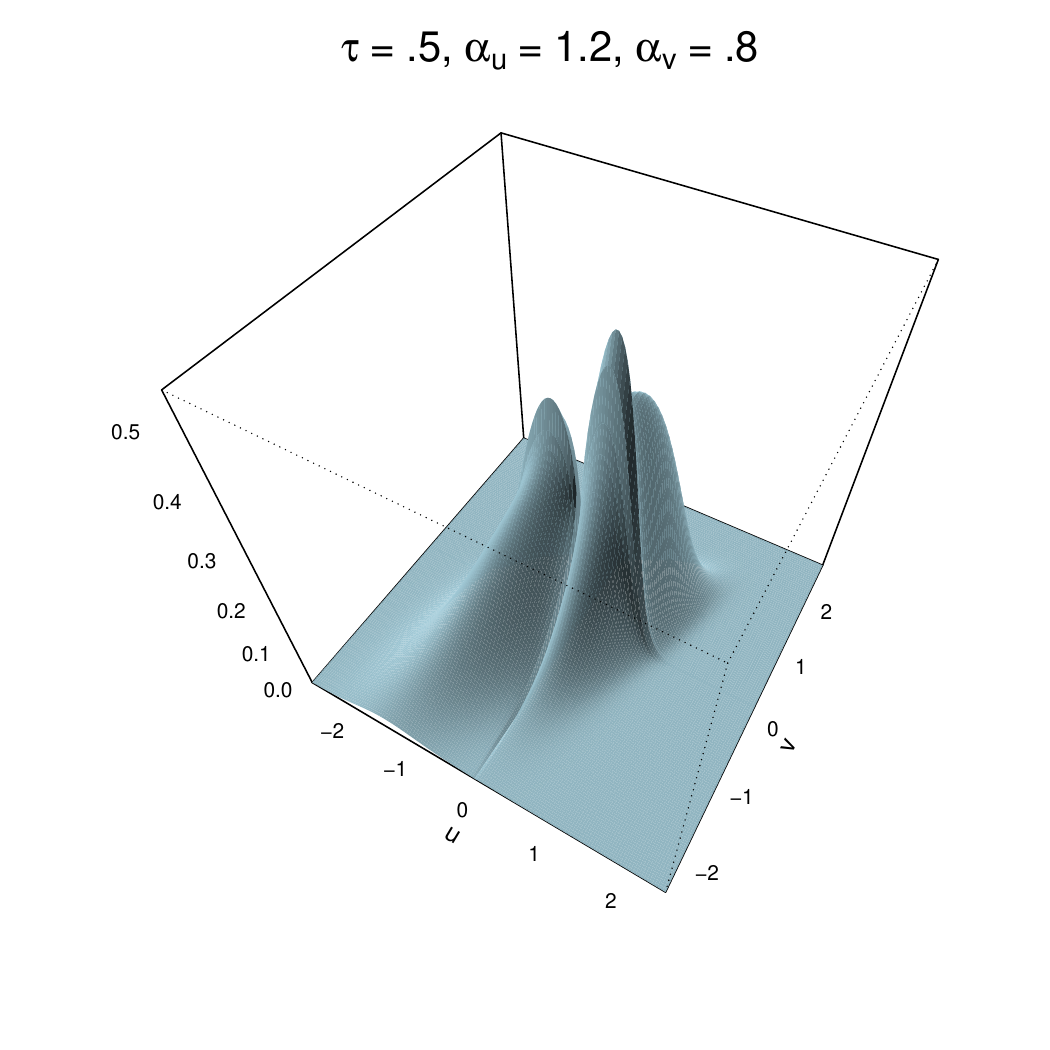}
    \includegraphics[height=.32\textwidth, width=0.32\textwidth]{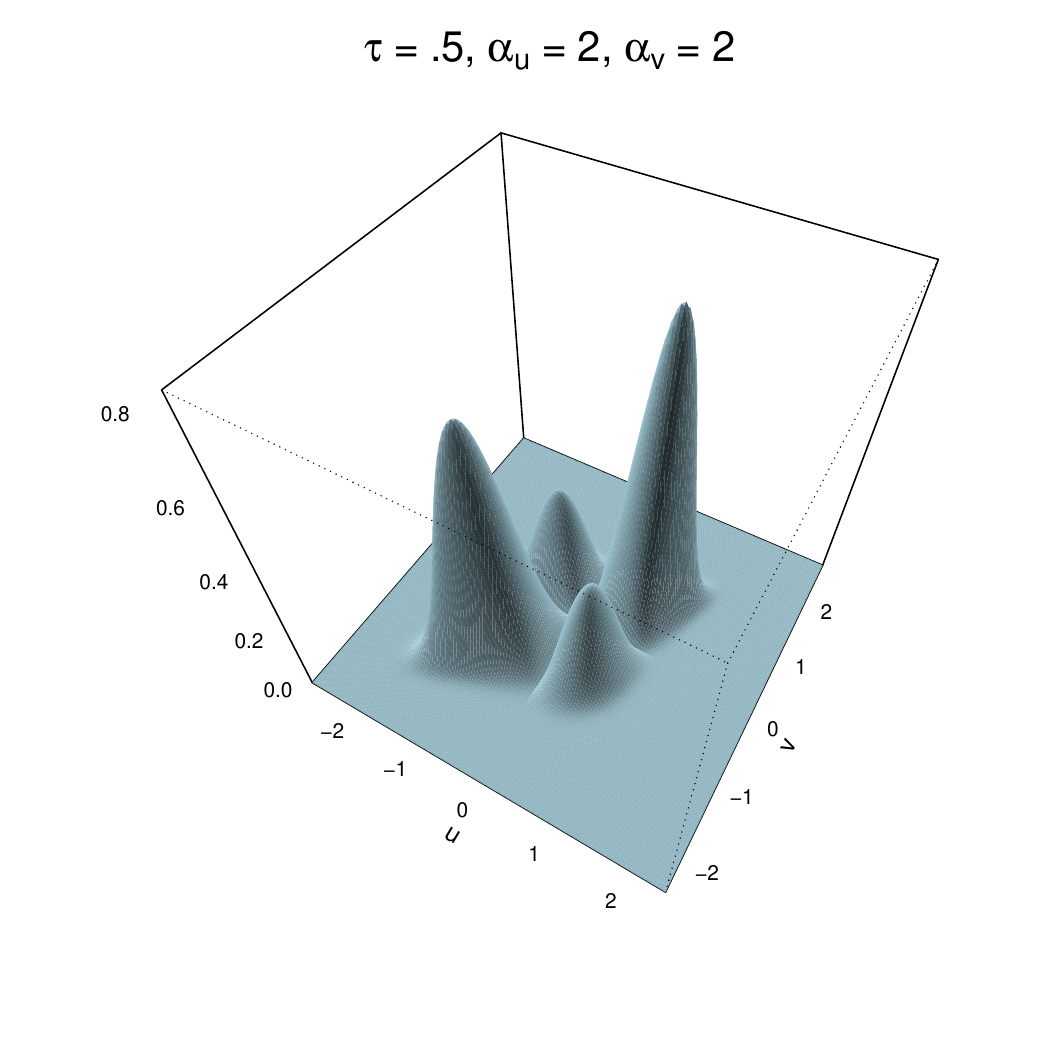}
    \caption{Beyond Gaussian: The richness of joint  density functions permitted by (\ref{paranorormal}).}
    \caption*{\small Notes. In the model (\ref{paranorormal}) the marginals $F_U$ and $F_V$ are nonparametric. To generate the panels, $F_U$ and $F_V$ are taken to be transformed power functions $u \mapsto \Phi(a \ (\mathrm{sign}(u)|u|^{\alpha} + \tau \ \max(0,u)^2)+b)$ for different choices of $\tau$ and $\alpha$, with $a$ and $b$ set such that the marginals have zero mean and unit variance; these examples extend those of \citen{wasserman-nonparanormal}. The correlation parameter $\rho$ is $.7$. The case $\tau=0$ and $\alpha=1$ for both $F_U$ and $F_V$ recovers the joint Gaussian density (not shown).}
    \label{fig:example}
\end{figure}

\paragraph{\textbf{The generalized HSM}}
The semi-parametric model (\ref{paranorormal}) presents an expressive generalization of the classical model, but for us it merely serves as a starting point to fix ideas.  We can get even more flexibility and expressivity while retaining the tractability of the identification and inference.  We further relax \eqref{eq:hsm}-\eqref{heckman}  by allowing the effect of $Z$ on $D^*$ to be non-additive and the dependence between $U$ and $V$ to be heterogeneous. The general form of our model specifies the joint distribution of 
latent outcome and selection drivers as\begin{equation}\label{general}
\Pr(Y^* \leq y, D^* \leq d \mid Z) = \Phi_2 ( \Phi^{-1}\left(F_{Y^*}(y)\right), \Phi^{-1}\left(F_{D^*}(d \mid Z)\right); \rho(y,d)),
\end{equation}
where the entire conditional distribution of $D^*$ can now depend on $Z$; 
  and $\rho(y,d)$ is a local correlation parameter that measures the \textit{strength of selection or sorting}.
The model imposes exclusion restrictions -- $Z$ affects the marginal distribution of $D^*$, but does not affect the marginal distribution of $Y^*$ nor the strength of selection. We show that this model admits identification and inference as convenient as the classical model, requiring again only that $Z$ takes on at least two values.  Moreover, the model is overidentified (hence testable) when $Z$ takes on more than two values.

Finally,  we incorporate exogenous covariates  yielding the model:
\begin{equation}\label{general2}
\Pr(Y^* \leq y, D^* \leq d \mid Z,X) = \Phi_2 ( \Phi^{-1}(F_{Y^*}(y\mid X)), \Phi^{-1}(F_{D^*}(d \mid Z,X)), \rho(y,d \mid X)),
\end{equation}
where the covariates $X$ can affect both the marginal conditional distributions of $Y^*$ and $D^*$ and  the sorting mechanism. This model expressly allows for  heterogenous effects of $X$'s on different parts of  distribution. The data on U.K. wages lends strong empirical support for this property, with marital status and other variables heterogenously
impacting different parts of the wage distribution. 
The heterogeneity in $X$ property also contrasts sharply with the classical model (\ref{heckman}) or its generalization (\ref{paranorormal}), which confine the covariates to shift the location of the distribution, but not other shape properties; see \citen{mr08} for a use of the Gaussian model where covariates shift both location and scale, but not other properties of the distribution.  The data firmly reject the location shift restrictions.

Our econometric development proceeds by 
specifying a flexible pointwise approximation to $\Pr(Y^* \leq y, D^* \leq d \mid X, Z)$ based upon the model's logic, akin to those used in univariate distribution regression (DR) models \cite{foresiperacchi95,chernozhukov+13inference}.\footnote{Flexible refers to the use of regressors constructed from original raw regressors by taking series transformations and interactions, for example.} This leads to a tractable Heckman-type approach to estimating the model parameter functions. The first step consists of a probit regression for the selection equation, as in the Heckman two-step method \cite{heckman79}. The second step estimates multiple bivariate probit regression with sample selection correction. We derive functional central limit theorems for all the estimators and their bootstrapped versions and use these results to perform uniform inference on function-valued parameters, for example, the distribution of the offered wages as well as counterfactual distributions induced by various hypotheses. This type of inference provides  simultaneous confidence bands and hypotheses tests about the functions of interest. Specifically, we rely on multiplier bootstrap \cite{gz84}, applied to the estimated influence functions as in \citen{lewbel95consistent}, \citen{hansen1996inference}, \citen{ch06}, \citen{ks12} and \citen{cck13}. 

We  utilize the proposed methods to analyze data on wages and employment in the UK, covering the period from 1978 to 2013. By estimating the conditional wage distributions for men and women and conducting wage decompositions that account for the endogeneity of employment selection, we shed light on the relationship between wage and employment in the UK. The results highlight the existence of positive sorting among single men and negative sorting among married women, which is consistent with the assortative matching hypothesis in the marriage market. Furthermore, the findings show that this difference in selection sorting explains a substantial portion of the gender wage gap at the upper end of the distribution, aligning with recent theories centered on glass ceiling. We still find that most of the gender gap in offered and observed wages can be attributed to differences in the wage structure that are often associated with gender discrimination in the labor market. The effect of education on wages is positive and increases with the distribution. In summary, our results demonstrate the significance of considering selection and flexible heterogeneity in empirical analysis and provide support for the use of the generalized Heckman selection model (\ref{general2}). Our results also align with those of \citen{mw18} which applied a quantile-regression based methodology of \citen{ab17} to U.S. data over a similar period.

This paper contributes to the extensive literature on sample selection in economics and statistics. The classical references in this field include works by \citen{gronau74}, \citen{heckman74}, \citen{lee82}, \citen{goldberger83}, \citen[Section 10.7]{amemiya85}, \citen[Section 9.4]{maddala86}, \citen{manski89}, \citen{manski94}, and \citen{vella98}. A commonly utilized approach to sample selection is the Heckman selection model developed by Heckman \citeyear{heckman74,heckman76,heckman79,heckman90}. Several extensions to the HSM have been made, including parametric extensions by \citen{lee83}, \citen{prieger02} and \citen{smith03} using different distributions, and a bivariate t-distribution extension by \citen{marchenko12} for heavy-tailed data. Semi-parametric versions have been developed by \citen{ahn93}, \citen{powell94}, \citen{andrews98}, and \citen{newey99}, while \citen{das03} introduced a nonparametric version with additive shocks, all of which focus on location effect models with homogeneous effects. None of these extensions can accommodate all sources of heterogeneity considered in our model. Previous studies on partial identification of unrestricted sample selection models include works by \citen{manski1990}, \citen{balke1994counterfactual}, \citen{manski94}, \citen{heckman2001instrumental}, and \citen{manski03}, among others. Our work contributes with new partial identification results to this development, with the main emphasis on achieving point identification.  In the online Supplemental material, we consider relaxed forms of the exclusion restriction on the sorting that obtain a nested sequence of bounds, which starts from a single point and ends with the (agnostic) \citen{balke1994counterfactual} bounds. This sequence provides a form of sensitivity analysis, analogous in spirit to \citen{christensen:sense}'s sensitivity analysis for structural models.

\citen[AB17]{ab17} proposed another extension of the HSM, which like our model accounts for multiple sources of heterogeneity. Their approach is based on quantile regression, which models the marginal distribution of the latent outcome, and a parametric copula model that links the latent selection and outcome variables. However, compared to our approach, this model requires the latent variables $(Y^*,D^*)$ to be continuous, while our approach accommodates mixed discrete-continuous distributions, which is important given that offered wages are often constrained to be above a minimum wage level, and observed wages have point masses at minimum wage and other levels due to wage rounding. Additionally, AB17 models the effect of covariates on the conditional quantile of the latent distribution, while we model the effect on the conditional latent distribution. The identification assumptions of the two models are also distinct, as our approach imposes more structure on the dependence between the outcome and selection processes, whereas AB17 requires continuous variation in the instrument $Z$ and real analyticity of the copula function governing the dependence of the shocks in the outcome and selection structures.  The analyticity condition implies extrapolability of the conditional distribution of outcome to the case where there is no selection, enabling identification at infinity.
 \ \ In this case, however, identification does not automatically imply consistent estimability, because the continuation of an analytical function is known to be an ill-posed problem.
 \ \ To overcome the latter problem, regularization must be used: AB17 impose parametric assumptions on the copula as a means of such regularization to carry-out estimation. 


Finally, there are relevant connections to the work of \citen{wasserman-nonparanormal} on the estimation of graphical models using the semi-parametric family (\ref{paranorormal}). While our approach considers the more general model (\ref{general2}), we contribute by offering a DR based approach, which can incorporate covariates $(X,Z)$ in a general yet tractable manner. Furthermore, our focus is on addressing the sample selection problem in the context of these models.

\paragraph{\textbf{Outline}}  Section \ref{sec:id}  examines the identification problem under sample selection using a new representation of a joint distribution. Section \ref{sec:model} introduces the DR model with selection and associated functionals,  estimators of the model parameters and functionals, and a multiplier bootstrap method to perform functional inference. Section \ref{sec:empirics} reports the results of the empirical application. The  proofs of the  results of Section \ref{sec:id} are gathered in Appendix \ref{app:id}. The online Supplemental Material (SM) contains deferred discussions of Sections \ref{sec:id}--\ref{sec:empirics}, the asymptotic theory for our estimation and inference methods, additional empirical results, a Monte Carlo simulation calibrated to the empirical application and other technical results.


\section{Local Gaussian Representation and Sample Selection}\label{sec:id}
\subsection{Local Gaussian Representation of a Joint Distribution}

Our first result shows that any 
joint distribution of two random variables has a local Gaussian representation (LGR).  This is a useful representation for econometric analysis, because it naturally nests the joint normal distribution and rich semi-parametric models that allow for nonparametric marginal distributions-- with much more room to spare. Indeed, we shall use the LGR to provide a new view of the  identification problem with sample selection and motivate our modeling choices later.

Let $Y^*$ and $D^*$ be two random variables with joint cumulative distribution function (CDF) $F_{Y^*,D^*}$ and marginal CDFs $F_{Y^*}$ and $F_{D^*}$. We label these variables with  asterisks because they will be latent variables when we introduce sample selection. 

\begin{lemma}[Local Gaussian Representation of Any Joint  Distribution]\label{lemma:lgr} The joint distribution $F_{Y^*,D^*}$ can be represented via a standard bivariate normal distribution at any point $(y,d)$ as 
\begin{equation}\label{eq:lgr}
F_{Y^*,D^*}(y,d) \equiv 
\Phi_2(\Phi^{-1}(F_{Y^*}(y)), \Phi^{-1}(F_{D^*}(d)); \rho(y,d)),
\end{equation}
where \ $\Phi_2(\cdot, \cdot; \rho)$ is the joint CDF of a standard bivariate normal random variable with parameter $\rho$; 
and $\rho(y,d) \in [-1,1]$ is the implied or local correlation parameter that depends on $(y,d)$, whose value is unique.
\end{lemma}

Lemma \ref{lemma:lgr} establishes that any bivariate CDF admits a unique pointwise representation by standard bivariate normal distributions. This result is related to Sklar's copula representation of the joint distributions but is different, especially in using the localized correlation.
Also this result is stronger than the comprehensive property of the Gaussian copula that establishes that this copula includes the two Frechet bounds and independent copula by suitable choice of the correlation parameter,  e.g., \citen{smith03}.  We also note that  Lemma \ref{lemma:lgr}  easily extends to CDFs conditional on covariates by making all the parameters dependent on the value of the covariates.

In what follows, it is convenient to define the LGR parameters:  $$\mu(y) := \Phi^{-1}(F_{Y^*}(y)),  \ \ \nu(d) := \Phi^{-1}(F_{D^*}(d)),$$ 
noting that
$\mu(y) \in \overline{\mathbb{R}}$ and  $\nu(d) \in \overline{\mathbb{R}}$, where $\overline{\mathbb{R}} := \mathbb{R} \cup \{-\infty,+\infty\}$ is the extended real number line.
By LGR, the mapping between $F_{Y^*,D^*}(y)$ and its LGR parameters $(\mu(y), \nu(d), \rho(d,y))$ is bijective.  Thus LGR carries the same information as the joint CDF.

The implied correlation $\rho(y,d)$ is a key parameter that measures local dependence. Indeed,  $\rho(y,d) = 0$ if and only if the distribution $F_{Y^*,D^*}$ factorizes at $(y,d)$ because 
$$
F_{Y^*,D^*}(y,d) = \Phi_2(\Phi^{-1}(F_{Y^*}(y)), \Phi^{-1}(F_{D^*}(d)); 0) =  F_{Y^*}(y) F_{D^*}(d),
$$
that is, $\rho(y,d) = 0$ if and only if the events $\{Y^*\leq y\}$ and $\{D^*\leq d\}$ are independent. Moreover, $\rho(y,d)$ is positive  if and only if correlation of 
$1(Y^*\leq y)$ and $1(D^* \leq d)$
is positive.
Thus, if $\rho(y,d)$ is positive  everywhere 
then $Y^*$ and $D^*$ are positively quadrant dependent \cite{lehmann66}.   We prove these assertions and provide additional discussion in Appendix \ref{app:lgr} of the SM.\footnote{See also \citen{survey18} for a recent review of copula-based measures of local dependence.}

\subsection{Identification of Sample Selection Model}
We consider now the sample selection problem where we observe two random variables $D$ and $Y$, which can be defined in terms of the latent variables  $D^*$ and $Y^*$ as
   \begin{eqnarray*}
    D &=& 1(D^* \leq 0),\\
    Y &=& Y^* \  \ \text{if} \ \ D=1, 
    \end{eqnarray*}
i.e., $D$ is an indicator for $D^* \leq 0$ and $Y^*$ is only observed when $D = 1$.  The goal is to identify features of the joint distribution of the latent variables from the joint distribution of the observed variables.  

We can write the distribution of the observed variables as
$$
\Pr(D = 1) = \Phi(\nu) \text{ and } \Pr(Y \leq y, D=1) = \Phi_2(\mu(y), \nu; \rho(y)),
$$
where $\mu(y)$, $\nu := \nu(0)$ and $\rho(y) := \rho(y,0)$ are the parameters of LGR for the latent distribution $F_{Y^*,D^*}$. The identified set for these parameters determine the identified set for $F_{Y^*}(y)$. Note that in particular,
$$
F_{Y^*}(y) = \Phi(\mu(y)).
$$
As shown below, $\mu(y)$ and $\rho(y)$ are partially identified. We proceed by characterizing the identified set for these parameters  and provide exclusion restrictions to achieve point identification.  We also examine the role of relaxed exclusion and other restrictions in reducing the size of the identified set in Appendix \ref{app:bounds} of the SM.  

To understand the source of partial identification under sample selection, note that there are two free probabilities,
$
\Pr(D = 1)$ and $\Pr(Y \leq y, D = 1)$, 
to identify three parameters,  $\mu(y)$, $\nu$ and $\rho(y)$. The selection probability pins down $\nu$:    $
   \nu = \Phi^{-1}(\Pr(D = 1)).
   $
The parameters $\mu(y)$ and $\rho(y)$ are partially identified by the set of  solutions in $(\mu,\rho)$ to
the equation
   $$
   \Pr(Y \leq y, D = 1) = \Phi_2(\mu,\Phi^{-1}(\Pr(D = 1)); \rho).
   $$
These solutions  form a one-dimensional manifold in $\mathbb{R} \times (-1,1)$ \cite{spivak65,munkres91}.\footnote{This is because $
\partial \Phi_2(\mu, \cdot; \rho)/\partial \mu  > 0
$, $
\partial \Phi_2(\cdot, \cdot; \rho)/\partial \rho > 0
$, and  $
\partial^2 \Phi_2(\cdot, \cdot; \rho)/\partial \mu \partial \rho > 0
$.}  
\subsection*{Exclusion Restrictions}To state the exclusion restrictions, let $Z$ be a candidate instrumental variable and $F_{Y^*,D^* \mid Z}$ be the joint CDF of $Y^*$ and $D^*$ conditional on $Z$. Then,  $F_{Y^*,D^* \mid Z}$ admits the LGR: 
  \begin{equation*}
  F_{Y^*,D^* \mid Z}(y, d \mid z) = \Phi_2(\mu(y \mid z), \nu(d \mid z); \rho(y,d \mid z)), 
  \end{equation*}
 with parameters $\mu(y \mid z) \in \overline{\mathbb{R}}$, $\nu(d \mid z) \in \overline{\mathbb{R}}$, and $\rho(y,d \mid z) \in [-1,1]$.  
The exclusion restrictions are the following.

\begin{assumption}[Exclusion Restrictions]\label{ass:er} There is a binary random variable $Z$  that satisfies:
   \begin{enumerate}
    \item Non-Degeneracy: $0 < \Pr(D = 1) < 1$ and $0 < \Pr(Z=1 \mid D = 1) <1$.
   \item Relevance: $
    \Pr(D = 1 \mid Z = 0) < \Pr(D = 1 \mid Z = 1) < 1.
   $
   \item Outcome exclusion: $\mu(y \mid z) = \mu(y)$ for all $y \in \mathbb{R}$ and $z \in \{0,1\}$.
   \item Selection Sorting exclusion: $\rho(y,0 \mid z) = \rho(y,0)$ for all $y \in \mathbb{R}$ and $z \in \{0,1\}$.
   \end{enumerate}
\end{assumption}
The condition that $Z$ is binary emphasizes that our identification strategy does not rely on large variation of $Z$. If $Z$ is not binary, we only require that Assumption \ref{ass:er} be satisfied for two values of $Z$. If it is satisfied for more than two values of $Z$, then the model is overidentified and the exclusion restrictions become testable.\footnote{We leave the development of such specification test to future research.}   Non-degeneracy states that there is sample selection and that $Z$ has variation in the selected population. Relevance requires that $Z$ affects the probability of selection. The condition $\Pr(D = 1 \mid Z = 1) < 1$ precludes identification at infinity, which we analyze separately in Remark \ref{remark:idi}. The sign of the first inequality can be reversed by relabelling the values of $Z$.   Outcome exclusion is a standard exclusion restriction, which is not sufficient for point identification in the presence of sample selection \cite{balke1994counterfactual,manski94,heckman2001instrumental,manski03}. It holds when $Y^*$ is independent of $Z$.\footnote{\citen{kitagawa2010} developed a test for the outcome exclusion.} Selection sorting exclusion requires the local correlation function to be independent of $Z$.\footnote{ \citen{torgovitsky2010identification} previously used this type of restriction to analyze identification of nonseparable models with continuous endogenous explanatory variables.}
AB17 consider an alternative condition to selection sorting exclusion based on real analyticity and continuous variation of $Z$. We refer to Appendices \ref{app:ab17} and \ref{app:analy} in the SM for a comparison with the analyticity approach and another alternative approach based on imposing semi-parametric structure on the sorting mechanism when $Z$ takes on more than 2 values.

We can get some intuition about the outcome and selection exclusion restrictions with the parametric and semi-parametric models of labor supply given in \eqref{eq:hsm}--\eqref{paranorormal}.  These models trivially satisfy the conditions stated above, because $Z$ only affects the latent disutility $D^*$ through the location, and the local correlation parameter is constant and does not depend on $Z$ (or $y$ for that matter). Yet the selection sorting exclusion does entail some loss of generality, which motivates forms of relaxed conditions on selection sorting that we consider in Appendix \ref{app:bounds} of the SM.

We now show how the presence of exclusion restrictions helps identify the parameters. Under Assumption \ref{ass:er} the conditional LGR at $d=0$ simplifies to
  \begin{equation}\label{eq:clgr}
  F_{Y^*,D^* \mid Z}(y, 0 \mid z) = \Phi_2(\mu(y), \nu(z); \rho(y)), \ \  z \in \{0,1\},
  \end{equation}
  where we simplify the notation $\nu(z) := \nu(0 \mid z) \text{ and } \rho(y) := \rho(y,0).$
We can relate this representation to the conditional distribution of  observed variables as
\begin{equation*}
\Pr(D = 1 \mid Z = z) =  \Phi(\nu(z)) ; \quad \Pr(Y \leq y, D=1 \mid Z=z) = \Phi_2(\mu(y), \nu(z); \rho(y)), \quad  z \in \{0,1\}.
\end{equation*}
 As before, $\nu(z)$ is identified from the conditional selection probability:
  \begin{equation}\label{eq:nu}
  \nu(z) = \Phi^{-1}\left( \Pr(D=1 \mid Z = z)\right), \quad z \in \{0,1\}.
  \end{equation}
Moreover, $\mu(y)$ and $\rho(y)$ are identified as the solution in $(\mu, \rho)$  to
  \begin{equation}\label{eq:murho}
  \Pr(Y \leq y, D=1 \mid Z=z) =  \Phi_2(\mu, \Phi^{-1}\left( \Pr(D=1 \mid Z = z)\right); \rho), \ \ z \in \{0,1\}.
 \end{equation}
This is a nonlinear system of two equations in two unknowns.  The result below states that the solution exists and is unique.

\begin{theorem}[Identification under Assumption 1]\label{theorem:id} Suppose that  Assumption \ref{ass:er} holds.  Suppose that the distribution of the observed variables $(Y,D,Z)$ implies $\rho(y)^2< 1$, then  $\mu(y)$ and $\rho(y)$ are point identified as the unique interior solution of \eqref{eq:murho} in $(\mu, \rho)$. 
\end{theorem}

The result follows by showing that  the Jacobian of the equations in \eqref{eq:murho} is a P-matrix for all $\mu \in \mathbb{R}$ and $\rho \in (-1,1)$, so uniqueness follows  by the global univalence result in Theorem 4 of \citen{gale65}.  
 We defer to Appendix \ref{app:full_thm} the analysis of the case $\rho(y)^2=1$ due to its more technical nature. This case deals with boundary situations that can occur at extreme values of $y$ or other non-typical circumstances.
In these situations, we can have either point or partial identification. They are easy to detect empirically.

\begin{remark}[Identification at Infinity]\label{remark:idi}  When the instruments are strong enough to set $$\Pr(D = 1 \mid Z = 1) = 1$$ and provided that the exclusion conditions holds $\mu(y \mid z) = \mu(y)$, the conditional LGR at $z=1$ gives $$\Pr(Y \leq y, D=1 \mid Z=1) = \lim_{\nu \nearrow + \infty} \Phi_2(\mu(y), \nu; \rho(y \mid 1)) = \Phi(\mu(y)),$$ which  identifies $\mu(y)$ by
   $$
  \mu(y) = \Phi^{-1}(\Pr(Y \leq y, D=1 \mid Z=1) ),
  $$
\textit{without the selection sorting exclusion}.\footnote{
Despite its extreme nature, this identification strategy is useful in empirical economics: e.g., \citen{mr08} use this reasoning to analyze wage penalty for highly educated women.}
This result is analogous to the identification at infinity of  \citen{chamberlain86} where $Z$ is continuous with unbounded support and 
  $
  \lim_{z \nearrow +\infty} \Pr(D = 1 \mid Z = z) = 1;
  $
  see also \citen{lewbel2007endogenous} for an alternative identification at infinity strategy using the special regressor approach. Interestingly we note that $\rho(y \mid 1)$ is not point-identified without further restrictions.  \qed
\end{remark}

\subsection{Selection Sorting Exclusion}\label{subsec:se}

We  provide a further interpretation of the selection exclusion restriction in a formulation of the sample selection problem in terms of the propensity score. The sample selection process is often represented as $D = 1\{V \geq \nu(Z)\}$, where $\nu(z) := \Phi^{-1} (p(Z))$, with $p(Z):=\Pr(D=1 \mid Z=z)$ being the propensity score, and $V \mid Z \sim N(0,1)$ is the unobserved selection normal score. Here we shall maintain Assumption \ref{ass:er}(1)--(3). The following theorem shows the condition that selection sorting exclusion  imposes on the relationship between $Y^*$ and $V$. Let $$F_{Y^*,V \mid Z}(y,v \mid z)= \Phi_2(\Phi^{-1}(F_{Y^*}(y)), v; \tilde{\rho}(y,v \mid z) )$$ be the LGR of the joint CDF of $(Y^*,V)$ conditional on $Z$, where we use that $Y^*$ is independent of $Z$ and $V \mid Z \sim N(0,1)$. 

\begin{theorem}[Interpretation of Selection Sorting Exclusion]\label{thm:se} Suppose Assumption \ref{ass:er}(1)--(3) hold. Then  Assumption \ref{ass:er}(4) holds if and only if $\tilde \rho(y,\nu(z) \mid z) = \tilde \rho(y)$ for any $z$ in the support of $Z$.\end{theorem}

Theorem \ref{thm:se} shows that selection exclusion holds whenever the implied correlation  $\tilde \rho(y,v \mid z)$ between $Y^*$ and $V$ does not depend on $v$ and $z$. A sufficient condition is that $(Y^*,V)$ are jointly independent of $Z$ and $\tilde \rho(y,v) = \tilde \rho(y)$ for all $v$ in the support of $\nu(Z)$,  where $\tilde \rho(y,v)$ is the local dependence parameter of the joint CDF of $(Y^*,V)$.  In the context of labor supply, $Y^*$ is latent wage and $V$ is the  ranking in the conditional distribution of utility/net benefit of employment.  The classical HSM in (\ref{heckman}) and its semi-parametric generalization in (\ref{paranorormal}) trivially satisfy this condition, because they have constant local correlation, $\tilde \rho(y, v)= \rho$, that is, the strength of selection does not vary with the level of offered wage or the employment ranking. 
Relative to such traditional models, the condition $\tilde \rho(y,v) = \tilde \rho(y)$ leaves some room for additionally flexibility -- namely, the strength of selection can vary with the level of offered wage.  For example, $y \mapsto \tilde \rho(y)$ increasing means that workers are more likely to select into the labor force if the offered wage's rank is higher (holding the employment ranking fixed). 

The selection exclusion $\tilde \rho(y,v) = \tilde \rho(y)$ is  essentially equivalent to the single index restriction: 
\begin{equation}\label{eq:index}
\Pr(Y^* \leq y \mid V=v) =  \Phi \left(a(y) + b(y) v\right),    
\end{equation}
where $y \mapsto a(y)$ and $y \mapsto b(y)$ are nonparametric functions linked to the LGR of $Y^*$ and $V$, $F_{Y^*,V}(y,v)= \Phi_2(\mu(y), v; \tilde{\rho}(y) )$, via $a(y) = \mu(y)/\sqrt{1-\tilde{\rho}(y)^2}$ and $b(y) = - \tilde \rho(y)/\sqrt{1-\tilde{\rho}(y)^2}$. This index form enhances interpretability and also connects to the econometric literature that utilizes the simplifying index restrictions; e.g., \citen{ichimura1993semiparametric},  \citen{klein1993efficient}, \citen{powell94} and \citen{vytlacil:equiv}. The proof of the equivalence in \eqref{eq:index} is given in Appendix \ref{app:index}.
Homogeneous models, such as the classical Heckman's labor supply model or the semi-parametric generalization (\ref{paranorormal}), impose the much stronger single index condition:
$$
\Pr(Y^* \leq y \mid V=v) =  \Phi \left(a(y) + b v\right),
$$
for all $y$, for $b = -\rho/\sqrt{(1- \rho^2)}$ being 
a rescaled correlation coefficient; note that the Gaussian restriction on $Y^*$ of the HSM corresponds to the further linearity restriction: $a(y) = a y$.    This connection highlights the generalization our model brings: in the general model $a(y) \neq a y$ allows for expressive departures from Gaussianity, and $b(y) \neq b$ allows for much richer patterns of dependence across the conditional distribution. The restriction \eqref{eq:index} is related to  a  linear model assumption on the expectation of $Y^*$ conditional on $V$ that was previously used in \citen{brinch2017beyond} and \citen{kowalski2016doing} to identify marginal treatment effects in treatment effects settings with endogenous treatment assignments and discrete instruments. One difference, however, is that our restriction is a single index restriction on the conditional distribution.\footnote{Single index restrictions on the conditional distribution like \eqref{eq:index} and linearity restrictions on the conditional expectation are not nested. This can be seen, for example, from the relationship
$$
\Ep[Y^* \mid V = v] = \int_{-\infty}^{\infty} [1(y > 0) - \Pr(Y^* \leq y \mid V=v)] \mathrm{d} y.
$$}
Moreover, \citen{brinch2017beyond} and \citen{kowalski2016doing} analysis imposes that $(Y^*,V)$ are jointly independent of $Z$. In Appendix \ref{app:index2}, we provide an alternative sufficient condition for the single index restriction that does not rely on this joint independence.

\begin{figure}
\includegraphics[width=.8\textwidth, height=3.5in]{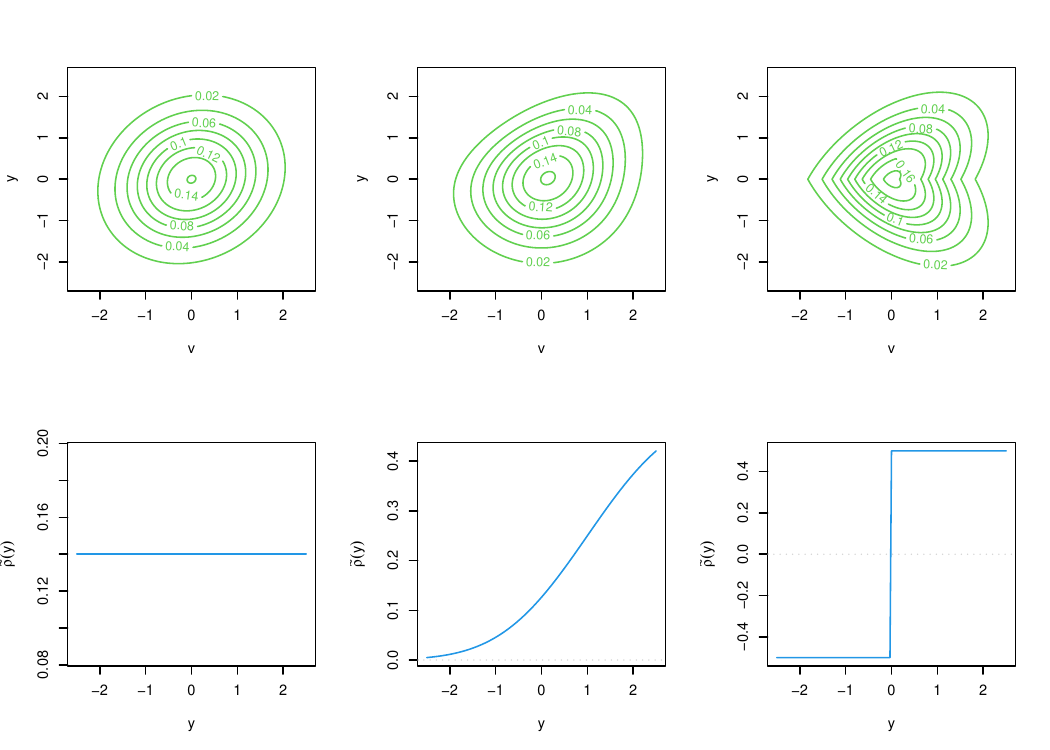}
    \caption{Joint CDF of $(Y^*,V)$ for different local correlation parameters $\tilde \rho(y)$; marginal CDFs of $(Y^*,V)$ are standard normal. }
    \caption*{\small Notes. Left panel: $\tilde \rho(y) = .14$; middle panel: $\tilde \rho(y) = \Phi(2(y-1)/3)/3$; right panel: $\tilde \rho(y) = 1(y \geq 0)/2 - 1(y < 0)/2$. The marginal distribution of $Y^*$ and $V$ are chosen as standard normal for simplicity.}
    \label{fig:example2}
    \end{figure}

We now elaborate on the value that the variation $\rho(y)$ with respect to $y$
brings to the table. What does this property mean in terms of the flexibility of dependence patterns between net offered wage $Y^*$ and employment ranking $V$? The standard
models that have homogeneous $\rho(y) = \rho$ or
\textit{equivalently} homogeneous $b(y) = b$, are only able to generate limited forms of dependence. The left panel of Figure \ref{fig:example2} presents an example of the contour
of the joint pdf of $(Y^*,V)$ when $\rho(y) = .14$.  In contrast, our model is able to generate richer forms
of dependence between $(Y^*, V)$ as illustrated in the middle and right panels. The middle panel has the same 
correlation between $Y^*$ and $V$ as the left panel, but we see that the
dependence increases from small values as we move towards the upper-right tail corner.  The right panel is another example,
where the correlation between $Y^*$ and $V$ is zero, but $(Y^*, V)$ are positively
dependent when $Y^*>0$
and negative dependent when $Y^*<0$. In all examples, the marginal distributions are fixed to be normal, but the joint distributions are not normal, except for the left panel. 


The above discussions aim to support the idea of using sorting exclusion, but one may and should challenge it in applications. To this end, we emphasize three points:
First,  as we mentioned in the discussion of Assumption  \ref{ass:er}, selection exclusion is testable when $Z$ takes on more than 2 values.
Second, one can work with the relaxed forms of exclusion restrictions, namely the sign
restrictions and r-relaxed exclusion restrictions and compute the bounds provided by Theorem \ref{theorem:ID2} in the SM.
Third, if richer instruments are available, one can still obtain results for identification under analyticity or other regularity assumptions on the local correlation function without selection sorting exclusion; see Appendix  \ref{app:analy} in the SM. 


\section{Econometrics of Distribution Regression Model with Sample Selection}\label{sec:model}
\subsection{The Model}
We consider a semi-parametric version of the LGR with covariates:
\begin{equation}\label{eq:drm}
    F_{Y^*,D^*}(y,0 \mid Z=z)   = \Phi_2(-x^\prime\beta(y), -z^\prime \pi;\rho(x'\delta(y))),
\end{equation}
where $Y^*$ is the latent outcome of interest, which can be either continuous, discrete or mixed;  $D^*$ is a latent variable that determines sample selection;   $X$ is a vector of covariates;  $Z=(Z_1,X)$; and $Z_1$ are excluded covariates, i.e., observed covariates that satisfy the exclusion restrictions.\footnote{It is understood that the models are flexible in the following sense. Given $x$ and $z$, we can generate constructed regressors  $t(x)$ and $b(z)$ as technical transformations of $x$ and $z$,  for example, by taking powers or splines of the components and their interactions. We then can reassign notation 
$x \leftarrow  t(x)$ and $z \leftarrow b(z)$. This convention entails no loss of generality provided that $t(x)$ and $b(z)$ contain the same information as  $x$ and $z$.  In this paper, we do not formally consider the case where the dimensions of $t$ and $b$ grow with the sample size, but this is possible along the lines of the series literature, with the inference being operationally equivalent to the case with the fixed number of series terms, provided that the number of series terms is small compared to the sample size  and that the approximation errors are negligible; e.g., \citen{newey:series} and \citen{chen:Chapter}.} The excluded covariates avoid reliance on functional form assumptions to achieve identification.

The model \eqref{eq:drm} is a semi-parametric version of \eqref{general2}, where we replace the components  $\Phi^{-1}(F_{Y^*}(y\mid X))$, $\Phi^{-1}(F_{D^*}(d \mid Z,X))$, and $\rho(y,d \mid X)$ by three indexes.\footnote{Here we include the covariates $X$ in $Z$ to lighten the notation and denote the covariates that satisfy the exclusion restriction by $Z_1$ instead of $Z$.} We shall refer to $-x^\prime\beta(y)$ as the outcome equation, to $-z^\prime \pi$ as the selection equation, and to $\rho(x'\delta(y))$ as the selection sorting equation.  We observe the selection indicator $D = 1(D^* > 0)$ and the outcome $Y=Y^*$ when $D=1$.\footnote{The minus signs  in \eqref{eq:drm} are included to take into account that the selection is defined by $D^* > 0$ instead of $D^* \leq 0$. We use this definition to facilitate the interpretation of the parameters and the comparison with the classical HSM; see Example \ref{ex:hsm}.}   In the empirical application that we consider below, $Y^*$ is offered wage, $D^*$ is the net utility from working, e.g., the difference between offered wage and reservation wage, $D$ is an employment indicator, $Y$ is the observed wage, $X$ includes labor market characteristics such as education, age, number of children and marital status, and $Z_1$ includes measures of out-of-work income. We shall discuss the validity of these measures as excluded covariates in Section \ref{sec:empirics}.

The model \eqref{eq:drm} is semi-parametric because $y\mapsto \beta(y)$ and $y\mapsto \delta(y)$ are unknown functions, i.e.  infinite dimensional parameters in general. This flexibility allows the effect of $X$ on the outcome and selection sorting equations to vary across the distribution. For example, it allows the return to education to vary across the distribution, the selection sorting to be different for high and low educated individuals, or to have positive selection sorting  at the upper tail and negative at the bottom tail or vice versa.\footnote{The parametric copula model of AB17 imposes that the sign of the sorting is the same across the latent wage distribution.} The function $u \mapsto \rho(u)$ is a known link  with range $[-1,1]$, e.g. the Fisher transformation \cite{fisher15}, $\rho(u) = \tanh(u)$. The corresponding distribution of $Y^*$ conditional on $Z$ is 
   $$F_{Y^*}(y\mid Z=z)= \lim_{\nu \nearrow + \infty} F_{Y^*,D^*}(y,v \mid Z=z)=\Phi(-x^\prime \beta(y)), \ \ z = (x,z_1).$$
The selection bias arises because this distribution is different from the distribution of the observed outcome $Y$, i.e.
    $F_{Y^*}(y\mid Z=z)\neq F_{Y}(y\mid Z=z,D=1)$

\begin{example}[Semi-Parametric version of Heckman Selection Model]\label{ex:hsm} Here we revisit the semi-parametric special case given in the Introduction, where
\begin{equation*}
D^* = \nu(Z) + V, \quad 
Y^* = \mu(X)+ U, 
\end{equation*}
where $(U,V)$ is  independent of $Z$ ($Z$ contains $X$ here) such that
$$
 F_{U,V}(u,v \mid Z=z) 
         = \Phi_2\left(\Phi^{-1}(F_{U}(u)), 
         \Phi^{-1}(F_V(v)); \rho\right).
         $$
Therefore, 
    $$
        F_{Y^*,D^*}(y,0 \mid Z=z) 
         = \Phi_2\left(\Phi^{-1}(F_{U}(y-\mu(x))), 
         \Phi^{-1}(F_V(-\nu(z))); \rho\right).
$$  
As noted in the Introduction, this (strictly) nests the classical Gaussian selection model where
$F_{U}(u) = \Phi(u/\sigma_U)$ and  $F_V(v)=  \Phi(v/\sigma_V)$, where $\sigma_U$ and $\sigma_V$ are the standard deviations of $U$ and $V$.  
For the econometric specification, we use parametrization:
$$
F_{U}(y-\mu(x)) = \Phi\left(-x'\beta(y)\right); \quad F_V(-\nu(z)) =\Phi(- z'\pi).
$$ \qed
\end{example}

 The parameters $\beta(y)$ and $\pi$ have the same interpretation as in probit models. Their signs are informative about the signs of the partial effects of the corresponding covariates on the conditional distribution of the outcome or probability of selection, and ratios of their components yield ratios of partial effects. Indeed, if $X$ is continuous, 
 $$
 \frac{\partial F_{Y^*}(y\mid Z=z)}{\partial x} = - \beta(y) \phi(-x^\prime \beta(y)),
 $$
 where $\phi$ is the standard normal PDF. The parameter $\delta(y)$ determines the sign of the effect of the covariates on the sorting because 
 $$
 \frac{\partial \rho(x'\delta(y))}{\partial x} = \delta(y) \dot{\rho}(x'\delta(y)),
 $$
 and  $\dot{\rho}(u) = \partial \rho(u)/ \partial u = 1 - \tanh(u)^2 > 0$. Appendix \ref{app:sect3} in the SM provides additional discussion on the interpretation of the model parameters.

\begin{example}[Data Generating Process] The model \eqref{eq:drm} has multiple data generating process representations as nonseparable systems. One  example is
\begin{eqnarray*}
       D^* &=& Z^\prime\pi + V, \ \ V\mid Z \sim \mathcal{N}(0,1), \\
        X^\prime \beta(Y^*) &=& \rho(X'\delta(Y^*))V+\sqrt{1-\rho(X'\delta(Y^*))^2} U, \ \  U\mid Z \sim \mathcal{N}(0,1), 
        \end{eqnarray*}
where $U$ and $V$ are independent. For example, in the wage application $V$ can be interpreted as unobserved utility from working (unobserved benefit of working for money-metric utility), net of what $Z$ already captures, and $U$ as unobserved skills or innate ability net of what $V$ and $X$ already capture. This representation is similar to the semi-parametric HSM in Example \ref{ex:hsm} with the difference that the equation for $Y^*$ is nonseparable. \qed
\end{example}

\subsection{Functionals: Factual, Counterfactual $\&$ Decomposition}
Several key functionals of the model's parameters \eqref{eq:drm} can be of interest. One is the marginal distribution of the latent outcome $Y^*$
   $$F_{Y^*}(y) = F_{Y^*}(y; \beta, F_X) := \int F_{Y^*}(y\mid Z=z)dF_Z(z)=\int \Phi(-x^\prime \beta(y))dF_X(x),$$
where $F_Z$ and $F_X$ are the marginal distributions of $Z$ and $X$, respectively.  In the case of the wage application, $F_{Y^*}$ corresponds to the distribution of the offered wage, which is a potential or latent outcome free of selection.   Using the formula above, we can also construct \textit{counterfactual} distributions by replacing  $\beta(y)$ and  $F_X$ by coefficients and distributions from different populations or groups, $\bar \beta(y)$ and $\bar F_X$. These distributions are useful to decompose the distribution of offered wages between females and males or between blacks and whites, which can be the basis to uncover discrimination in the labor market.  Another functional is the probability of selection
$$
\Pr(D=1) =  \Pr(D = 1; \pi, F_Z) := \int \Pr(D=1 \mid Z=z) dF_Z(z) = \int \Phi(z^\prime \pi)dF_Z(z),
$$
which we can also use to define counterfactuals and employ them to decompose differences in employment rates between employment structure effects, $\pi$, and composition effects, $F_Z$.
 
We can also use the model to construct distributions for the observed outcome using that
    \begin{align*}
    F_{Y \mid D}(y \mid 1) = F_Y(y; \beta, \pi, \delta, F_Z) 
    &:=
    \int \frac{\Phi_2\left( -x^\prime\beta(y), z^\prime \pi; -\rho(x'\delta(y)) \right)}{\Phi(z^\prime \pi)}dF_Z(z\mid D=1) \\
    &= \frac{\int \Phi_2\left( -x^\prime\beta(y), z^\prime \pi; -\rho(x'\delta(y)) \right) dF_Z(z)}{\int \Phi(z^\prime \pi) dF_Z(z)},
    \end{align*}
where the second equality follows from the Bayes rule. We can again construct counterfactual distributions by changing $\beta(y)$, $\pi$, $\delta(y)$ and $F_Z$. In the wage application, we will decompose the differences in the wage distribution between genders or across time into changes in the worker composition  $F_Z$, wage structure $\beta(y)$, selection structure $\pi$, and selection sorting $\delta(y)$. Both selection effects are new to this model.

Quantiles and other functionals of the distributions of latent and observed outcomes can be constructed by applying the appropriate operator. For example, the $\tau$-quantile of the latent outcome is  $Q_{Y^*}(\tau) = \mathbf{Q}_{\tau}(F_{Y^*})$, where $\mathbf{Q}_{\tau}(F) := \inf\{y \in \mathbb{R} : F(y) \geq \tau\}$ is the quantile or left-inverse operator.


\subsection{Estimation} To estimate the model parameters and functionals of interest, we assume that we have a random sample of size $n$ from $(D,DY,Z)$, $\{(D_i, D_i Y_i,Z_i)\}_{i=1}^n$,  where we use $D Y$ to indicate that we only observe $Y$ when $D=1$.  

Before describing the estimators, it is convenient to introduce some notation.   Let  $\mathcal{Y}$ be the region of interest of $Y$, and denote $\theta_y := (\beta(y), \delta(y))$, where we replace the arguments in $y$ by subscripts to lighten the notation.\footnote{If the support of $Y$ is finite,  $\mathcal{Y}$ can be the entire support, otherwise $\mathcal{Y}$ should be a subset of the support excluding low density areas such as the tails.}

The estimation relies on the relationship between conditional distributions and binary regressions. Thus, the CDF of $Y$ at a point $y$ conditional on $X$ is the expectation that an indicator that $Y$ is less than $y$ conditional on $X$,
$$
F_{Y \mid X}(y \mid x) = \Ep[1(Y \leq y) \mid X = x].
$$
 To implement this idea,  we construct the set of indicators for the selected observations 
$$
I_{yi} = 1(Y_i\leq y) \text{ if } D_i = 1,
$$
for each $y \in \mathcal{Y}$. In the presence of sample selection, we cannot just run a probit binary regression of $I_{yi}$ on $X_i$ to estimate the parameter $\beta(y)$ as in \citen{foresiperacchi95} and \citen{chernozhukov+13inference}. The problem is similar to running least squares in the HSM. Instead, we use that 
 \begin{align*}
      \ell_i(\pi,\theta_y)=\left[1-\Phi(Z_i^\prime\pi)\right]^{1-D_i}&\times \Phi_2(-X_i^\prime\beta(y),Z_i^\prime\pi;-\rho(X_i'\delta(y)))^{D_iI_{yi}} \\ &\times\Phi_2(X_i^\prime\beta(y),Z_i^\prime\pi;\rho(X_i'\delta(y)))^{D_i(1-I_{yi})}
      \end{align*}
is the likelihood of $(D_i, I_{yi})$ conditional on $Z_i$. This likelihood is the same as the likelihood of a bivariate probit model or more precisely a probit model with sample selection \cite{zellner65,poirier80,vandeven81}. 

We estimate the model parameters using a computationally attractive two-step method to maximize the average log-likelihood, similar to the Heckman two-step method. The first step is a probit regression for the probability of selection to estimate $\pi$, which is identical to the first step in the Heckman two-step method. The second step consists of multiple distribution regressions (DRs) with sample selection corrections to estimate $\beta(y)$ and $\delta(y)$ for each value of $y \in \mathcal{Y}$. These steps are summarized in the following algorithm:

\begin{algorithm}[Two-Step DR Method]\label{alg:tsdr}   (1) Run a probit for the selection equation to estimate $\pi$:
(2) Run multiple DRs with sample selection correction to estimate $\theta_y$. That is, 
\begin{align*} 
& (1) \quad \hat{\pi} =  \argmax_{c \in \mathbb{R}^{d_{\pi}} } \frac{1}{n} \sum_{i=1}^n \left[D_{i}\log{\Phi(Z_i^\prime c)}+(1-D_{i})\log{\Phi(-Z_i^\prime c)}\right],\\ 
& (2) \quad \hat{\theta}_y =  \argmax_{t=(b,d) \in \Theta} \frac{1}{n} \sum_{i=1}^n D_{i}  \left[ \right.I_{yi} \log{\Phi_2\left(-X_i^\prime b,Z_i^\prime \hat{\pi}; -\rho(X_i'd)\right)} \\
  &  \quad \quad \quad \quad + \left.(1-I_{yi}) \log{\Phi_2\left(X_i^\prime b,Z_i^\prime \hat{\pi}; \rho(X_i'd) \right)} \right], \ \ y \in \mathcal{Y}.
\end{align*}
where $I_{yi} = 1(Y_i \leq y)$ and $\Theta \in \mathbb{R}^{d_{\theta}}$ is a compact parameter set, and 
  $$
  d_\pi := \dim \pi, \ d_{\theta} := \dim \theta_u, \ \ \rho(u) := \tanh(u) = \frac{e^u - e^{-u}}{e^u + e^{-u}} \in [-1,1].
  $$
\end{algorithm}

In practice we replace the set  $\mathcal{Y}$ by a finite grid $\bar{\mathcal{Y}}$ if $\mathcal{Y}$ contains many values. 

The estimators of the functionals of interest are constructed from the estimators of the parameters using the plug-in method. For example, the estimator of the distribution of the latent outcome and the estimator of the probability of selection are
\begin{equation}\label{eq:ldisthat}
\hat{F}_{Y^*}(y):= F_{Y^*}(y; \hat \beta, \hat F_X)  = \frac{1}{n}\sum_{i=1}^n\Phi(-X_i^\prime \hat{\beta}(y)),
\quad 
\hat{\Pr}(D = 1) :=  \Pr(D = 1; \hat \pi, \hat F_Z) = \frac{1}{n}\sum_{i=1}^n \Phi(z^\prime \hat \pi),
\end{equation}
and the estimators of the counterfactual distributions of the observed outcome are constructed from
\begin{equation}\label{eq:odisthat}
\hat{F}_{Y \mid D}(y \mid 1) := F_Y(y; \hat \beta, \hat \pi, \hat \delta, \hat F_Z)  =  \frac{\sum_{i=1}^n \Phi_2(-X_i^\prime \hat{\beta}(y),Z_i^\prime \hat{\pi};-\rho(X_i'\hat{\delta}(y)))}{\sum_{i=1}^n \Phi(Z_i^\prime \hat{\pi})},
\end{equation}
by choosing the estimators of $ \hat{\beta}(y)$, $\hat{\pi}$, and $\hat{\delta}(y)$ and the sample values of $Z$ appropriately.  Estimators of quantiles and other functionals of these distributions are obtained by applying the operators that define the functionals to the estimator of the distribution. See Section \ref{sec:theory} of SM for details. 


\subsection{Inference on Functional Parameters} 
The model parameters and functionals of interest are generally function-valued. We show how to construct confidence bands for them that can be used to test functional hypotheses such as the entire function being zero, non-negative or constant. To explain the construction, consider the case where the functional of interest is a linear combination of the model parameter $\theta_y$, that is the function $y\mapsto c'\theta_y,$ $y\in\mathcal{Y}$, where $c \in \mathbb{R}^{d_{\theta}}$.  The set $CB_p(c'\theta_y)$ is an asymptotic $p$-confidence band for $c'\theta_y$ if it satisfies 
  $$
  \Pr\left[c'\theta_y \in CB_p(c'\theta_y), \text{ for all } y \in\mathcal{Y}\right] \to p.
  $$
We form  $CB_p(c'\theta_y)$  as
  $CB_p(c'\theta_y):=c'\hat{\theta}_y\pm cv(p)SE(c'\hat{\theta}_y),$
  where $\hat{\theta}_y$ is the estimator of $\theta_y$ defined in Algorithm \ref{alg:tsdr},  $SE(c'\hat{\theta}_y)$ is the standard error of $c'\hat{\theta}_y$, and $cv(p)$ is a critical value, i.e. a consistent estimator of the $p$-quantile of the statistic
  $$t_{\mathcal{Y}} = \sup_{y\in \mathcal{Y}} \frac{|c'\hat{\theta}_y-c'\theta_y|}{SE(c'\hat{\theta}_y)}.$$

We obtain the standard error and critical value from the limit distribution of the stochastic process $y \mapsto \hat{\theta}_y$  derived in Section \ref{sec:theory} of the SM. In practice, it is convenient to estimate the critical value using resampling methods. Multiplier bootstrap is computationally attractive in our setting because it does not require parameter re-estimation and therefore avoids the nonlinear optimization in both steps of Algorithm \ref{alg:tsdr}. The multiplier bootstrap is implemented using the following algorithm:
\begin{algorithm}[Multiplier Bootstrap]\label{alg:mb} (i) For $b \in 1,\ldots, B$ and the finite grid $\bar{\mathcal{Y}} \subseteq \mathcal{Y}$, repeat the steps: (1) Draw the bootstrap multipliers $\{\omega_{i}^b : 1 \leq i \leq n \}$ independently from the data and normalized them to have zero mean,
   $
    \omega_i^b = \tilde \omega_i^b - \sum_{i =1}^n \tilde  \omega_i^b/n, \ \ \tilde \omega_i^b \sim \text{ i.i.d. } \mathcal{N}(0,1).
    $
(2) Obtain the bootstrap estimator of the model parameter 
    $$
    \hat \theta^b_y = \hat \theta_y + n^{-1} \sum_{i = 1}^n \omega_i^b \ \widehat \psi_i(\hat \theta_y, \hat \pi),
    $$
    where $\widehat \psi_i(\hat \theta_y, \hat \pi)$ is an estimator of the influence function of $\hat{\theta}_y$ given in equation \eqref{eq:if} of the SM. 
(3) Construct bootstrap realization of maximal t-statistic $t_{\mathcal{Y}}$ for the functional of interest,
    $$
    t_{\mathcal{Y}}^b = \max_{y \in \bar{\mathcal{Y}}} \frac{|c'\hat \theta^b_y - c'\widehat \theta_y|}{SE(c'\hat{\theta}_y)}, \ \ SE(c'\hat{\theta}_y) = \sqrt{c' \widehat \Sigma_{\theta_y\theta_y} c},
    $$
    where $\widehat \Sigma_{\theta_y\theta_y}$ is an estimator of the asymptotic variance-covariance matrix of $\hat{\theta}_y$ given in equation \eqref{eq:se} of the SM.
(ii) Compute the critical value $cv(p)$ as the simulation $p-$quantile of $t_{\bar{\mathcal{Y}}}^b$,
$
cv(p) =  p-\text{quantile of }  \{t_{\mathcal{Y}}^b : 1 \leq b \leq B\}
$
\end{algorithm}

 The centering of the multipliers in step (i1) of the algorithm is a finite sample adjustment. Confidence bands for other functionals of the model parameter can be constructed using a similar bootstrap method.

\section{Wage Decompositions in the UK}\label{sec:empirics}  We apply the DR model with sample selection to carry out wage decompositions accounting for endogenous employment participation using data from the United Kingdom.

\subsection{Data} The data come from the U.K. Family Expenditure Survey (FES) for the years 1978 to 2001, Expenditure and Food Survey (EFS) for the years 2002 to 2007, and Living Costs and Food Survey (LCFS) for the years 2008 to 2013. Despite the differences in the name, these surveys contain comparable information. Indeed, the FES was combined to the National Food Survey to form the EFS, which was renamed LCFS when it became a module of the Integrated Household Survey. The data from the FES has been previously used by \citen{gosling00}, \citen{brs03}, \citen{bgim07} and \citen{ab17} to study wage equations in the U.K. labor market. We are not aware of any previous use of the data from the EFS and LCFS for this purpose.\footnote{See \citen{rv18} for another recent application of the data to the analysis of female labor force participation. } 
The three surveys contain repeated cross-sectional observations for women and men.  The selection of the sample is similar to the previous work that used the FES. Thus, we keep individuals with ages between 23 to 59 years, and drop full-time students, self-employed workers, those married with spouse absent, and those with missing education or employees whose wages are missing. This leaves a sample of 258,900 observations, 139,504 of them correspond to women and 119,396 to men. The sample size per survey year and gender ranges from 2,197 to 4,545.  

The outcome of interest, $Y$, is the logarithm of real hourly wage rate. We construct this variable as the ratio of the weekly usual gross main nominal earning to the weekly usual working hours, deflated by the U.K. quarterly retail price index. The selection variable, $D$, is an indicator for being employed.\footnote{For data before 1990, $D=0$ if the individual is in one of the following status: seeking work, sick but seeking work, sick but not seeking work, retired and unoccupied. For those in and after 1990, $D=0$ if the individual is seeking work and available, waiting to start work, sick or injured, retired or unoccupied.} The covariates, $X$, include 5  indicators for age when ceasing school ($\leq$15, 16, 17--18, 19--20, 21--22 and $\geq$ 23), a quartic polynomial in age, an indicator of being married or cohabiting, 6 variables with the number of kids by age categories (1, 2, 3--4, 5--10, 11--16,  and 17-18), 36 survey year indicators, and 11 region indicators (Northern 5.48\%, Yorkshire 9.56\%, North Western 10.20\%, East Midlands 7.36\%, West Midlands 9.13\%, East Anglia 5.31\%, Greater London 10.06\%, South Eastern 16.82\%, South Western 7.94\%, Wales 4.99\%, Scotland 8.92\%, and Northern Ireland 4.23\%).\footnote{In the rest of the paper we shall refer to an individual being married or cohabiting as married.} We provide descriptive statistics of the variables and   some background on the U.K. labor market using our data in Section \ref{app:empirics}  of the SM.

The excluded covariate, $Z_1$,  is a potential out-of-work income benefit interacted with the marital status indicator used before in  \citen{brs03} and \citen{bgim07}. This benefit is constructed with the Institute for Fiscal Studies (IFS) tax and welfare-benefit model (TAXBEN). TAXBEN is a static tax and benefit micro-simulation model of taxes on personal incomes, local taxes, expenditure taxes, and entitlement to benefits and tax credits that operates on large-scale, representative household surveys \cite{b09}. It is designed to calculate the income of a tax unit if the individual was considered out-of-work.\footnote{Our definition of the out-of-work benefit income is slightly different from  the definition of \citen{brs03} and \citen{bgim07}. They calculated it as the income of a tax unit if all the individuals within the tax unit were out of work. In our view our definition might better reflect the opportunity cost or outside value option of working that the individual faces.} It is composed of eligible unemployment and housing benefits, which are determined by the demographic composition of the tax unit and the housing costs that the tax unit faces. These costs vary by region and over time due to numerous policy changes that have occurred over time. There is no consensus in the literature about the validity of this variable as an excluded covariate. In our case, the outcome and selection sorting exclusions imply that, conditional on the observed covariates, the offered wage and dependence between offered wage and net reservation wage do not depend on the level of the benefit.  We shall assume that the exclusion restrictions are satisfied  and refer to \citen{brs03} and \citen{bgim07} for a discussion on the plausibility of the outcome restriction.  In the Introduction, we stated a rich semi-parametric generalization of Heckman's labor supply model that trivially satisfies the selection sorting exclusion, motivating their use in our analysis.

\subsection{Empirical Specifications} We estimate the DR model for different samples and carry out several wage decompositions where we compare the distributions of men and women, or the distributions over time within genders. The specifications of the selection and outcome  equations include all the covariates  described above except for the excluded covariates in the outcome equation. The parameter of the selection sorting function is notoriously more difficult to estimate than the parameters of the selection and outcome equations. We consider four simplified specifications of the sorting function where  the covariates included in the index $X'\delta(y)$ are:
\begin{itemize}
\item Specification 1: a constant.
\item Specification 2: a constant and the marital status indicator.
\item Specification 3: a constant and a linear trend on the year of the survey.
\item Specification 4: a constant and a linear trend on the year of the survey interacted with the marital status indicator.  
\end{itemize}
We also experimented with other specifications that include the education indicators, indicators of survey year, or age.  We do not report these results because they do not show any clear pattern mainly due to imprecision in the estimation of the parameter $\delta(y)$.\footnote{The main results on the   wage decompositions presented below are not sensitive to the specification of the sorting equation.}

\subsection{Selection Sorting}
We report point estimates and 95\% confidence bands for the local correlation function $y \mapsto \rho(x'\delta(y))$ in the selection sorting equation. Estimates and 95\% confidence bands for the coefficients of the selection, outcome and selection sorting equations are given in Section \ref{app:empirics} of the SM. The estimates are obtained with Algorithm \ref{alg:tsdr} replacing $\mathcal{Y}$ by a finite grid containing the sample quantiles of log real hourly wage with indexes $\{0.10, 0.11, \ldots, 0.90 \}$ in the pooled sample of men and women. We report all the estimates as a function of the quantile index. The confidence bands are constructed by Algorithm \ref{alg:mb} with $B=500$ bootstrap repetitions and the same finite grid  as for the estimates. We also report  estimates from the HSM of Example \ref{ex:hsm} with dashed lines as a benchmark of comparison.

\begin{figure}
    \includegraphics[height=0.3\textwidth, width=0.49\textwidth,page=2]{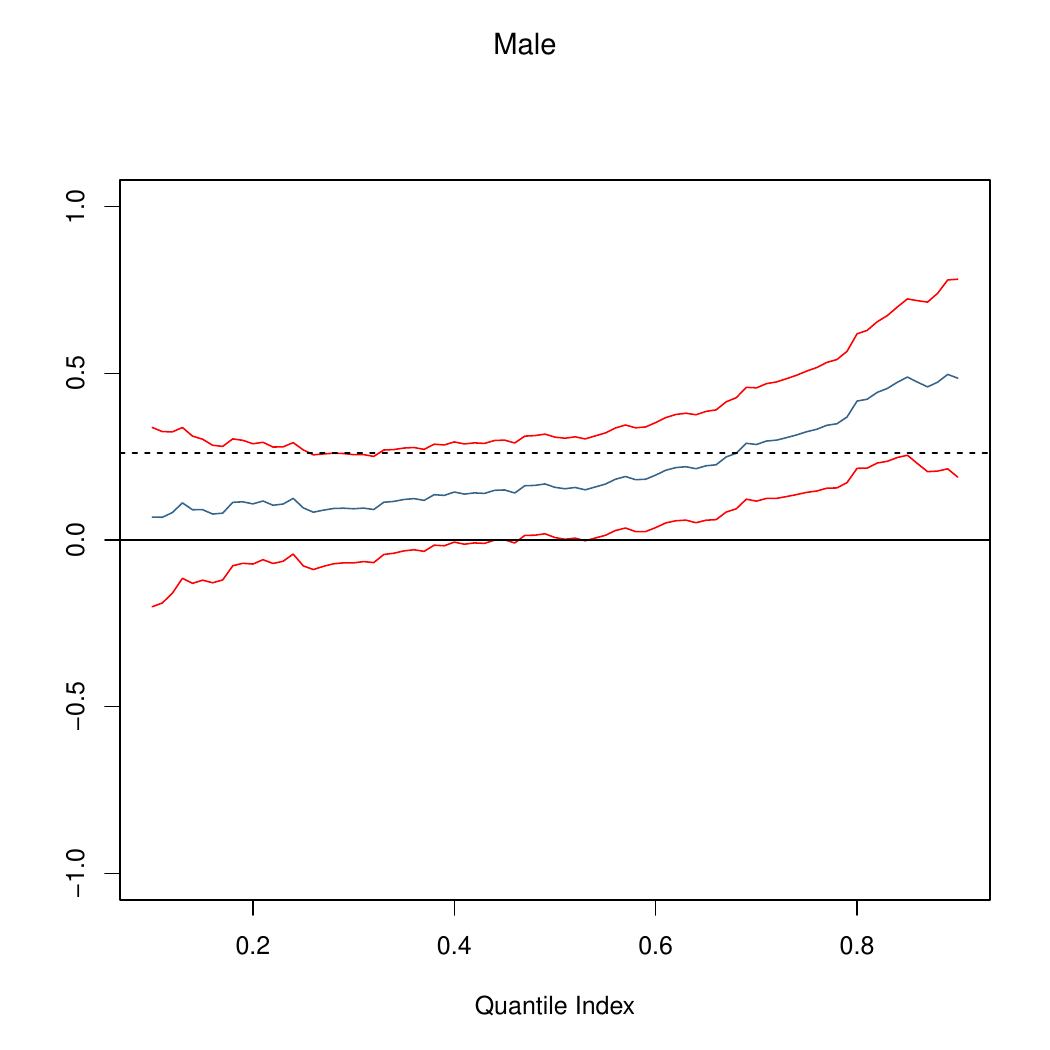}
    \includegraphics[height=0.3\textwidth, width=0.49\textwidth,page=4]{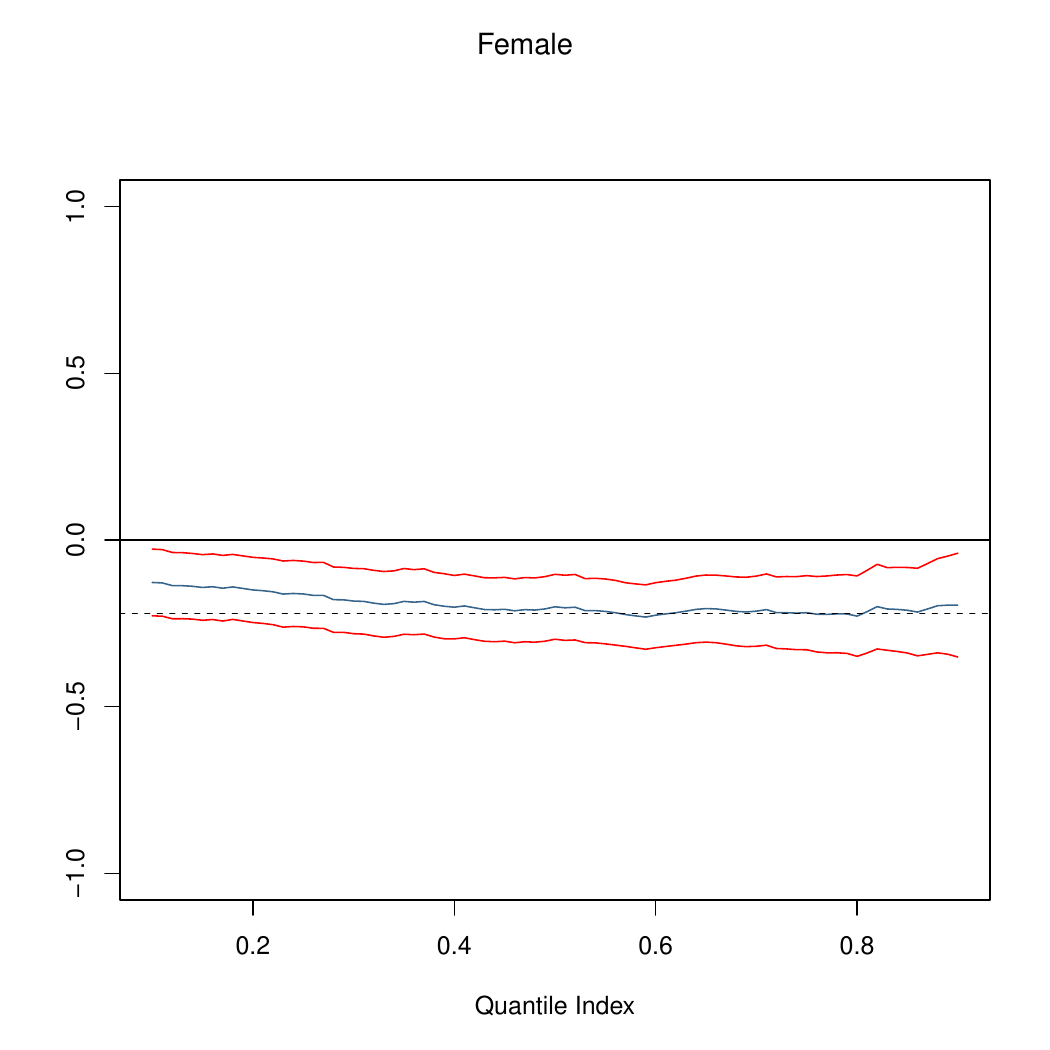}
    \caption{Estimates and 95\% confidence bands for the selection sorting function: specification 1}\label{fig:rho_men_spec1}
\end{figure}

\begin{figure}
    \includegraphics[height=.6\textwidth, width=0.4\textwidth,page=2]{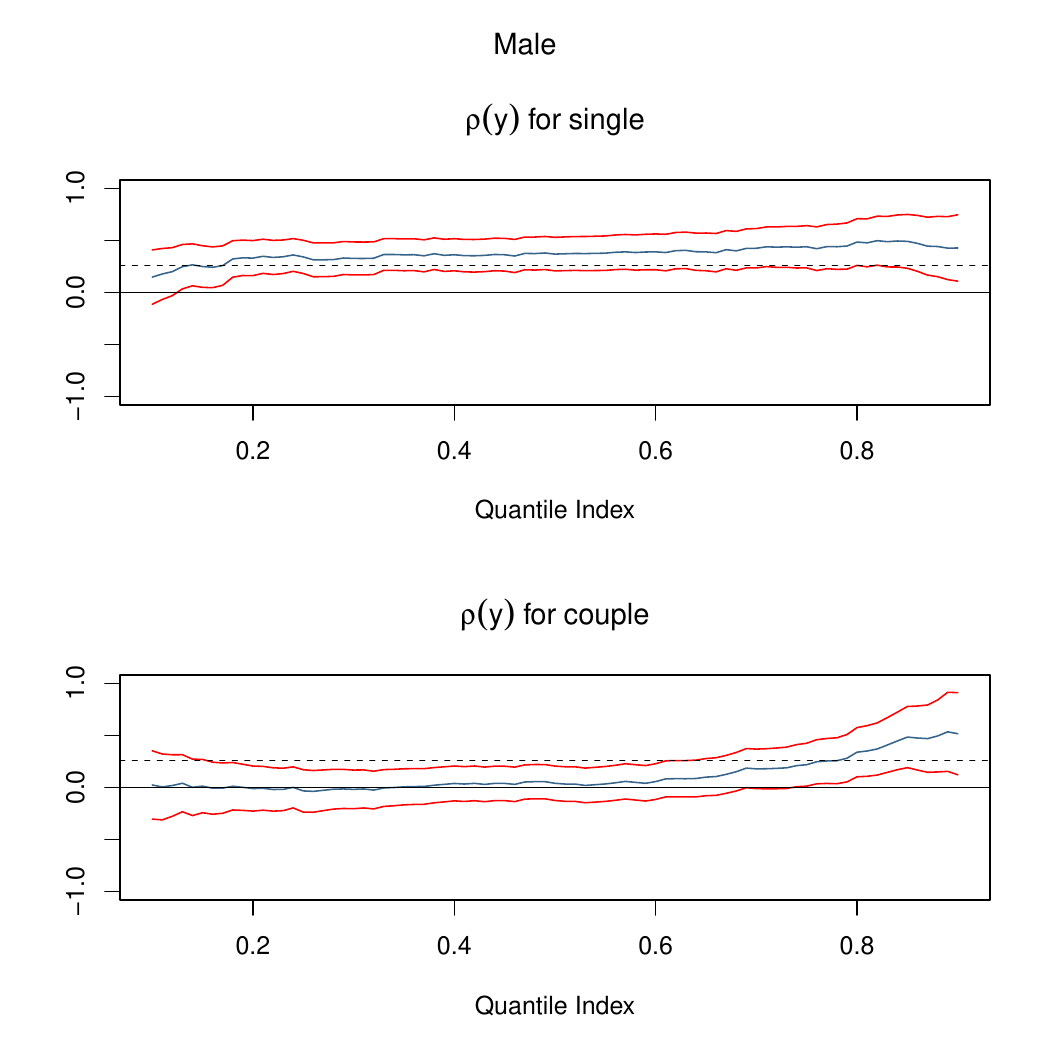}
    \includegraphics[height=.6\textwidth, width=0.4\textwidth,page=4]{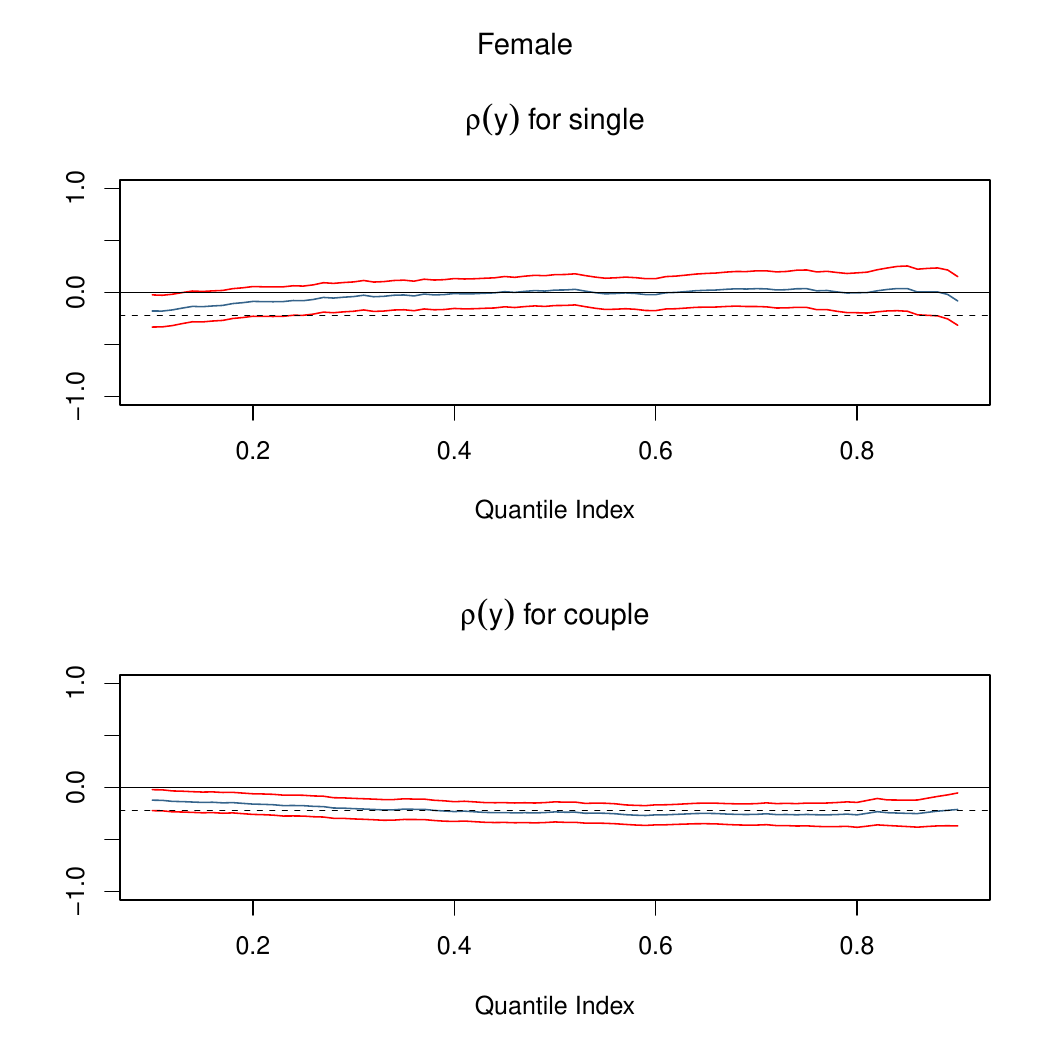}
    \caption{Estimates and 95\% confidence bands for the selection sorting function: specification 2}\label{fig:rho_men_spec2}
\end{figure}

\begin{figure}
    \includegraphics[height=0.5\textwidth, width=0.49\textwidth,page=2]{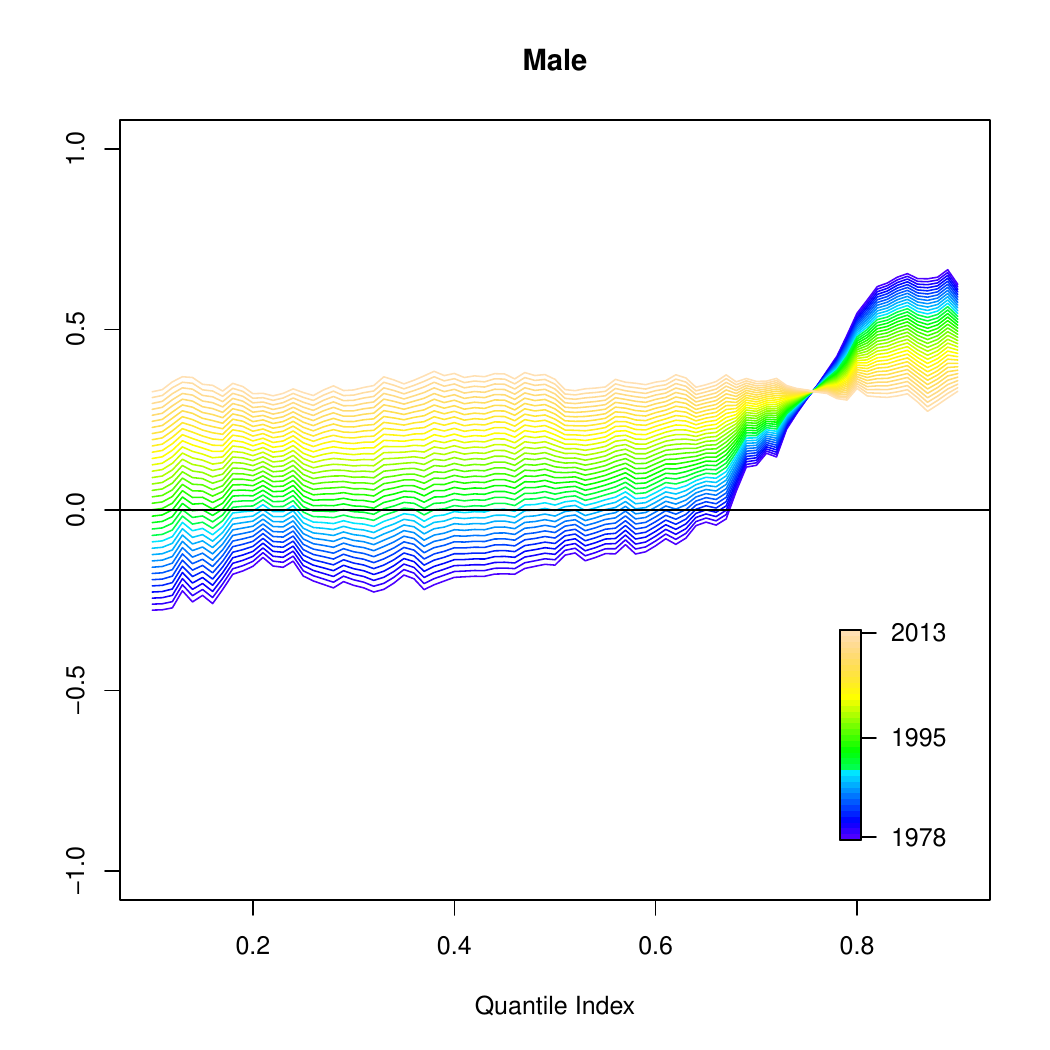}
    \includegraphics[height=0.5\textwidth, width=0.49\textwidth,page=5]{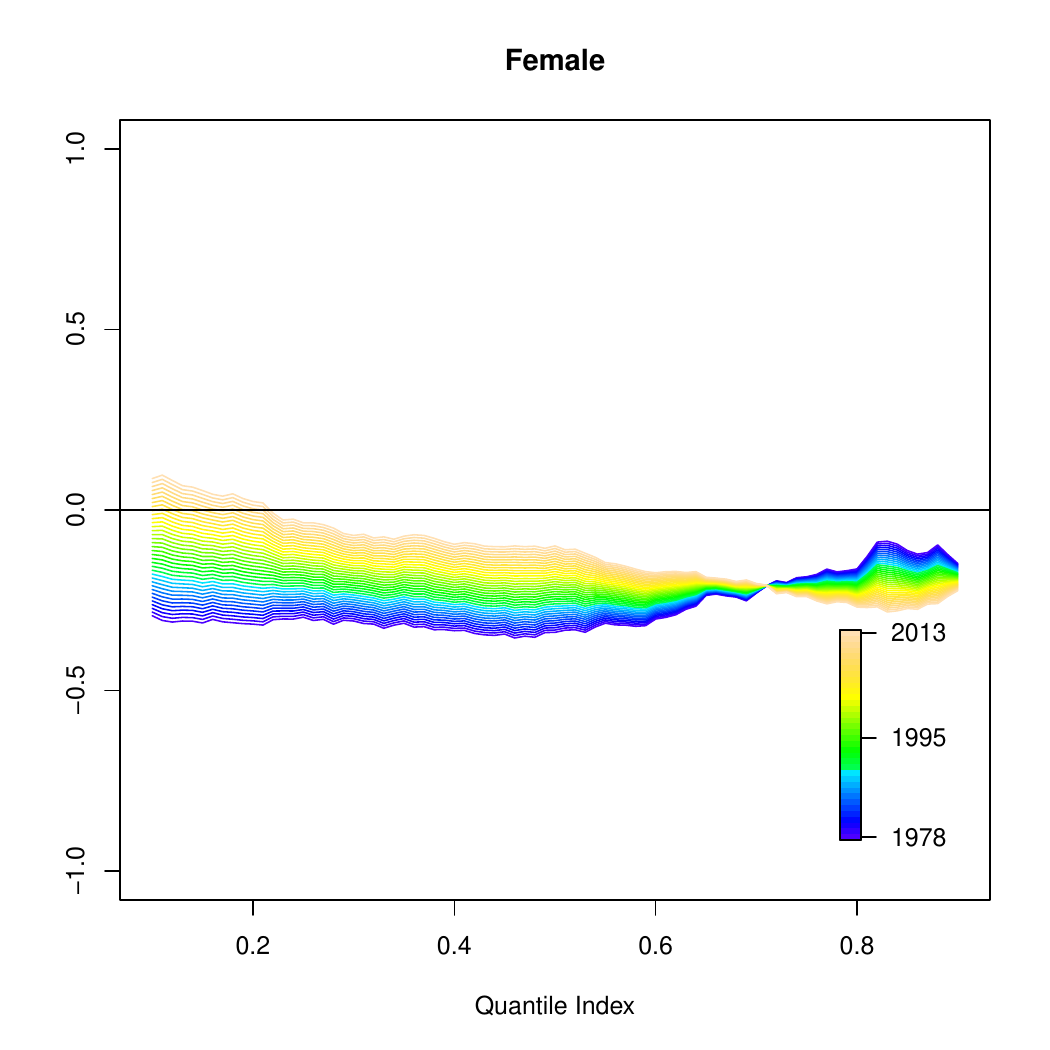}
    \caption{Estimates of the selection sorting function: specification 3}\label{fig:rho_men_spec3}
\end{figure}

Figures \ref{fig:rho_men_spec1}--\ref{fig:rho_men_spec3} display the estimates of the sorting functions for specifications 1--3, respectively.  Figure \ref{fig:rho_men_spec1} shows  positive selection sorting for men and negative selection sorting for women.  In both cases we cannot reject that the sorting is constant across the distribution. This finding is refined in Figure \ref{fig:rho_men_spec2}, where we  uncover that the positive male sorting comes mainly from bachelors, whereas the negative female sorting comes from married women.  This pattern is consistent with a marriage market where there is assortative matching in offered wages given observable characteristics, where women with high potential wages are married to highly paid working men and decide not to work \cite{neal04}. Figure \ref{fig:rho_men_spec3} shows that the sorting homogeneity found in the pooled sample hides some heterogeneity across time. Thus, we find that the male sorting is heterogeneous in the early years,  negative at the bottom and positive  at the top  of the distribution, and progressively becomes homogenous. The female sorting is more homogenous over time, but also displays a positive trend, especially at the bottom of the distribution. Figure \ref{fig:rho_men_spec4} in the SM shows that the trends in sorting are driven by married individuals at the bottom of the distribution and single individuals at the top of the distribution.\footnote{We do not report confidence bands for specifications 3 and 4 to avoid cluttering. The confidence bands for the coefficients of the selection sorting function $\delta(y)$ in the SM show that the results on the trends are statistically significant.}

\subsection{Distributions of Offered and Observed Wages, and Wage Decompositions}
\begin{figure}
    \includegraphics[height=0.49\textwidth, width=0.49\textwidth,page=1]{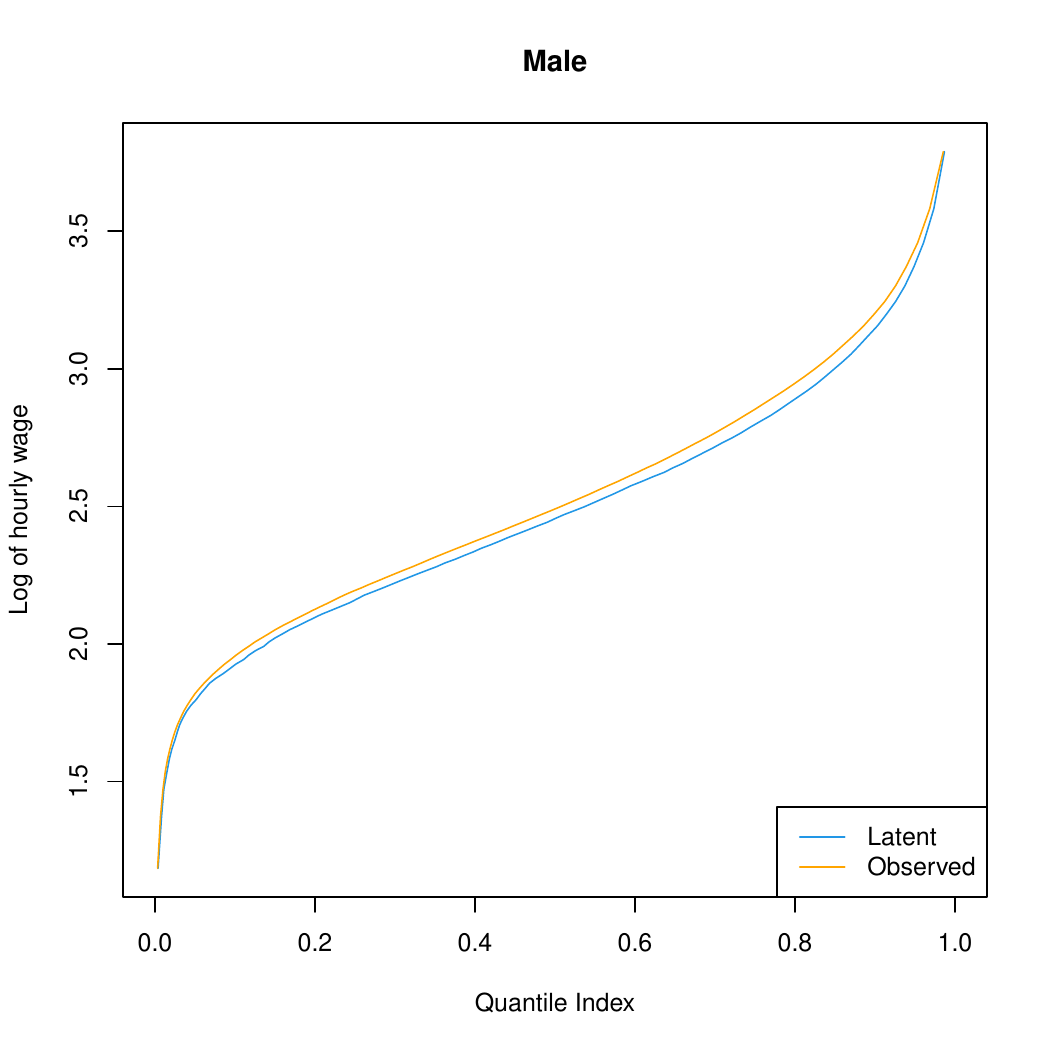}
    \includegraphics[height=0.49\textwidth, width=0.49\textwidth,page=2]{Figures/Comp_Dist_11378_lintimexcouple_500.pdf}
    \caption{Estimates of the quantiles of observed and offered (latent) wages: specification 4}\label{fig:latent}
\end{figure}

Figure \ref{fig:latent} shows point estimates of the quantiles of offered and observed wages for men and women based on specification 4. Estimates for the other specifications and confidence bands for all the specifications are given in the SM.  The offered wage is a latent variable defined for all the individuals that is free of sample selection. As we showed in Section \ref{sec:model}, the distributions of both types of wages can be expressed as functionals of the model parameters, and estimated using the plug-in estimators \eqref{eq:ldisthat} and \eqref{eq:odisthat}.\footnote{The model-based estimator of the observed distribution in  \eqref{eq:odisthat} produces almost identical estimates to the empirical distribution of the observed wages.} We find different sample selection biases for men and women.  Thus, the quantiles of the observed wages are below the quantiles of latent wages for men, consistent with the positive selection in the sorting equation, whereas they are similar for women. In this case, the bias coming from the negative sorting is almost exactly offset by the difference in the composition between working and non-working women. 

\begin{figure}
    \includegraphics[height=0.49\textwidth, width=0.49\textwidth,page=17]{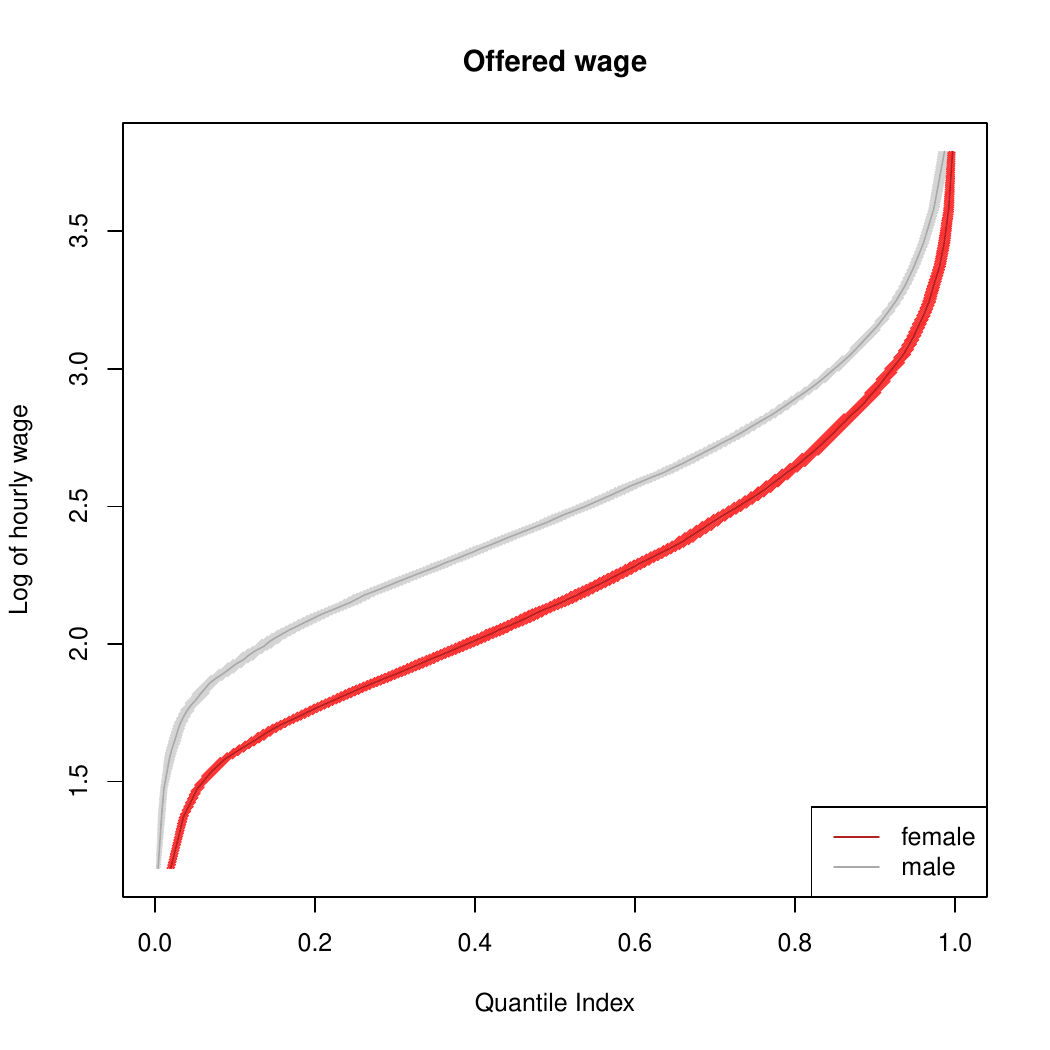}
    \includegraphics[height=0.49\textwidth, width=0.49\textwidth,page=21]{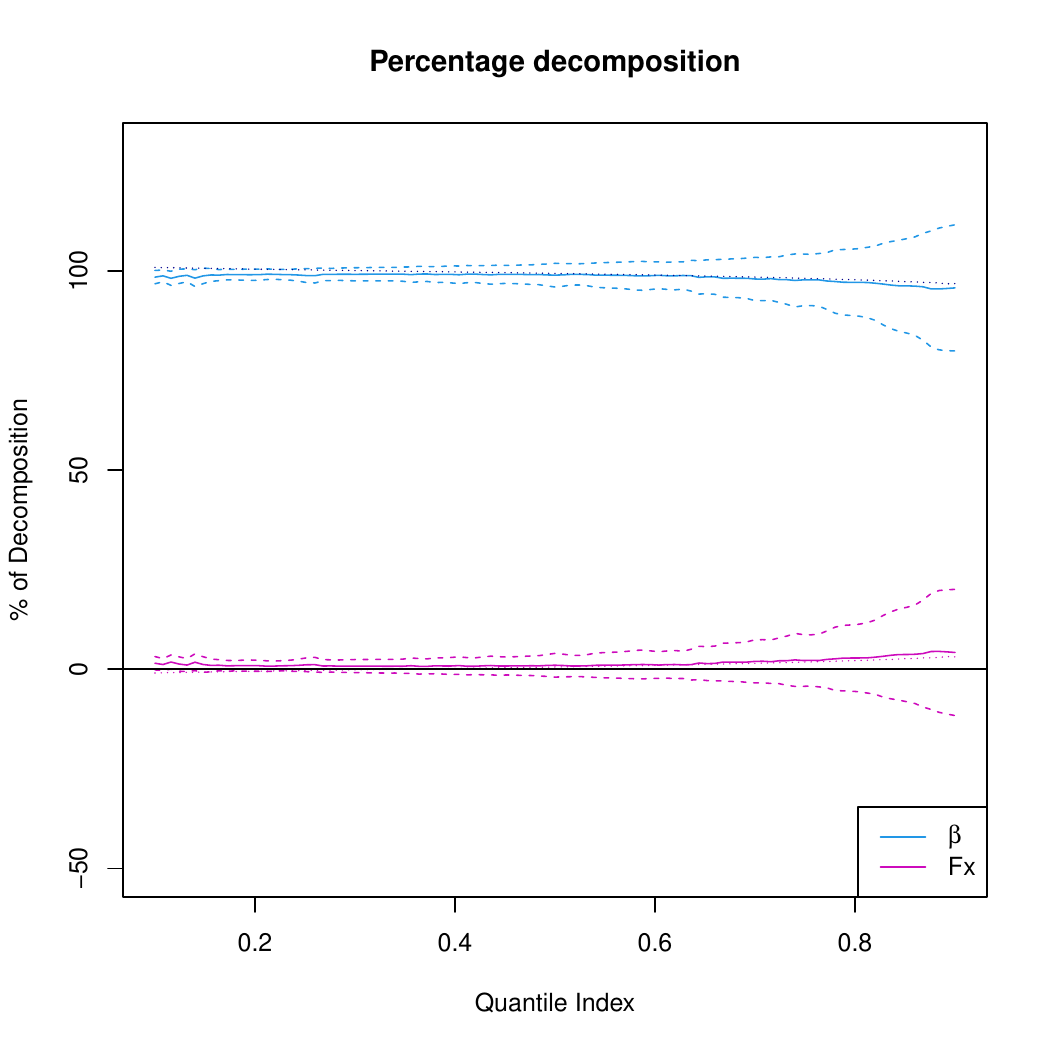}
    \caption{Estimates and 95\% confidence bands for the quantiles of offered (latent) wages and decomposition between women and men: specification 4}\label{fig:latent2}
\end{figure}

 Figure \ref{fig:latent2} compares the quantile function of offered wages  between men and women and carries out a gender wage gap analysis  based on specification 4. The estimates and 95\% confidence bands for the other specifications are reported in the SM. The gender wage gap analysis is based on the counterfactual distributions 
$$
F_{Y^* \langle j, k \rangle}(y) = F_{Y^*}(y; \beta_j, F_{X_k}) = \int \Phi(-x'\beta_j(y)) dF_{X_k}(x),
$$
where $\beta_j(y)$ is the coefficient of the wage equation in group $j$, $F_{X_k}$ is the distribution of the characteristics in group $k$, and $j$ and $k$ are group indices for women and men.  $F_{Y^* \langle j, k \rangle}$ corresponds to the distribution of offered wages that we would observed when the wage structure is as in group $j$ and the distribution of  characteristics is as in group $k$. We decompose the  difference in the quantile functions of the latent wages between women (group 1) and men (group 0) using the counterfactual distributions as
$$
F_{Y^* \langle 1, 1 \rangle} -  F_{Y^* \langle 0, 0\rangle}  = [F_{Y^* \langle 1, 1 \rangle} -  F_{Y^* \langle 0, 1 \rangle}] + [F_{Y^* \langle 0, 1 \rangle} -  F_{Y^* \langle 0, 0 \rangle}],
$$
where the first term is the wage structure or discrimination effect and the second term is the composition effect. We obtain estimates of the counterfactual distributions and quantile functions using the plug-in estimator in \eqref{eq:ldisthat} and the quantile and increasing rearrangement operators, see Section \ref{sec:theory} of the SM. 
We find that the wages offered to women are between $21$ and $40\%$ lower than the wages offered to men at the same quantile index. The majority of this difference  is explained by differences in the wage structure, $\beta(y)$, whereas differences in composition, $F_X$, have very little explanatory power. This result can be interpreted as evidence of gender discrimination in the labor market.


 We next use the DR model to decompose changes in the distribution of the observed wage between women and men, and between the first and second halves of the sample period for each gender. We extract four components that correspond to different inputs of the DR model: (1) selection (employment) sorting: $\delta(y)$; (2) selection (employment) structure: $\pi$; (3) outcome (wage) structure: $\beta(y)$; and (4) composition: $F_Z$. 
To define the effects of these components, let $F_{Y\langle t, s, r, k\rangle}$ be the counterfactual distribution of wages when the sorting is as in group $t$, the employment structure is as in group $s$, the wage structure is as  in group $r$, and the composition of the population is as in group $k$. The actual distribution in group $t$ therefore corresponds to $F_{Y\langle t, t, t, t\rangle}$.  We assume that there are two groups indexed by $0$ and $1$ that correspond to demographic populations such as men and women, or time periods such as the first and second halves of the sample years. Then, we can decompose the distribution of observed wage between group $1$ and group $0$ as:
\begin{multline*}
F_{Y\langle 1, 1, 1, 1 \rangle} -  F_{Y\langle 0, 0, 0, 0 \rangle}  = [F_{Y\langle 1, 1, 1, 1 \rangle} -  F_{Y\langle 0, 1, 1, 1 \rangle}] + [F_{Y\langle 0, 1, 1, 1 \rangle} -  F_{Y\langle 0, 0, 1, 1 \rangle}] \\ + [F_{Y\langle 0, 0, 1, 1 \rangle} -  F_{Y\langle 0, 0, 0, 1 \rangle}] + [F_{Y\langle 0, 0, 0, 1 \rangle} -  F_{Y\langle 0, 0, 0, 0 \rangle}],
\end{multline*}
where the first term in square brackets of the right hand side is a sorting effect, the second an employment structure effect, the third a wage structure effect, and the forth a composition effect. This is a distributional version of the classical Oaxaca-Blinder decomposition that accounts for sample selection \cite{kitagawa55,oaxaca73,blinder73}. It is well-known that the order of extraction of the components in this type of decompositions might matter. As a robustness check, we estimate the decomposition changing the ordering of the components. In results not reported, we find that the main findings are not sensitive to the change of ordering. 


In terms of the DR model, the counterfactual distribution can be expressed as the functional
$$
F_{Y\langle t, s, r, k\rangle}(y) = F_Y(y; \beta_r, \pi_s, \delta_t, F_{Z_k}) = \frac{\int \Phi_2\left( -x^\prime\beta_r(y), z^\prime \pi_s; -\rho(x'\delta_t(y)) \right) dF_{Z_k}(z)}{\int \Phi(z^\prime \pi_s) dF_{Z_k}(z)},
$$
where $\delta_t$ is the coefficient of the sorting function in group $t$, $\pi_s$ is the coefficient of the employment equation in group $s$, $\beta_r$ is the coefficient of the wage equation in group $r$, and $F_{Z_k}$ is the distribution of characteristics in group $k$. Given random samples for groups $0$ and $1$, we construct a plug-in estimator of $F_{Y\langle t, s, r, k\rangle}$ by suitably combining the estimators of the model parameters and distribution of covariates from the two groups.

\begin{remark}[Selection Effects]\label{remark:se} To interpret the selection effects, it is useful to consider a simplified version of the model without covariates where 
$    F_Y(y;\pi,\rho)
= \Phi_2\left( -\beta, \pi; -\rho \right)/\Phi (\pi).
$
Here we drop the dependence of $\beta$ and $\rho$ on $y$ to lighten the notation, and make explicit the dependence of $F_Y$ on the selection parameters $\pi$ and $\rho$ to carry out comparative statics with respect to them. Then, by the properties of the normal distribution
$$
    \frac{\partial  F_Y(y;\pi,\rho)}{\partial \rho} = - \frac{\phi_2(-\beta,\pi; -\rho)}{ \Phi(\pi)} < 0,
$$
and
$$
    \frac{\partial F_Y(y;\pi,\rho)}{\partial \pi} \propto \Phi\left(\frac{-\beta + \rho \pi}{\sqrt{1-\rho^2}}  \right) \Phi(\pi) - \int_{-\infty}^{\pi} \Phi\left(\frac{-\beta + \rho x}{\sqrt{1-\rho^2}}  \right) \phi(x) dx \left\{\begin{array}{c} < 0 \text{ if } \rho < 0, \\  = 0 \text{ if } \rho = 0, \\ > 0 \text{ if } \rho > 0, \end{array}\right.
$$ 
where $\Phi$ and $\phi$ are the standard normal CDF and PDF, and  $\phi_2(\cdot, \cdot; \rho)$ be the joint PDF of a standard bivariate normal random variable with parameter $\rho$.\footnote{To obtain the  derivative we use that $\Phi_2\left( -\beta, \pi; -\rho \right) = \int_{-\infty}^{\pi} \Phi\left( \frac{-\beta + \rho x}{\sqrt{1-\rho^2}}\right) \phi(x) dx$.} 
Increasing $\rho$, therefore, shifts the distribution to the right (increases quantiles) because it makes selection sorting more positive while the size of the selected population is fixed. The effect of increasing $\pi$ is more nuanced and depends on the sign of $\rho$. Intuitively, $\pi$ affects the size of the selected population and the relative importance of observables and unobservables in the selection. For example, when selection sorting is negative, increasing the size of the selected population by increasing $\pi$ shifts the distribution of the right (increases quantiles) because the newly selected individuals have smaller (more negative) selection unobservables that correspond to larger (more positive) outcome unobservables. In other words, the newly selected individuals are relatively less adversely selected. The sign of the selection effects might be different in the presence of covariates if the parameter variation changes the composition of the selected population. We provide an example of this sign reversal in Appendix \ref{app:sign} of the SM. \qed
\end{remark}

\begin{figure}
    \includegraphics[height=.49\textwidth, width=.49\textwidth,page=14]{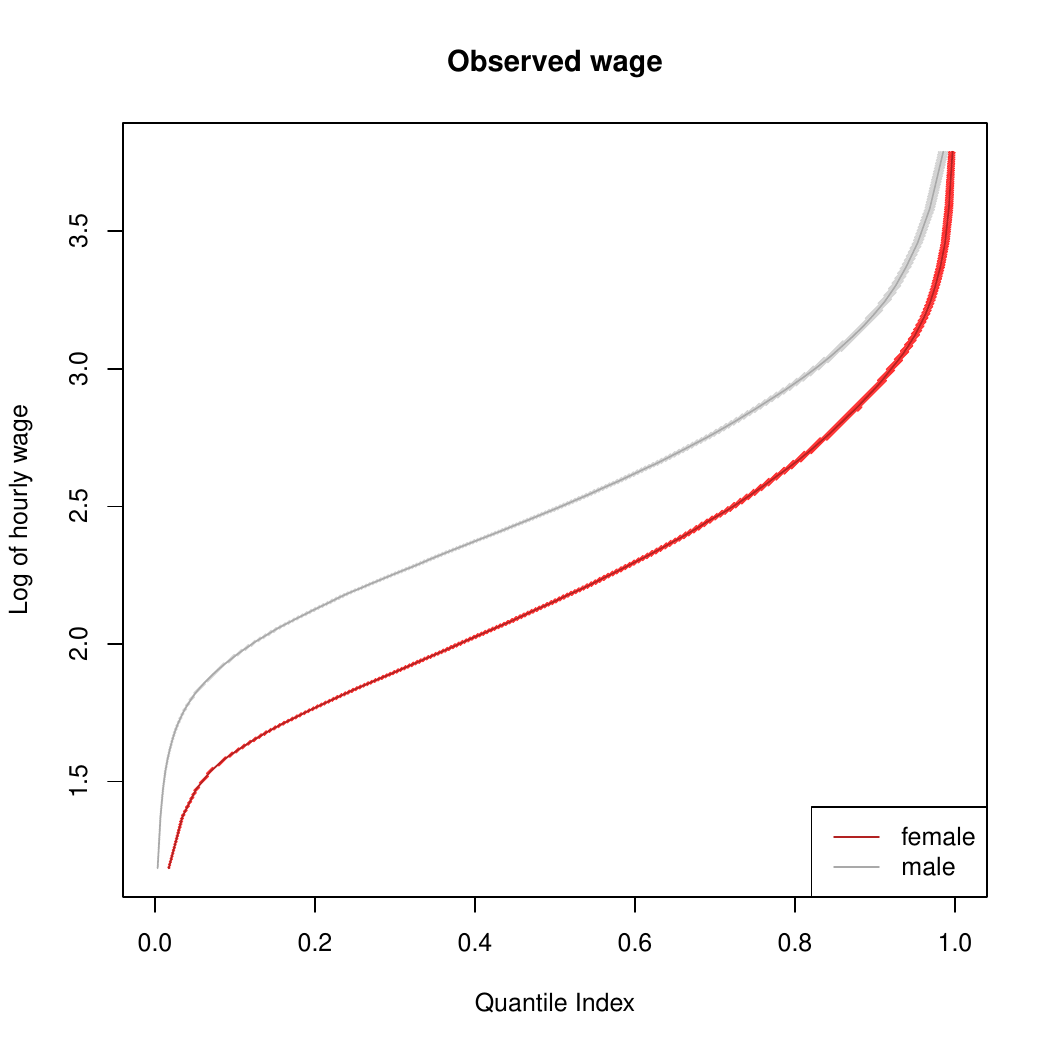}
        \includegraphics[height=.49\textwidth, width=.49\textwidth,page=43]{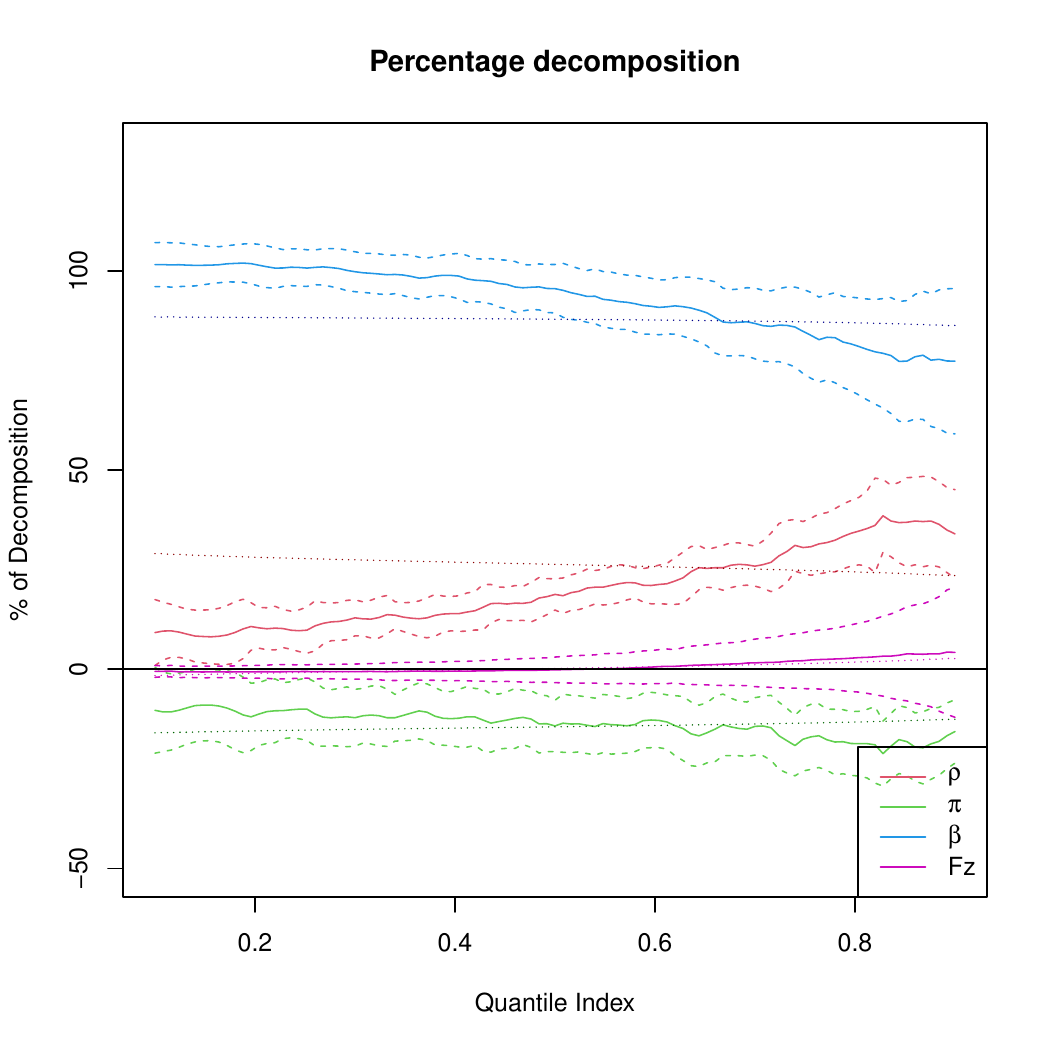}
    \caption{Estimates and 95\% confidence bands for the quantiles of observed wages and decomposition between men and women in specification 4}\label{fig:dec_spec1}
\end{figure}

Figure \ref{fig:dec_spec1} reports estimates of the quantile functions of observed wages for men and women, together with the relative contributions of each component to the decomposition between men (group $0$) and women (group $1$) based on specification 4. The bands for the contributions are joint for all the components and rely on the delta method; see Remark \ref{remark:qinfer} in the SM. Estimates of the components of the decomposition and the analysis based on specifications 1--3 are given in the SM.  The distribution for men first order stochastically dominates the distribution for women.  Most of this gender wage gap is explained by differences in the wage structure, i.e. differences in the returns to observed characteristics.
However,  differences in sorting and employment structure also account for an important percentage of the gap, especially at the top of the distribution. Thus, we uncover that the negative female sorting explains about 30--40\% of the gap at the top of the distribution. A possible explanation is that women with very high potential wages decide not to work  because there are no high-paid jobs available to them due to glass ceiling \cite{abv03}. The negative contribution of the employment structure can be explained by the order of the decomposition where we are applying the male employment structure to the female distribution with positive male sorting.\footnote{While the sign of the employment structure contribution changes with the order of the decomposition, neither its importance nor the significance of the contributions of the other components are sensitive to this order.} In this case we are increasing the proportion of employed women, where the added women come from a pool with lower positive selection, and this negative effect is not reversed by a change in the composition of the working women; see Remark \ref{remark:se} for more details. The aggregate selection effect, defined as the sum of the selection sorting and selection structure effects, is positive and statistically significant at the top of the distribution; see Figure \ref{fig:adec} in the SM. Differences in the composition of the characteristics contribute very little to explain the gender gap.  Finally, the estimates from the HSM in dotted lines pick up the average contributions of the components, but miss all the heterogeneity across the distribution.

\begin{figure}
    \includegraphics[height=.49\textwidth, width=.49\textwidth,page=14]{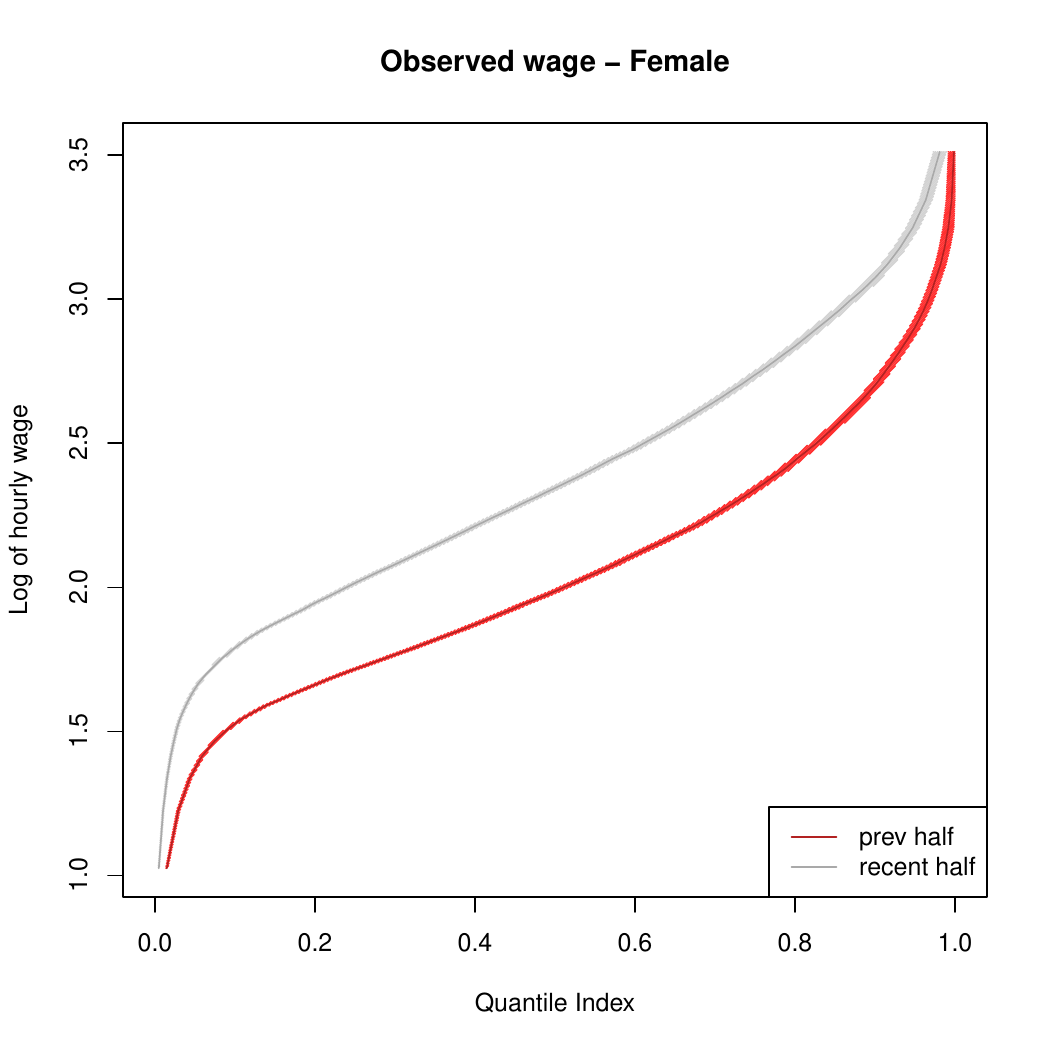}
        \includegraphics[height=.49\textwidth, width=.49\textwidth,page=43]{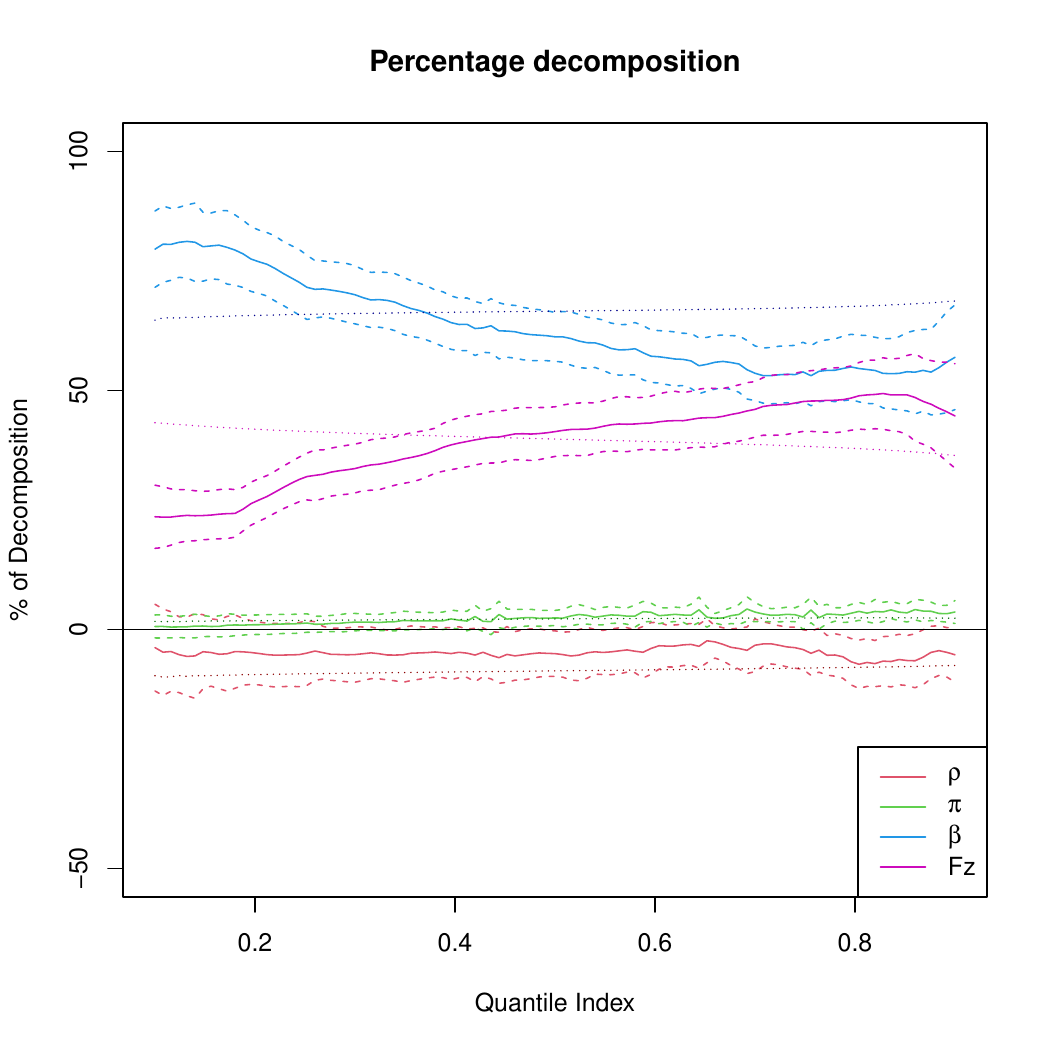}

    \caption{Estimates and 95\% confidence bands for the quantiles of observed wages and decomposition between first and second half of the sample period for women in specification 2}\label{fig:dec2_women_spec1}
\end{figure}

\begin{figure}
    \includegraphics[height=.49\textwidth, width=.49\textwidth,page=14]{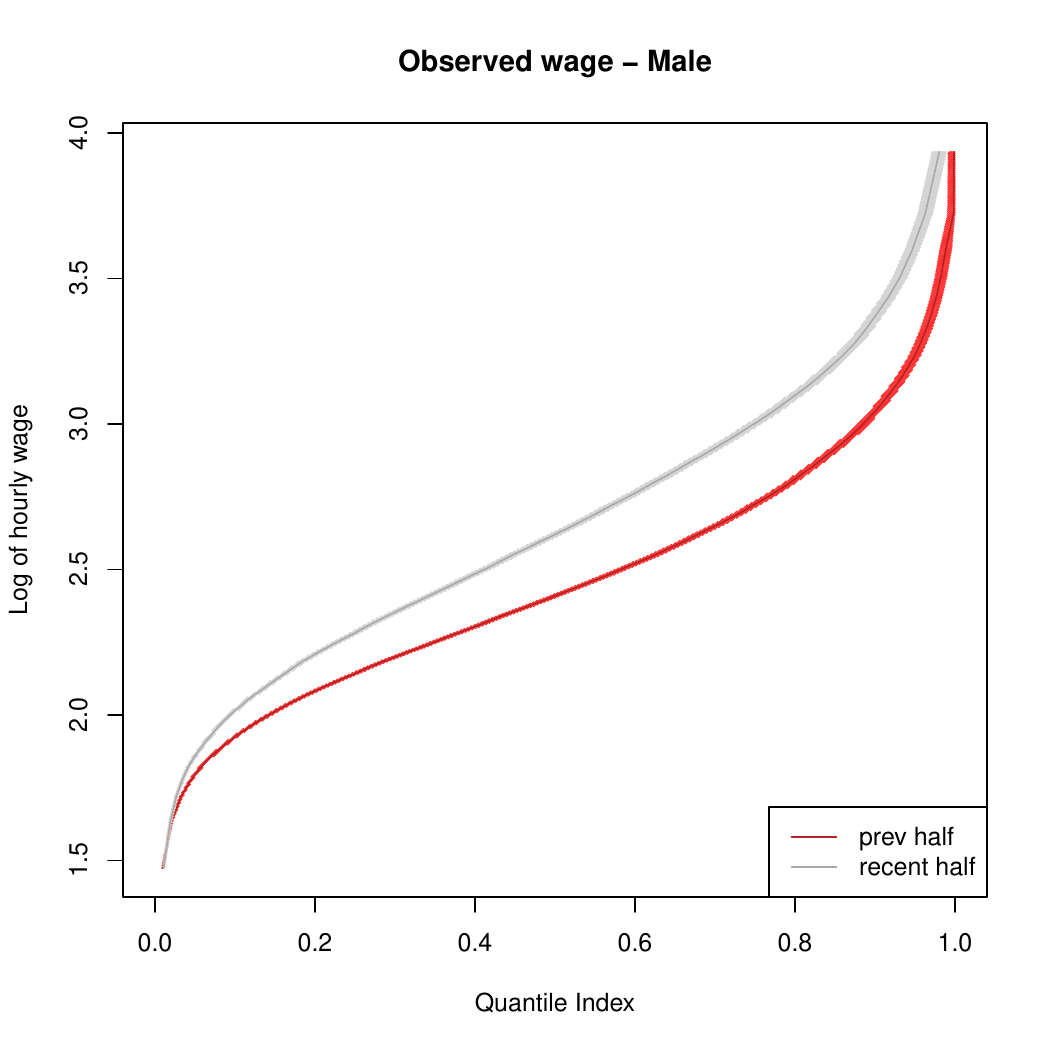}
        \includegraphics[height=.49\textwidth, width=.49\textwidth,page=43]{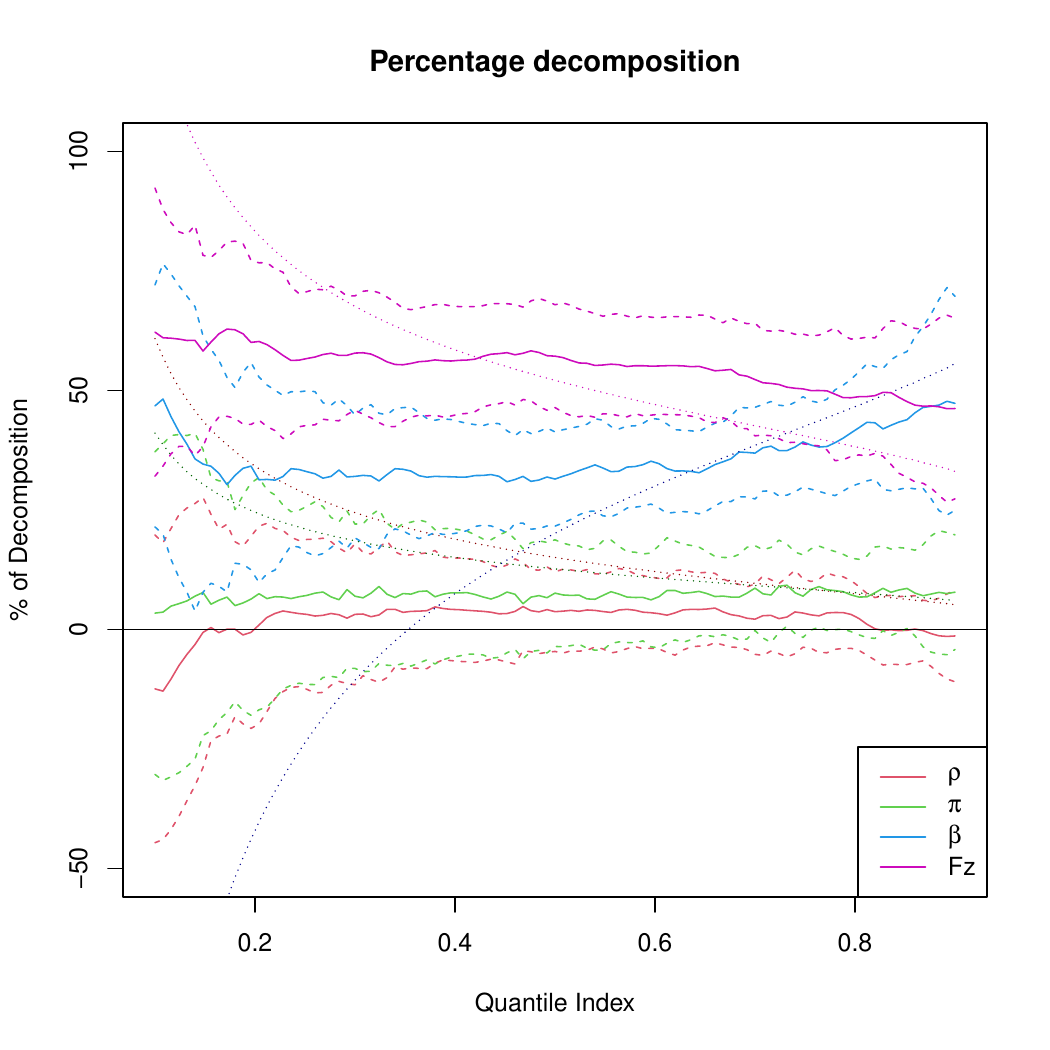}

    \caption{Estimates and 95\% confidence bands for the quantiles of observed wages and decomposition between first and second half of the sample period for men in specification 2}\label{fig:dec2_men_spec1}
\end{figure}

Figures \ref{fig:dec2_women_spec1} and \ref{fig:dec2_men_spec1} report estimates of the quantile functions of observed wages for the first and second halves of the sample period, together with the relative contributions of each component to the decomposition between second half (group $0$) and first half (group $1$) based on specification 2 for women and men, respectively. Estimates of the components of the decompositions  are given in the SM. The distribution for the second half first order stochastically dominates the distribution for the first half in both cases. For women, the most important components are the wage structure and composition effects in this order. The importance of the wage structure is decreasing along the distribution, whereas the importance of the composition is increasing. Composition and wage structure  are also the most important components for men. The small contributions of the selection sorting component to the change in the distribution of wages between the two time period for both genders seem to contradict the linear time trends that we found in the coefficient of the sorting selection function. This might be explained by the inability of a coarse partition of the sample into two halves to capture the gradual increase in selection sorting, together with the changes in the composition.

\subsection{Discussion}
The main findings can be summarized as: (1) positive sorting for men and negative sorting for women driven by single men and married women, which is consistent with assortative matching in the marriage market; (2) heterogeneity in selection sorting decreases gradually over time; (3) differences in returns to characteristics in the wage equation, which might  be associated with gender discrimination in the labor market,  account for most of the gender wage gap; (4) selection sorting on unobservables explains up to 39\% of the gender wage gap at the top of the distribution, which can be taken as evidence of glass ceiling; and (5) changes in the structure of the wage equation and composition of the characteristics account for most of the differences in the wage distribution between the two halves of the sample period within each gender.

We  compare and contrast these findings with previous results from the  literature that studied similar issues. These results were obtained from different data and/or using different methodology. \citen{bgim07} applied a bound approach that does not require of exclusion restrictions to study the evolution of wage inequality using the FES data for the period 1978--2000.  They assumed positive sorting for men and women in some of their estimates  to make the bounds more informative.  Interestingly, they mentioned the possibility that the assumption is violated for married women due to assortative matching in the marriage market.\footnote{In results not reported, we find that the negative sorting for married women is robust to the definition of the out-of-work benefit income variable. Thus, we find similar estimates using the income of a tax unit if all the individuals within the tax unit were out of work as the excluded covariate, as in \citen{bgim07}.} They also found evidence against the validity of out-of-work benefit income as a valid excluded covariate for men.  AB17 using the same data from the FES, also found positive sorting for men, stronger  for single than for married men, but they employed an alternative methodology that combines quantile regression for the marginal distributions with a parametric model for the copula. Contrary to our findings, they also found positive selection for women, which is statistically significant only for married women.  \citen{mr08} estimated a HSM using data from the US-CPS for the periods 1975-1979 and 1995-1999. They found that the selection sorting for women shifted from negative to positive between the two periods. We also find for the UK that the sorting for most women has a  positive trend over time, but remains negative even in 2013 for most of the distribution.  \citen{mw18} applied the methodology of AB17 to data from the US-CPS for the period 1976--2014. They also found negative sorting for women at the beginning of the sample period that became positive during the 90s, and positive sorting for men throughout the entire period. \citen{bertrand17} pointed out multiple possible explanations for the glass ceiling based on the field of education, psychological attributes or preferences for job flexibility that are compatible with our finding on the importance of sorting on unobservables at the top of the distribution. None of the previous papers distinguished between the selection sorting and selection structure effects.

One limitation of our dataset is that it does not contain a direct measure of work experience. As a final robustness check, we find that the results are not sensitive to the exclusion of college graduates from the sample by redoing the analysis excluding all the individuals who cease school after age 18. This is a relevant exclusion because work experience is a more relevant determinant of wage for highly educated workers.\footnote{These results are available from the authors upon request.}

\section{Conclusion}

 We develop a distribution regression model with sample selection that accommodates rich patterns of heterogeneity in the effects of covariates on outcomes and selection.  The model is semi-parametric in nature, as it has function-valued parameters, and is able to considerably generalize the classical selection model of \citen{heckman74}. Furthermore, the model accounts for richer covariate effects than the previous semi-parametric generalizations which allowed only the location effects for covariates.  We propose to  estimate the model by a process of bivariate probit regressions, indexed by threshold-dependent parameters. We show that the resulting estimators of the function-valued parameters are approximately Gaussian and concentrate in a $1/\sqrt{n}$ neigborhood of the true values.  We present an extensive wage
decomposition analysis for the U.K. using new data, generating both new findings and demonstrating the power of the method. Our identification approach is especially designed for the sample selection problem but can be applied to other settings. In work in progress, we show how the selection exclusion can be used to identify causal effects in treatment effects models with endogeneity in the presence of an instrumental variable that can be binary.

\bibliographystyle{econometrica}
\bibliography{bibliography,mybibVOLUME}

\appendix

\section{Proofs of Results in Section \ref{sec:id}}\label{app:id}

\subsection{Proof of Lemma \ref{lemma:lgr}}
Let $\phi_2(\cdot, \cdot; \rho)$ be the joint probability density function (PDF) of a standard bivariate normal random variable with parameter $\rho$. The representation \eqref{eq:lgr} exists and is unique by the following properties of the standard bivariate normal distribution:
\begin{enumerate}
\item $\rho \mapsto \Phi_2(\cdot, \cdot; \rho)$ is continuously differentiable and  $
\partial \Phi_2(\cdot, \cdot; \rho)/\partial \rho = \phi_2(\cdot, \cdot; \rho) > 0
$ \cite{sibuya59,sungur90};
\medskip
\item $
\lim_{\rho \nearrow 1} \Phi_2(x, y; \rho) = \min[\Phi(x),\Phi(y)];
$
\medskip
\item $
\lim_{\rho \searrow -1} \Phi_2(x, y; \rho) = \max[ \Phi(x) + \Phi(y) -1, 0];
$
\end{enumerate}
together with the Frechet-Hoeffding bounds
$$
\max[ \Phi(\mu(y)) + \Phi(\nu(d)) - 1, 0]  \leq F_{Y^*,D^*}(y,d) \leq   \min[\Phi(\mu(y)),\Phi(\nu(d))]. 
$$\qed


\subsection{Full Statment and Proof of Theorem \ref{theorem:id}}\label{app:full_thm} Let $$ F_{Y,D \mid Z} (y,1 \mid z) := \Pr(D=0 \mid Z=z) + \Pr(Y\leq y, D=1 \mid Z=z).$$ 

The full statement of the theorem is:
\begin{enumerate}
  \item If $ F_{Y,D \mid Z} (y,1 \mid z)  = 1$, $z \in \{0,1\}$, then,
  $$
  \rho(y) = 1, \ \   \mu(y) \in [\Phi^{-1}(\Pr(D=1 \mid Z=1)), + \infty).
  $$ 
  \item If $F_{Y,D \mid Z} (y,1 \mid z) = \Pr(D=0 \mid Z=z)$, $z \in \{0,1\}$, then,
  $$
\rho(y) = - 1, \   \mu(y) \in (-\infty, \Phi^{-1}\left(\Pr(D=0 \mid Z=1)\right)].
  $$ 
  \item  If $F_{Y,D \mid Z} (y,1 \mid 1) - \Pr(D=0 \mid Z=1) = F_{Y,D \mid Z} (y,1 \mid 0) - \Pr(D=0 \mid Z=0) > 0$,  then,
  $$
  \rho(y) = 1, \ \ \mu(y) = \Phi^{-1}\left(F_{Y,D \mid Z} (y,1 \mid 1) - \Pr(D=0 \mid Z=1) \right).
   $$
  \item  If $ F_{Y,D \mid Z} (y,1 \mid 1)  <  1$  and $F_{Y,D \mid Z} (y,1 \mid 0) = 1$, then,
  $$
   \rho(y) = 1, \ \ \mu(y) = \Phi^{-1}\left(F_{Y,D \mid Z} (y,1 \mid 1) - \Pr(D=0 \mid Z=1) \right).
  $$
  \item  If $F_{Y,D \mid Z} (y,1 \mid 1) > \Pr(D=0 \mid Z=1)$ and $F_{Y,D \mid Z} (y,1 \mid 0) = \Pr(D=0 \mid Z=0)$, then,
$$
 \rho(y) = - 1, \  \mu(y) =  \Phi^{-1}\left(F_{Y,D \mid Z} (y,1 \mid 1) \right). 
  $$
  \item  If  $F_{Y,D \mid Z} (y,1 \mid 1) =  F_{Y,D \mid Z} (y,1 \mid 0) < 1$, then,
\begin{eqnarray*}
& \rho(y) = - 1, \  \mu(y) = \Phi^{-1}\left(F_{Y,D \mid Z} (y,1 \mid 1) \right).
\end{eqnarray*}
\item Otherwise, $\mu(y)$ and $\rho(y)$ are point identified as the solution in $(\mu, \rho)$  to \eqref{eq:murho}. This solution exists and is unique.
\end{enumerate}

\textbf{Proof.} The identification of $\nu(z)$ follows from equalizing the marginals of $F_{Y^*,D^* \mid Z}$ with respect to $D^*$  and the conditional LGR at $D^*=0$. Since $\nu(z)$ is identified, we shall use 
$$\Phi(\nu(z)) = \Pr(D=1 \mid Z=z)  \text{ and } \bar{\Phi}(\nu(z)) = \Pr(D=0 \mid Z=z),  z \in \{0,1\},$$ to lighten the notation. We also recall the Frechet-Hoeffding bounds  in our setting, as they will be extensively used in the proof:
\begin{equation}\label{eq:fhb}
\max[ \Phi(\mu(y)) - \bar{\Phi}(\nu(z)), 0]  \leq  \Phi_2(\mu(y), \nu(z); \rho(y)) \leq   \min[\Phi(\mu(y)),\Phi(\nu(z))], z \in \{0,1\},\tag{\text{FHB}}
\end{equation}
where the upper bound is only attained at $\rho(y) =1$ and the lower bound at $\rho(y) = -1$.

\paragraph{\textbf{Cases (1)--(2)}} We consider first the partially identified cases. In case (1), 
$$ \Phi_2(\mu(y), \nu(z); \rho(y)) = F_{Y,D \mid Z} (y,1 \mid z) - \bar{\Phi}(\nu(z)) = \Phi(\nu(z)), z \in \{0,1\},$$  implies that $\rho(y) = 1$ by the upper \ref{eq:fhb}. The identified set for $\mu(y)$ is obtained from $$\Phi(\mu(y)) \geq \max[\Phi(\nu(0)), \Phi(\nu(1))] = \Phi(\nu(1))$$ by the upper \ref{eq:fhb}
 and Assumption \ref{ass:er}(2). In case (2), $$ \Phi_2(\mu(y), \nu(z); \rho(y)) = F_{Y,D \mid Z} (y,1 \mid z) - \bar{\Phi}(\nu(z)) = 0, z \in \{0,1\},$$  implies that $\rho(y) = -1$ by the lower \ref{eq:fhb}. The identified set for $\mu(y)$ is obtained from $$\Phi(\mu(y)) \leq \min[\bar{\Phi}(\nu(0)), \bar{\Phi}(\nu(1))] = \bar{\Phi}(\nu(1))$$ by the lower \ref{eq:fhb} and Assumption \ref{ass:er}(2).

In the rest of the proof we can assume that the conditions that define cases (1)--(2) do not hold.
\paragraph{\textbf{Cases (3)--(4)}} These boundary cases  correspond to $\rho(y) = 1$. They are identifiable because $\rho(y) = 1$ if and only if $F_{Y,D \mid Z} (y,1 \mid 1) - \bar{\Phi}(\nu(1))= F_{Y,D \mid Z} (y,1 \mid 0) - \bar{\Phi}(\nu(0)) $ or $F_{Y,D \mid Z} (y,1 \mid 0) = 1$. The only if part follows from the upper \ref{eq:fhb} 
and Assumption \ref{ass:er}(2).  Indeed, there are 2 cases depending on the values of $\Phi(\nu(0))$ and $\Phi(\mu(y))$. If $\Phi(\mu(y)) < \Phi(\nu(0))$,  then $ \min[\Phi(\nu(z)),\Phi(\mu(y))] = \Phi(\mu(y))$ for $ z \in \{0,1\}$ so that $F_{Y,D \mid Z} (y,1 \mid 1) - \bar{\Phi}(\nu(1))= F_{Y,D \mid Z} (y,1 \mid 0) - \bar{\Phi}(\nu(0))$. If $\Phi(\nu(0)) \leq \Phi(\mu(y))$, then $\Phi_2(\mu(y), \nu(0); 1)  =  \Phi(\nu(0))$ so that $F_{Y,D \mid Z} (y,1 \mid 0) = 1$.  
For the case $F_{Y,D \mid Z} (y,1 \mid 1) - \bar{\Phi}(\nu(1))= F_{Y,D \mid Z} (y,1 \mid 0) - \bar{\Phi}(\nu(0))$, the if part follows because $\nu \mapsto \Phi_2(\cdot, \nu; \rho)$ is strictly monotonic when $\rho \in (-1,1)$ and $\nu(1) > \nu(0)$ by Assumption \ref{ass:er}(2) so that $\Phi_2(\mu(y), \nu(0); \rho(y)) \neq \Phi_2(\mu(y), \nu(1); \rho(y))$. This shows that $\rho(y) \not\in (-1,1)$.  Moreover, this case is ruled out when $\rho(y) = -1$ by the lower \ref{eq:fhb},
$F_{Y,D \mid Z} (y,1 \mid 1) >  \bar{\Phi}(\nu(1))$, and Assumption \ref{ass:er}(2).\footnote{Note that in case (4), $F_{Y,D \mid Z} (y,1 \mid 1) -  \bar{\Phi}(\nu(1)) = \max[ \Phi(\mu(y)) - \bar{\Phi}(\nu(1)), 0] \geq  \max[ \Phi(\mu(y)) - \bar{\Phi}(\nu(0)), 0] = \Phi(\nu(0)) > 0$ when $\rho(y) = -1$.} The case $F_{Y,D \mid Z} (y,1 \mid 0) = 1$ implies that $\Phi_2(\mu(y), \nu(0); \rho(y)) = \Phi(\nu(0))$, which is only possible when $\rho(y) = 1$ by the upper \ref{eq:fhb}.

Now, we can analyze the identification of $\mu(y)$ using the upper \ref{eq:fhb}. Case (3) corresponds to $F_{Y,D \mid Z} (y,1 \mid z) - \bar{\Phi}(\nu(z))= \Phi(\mu(y))$, $z \in \{0,1\},$ which identify $\mu(y)$. Case (4) corresponds to $F_{Y,D \mid Z} (y,1 \mid 0) = 1$ and $F_{Y,D \mid Z} (y,1 \mid 1) - \bar{\Phi}(\nu(1))= \Phi(\mu(y))$. The second equation identifies $\mu(y)$. 

\paragraph{\textbf{Cases (5)--(6)}} These boundary cases correspond to $\rho(y) = -1$. They are identifiable because $\rho(y) = -1$ if and only if $F_{Y,D \mid Z} (y,1 \mid 0) = F_{Y,D \mid Z} (y,1 \mid 1)$ or $F_{Y,D \mid Z} (y,1 \mid 0) =  \bar{\Phi}(\nu(0))$. Symmetrically to $\rho(y) = 1$, the only if part follows from the lower \ref{eq:fhb}  and Assumption \ref{ass:er}(2), whereas the  if part for the case $F_{Y,D \mid Z} (y,1 \mid 0) = F_{Y,D \mid Z} (y,1 \mid 1)$ follows from the upper \ref{eq:fhb},   strict monotonic of $\nu \mapsto \Phi_2(\cdot, \nu; \rho)$ when $\rho \in (-1,1)$, and Assumption \ref{ass:er}(2).\footnote{Indeed, this result follows from a similar argument to the case $F_{Y,D \mid Z} (y,1 \mid 1) - \bar{\Phi}(\nu(1))= F_{Y,D \mid Z} (y,1 \mid 0) - \bar{\Phi}(\nu(0))$ using that $\Phi_2(\mu(y), \nu(z); \rho(y)) + \bar{\Phi}(\nu(0)) = \Phi_2(-\mu(y), \nu(z); -\rho(y)) $.}  The  if part for $F_{Y,D \mid Z} (y,1 \mid 0) =  \bar{\Phi}(\nu(0))$ follows because this case implies that $\Phi_2(\mu(y), \nu(0); \rho(y)) = 0$, which is only possible when $\rho(y) = -1$ by the lower \ref{eq:fhb}.

Now, we can analyze the identification of $\mu(y)$ using the lower \ref{eq:fhb}. Case (5) corresponds to $F_{Y,D \mid Z} (y,1 \mid 1) = \Phi(\mu(y))$, which identifies $\mu(y)$. Case (6) corresponds to $F_{Y,D \mid z} (y,1 \mid z) =  \Phi(\mu(y))$, $z \in \{0,1\}$. Both of these equations have the same solution that identifies $\mu(y)$. 

\paragraph{\textbf{Case (7)}}  Finally, consider now the non-boundary case where $\rho(y) \in (-1,1)$.   The parameters $\mu(y)$ and $\rho(y)$ are identified as the solution in $(\mu,\rho)$ to \eqref{eq:murho}.  
This nonlinear system of 2 equations has unique solution under Assumption \ref{ass:er}(2). This result follows from Theorem 4 of \citen{gale65}, after showing that the Jacobian of the system \eqref{eq:murho} is a P-matrix when $\rho(y) \in (-1,1)$, which is what we demonstrate in the remainder of the proof.

Let  $\partial_{\mu}\Phi_2(\mu,\nu;\rho) = \partial \Phi_2(\mu,\nu;\rho)/\partial \mu $ and $\partial_{\rho}\Phi_2(\mu,\nu;\rho) = \partial \Phi_2(\mu,\nu;\rho)/\partial \rho $. The Jacobian matrix of the system,
$$
J(\mu(y),\rho(y)) =  \left(\begin{array}{cc}\partial_{\mu} \Phi_2(\mu(y), \nu(1); \rho(y)) & \partial_{\rho} \Phi_2(\mu(y), \nu(1); \rho(y)) \\
\partial_{\mu} \Phi_2(\mu(y), \nu(0); \rho(y)) & \partial_{\rho} \Phi_2(\mu(y), \nu(0); \rho(y)) \end{array}\right),
$$
is a P-matrix for all $\mu(y) \in \mathbb{R}$ and $\rho(y) \in (-1,1)$ because by the properties of the bivariate normal CDF:
$$
\partial_{\mu} \Phi_2(\mu(y), \nu(1); \rho(y)) = \Phi\left( \frac{\nu(1) - \rho(y) \mu(y)}{\sqrt{1-\rho(y)^2}}\right) \phi(\mu(y)) > 0,
$$
$$
\partial_{\rho} \Phi_2(\mu(y), \nu(0); \rho(y)) =  \phi_2(\mu(y), \nu(0); \rho(y)) > 0,
$$
and
$$
\det(J(\mu(y),\rho(y))) = \phi(\mu(y))^2 \left[\Phi\left( \tilde \nu(1,y) \right) \phi\left( \tilde \nu(0,y)\right) - \Phi\left( \tilde \nu(0,y)\right) \phi\left( \tilde \nu(1,y)\right) \right] > 0,
$$
where $\tilde{\nu}(0,y) = [\nu(0) - \rho(y) \mu(y)]/\sqrt{1-\rho(y)^2}$ and  $\tilde{\nu}(1,y) = [\nu(1) - \rho(y) \mu(y)]/\sqrt{1-\rho(y)^2}$. 
In the last result we use that, by the properties of the normal distribution,  $$\phi_2(\mu, \nu; \rho) = \phi\left( [\nu - \rho \mu]/\sqrt{1-\rho^2}\right)  \phi(\mu)$$ and  the inverse Mills ratio $\nu \mapsto \lambda(\nu) := \phi(\nu)/\Phi(\nu)$ is strictly decreasing in $\mathbb{R}$, so that
$$
\Phi\left( \tilde \nu(1,y)\right) \phi\left( \tilde \nu(0,y)\right) - \Phi\left( \tilde \nu(0,y)\right) \phi\left( \tilde \nu(1,y)\right) > 0,
$$
since $ \tilde \nu(0,y) <  \tilde \nu(1,y)$. \qed

\subsection{Proof of Theorem \ref{thm:se}}  Let $D^* = V - \nu(Z)$ and $V \mid Z \sim N(0,1)$ such that $p(z) = F_{D^* \mid Z}(0 \mid z) = \Phi(\nu(z))$.\footnote{Note that any definition of $D^*$ that yields the same probability of $D=1$ conditional on $Z$ is observationally equivalent. In particular, we can choose without loss of generality $D^*$ with continuous CDF.} Note that $\Phi^{-1}(F_{Y^*}(y)) = \mu(y) $, $ \Phi^{-1}(F_{D^* \mid Z}(0 \mid z) ) = \nu(z)$, and
\begin{multline*}
\Phi_2(\mu(y), \nu(z); \tilde{\rho}(y,\nu(z) \mid z) ) = F_{Y^*,V \mid Z}(y,\nu(z) \mid z) 
  = \Pr(Y^* \leq y, V \leq \nu(z) \mid Z = z) \\ =  \Pr(Y^* \leq y, D^* \leq 0 \mid Z = z) = F_{Y^*,D^* \mid Z}(y,0 \mid z) 
  = 
  \Phi_2(\mu(y), \nu(z); \rho(y \mid z)),
\end{multline*} 
by definition of the LGR, such that $\tilde{\rho}(y,\nu(z) \mid z) =  \rho(y \mid z)$
The result holds because $\tilde{\rho}(y,\nu(z) \mid z) =  \rho(y)$ if $ \rho(y \mid z) = \rho(y)$ and $\rho(y \mid z) = \tilde{\rho}(y)$ if $\tilde{\rho}(y,\nu(z) \mid z) = \tilde{\rho}(y)$. \qed

\subsection{Proof of Equivalence \eqref{eq:index}}\label{app:index}
Assume joint independence, $(Y^*,V) \indep Z$.  Then,  $\tilde \rho(y,v) = \tilde \rho(y)$ for all $v \in \mathcal{V}_0$, where $\mathcal{V}_0$ is an open interval containing the support of $\nu(Z)$, implies the single index property because
$$
\Pr(Y^* \leq y \mid V=v) = \frac{(\partial/\partial v) F_{Y^*,V}(y,v)}{(\partial/\partial v) F_{V}(v)} = \frac{(\partial/\partial v) \Phi_2(\mu(y), v; \tilde{\rho}(y) )}{(\partial/\partial v) \Phi(v)} = \Phi \left(a(y) + b(y) v\right),
$$
for $a(y) = \mu(y)/\sqrt{1-\tilde{\rho}(y)^2}$ and $b(y) = - \tilde \rho(y)/\sqrt{1-\tilde{\rho}(y)^2}$. Conversely, assuming that the index restriction holds for all $v \in \mathcal{V}_0$, the partial differential equation
 $$
 (\partial / \partial v) F_{Y^*,V}(y,v) = \Phi \left(a(y) + b(y) v \right) (\partial/\partial v) \Phi(v)
 $$
 has a solution at the LGR, $F_{Y^*,V}(y,v) =  \Phi_2\left(\mu(y), v; \tilde \rho(y) \right)$, for all $v \in \mathcal{V}_0$, 
 where $\mu(y) = a(y) / \sqrt{1 + b(y)^2}$ and $\tilde \rho(y) = - b(y) / \sqrt{1 + b(y)^2}$. In both cases we use that $V \sim N(0,1)$.

\subsection{Sufficient Condition for Index Restriction Without Joint Independence}\label{app:index2}
Assume only marginal outcome independence, $Y^* \indep Z$ ,and note that $V \indep Z$ follows by construction.  Then,   $\tilde \rho(y,\nu(z) \mid z) = \tilde \rho(y)$  and $\partial \tilde \rho(y,\nu(z) \mid z)/\partial v = 0$ for $z \in \{0,1\}$,\footnote{For example, $
\tilde \rho(y, v | z) = \tanh(y + (v - \nu(z))^2)
$ satisfies these restrictions. We are grateful to an anonymous referee for a comment that motivated these sufficient condition and  example.} imply the single index property because
\begin{multline*}
   \Pr(Y^* \leq y \mid Z=z, V=v) = \frac{(\partial/\partial v) F_{Y^*,V \mid Z}(y,v \mid z)}{(\partial/\partial v) F_{V}(v)} = \frac{(\partial/\partial v) \Phi_2(\mu(y), v; \tilde{\rho}(y,v\mid z) )}{(\partial/\partial v) \Phi(v)} \\
   = \Phi \left(a(y,v\mid z) + b(y,v \mid z) v\right) + \phi \left(a(y,v\mid z) + b(y,v \mid z) v\right) \frac{\partial \tilde \rho(y,v \mid z)}{\partial v},
\end{multline*}
for $a(y,v\mid z) = \mu(y)/\sqrt{1-\tilde{\rho}(y,v\mid z)^2}$ and $b(y,v\mid z) = - \tilde \rho(y,v \mid z)/\sqrt{1-\tilde{\rho}(y,v \mid z)^2}$, and the conditions on $\tilde \rho(y,v \mid z)$ and  $\partial \tilde \rho(y,v \mid z)/\partial v$ yield
$$
 \Pr(Y^* \leq y \mid Z=z, V=\nu(z)) = \Phi \left(a(y) + b(y) \nu(z)\right), \quad z \in \{0,1\},
$$
for $a(y) = \mu(y)/\sqrt{1-\tilde{\rho}(y)^2}$ and $b(y) = - \tilde \rho(y)/\sqrt{1-\tilde{\rho}(y)^2}$.


\newpage

\appendix 



\begin{center}
\Large{Supplementary Material to ``Distribution Regression with Sample Selection, with an Application to Wage Decompositions in the UK ''}

\setcounter{footnote}{0}
\setcounter{page}{1}
\setcounter{figure}{0}
\setcounter{table}{0}
\renewcommand{\thetable}{S\arabic{table}}   
\renewcommand{\thefigure}{S\arabic{figure}}
\thispagestyle{plain}
\renewcommand{\thesection}{S\arabic{section}}

\normalsize{Victor Chernozhukov, Iv\'an Fern\'andez-Val, and Siyi Luo}
\end{center}

\begin{abstract}
The supplementary material includes seven appendices. Appendices \ref{app:id}--\ref{app:sign} contain deferred discussions of Sections \ref{sec:id}--\ref{sec:empirics}. They include properties of the LGR, comparison with AB17, identification approaches with rich instruments, bound analysis with an example, models with homogeneous parameters and an example of sign reversal in the selection effects. Appendices \ref{sec:theory} and \ref{app:proofs} contain the asymptotic theory for the estimation and inference methods, and the corresponding proofs. Appendix \ref{app:mc} reports the results of a 
Monte Carlo simulation calibrated to the empirical application. Appendix \ref{app:empirics} shows additional results of the empirical application. They include descriptive statistics and background on the U.K. labor market, a model of offered and reservation wages,  estimates of the coefficients of the employment (selection) equation,  estimates and 95\% confidence intervals for the coefficients of the wage equation, estimates and 95\% confidence intervals for the sorting equation not reported in the main text, estimates and 95\% confidence intervals for the decomposition of the employment rate, estimates and 95\% confidence bands for the components of the wage decomposition of observed wages in the specification 4,  estimates and 95\% confidence bands for the offered and observed wages and their decompositions for the specifications 1--3, and estimates and 95\% confidence bands for the wage decomposition between first and second half of the sample period for men and women in specification 2.

\end{abstract}

\newpage

\newpage
\section{Deferred Discussions of Section \ref{sec:id}}\label{app:sect2}

\subsection{Discussion of the LGR and Local Correlation}\label{app:lgr}
In the LGR, the marginal CDFs of $Y^*$ and $D^*$ are represented by local Gaussian links
$$
F_{Y^*}(y) = \Phi(\mu(y)), \quad F_{D^*}(d) = \Phi(\nu(d)),
$$
and the copula of $Y^*$ and $D^*$ is represented by a local Gaussian copula
\begin{equation}\label{LGC}
\begin{array}{l}
 C_{Y^*,D^*}(u,v) =   \Phi_2(\Phi^{-1}(u), \Phi^{-1}(v); \rho(y_u,d_v)), \\
 \forall (u,v) \in [0,1]^2 : \exists y_u \in \mathbb{R}: F_{Y^*}(y_u) = u, \quad \exists d_v  \in \mathbb{R}: F_{D^*}(d_v) = v.
\end{array}
\end{equation}

\citen{kolev06} developed a closely related result to (\ref{LGC}) for the copula. They established that the copula of any bivariate distribution can be represented by the bivariate Gaussian copula with a local correlation parameter. Like the copula, the LGR is convenient because it separates $\mu(y)$ and $\nu(d)$ as two parameters determining the marginals of $Y^*$ and $D^*$ from  $\rho(y,d)$ as a parameter determining the dependence between $Y^*$ and $D^*$.  
  Unlike the copula, the arguments of the LGR are the same as the arguments of the joint CDF and  the domain of the LGR is therefore $\mathbb{R}^2$.  
  



Here we present additional discussion of the local correlation $\rho(y,d)$.\footnote{The parameter $\rho(y,d)$ is the tetrachoric correlation coefficient between $\{Y^*\leq y\}$ and $\{D^*\leq d\}$ \cite{pearson1900}.} 
In general, the sign of $\rho(y,d)$ determines the sign of the local dependence between $Y^*$ and $D^*$ at $(y,d)$ as measured by the covariance between $1\{Y^*\leq y\}$ and $1\{D^*\leq d\}$.
\begin{lemma}[Correlation interpretation for $\rho(y,d)$]\label{lemma:corr} For $(Y^*, D^*)$ having LGR with local correlation parameter $\rho(y,d)$, the map
$\rho(y,d) \mapsto \mathrm{Cov}(1\{Y^*\leq y\}, 1\{D^*\leq d\})$, mapping $[-1,1]$ to the real line, is a strictly increasing (hence bijective) function.  Moreover, $\mathrm{Cov}(1\{Y^*\leq y\}, 1\{D^*\leq d\})$ and $\rho(y,d)$ have the same sign (positive, negative, or 0).

\end{lemma}

It follows that $\rho(y,d)$ is positive (resp. negative) everywhere if and only if $Y^*$ and $D^*$ are positively (resp. negatively) quadrant dependent in the sense defined by \cite{lehmann66}. The local dependence measure $ \text{Cov}(1\{Y^*\leq y\}, 1\{D^*\leq d\})$ is called Laplace covariance kernel by \citen{dhkv15}. 


 When $\rho(y,d) = \rho(y)$  we can further characterize the local dependence in terms of the normal scores $Y^s := \Phi^{-1}\left( F_{Y^*}(Y^*)  \right)$ and $D^s := \Phi^{-1}\left( F_{D^*}(D^*)  \right)$. Let $F_{Y^s,D^s}(y,d) = \Phi_2(\mu_s(y), \nu_s(d); \rho_s(y,d))$ be the LGR of the distribution of $(Y^s,D^s)$.
\begin{lemma}[Other Properties of $\rho(y,d)$]\label{lemma:lgr2}  If $\rho(y,d) = \rho(y)$ and $d \mapsto F_{D^*}(d)$ is continuous on $\mathbb{R}$,
$$
\rho_s(y) = - \frac{\mathrm{Cov}(1\{Y^s\leq \mu_s(y)\}, D^s)}{\phi(\mu_s(y))},
$$
where
$
\mathrm{Cov}(Y^s,D^s) = \Ep[\rho_s(Y^s)],
$
provided that $\mathrm{Cov}(Y^s,D^s)$ exists, where $\mu_s(y) = \sup \{y^* \in \mathbb{R} : F_{Y^*}(y^*) \leq \Phi(y) \}$, 
and $ \rho_s(y) = \rho(y^*)$ for $y^*$ such that $ F_{Y^*}(y^*) = \Phi(\mu_s(y))$.
\end{lemma} 

Note that for the points $y$ in the support of $Y^s$, 
$
\rho_s(y) = -{\mathrm{Cov}(1\{Y^s\leq y\}, D^s)}/{\phi(y)}.
$
Lemma \ref{lemma:lgr2} therefore characterizes  $\rho_s(y)$,  and therefore also  $\rho(y)$, as a measure of local dependence between $Y^s$ and $D^s$.    It also shows that the covariance between these normal scores, a measure of global dependence, can be expressed as  the average of the local dependence parameter.  The case considered with $\rho(y,d) = \rho(y)$ and continuous $D^*$ is relevant for the discussion of the  sample selection problem in Section \ref{subsec:se}.

\

\subsection{Comparison with AB17.}\label{app:ab17} We compare our identification conditions with  AB17.   We start by restating AB17 assumptions in terms of the LGR   of the joint distribution of $(Y^*,V)$. In addition to Assumption \ref{ass:er}(1)--(3),  AB17 assumed that (i) $(Y^*,V)$ are jointly independent of $Z$, (ii) $v \mapsto \tilde \rho(y,v)$ is real analytic on the unit interval,  (iii) the support of  $p(Z)$ contains an open interval, and (iv) $y \mapsto F_{Y*}(y)$ is continuous and strictly increasing.\footnote{\citen{hs84} also exploited a real analyticity assumption to solve a censoring problem in duration analysis.}   Assumption \ref{ass:er} neither implies nor is implied by conditions (i)--(iv). Thus, condition (iii) requires  $Z$ to have continuous variation and is therefore more restrictive than our assumption that $Z$ can be binary.  Unlike (iv), we do not require the CDF of $Y^*$ to be continuous. Our identification result  applies to discrete and mixed outcomes. However,  under condition (i), Theorem \ref{thm:se} shows that selection exclusion requires that $v \mapsto \tilde \rho(y,v) $ is constant, which is stronger than (ii). In other words, we impose stronger restrictions in the dependence between the latent outcome and unobserved selection ranking, but require less variation in the excluded covariate $Z$ and outcome $Y^*$. 

More specifically,  let  $p(z) = \Pr(D=1 \mid Z=z)$ and $V = F_{D^*\mid Z}(D^* \mid Z) $ such that $V \mid Z \sim U(0,1)$.\footnote{We assume that $D^*$ is absolutely continuous with strictly increasing distribution. This assumption is without loss of generality because the distribution of $D^*$ is only identified at $D^*=0$.} AB17 assumed that (i) $(Y^*,V)$ are jointly independent of $Z$, (ii) $v \mapsto C_{Y^*,V}(\cdot,v)$ is real analytic on the unit interval, where $C_{Y^*,V}$ is the copula of $(Y^*,V)$, and (iii) the support of  $p(Z)$ contains an open interval. 
We now show that our selection sorting exclusion neither implies nor is implied by conditions (i) and (ii).   Selection sorting exclusion implies that for any $u \in [0,1]$ that satisfies $F_{Y^*}(y_u) = u$ for some $y_u$,
$$
C_{Y^*,V \mid Z}(u,p(z) \mid z) = C_{Y^*,D^* \mid Z}(u,p(z)  \mid z) = \Phi_2(\Phi^{-1}(u), \Phi^{-1}(p(z)); \rho(y_u,0)) = C_{Y^*,V}(u,p(z)),
$$
since $p(z) = F_{D^* \mid Z}(0 \mid z)$. This implication is weaker than condition (i) but it suffices for the identification argument in AB17. However, it only guarantees that $v \mapsto C_{Y^*,V}(\cdot,v)$ is real analytic on the support of $p(Z)$.\footnote{Note that $v \mapsto \Phi_2(\cdot, \Phi^{-1}(v); \rho(\cdot,0))$ is a real analytic function.}  Therefore, we conclude that selection exclusion  implies conditions (i) and (ii) only if the support of $p(Z)$ is the unit interval. To verify that the converse is also not true, note that the LGR of $(Y^*,V)$ conditional on $Z$ under condition (i) is
$$
F_{Y^*,V \mid Z}(y,v \mid z) = \Phi_2(\tilde \mu(y), \tilde \nu(v); \tilde \rho(y,v)). 
$$
This, together with $\tilde \mu(y) = \mu(y)$ and $\tilde \nu(p(z)) = \nu(z)$, imply that
$$
F_{Y^*,D^* \mid Z}(y,0 \mid z) =  \Phi_2(\tilde \mu(y), \tilde \nu(p(z)); \tilde \rho(y,p(z))) =  \Phi_2(\mu(y),  \nu(z); \tilde \rho(y,p(z))),
$$
which satisfies the selection exclusion only if $\tilde \rho(y,v) = \tilde \rho(y)$ for all $v$ in the support of $p(Z)$, i.e. the local dependence between $Y^*$ and $V$ does not vary with the value of $V$ in this region.
We finally note that condition (i) together with $\tilde \rho(y,v) = \tilde \rho(y)$ for all $v$ in the unit interval imply condition (ii) because
$$
C_{Y^*,V}(\cdot,v) = \Phi_2(\cdot, \Phi^{-1}(v); \tilde \rho(\cdot))
$$
is a real analytic function with respect to $v$ in the unit interval. Alternatively, condition (ii) is equivalent to $v \mapsto \tilde \rho(\cdot,v)$ being real analytic, which is weaker than $\tilde \rho(y,v) = \tilde \rho(y)$.

\subsection{Further Identification Results with Richer Instruments}\label{app:analy} Suppose that $(Y^*, V)$ are jointly independent of $Z$:\begin{equation}\label{ass:completeIndep}
(Y^*,V) \indep Z.
\end{equation}
This assumption implies that the sorting mechanism depends only on the instrumental values through the propensity score
\begin{eqnarray}\label{ass:completeIndep2}
 \Pr (Y^* \leq y, D^* \leq 0 \mid Z=z) & = & \Phi_2( \mu(y), \nu(z); \rho(y, \nu(z)) \notag \\ 
& = & \Pr (Y^* \leq y, D^* \leq 0 \mid \nu(Z) =\nu(z)), \quad z \in \mathcal{Z},
\end{eqnarray}
 where $\nu(z)$ is the transformed propensity score $\Phi^{-1} (p(z))$, and $\mathcal{Z}$ is the support of $Z$.   

Using \eqref{ass:completeIndep2},  the system of equations for the identification analysis can be written as:
\begin{eqnarray}\label{eq:defF}
 \mathsf{R} (y, v):= \Pr (Y^* \leq y, D^* \leq 0 \mid \nu(Z) = v)  = \Phi_2( \mu(y), v; \rho(y, v)) \quad v \in \mathcal{V},
\end{eqnarray}
where $\mathcal{V}$ is the support of $\nu(Z)$.  Intuitively, if the propensity scores have enough variation, the equations above should contain identifying information. Still, considerable care is required to turn this information into point identification and estimability.

\paragraph{\textbf{Real Analyticity Approach.}}  This approach imposes real-analyticity assumptions on $v \mapsto \rho(y,v)$ over $\Bbb{R}$, but avoids making parametric assumptions.  Specifically, analyticity means that over each open neighborhood the mapping can be represented by a convergent Taylor expansion around a point in that neighborhood \cite[Def. 1.1.5]{krantz2002primer}. For example, a continuously infinite-differentiable function with derivatives bounded by a common constant has this property. The key property of real analytical functions is the continuation:  knowing a function over an open region, implies we know the function over entire $\Bbb{R}$ \cite[Cor. 1.2.6]{krantz2002primer}. Thus, intuitively, such assumption rules out "wilder" forms of $\rho(y,v)$ that cause identification failure and attain the agnostic Balke-Pearl bounds.\footnote{The Balke-Pearl bounds are attained by having $\rho(y,v)$ vary between +1 or -1 over each small neigborhood of a given point $v_0$, to trace the smallest root and the maximal root as bounds for $\mu(y)$, then taking the smallest upper bound and taking the largest lower bound across instrumental values $v$. Such oscillatory (and unreasonable) behavior of the sorting function $\rho$ can be ruled out by a variety of means, which underlie but don't exhaust all of the point identification results in this paper.}

\begin{lemma}[Identification Under Real Analyticity]\label{lemma:RA} Fix a value of $y$ in what follows.
Suppose that $v \mapsto \rho(y, v)$ is real analytic on $\Bbb{R}$, mapping into $[-1,+1]$. Also  assume  that $v \mapsto \rho(y, v)$ 
can be continuously extended to $\overline{\Bbb{R}}$, and the propensity score variable $p(Z)$ is continuous on an open subinterval of $(0,1)$. Then the pair $(\mu(y), \rho(y,v))$ are identified: $\mu(y)$ is identified at infinity as 
$$\mu(y) = \lim_{v \to \infty}\Phi^{-1}(\mathsf{R}(y,v))$$ and $\rho(v)$ is identified as the unique solution to $\mathsf{R}(y,v)= \Phi_2( \mu(y), v, \rho(y, v))$ for any $v \in \mathcal{V}$.
\end{lemma}

The lemma above extends the argument in AB17 to the present LGR-based approach. The drawback of the analyticity approach is its restrictive nature; for example, functions generated from finite spline sieves are not real analytical.  Moreover, identification does not immediately translate into estimability due to the ill-conditioned nature of the analytical continuation, especially at "infinity" -- see \citen{trefethen2020quantifying} for further discussion. Imposing parametric functional forms on $v \mapsto \rho(y,v)$ provides necessary regularization, as used, for example, in AB17 for estimation purposes.\footnote{See \citen{trefethen2020quantifying} for making this point explicit through the use of polynomials.} But once we make the parametric assumptions, we can also get identification and estimability without requiring continuous instruments and without relying on the identification at infinity arguments.

\paragraph{\textbf{Semi-Parametric Approach}} We assume that $\rho(y,v) = \rho(y,v; \theta_0(y))$
for $\theta_0(y) \in \Theta \subset \Bbb{R}^d$, where $\rho(y,v; \theta)$ is a known function up to the parameter $\theta$ and  $\Theta$ is a compact region 
with non-empty open interior in $\Bbb{R}^d$; and that the true value of $\rho(y,v)$ is induced by some value $\theta_0(y)$ in the interior of $\Theta$. Let
$$
\mathsf{L}_y(m, v; \theta) : = \Phi_2( m, v; \rho(y, v; \theta)).
$$
Moreover, we assume that $\theta \mapsto \rho(y, v; \theta)$ is continuously differentiable at each $v$ and $y$ of interest. 
Define the map $$ (m, \theta) \mapsto \mathsf{L}_y(m, \theta) := \left\{\mathsf{L}_y(m, v; \theta), v \in \mathcal{V}\right\},$$ where $\mathcal{V}$ is the support of $\nu(Z)$, which we assume is a finite set.
Define the  Jacobian of this map as $\mathsf{JL}_y(m, \theta)$. The first row of $\mathsf{JL}_y(m, \theta)$ is  $\{ (\partial/\partial m) \mathsf{L}_y(m, \theta) \}' $ and the $(k+1)$-st row is
$\{ (\partial/\partial \theta_k) \mathsf{L}_y(m, \theta) \}'$, $k=1,\ldots,d$, where $\theta_k$ is the $k$th component of $\theta$.

We have the system of equations
\begin{equation}\label{eq:system}
\mathsf{R}(y, v) = \mathsf{L}_y(m, v; \theta), \quad v \in \mathcal{V}.
\end{equation}
We shall apply univalence results  to deduce global and local identifiability. 

 To proceed, we assume that $v \mapsto \rho(y,v;\theta)$ is non-redundantly parameterized, namely, each set of values of this function over the set $\mathcal{V}$ can only correspond to one value of $\theta$.  Define $\mathcal{M}$ as a compact interval in $\Bbb{R}$ containing
$\mu(y)$ in its interior.
 
\begin{lemma}[Identification with Discrete Instruments]\label{lemma:parametric} Given the set-up in this subsection, assume that $\mathcal{V}$, the support of $\nu(Z)$, is a finite set with  cardinality $|\mathcal{V} |$ at least $d + 1$.  Then the system of equations (\ref{eq:system}) 
has the unique local solution at $(m, \theta) = (\mu(y), \theta_0(y)) $ if the Jacobian 
$\mathsf{JL}_y(\mu(y), \theta_0(y))$ has rank $d+1$. Moreover, if \ 
$\Theta$  is rectangular and  $\mathsf{JL}_y(m, \theta)$  is full rank and $R \mathsf{JL}(m, \theta)$ is a P-matrix over 
$\mathcal{M} \times \Theta$, for some $(d+1) \times |\mathcal{V}|$ matrix $R$, then $(\mu(y), \theta_0(y))$ are 
globally identified. More generally if \ $\Theta$ is a convex polyhedron or a set with $C^1$ boundary, with non-empty 
interior, and if the generalized P-matrix or positive quasi-definite matrix conditions of 
\citen{mas1979homeomorphisms} in Theorem 1 and Theorem 2 are satisfied for the projection matrix 
$R \mathsf{JL}_y(m, \theta)$ for some $(d+1) \times |\mathcal{V}|$ matrix $R$, then  $(\mu(y), \theta_0(y))$ are 
globally identified.
\end{lemma}


Unlike Lemma \ref{lemma:RA}, Lemma \ref{lemma:parametric} requires only that the instrument and the induced propensity score take on $d+1$ values, because it does not rely on extrapolation or identification  at infinity (the parametric function needs to be only defined over $\mathcal{V}$). The conditions for identification above for $d=1$ have particularly interpretable form, leading to Theorem \ref{theorem:id} in the main text.

\paragraph{\textbf{Intuition}} A simple geometrical insight that helps interpret the local identification conditions is that in order for $\mu(y)$ to be a non-unique root locally, we need to be able to move the path of roots for $m$, $m_t$, along a time axis with constant non-zero velocity $c \neq 0$, that is $\dot m_t = c$, starting from $m_0=\mu(y)$. Since we can rescale time, without loss of generality, set $|c|=1$. Then, this move is feasible if we can find a corresponding move $\dot \theta_t= \delta_t$ that can implement $m_t = m_0 + ct$ as a root, starting at $\theta_0$.  In summary, a path of local alternative solutions can be potentially generated by solving the following differential equation system: for all $t \in (0, \epsilon)$ 
$$m_t (y) = m_0 + c t,  \ \ \theta_t = \theta_0 + \int_0^{t} \delta_s ds$$ 
such that
$$
c = \dot m_t(y)   = \frac{ (\partial/\partial \theta_t') \mathsf{L}_y(m_t, v; \theta_t)  }{(\partial/\partial m_t) \mathsf{L}_y(m_t, v; \theta_t)} \delta_t, \quad \forall v \in \mathcal{V},
$$
and subject to the boundary conditions $\rho(y, v ; \theta_t)  \in [-1,1]$. If this path exists, then we have non-identification of $\mu(y)$.  Otherwise, we have local identification of $\mu(y)$.

 Since $\epsilon>0$ is arbitrary and the problem is smooth, it suffices to  check local identification by just examining the failure of local identification at the origin $t=0$. If the instrument is not rich enough, namely, the cardinality of  $\mathcal{V}$ is smaller than $d+1$, a local path direction $\delta_0$ always exists. More generally,  $\delta_0$ exists if the matrix $M$ with rows 
$$
\left.\frac{ (\partial/\partial \theta_t') \mathsf{L}_y(m_t, v; \theta_t)  }{(\partial/\partial m_t) \mathsf{L}_y(m_t, v; \theta_t)} \right|_{t=0}, \quad v \in \mathcal{V},
$$
 linearly spans the vector $c  \boldsymbol{1}$, where $\boldsymbol{1}$ is a vector of ones.  If the spanning condition does not hold, $\delta_0$ does not exist, leading to local identification. For example, if $d=1$, and the matrix $M$  has two scalar entries $M_1 \neq M_2$ in its rows, then solving  $M_1 \delta_0 =c, M_2 \delta_0 =c$  for the same $\delta_0$ is impossible, yielding local identification. This case corresponds to Theorem \ref{theorem:id}.  This view of local identification is equivalent to local identification based on the Jacobian $\mathsf{JL}_y (\mu(y), \theta_0)$ having full rank, but it is perhaps more illuminating.

\subsection{Relaxing Selection Sorting Exclusion}\label{app:bounds}

We next consider replacing the selection sorting exclusion with weaker conditions.  While the key motivating semi-parametric example (\ref{paranorormal}) easily meets the exclusion restriction, we analyze how deviations from 
such structures can impact the identification results. 
This also helps us connect the results to the Balke-Pearl bounds that only rely on the outcome exclusion. We provide additional theoretical discussions for motivating this investigation in Section \ref{subsec:se}

\begin{assumption}[Relaxed Selection Restrictions]\label{ass:rel}Suppose Assumption 1 (1)-(3) hold, namely non-degeneracy, relevance, and outcome exclusion, and in addition either: 
   \begin{enumerate}
\item[(5)] r-Relaxed Selection Exclusion: $|\rho(y \mid 0)- \rho(y\mid 1)| \leq r$ for some $0<r\leq 2$; or
\item[(6)] Positive Selection: 
$\rho(y\mid z)\geq 0$ for each $z \in \{0,1\}$.
   \end{enumerate}
\end{assumption}
The first condition allows the strength of selection to respond to $Z$, but in a limited fashion governed by the parameter $r$. As $r$ varies from $0$ to $2$, we can trace a nested sequence of identification regions, starting with a point and ending with the Balke-Pearl bounds resulting from outcome exclusion. 

The second condition restricts the sign of the implied correlation to be positive (the sign in this condition can be reversed by changing the sign of outcome). This also leads to tighter bounds relative to the case with just outcome exclusion.
It is useful to note that the implied correlation is positive, $\rho(y)>0$, if and only if $1(Y^* > y)$ and $1(D^* >0)$
are positively correlated for all $y$. 
In the context of labor supply,
this condition means that there is positive sorting into employment throughout the offered wage distribution.
Therefore, the positivity condition has a rather natural interpretation in terms of intrinsic economic quantities.

With this relaxed condition, we work with the two equations:
\begin{equation}\label{eq:2eqs}
  \Pr(Y \leq y, D=1 \mid Z=z) =  \Phi_2(\mu, \Phi^{-1}\left( \Pr(D=1 \mid Z = z)\right); \rho_z), \ \ z \in \{0,1\},
 \end{equation}
 that share one unknown $\mu$, and each has its own unknown $\rho_z$.  The following statement is straightforward, but we record it formally nonetheless. In the statement we let the parameter $(\mu, \rho_0, \rho_1)$ range over the parameter set $\overline{\Bbb{R}} \times [-1,1] \times [-1,1]$.

 \begin{theorem}[Identification under Assumption 2]\label{theorem:ID2}
 Let $S_z= \{ (\mu_z, \rho_z)\}$ denote all solutions to equation (\ref{eq:2eqs}) for $z=0$ and $z=1$. Each $S_z$ is not empty, and, restricted to the non-degenerate case $\rho^2_z< 1$, is a one-dimensional manifold.
 Under outcome exclusion, the identification region for $(\mu(y), \rho(y \mid 0), \rho(y \mid 1))$ is $R_0 = \{ (\mu, \rho_0, \rho_1) : (\mu, \rho_z) \in S_z, \text{ for each } z \in \{0,1\}\}.$ Under outcome exclusion and  the relaxed selection, the identification region is $R_0(r)= R_0 \cap 
 \{(\mu, \rho_0, \rho_1): |\rho_0 - \rho_1| \leq r\}.$
Under outcome exclusion and  positive selection, the region is $R_0^+ = R_0 \cap \{(\mu, \rho_0, \rho_1): \min(\rho_0, \rho_1) \geq 0\}.$
The identification regions for $\mu(y)$ are given by applying the projection operator $\pi_1$ taken with respect to the first coordinate: this gives regions $\pi_1 R_0$, $\pi_1 R_1(r)$, and $\pi_1 R_0^+$, respectively.
\end{theorem}

We observe that that
the identification region $\pi_1 R_0$ is (equivalent to) the Balke-Pearl bound. The regions $R_1(r)$ and  $R_0^+$ tighten the bounds: by construction, these sets are weakly smaller than $R_0$, and the example below demonstrates they can be much smaller. In fact, $R_1(r)$ converges to a singleton as $r \searrow 0$. We can also intersect the two sets if both forms of relaxation are assumed to hold.


\begin{figure}[ht]
    \includegraphics[height=.6\textwidth, width=\textwidth]{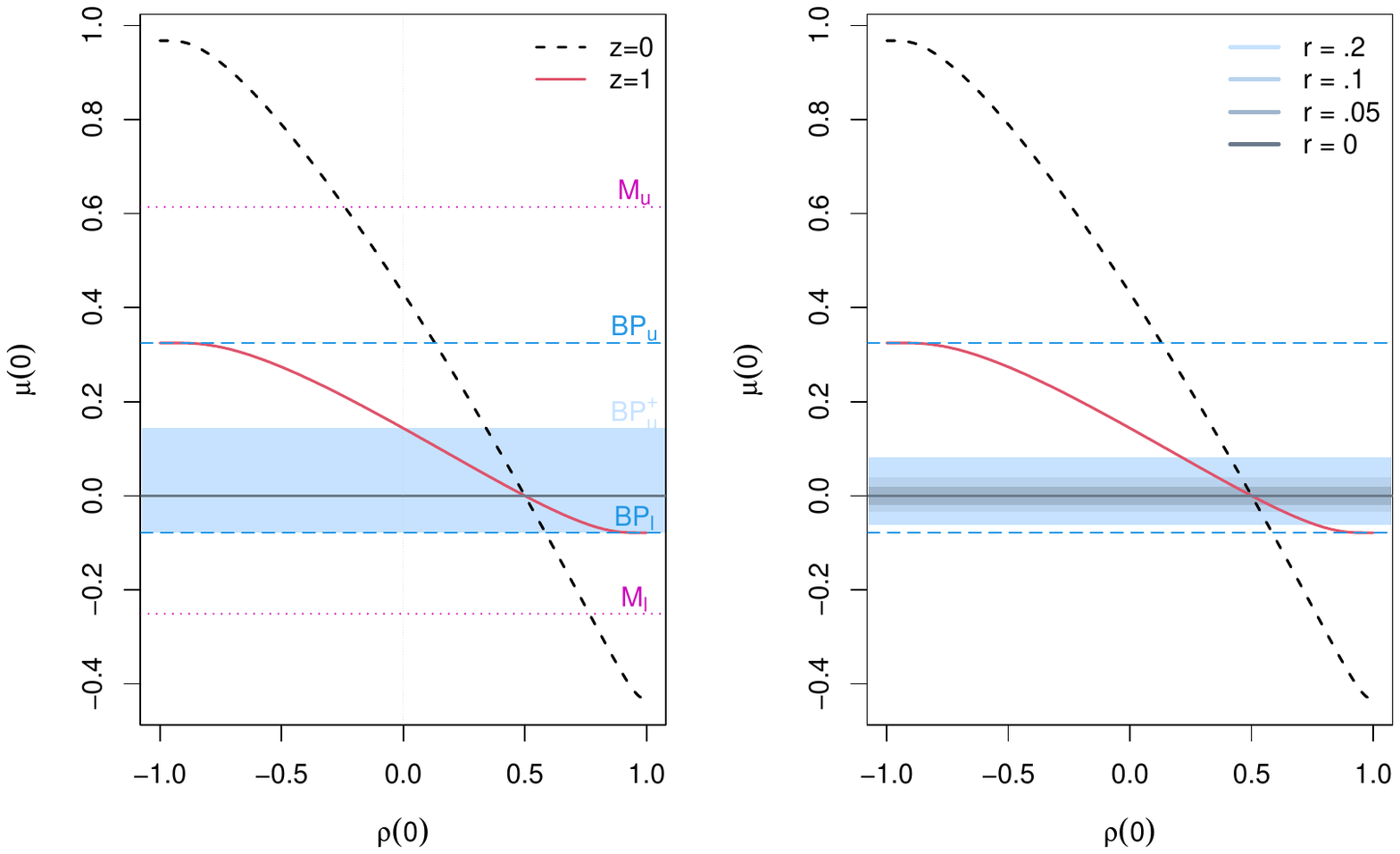}
    \caption{Illustration of identification analysis via exclusion restrictions.}
    \caption*{\footnotesize Notes. The curves represent solutions in $(\mu,\rho)$ to $ \Phi_2(0,z; 0.5) =  \Phi_2(\mu, z; \rho)$ for $z \in \{0,1\}$.  The point identification $\mu(0)=0$ occurs at the intersection of the curves. The left panel also demonstrates the construction of identified sets for $\mu(0)$ when the selection exclusion is removed ( $[BP_l, BP_u]$) or replaced by positive sorting
    ( $[BP_l, BP^+]$). The right panel demonstrates the construction of identified sets when the selection takes the relaxed form of Assumption \ref{ass:rel}(5).}\label{fig:example}
\end{figure}

\subsection{The role of selection exclusions and their relaxed forms in identification.}\label{sec:exampleID} We illustrate the role of exclusion restrictions or their relaxed versions, we consider a simple example.\footnote{This example is inspired by an example graciously provided to us by one of the referees.} Assume that $Y^*$ is binary with $Y^*=1\{U > 0 \}$, $D = 1\{V \leq Z\}$, $(U,V)$ are standard bivariate normal with parameter $\rho=0.5$, $\Pr(D=1 \mid Z=z) = \Phi(z)$ and $\Pr(Z = 1) = 1/2$.   The distribution of $Y^*$ is fully characterized by $F_{Y^*}(0) = \Pr(Y^* = 0)$ or equivalently by $\mu(0) = \Phi^{-1}\left(F_{Y^*}(0)\right)$.  In what follows, none of the bounds to be constructed  make use of the parametric distribution of $Y^*$ and $D$, other than to compute $\Pr(Y^* \leq 0, D = 1 \mid Z=z)$. 

Figure \ref{fig:example} plots the solutions in $(\mu,\rho)$ to \eqref{eq:murho} at $y=0$ for each value of $z$ without  imposing any exclusion restriction, together with bounds for $\mu(0)$ after sequentially imposing exclusion restrictions. In the right panel, the solid and dashed curves are the solutions in $(\mu,\rho)$ to  $ \Phi_2(0,z; 0.5) =  \Phi_2(\mu, z; \rho)$ for $z \in \{0,1\}$, respectively.  The  horizontal  lines  indicate the construction of bounds on $\mu(0)$ by projection. 
Without imposing any exclusion restriction the identified set is  $[M_l,M_u]=[-.25,.61]$. This set is obtained as the mean of the maximum/minimum of the curves \cite{manski1990}. 
When we impose the outcome exclusion we obtain the identified set $\pi_1 R_0 = [BP_l, BP_u]=[-.08,.32]$.  This set is obtained as the minimum/maximum of the maximum/minimum of the curves  \cite{balke1994counterfactual}.  When we impose
both outcome and exclusion restrictions, we obtain $\pi_1 R_1(0)=\{\mu(0)\}= \{0\}$ as the identified set; that is, we point-identify the parameter value $\mu(0)$. Graphically, we obtain this set by intersecting the two curves and projecting the intersection point on the vertical axis.

In the left panel, we also show how the positive selection restriction in Assumption \ref{ass:rel}(6) helps tighten the bound. Indeed, the resulting identified set $\pi_1 R_0^+$ is $[BP_l,BP^+]=[-.08,.14]$, which is tighter than the Balke-Pearl set $\pi_1 R_0$.  Finally, the right panel shows bounds imposing the relaxed selection restriction in Assumption \ref{ass:rel}(5) for $r \in \{0,.05, .10, .20\}$. The resulting identified sets are $\pi_1 R_1(0) = \{0\}$, $\pi_1 R_1(.05) = [-.02,.02]$, $\pi_1 R_1(.10) = [-.03,.04]$, and $\pi_1 R_1(.20) = [-.06,.08]$. The set $\pi_1 R_0$ can be obtained by setting $r$ sufficiently large (in particular, $r=1.13$ is sufficient).

\subsection{Proof of Lemma \ref{lemma:corr}} We have that
\begin{multline*}
\mathrm{Cov}(1\{Y^*\leq y\}, 1\{D^*\leq d\}) = \Phi_2(\mu(y), \nu(d); \rho(y,d)) -  \Phi(\mu(y)) \Phi(\nu(d)) \\ =  \Phi_2(\mu(y), \nu(d); \rho(y,d)) -  \Phi_2(\mu(y), \nu(d); 0),
\end{multline*}
and $\rho \mapsto \Phi_2(\cdot, \cdot; \rho)$ is strictly increasing. The conclusion about signs follows.\qed

\subsection{Proof of Lemma \ref{lemma:lgr2}}  We start by characterizing the marginal and joint CDFs of $Y^s=\Phi^{-1}\left( F_{Y^*}(Y^*)  \right)$ and $D^s=\Phi^{-1}\left( F_{D^*}(D^*)  \right)$. Since $d \mapsto F_{D^*}(d)$ is continuous, $F_{D^*}(D^*) \sim U(0,1)$ by the probability integral transform. Hence,
$$
F_{D^s}(d) = \Pr(F_{D^*}(D^*)  \leq \Phi(d)  ) = \Phi(d).
$$
Let $\mathcal{Y}^s = \{ y \in \overline{\mathbb{R}}: \Phi(y) = F_{Y^*}(y^*) \text{ for some } y^* \in \mathbb{R}\} $  be the support of $Y^s$. For any $y \in \mathcal{Y}^s$,
$$
F_{Y^s}(y) = \Pr(\Phi^{-1}\left( F_{Y^*}(Y^*) \right) \leq \Phi^{-1} \left( F_{Y^*}(y^*) \right) ) = F_{Y^*}(y^*) = \Phi(y).
$$
The CDF extends to all $y \in \overline{\mathbb{R}}$ by
$$
F_{Y^s}(y) = F_{Y^s}(\mu_s(y)) = \Phi(\mu_s(y)), \ \ \mu_s(y) = \sup \{y^* \in \overline{\mathbb{R}} : F_{Y^*}(y^*) \leq \Phi(y) \}.
$$
Hence, the LGR of the distribution of $(Y^s,D^s)$ is
$$
F_{Y^s,D^s}(y,d) = \Phi_2(\mu_s(y), d; \rho_s(y,d)),
$$
where $\rho_s(y,d) = \rho_s(\mu_s(y), d)$ because $F_{Y^s,D^s}(y,d) =  F_{Y^s,D^s}(\mu_s(y), d)$. 

Next, we show that for any integrable random variable $X$ and random variable $Z$,
\begin{equation}\label{eq:cov1}
\mathrm{Cov}(1\{Z \leq z\},X) = - \int_{-\infty}^{\infty}  [F_{X,Z}(x,z) - F_{X}(x) F_{Z}(z)] \mathrm{d} x
\end{equation}
and 
\begin{equation}\label{eq:cov2}
\mathrm{Cov}(Z,X) = - \int_{-\infty}^{\infty} \mathrm{Cov}(1\{Z \leq z\},X) \mathrm{d} z.
\end{equation}
Given \eqref{eq:cov1}, \eqref{eq:cov2} follows from \citen{hoffding-40} and \citen{lehmann66}. To show \eqref{eq:cov1},   let $(X_1,Z_1)$ and $(X_2,Z_2)$ be two independent copies of $(X,Z)$. Then,
$$
\mathrm{Cov}(1\{Z \leq z\},X) = \frac{1}{2} \Ep[(1\{Z_1 \leq z\} - 1\{Z_2 \leq z\})(X_1-X_2)].
$$ 
The result follows from $X_1 - X_2 = \int_{-\infty}^{\infty} [1\{X_2 \leq x\} - 1\{X_1 \leq x\}] \mathrm{d} x$ and interchanging the expectation and integral.

Finally, the results follow from applying \eqref{eq:cov1} and \eqref{eq:cov2} to $X=D^s$ and $Z = Y^s$, together with $\rho_s(y,d) = \rho_s(y)$ because $\rho(y,d) = \rho(y)$ by assumption and 
\begin{multline*}
\Phi_2(\mu_s(y), d; \rho_s(y,d)) = F_{Y^s,D^s}(y,d) =  F_{Y^s,D^s}(\mu_s(y), d) = F_{Y^*,D^*}(y^*,d^*) \\ = \Phi_2(\mu(y^*), \nu(d^*); \rho(y^*,d^*)) = \Phi_2(\mu_s(y), d; \rho(y^*,d^*)),
\end{multline*}
where $y^*$ and $d^*$ are the solution to $F_{Y^*}(y^*) = \Phi(\mu_s(y))$ and $F_{D^*}(d^*) = \Phi(d)$.  Indeed, by \eqref{eq:cov1},
\begin{multline*}
\mathrm{Cov}(1\{Y^s \leq y\},D^s)  = - \int_{-\infty}^{\infty}  [\Phi_{2}(\mu_s(y), d; \rho_s(\mu_s(y))) - \Phi_{2}(\mu_s(y), d; 0) ] \mathrm{d} d \\ = - \int_{-\infty}^{\infty} \int_{0}^{\rho_s(\mu_s(y))} \phi_{2}(\mu_s(y), d; r) \mathrm{d} r \mathrm{d} d = - \rho_s(\mu_s(y)) \phi(\mu_s(y)),
\end{multline*}
by $\Phi_{2}(\mu_s(y), d; 0) = \Phi(\mu_s(y)) \Phi(d)$, $\phi_2(\cdot, \cdot; r) = \partial \Phi_2(\cdot, \cdot; r)/\partial r$  \cite{sibuya59,sungur90}, and $$\int_{-\infty}^{\infty} \phi_2(\mu_s(y), d; r) \mathrm{d} d =  \phi(\mu_s(y))$$ after interchanging the order of the integrals.  Then, by \eqref{eq:cov2},
$$
\mathrm{Cov}(Y^s,D^s) =  \int_{-\infty}^{\infty} \rho_s(\mu_s(y)) \phi(\mu_s(y)) \mathrm{d} y =  \int_{-\infty}^{\infty} \rho_s(y)  \mathrm{d} F_{Y^s}(y) = \Ep[\rho_s(Y^s)].
$$
 \qed

\subsection{Proof of Lemma \ref{lemma:RA}} The result is immediate from $y \mapsto \mathsf{R}(y,v)$ being real analytical on $\Bbb{R}$, which pins down its value on the entire $\Bbb{R}$ via analytical continuation \cite[cor. 1.2.6]{krantz2002primer}, including the limit value which exists by assumption. This gives the identification of $\mu(y)$ and the identification of $\rho(y,v)$ follows using the same argument as in the proof of Lemma \ref{lemma:lgr}. \qed

\subsection{Proof of Lemma \ref{lemma:parametric}} The local identification result is an immediate consequence of Rothenmberg \cite{rothenberg1971identification}, and the global result follows from \citen{mas1979homeomorphisms}, Theorem 1 or Theorem 2. \qed

\subsection{Proof of Theorem \ref{theorem:ID2}} The sets $S_0$ and $S_1$ are not empty by Lemma \ref{lemma:lgr} applied to the conditional CDFs $F_{Y^*,D^* \mid Z}(y,0 \mid z),$ $z \in \{0,1\}$.  They are one-dimensional manifolds by the implicit function theorem, because $(\mu,\rho) \mapsto   \Phi_2(\mu, \Phi^{-1}\left( \Pr(D=1 \mid Z = z)\right); \rho) - \Pr(Y \leq y, D=1 \mid Z=z)$ is continuously differentiable and surjective for $z \in \{0,1\}$.  The rest of the claims follow by definition. \qed


\section{Deferred Discussions of Section \ref{sec:model}}\label{app:sect3}

 \begin{remark}[Homogeneous parameters] One important case arises when some of the components of $\beta(y)$ do not vary with $y$. This case is related with homogeneity on quantile effects with respect to the covariates and quantile index. Assume that $y \mapsto \Phi(-x^\prime \beta(y))$ is one-to-one with inverse (quantile) function $u \mapsto Q_{Y^*}(u \mid x)$ . Then, $\Phi(-x^\prime \beta(y)) = \Phi(-x_1^\prime \beta_1(y) - x_2^\prime \beta_2)$ for all $y \in \mathbb{R}$ if and only if $\partial Q_{\widetilde{Y}^*}(u \mid x) / \partial x_2 = \beta_2$ for all $u \in [0,1]$, where $Q_{\widetilde{Y}^*}(u \mid x)$ is the $u$-quantile of $\widetilde{Y}^* = -X_1^\prime \beta_1(Y)$ conditional on $X=x$.  The only if part follows from 
$
\Pr(\widetilde{Y}^* \leq \widetilde y \mid X = x) = \Phi(\widetilde y - x_2^\prime \beta_2)
$ as $\Phi(-X_1^\prime \beta_1(Y) - X_2^\prime \beta_2) \mid X \sim U(0,1)$, so that $Q_{\widetilde{Y}^*}(u \mid x)  = \Phi^{-1}(u) +  x_2^\prime \beta_2$. The if part follows from $Q_{\widetilde{Y}^*}(u \mid x) = x_2^\prime \beta_2 + \Phi^{-1}(u)$, so that $\Pr(Y \leq y \mid X=x) = \Pr(\widetilde{Y}^* \leq -X_1^\prime \beta_1(y) \mid X = x) = \Phi(-x_1^\prime \beta_1(y) - x_2^\prime \beta_2)$. For example, when $X_1$ only includes a constant, then the DR model corresponds to the transformation model $-\beta_1(Y) = X_2^\prime \beta_2 + \Phi^{-1}(U)$, $U \mid X \sim U(0,1)$, where the covariates $X_2$ have homogeneous effects on the quantiles of $-\beta_1(Y)$.  The HSM is a special case of the transformation model with $y \mapsto -\beta_1(y)$ being linear. \qed
\end{remark}

\section{Deferred Discussions of Sections \ref{sec:empirics}}\label{app:sign}

\begin{example}[Sign Reversal in Selection Effects]
The sign of the selection effects might be different in the presence of covariates if the parameter variation changes the composition of the selected population. 
Consider the following simple example with only one covariate based on the wage application. Let the covariate be an indicator for high skills. Assume that high-skilled workers are relatively more likely to participate than low-skilled workers, there is no selection sorting on unobservables, which corresponds to $\rho(x'\delta(y)) =0$ in the model, and the distribution of offered wages for high-skilled workers first-order stochastically dominates the same distribution for low-skilled workers. In this case increasing the probability of participation for high-skilled workers, which corresponds to increasing the component of $\pi$ associated with the high-skill indicator in the model, both increases the overall probability of participation and shifts the distribution of observed wages to the right (increases quantiles), despite the lack of selection sorting. Intuitively, the distribution of observed wages is a mixture of the distribution of wages for employed high-skilled and low-skilled workers, and we are increasing the relative proportion of employed high-skilled workers. The opposite holds if the distribution of offered wages for high-skilled workers is first-order stochastically dominated by the same distribution for low-skilled workers. \qed
\end{example}

\section{Asymptotic Theory}\label{sec:theory}
 We derive asymptotic theory for the estimators of the model parameters and functionals of interest.

\subsection{Limit distributions} We first introduce some notation that is useful to state the assumptions that we make to derive the limit distribution of the estimators. Let  $\widetilde S_1 :=   \partial_{\pi} L_1(\pi)$ and  $\widetilde S_{2y} :=  \partial_{\theta_y} L_2(\theta_y, \pi)$ be the scores of the first and second steps   in Algorithm \ref{alg:tsdr} evaluated at the true parameter values, and  $H_1 := \Ep \left[ \partial_{\pi \pi'} L_1(\pi) \right]$ and $H_{2y} := \Ep \left[ \partial_{\theta_y \theta_y} L_2(\theta_y, \pi) \right]$ be the corresponding expected Hessians. Let 
\begin{equation}\label{eq:cf}
\Sigma_{\theta_y\theta_{\tilde y}} := H_{2y}^{-1} \left\{n \Ep \left[ \widetilde S_{2y} \widetilde S_{2\tilde y}' \right] - J_{21y} H_1^{-1} J_{21\tilde y}' \right\} H_{2\tilde y}^{-1},
\end{equation}
where  $J_{21y} := \Ep \left[ \partial_{\theta_y \pi'} L_2(\theta_y, \pi) \right]$, $d_{\pi} := \dim \pi$,  and $d_{\theta} := \dim \theta_y$.

\begin{assumption}[DR Estimator with Sample Selection]\label{ass:fclt} (1) Random sampling: $\{(D^*_i, Y^*_i,Z_i)\}_{i=1}^n$ is a sequence of independent and identically distributed copies of $(D^*,Y^*,Z)$. We observe $D = 1(D^* > 0)$ and $Y = Y^*$ if $D=1$.
(2) Model: the distribution of $(D^*,Y^*)$ conditional on $Z$ follows the DR model \eqref{eq:drm}. (3) The support of $Z$, $\mathcal{Z}$,  is a  compact set.  (4) The set $\mathcal{Y}$
is either finite or a bounded interval. In the second case, the density function of $Y$ conditional on $X$ and $D=1$, $f_{Y \mid X,D}(y \mid x, 1)$, exists, is uniformly bounded above, and is uniformly continuous in $(y,x)$ on $\mathcal{Y}\times \mathcal{X}_1$, where $\mathcal{X}_1$ is the support of $X$ conditional on $D=1$.
(5) Identification and non-degeneracy: the equations $\Ep[\partial_{\pi} L_1(\tilde \pi)] = 0$ and $\Ep[\partial_{\theta_y} L_2(\tilde \theta_y, \tilde \pi)] = 0$ posses a unique solution at $(\tilde \pi, \tilde \theta_y) = (\pi,\theta_y)$ that lies in the interior of a compact set $\Pi \times \Theta \subset \mathbb{R}^{d_{\pi} + d_{\theta}}$ for all $y \in \mathcal{Y}$; and the matrices $H_1$, $H_{2y}$  and $\Sigma_{\theta_y\theta_{y}}$ are nonsingular for each $y \in \mathcal{Y}$.
\end{assumption}

Part (1) is a standard condition about the sampling and selection process, which is designed for cross sectional  data.  Part (2) imposes the semi-parametric DR model on the LGR of the conditional distribution of $(D^*,Y^*)$ at $d=0$.  Part (3) imposes some compactness conditions, which can be generalized at the cost of more complicated proofs.  Part (4) covers continuous, discrete and mixed continuous-discrete outcomes. Part (5) imposes directly identification and that the variance-covariance matrix of the first-step estimator and the covariance function of the second-step estimator are well-behaved. Note that $H_1$, $H_{2y}$ and $J_{21y}$ are finite by Part (3). 
More primitive conditions for part (5) can be found in the conditional maximum likelihood literature, e.g.,  \citen{newey94}. 

The main result of this section is a functional central limit theorem for $\hat{\theta}_y$.  Let $\ell^{\infty}(\mathcal{Y})$ be the set of bounded functions on $\mathcal{Y}$, and $\leadsto$ denote weak convergence (in distribution). 
\begin{theorem}[FCLT for $\hat{\theta}_y$]\label{thm:fclt} Under Assumption \ref{ass:fclt}, 
$$
\sqrt{n} (\hat \pi - \pi) = -H_1^{-1} \widetilde{S}_1 + o_P(1) \leadsto Z_{\pi} \sim \mathcal{N}(0, -H_1^{-1}), \text{ in } \mathbb{R}^{d_{\pi}}
$$
and 
$$
\sqrt{n} (\hat \theta_y - \theta_y)= - H_{2y}^{-1} \sqrt{n} \left( \widetilde S_{2y} - J_{21y} H_1^{-1} \widetilde S_1 \right) + o_P(1) \leadsto Z_{\theta_y} \text{ in } \ell^{\infty}(\mathcal{Y})^{d_{\theta}},
$$
where  $y \mapsto Z_{\theta_y}$ is a zero-mean Gaussian process with uniformly continuous sample paths and covariance function $
\Sigma_{\theta_y\theta_{\tilde y}}$,  $y,\tilde y \in \mathcal{Y}$, defined in \eqref{eq:cf}. 
\end{theorem}

\begin{remark}[Comparison with \citen{chernozhukov+13inference}] The asymptotic distribution in Theorem  \ref{thm:fclt} for $\widehat \theta(y)$ does not follow from the theory of \citen{chernozhukov+13inference} for DR-estimators without sample selection. This theory does not cover two-step M-estimators with an objective function that is not concave. We rely on compactness of the parameter space instead of concavity. \qed
\end{remark}

The first order term in the limit of $\sqrt{n} (\hat \theta_y - \theta_y)$ is the sample average of the influence function of $\hat \theta_y$. We  construct an estimator of the covariance function $
\Sigma_{\theta_y\theta_{\tilde y}}$ based on this function. Thus, we form
\begin{equation}\label{eq:se}
\widehat \Sigma_{\theta_y\theta_{\tilde y}} =  n^{-2} \sum_{i=1}^{n}  \widehat \psi_{i}(\hat \theta_y, \hat \pi) \widehat \psi_{i}(\hat \theta_{\tilde y}, \hat \pi)'.
\end{equation}
Here, $\widehat \psi_{i}$ is an estimator of the influence function of $\hat \theta_y$,
\begin{equation}\label{eq:if}
\widehat \psi_{i}(t,c) = - \widehat H_{2y}(t,c)^{-1} \left( S_{2yi}(t,c) - \widehat J_{21y}(t,c) \widehat H_1(c)^{-1} S_{1i}(c) \right),
\end{equation}
where $S_{1i}(c)$ and $S_{2iy}(t,c)$ are the individual scores of the first and second steps of Algorithm \ref{alg:tsdr},
\begin{eqnarray*}
S_{1i}(c) &:=&  \partial_{c} L_{1i}(c), \  L_{1i}(c) :=  D_{i}\log{\Phi(Z_i^\prime c)}+(1-D_{i})\log{\Phi(-Z_i^\prime c)}, \\ S_{2yi}(t,c) &:=& \partial_{t} L_{2yi}(t,c), \  t=(b,d)\\
L_{2yi}(t,c) &:= & D_{i}  \left[ I_{yi} \log{\Phi_2\left(-X_i^\prime b,Z_i^\prime c; -\rho(x'd)\right)}
 +  (1-I_{yi}) \log{\Phi_2\left(X_i^\prime b,Z_i^\prime c; \rho(x'd) \right)}\right], 
\end{eqnarray*}
and
\begin{eqnarray*}
\widehat H_1(c) :=  \partial_{c c'} L_1(c), \ \widehat H_{2y}(t,c) :=  \partial_{t t'} L_2(t, c),  \  \widehat J_{21y}(t,c) :=  \partial_{t c'} L_2(t, c), 
\end{eqnarray*}
are estimators of $H_1$, $H_{2y}$, and $J_{21y}$ when evaluated at $c = \hat \pi$ and $t = \hat \theta_y$.

We now establish a functional central limit theorem for the estimators of functionals of the model parameters. This result is based on expressing the functional as a suitable operator of the model parameters and using the functional delta method \cite[Chapter 3.9]{vdV-W}. To present the result in a concise manner, we consider a generic functional $$u \mapsto \Delta_u = \varphi_u(\pi, \theta_\cdot, F_Z),$$ 
where $u \in \mathcal{U}$, a totally bounded metric space, and $\varphi_u$ is an operator that maps $\mathbb{D}_{\Delta}$ to the set $\ell^\infty(\mathcal{U})$, where $\Delta_\cdot$ takes values. Here $\mathbb{D}_{\Delta}$ denotes the space for  the parameter tuple $(\pi, \theta_\cdot, F_Z)$; this space is not stated here explicitly, but is restricted by the regularity conditions of the previous section.  Here
we identify $F_Z$ with an integral operator $f \mapsto \int f(z) d F_Z(z)$ taking values in $\ell^\infty(\mathcal{F})$ that acts on a Donsker set of bounded measurable functions $\mathcal{F}$, which includes indicators of rectangular sets;  see \citen{chernozhukov+13inference} and examples below.
The parameter space $\mathbb{D}_{\Delta}$ is a subset of a normed space $\mathbb{D}:=\mathbb{R}^{d_{\pi}} \times \ell^{\infty}(\mathcal{Y})^{d_{\theta}} \times \ell^{\infty}(\mathcal{F}) $.  In this notation, the plug-in estimator of the functional $\Delta_u$ is $$\hat \Delta_u = \varphi_u(\hat \pi, \hat \theta_y, \hat F_Z),$$ 
where $\hat \pi$ and $\hat \theta_y$ are the estimators of the parameters defined in Algorithm \ref{alg:tsdr} and $\hat F_Z$ is the empirical distribution of $Z$. 

We provide some examples.  The distribution of the latent outcome is given by:
$$
F_{Y^*}(y) = \varphi_y(\pi, \theta_y, F_Z) = \int \Phi(-x'\beta_y) dF_Z(z),
$$
$\mathcal{F}$  contains $\{\Phi(-\cdot'\beta_y) : y \in \mathcal{Y} \}$ as well as the indicators of all rectangles in $\overline{\mathbb{R}}^{d_z}$, $\overline{\mathbb{R}} := \mathbb{R} \cup \{-\infty, +\infty\}$, $d_z = \dim Z$,  and $\mathcal{U} = \mathcal{Y}$. 
The quantile function of the latent outcome is $$Q_{Y^*}(\tau) = \varphi_{\tau}(\pi, \theta_y, F_Z) =   \mathbf{Q}_{\tau} \mathbf{R} F_{Y^*},$$ $\mathcal{F}$  is the same as for the distribution of the latent outcome, $\mathcal{U}$ is a closed subset of $(0,1)$ including the quantile indexes of interest, $\mathbf{R}$ is the non-decreasing rearrangement operator, and $\mathbf{Q}_{\tau}$ is the left-inverse (quantile) operator. The distribution of the observed outcome is given by:
$$
   F_Y(y \mid D=1) = \varphi_y(\pi, \theta_y, F_Z) = \frac{\int \Phi_2\left( -x^\prime\beta(y), z^\prime \pi; -\rho(x'\delta(y)) \right) dF_Z(z)}{\int \Phi(z^\prime \pi) dF_Z(z)},
$$
$\mathcal{F}$  contains $\{\Phi_2\left( -\cdot^\prime\beta(y), \cdot^\prime \pi; -\rho(\cdot'\delta(y)) \right) : y \in \mathcal{Y} \}$ as well as the indicators of all rectangles in $\overline{\mathbb{R}}^{d_z}$,  and $\mathcal{U} = \mathcal{Y}$. 


The following result is a corollary of Theorem \ref{thm:fclt} by the functional delta method. Let $UC(\mathcal{Y},\xi)$ be the set of functions on $\mathcal{Y}$ that are uniformly continuous with respect to $\xi$, a standard metric on $\mathbb{R}$, and $UC(\mathcal{F},\lambda)$ be the set of functionals on $\mathcal{F}$ that are uniformly continuous with respect to $\lambda$, where $\lambda(f,\tilde f) = [\Pr(f - \tilde f)^2]^{1/2}$ for any $f, \tilde f \in \mathcal{F}$.

\begin{corollary}[FCLT for $\hat \Delta_u$]\label{cor:fclt} Suppose that Assumption \ref{ass:fclt} holds, and $(p, t_y, F) \mapsto \varphi_\cdot(p, t_y, F)$, from $\mathbb{D}_{\Delta} \subset \mathbb{D}$ to $ \ell^\infty(\mathcal{U})$  is Hadamard differentiable at $(\pi, \theta_y, F_Z)$, tangentially to $\mathbb{R}^{d_{\pi}} \times UC(\mathcal{Y},\xi)^{d_{\theta}} \times UC(\mathcal{F},\lambda)$ with derivative $(p, t_y, F) \mapsto \varphi_\cdot'(p, t_y, F)$ that is defined and continuous on $\mathbb{R}^{d_{\pi}} \times \ell^{\infty}(\mathcal{Y})^{d_{\theta}} \times \ell^{\infty}(\mathcal{F})$. Then,
$$
\sqrt{n}(\hat \Delta_u -  \Delta_u) \leadsto Z_{\Delta_u} := \varphi_u'(Z_{\pi}, Z_{\theta_y}, Z_{F}) \text{ in } \ell^{\infty}(\mathcal{U}),
$$
where $Z_{\pi}$  and $Z_{\theta_y}$ are the random limits in Theorem \ref{thm:fclt}, $Z_{F}$ is a tight $F_Z$-Brownian bridge, and $u \mapsto  Z_{\Delta_u}$ is a tight zero-mean Gaussian process. 
\end{corollary}

\begin{remark}[Hadamard Differentiable Functionals]\label{remark:hd} The distributions of the latent and observed outcome together with counterfactual distributions constructed thereof are examples of Hadamard differentiable functions. In the case of the latent outcome, the result follows from the Hadamard differentiability of the counterfactual operator in \citen{chernozhukov+13inference}. In the case of the observed outcome, the result follows from the differentiability of the counterfactual operator and the composition rule for Hadamard derivatives applied to the ratio of two functions. Quantile (left-inverse) functionals of these distributions are Hadamard differentiable under additional conditions that guarantee that the quantile operator is Hadamard differentiable. These include that the outcome variable be continuous with density bounded above and away from zero \cite{cfg10}. Then the Hadamard differentiability of the quantile function follows from the composition rule for Hadamard derivatives.\qed
\end{remark}

\begin{remark}[Inference on Quantile Functions]\label{remark:qinfer} There are two alternatives to construct confidence bands for quantile functions. The first approach is the standard method based on characterizing the limit distribution of the estimator of the quantile function using the delta method, which relies on the Hadamard differentiability of the inverse operator. As we mention in Remark \ref{remark:hd}, this differentiability requires additional conditions including that the outcome variable be continuous. The second approach applies to any type of outcome variable. It is based on the generic method of \citen{cfmw16} that inverts confidence bands for distribution functions into confidence bands for quantile function. This method does not rely on the delta method and is therefore more robust to modeling assumptions and widely applicable.  It  has the  shortcoming, however,  that the bands might not be centered at the point estimate of the quantile function. We apply the second method to obtain most of the results in the empirical application.\qed
\end{remark}

\subsection{Multiplier Bootstrap} We make the following assumption about the bootstrap multipliers of Algorithm \ref{alg:mb}:
\begin{assumption}[Multiplier Bootstrap]\label{ass:mb} The multipliers  $(\omega_{1},...,\omega_{n})$ are i.i.d. draws from a
random variable $\omega \sim \mathcal{N}(0,1)$, and are
independent of  $\{(D^*_i, Y^*_i,Z_i)\}_{i=1}^n$ for all $n$.
\end{assumption}

Let 
$$
\hat{\theta}_y^b  = \hat \theta_y + n^{-1} \sum_{i = 1}^n \omega_i \ \widehat \psi_i(\hat \theta_y, \hat \pi)
$$
be the multiplier bootstrap version of  $\hat{\theta}_y$. We establish a functional central limit theorem for the bootstrap for  $\hat{\theta}_y$. 
  Here we use $\leadsto_{\Pr}$ to denote
bootstrap consistency, i.e. weak convergence conditional on the data in
probability, which is formally defined in Appendix \ref{app:notation}.

\begin{theorem}[Bootstrap FCLT for $\hat{\theta}_y$]\label{thm:bfclt} Under the conditions of Theorem \ref{thm:fclt} and Assumption \ref{ass:mb}, 
$$
\sqrt{n} (\hat \theta_y^b - \hat \theta_y) \leadsto_{\Pr} Z_{\theta_y} \text{ in } \ell^{\infty}(\mathcal{Y})^{d_{\theta}},
$$
where  $y \mapsto Z_{\theta_y}$ is the same Gaussian process as in Theorem \ref{thm:fclt}. 
\end{theorem}

The following result is a corollary of Theorem \ref{thm:bfclt} by the functional delta method for the bootstrap \cite[Chapter 3.9]{vdV-W}. Let $\hat \Delta^b_u = \varphi_u(\hat \pi^b, \hat \theta^b_y, \hat F^b_Z)$, be the multiplier bootstrap version of $\hat \Delta_u$ where $$\hat \pi^b = \hat \pi - n^{-1} \sum_{i=1}^n \omega_i  \ \hat{H}_1(\hat \pi)^{-1} S_{1i}(\hat \pi),$$ and $\hat F^b_Z$ is the weighted empirical distribution of $Z$  that uses  $(1+\omega_1, \ldots, 1+\omega_n)$ as sampling weights. 

\begin{corollary}[Bootstrap FCLT for $\hat \Delta_u$] Suppose that the conditions of Corollary \ref{cor:fclt} and Assumption \ref{ass:mb} hold. Then,
$$
\sqrt{n}(\hat \Delta^b_u - \hat \Delta_u) \leadsto_{\Pr} Z_{\Delta_u} \text{ in } \ell^{\infty}(\mathcal{U}),
$$
where $Z_{\Delta_u}$ is the same process as in Corollary \ref{cor:fclt}. 
\end{corollary}

\section{Proofs of Section \ref{sec:theory}}\label{app:proofs}

\subsection{Notation}\label{app:notation}
We adopt the standard notation in the empirical process literature, e.g. \citen{vdV-W}, 
\begin{equation*}
{\mathbb{E}_n}[f]={\mathbb{E}_n}[f(A)]=n^{-1}\sum_{i=1}^{n}f(A_{i}),
\end{equation*}
and 
\begin{equation*}
\mathbb{G}_n[f]=\mathbb{G}_n[f(A)]=n^{-1/2}\sum_{i=1}^{n}(f(A_{i})-\Ep[f(A)]).
\end{equation*}
When the function $\widehat{f}$ is estimated, the notation should
interpreted as: 
\begin{equation*}
\mathbb{G}_n[\widehat{f}\ ]=\mathbb{G}_n[f]\mid_{f=\widehat{f}}\text{\ and \ 
}\Ep[\widehat{f}\ ]=\Ep[f]\mid_{f=\widehat{f}}.
\end{equation*}
%
We also follow the notation and definitions in \citen{vdV-W}
of bootstrap consistency. Let $D_{n}$ denote the data vector and $E_{n}$ be
the vector of bootstrap weights. Consider the random element $%
Z_{n}^{b}=Z_{n}(D_{n},E_{n})$ in a normed space $\mathbb{Z}$. We say that
the bootstrap law of $Z_{n}^{b}$ consistently estimates the law of some
tight random element $Z$ and write $Z_{n}^{b}\rightsquigarrow_{\Pr}Z$ in $%
\mathbb{Z}$ if 
\begin{equation}
\begin{array}{r}
\sup_{h\in\text{BL}_{1}(\mathbb{Z})}\left|\Ep^{b}h%
\left(Z_{n}^{b}\right)- \Ep h(Z)\right|\rightarrow_{\Pr^{*}}0,%
\end{array}
\label{boot1}
\end{equation}
where $\text{BL}_{1}(\mathbb{Z})$ denotes the space of functions with
Lipschitz norm at most 1, ${\Ep}^{b}$ denotes the conditional
expectation with respect to $E_{n}$ given the data $D_{n}$, and $%
\rightarrow_{\Pr^{*}}$ denotes convergence in (outer) probability.

We use the $Z$-process framework described in Appendix E.1 of \citen{chernozhukov+13inference}. To set-up the problem in terms of this framework, we need to introduce some notation. Let $W := (Z,D,YD)$ denote all the observed variables and $\xi_y := (\pi',\theta_y')'$ be a vector with the model parameters of the first and second steps.  Let
$$
\varphi_{y,\xi}(W) := 
\left[
\begin{array}{c}
  S_{1,\xi}(W)   \\
  S_{2y,\xi}(W) 
\end{array}
\right] = \left[
\begin{array}{c}
  \frac{\partial \ell_{1,\xi}(W)}{\partial \pi}   \\
  \frac{\partial \ell_{2y,\xi}(W)}{\partial \theta_y} 
\end{array}
\right]
$$
where 
\begin{eqnarray*}
\ell_{1,\xi}(W) &:=& D\log{\Phi(Z^\prime \pi)}+(1-D)\log{\Phi(-Z^\prime \pi)}, \\
\ell_{2y,\xi}(W) &:=& D [I_{y} \log{\Phi_2\left(-X^\prime \beta(y),Z^\prime \pi; -\rho(X'\delta(y))\right)} +  (1-I_{y}) \log{\Phi_2\left(X^\prime \beta(y),Z^\prime \pi; \rho(X'\delta(y)) \right)} ], 
\end{eqnarray*}
be the scores of the first and second steps; and 
\begin{equation}\label{eq:hess}
J(y) = \Ep \left[ \frac{\partial \varphi_{y,\xi}(W)}{\partial \xi'}\right] = \left[
\begin{array}{cc}
  H_1   & 0  \\
  J_{21y}   & H_{2y}   
\end{array}
\right]
\end{equation}
be the expected Hessian evaluated at the true value of $\xi_y$. We provide more explicit expressions for the score and expected Hessian in Appendix \ref{app:scores}. Note that 
\begin{equation}\label{eq:invhess}
J^{-1}(y) = 
\left[
\begin{array}{cc}
  H_1^{-1}   & 0  \\
  - H_1^{-1} J_{21y} H_{2y}^{-1}   & H_{2y}^{-1}   
\end{array}
\right]
\end{equation}
by the inverse of the partitioned inverse formula, and 
\begin{equation}\label{eq:vscore}
\Ep[\varphi_{y,\xi}(W) \varphi_{\tilde y,\xi}(W)'] = \left[
\begin{array}{cc}
  \Ep[S_{1,\xi}(W) S_{1,\xi}(W)']    & 0  \\
  0   & \Ep[S_{2y,\xi}(W) S_{2\tilde y,\xi}(W)']    
\end{array}
\right]
\end{equation}
because $\Ep[S_{1,\xi}(W) S_{2y,\xi}(W)'] = 0$ for all $y \in \mathcal{Y}$. 


\subsection{Auxiliary Results}
We start by providing sufficient conditions that are useful to verify Condition Z in  \citen{chernozhukov+13inference}. They are an alternative to  Lemma E.1 of \citen{chernozhukov+13inference}, where we replace the requirement that the function $\xi \mapsto \Psi(\xi,y) := \Ep[\varphi_{y,\xi}(W)]$ is the gradient of a convex function by compactness of the parameter space for $\xi_y$ and an identification condition.\footnote{We adapt the notation of \citen{chernozhukov+13inference} to our problem by using $y$, $\mathcal{Y}$, $\xi_y$, $d_{\xi}$ and $\Xi$ in place of $u$, $\mathcal{U}$, $\theta_0(u)$, $p$, and $\Theta$.}

\begin{lemma}[Simple sufficient condition for Z]
\label{cont inverse} Suppose that $\Xi$ is a compact subset of $\mathbb{R}^{d_{\xi}}$, and $\mathcal{Y}$ is a compact interval in $\mathbb{R}$. Let $\mathcal{I}$ be an open set
containing $\mathcal{Y}$. Suppose that (a) $\Psi :\Xi \times \mathcal{I}
\mapsto \mathbb{R}^{d_{\xi}}$ is continuous, and $\xi \mapsto \Psi (\xi ,y) $
possesses a unique zero at $\xi_y$ that is in the interior of $\Xi$ for each $y\in \mathcal{Y}$%
, (b) for each $y\in \mathcal{Y}$, $\Psi (\xi_y,y)=0$, (c) $\frac{%
\partial }{\partial (\xi ^{\prime },y)}\Psi (\xi ,y)$ exists at $%
(\xi_y,y)$ and is continuous at $(\xi_y,y)$ for each $y\in 
\mathcal{Y}$, and $\dot{\Psi}_{\xi_y,y}:=\frac{\partial }{\partial
\xi ^{\prime }}\Psi (\xi ,y)|_{\xi_y}$ obeys $\inf_{y\in 
\mathcal{Y}}\inf_{\Vert h\Vert =1}\Vert \dot{\Psi}_{\xi_y,y}h\Vert
>c_{0}>0$. Then Condition Z of \citen{chernozhukov+13inference} holds and $y \mapsto \xi_y$ is
continuously differentiable.
\end{lemma}

\noindent \textbf{Proof of Lemma \ref{cont inverse}. }  We restate the statement of Condition Z of \citen{chernozhukov+13inference} with our notation for the reader's reference.

\textsc{\ Condition Z.} \textit{\ Let $\mathcal{Y}$ be a compact set of some
metric space, and $\Xi $ be an arbitrary subset of $\mathbb{R}^{d_{\xi}}$.
Assume (i) for each $y\in \mathcal{Y}$, $\Psi (\cdot ,y):\Xi \mapsto 
\mathbb{R}^{d_{\xi}}$ possesses a unique zero at $\xi_y$, and, for some $%
\delta >0$, $\mathcal{N}:=\cup _{y\in \mathcal{Y}}B_{\delta }(\xi_y)
$ is a compact subset of $\mathbb{R}^{d_{\xi}}$ contained in $\Xi $, (ii) the
inverse of $\Psi (\cdot ,y)$ defined as $\Psi ^{-1}(x ,y) := \{ \xi \in
\Xi: \Psi(\xi,y) =x \}$ is continuous at $x=0$ uniformly in $y\in 
\mathcal{Y}$ with respect to the Hausdorff distance, (iii) there exists $%
\dot{\Psi}_{\xi_y,y}$ such that $\lim_{t \searrow 0}\sup_{y\in {%
\mathcal{Y}},\Vert h\Vert=1}|t^{-1}[\Psi(\xi_y+th,y) -
\Psi(\xi_y,y)] -\dot{\Psi}_{\xi_y,y}h|=0$, where $%
\inf_{y\in \mathcal{Y}}\inf_{\Vert h\Vert =1}$ $\Vert \dot{\Psi}_{\xi_y,y}h\Vert >0$, and (iv) the maps $y \mapsto \xi_y$ and $y
\mapsto \dot{\Psi}_{\xi_y,y}$ are continuous.}

The first part of Z(i) follows immediately from condition
(a). The verifications of the second part of Z(i), Z(iii) and Z(iv) are omitted because they follow by the same argument as in the proof of Lemma E.1 of \citen{chernozhukov+13inference}. 

To show Condition Z(ii), we need to verify that for any $x_t \to 0$ such
that $x_t \in \Psi(\Xi,y)$,  we have that $d_H(\Psi^{-1}(x_t,y) , \Psi^{-1}(0, y)) \to
0 $, where $d_H$ is the Hausdorff distance, uniformly in $y \in \mathcal{Y}$%
. Suppose by contradiction that this is not true, then there is $(x_t, y_t)$
with $x_t \to 0$ and $y_t \in \mathcal{Y}$ such that $d_H(\Psi^{-1}(x_t,y_t)
, \Psi^{-1}(0, y_t)) \not \to 0.$ By compactness of $\mathcal{Y}$, we can
select a further subsequence $(x_k,y_k)$ such that $y_k \to y$, where $y \in 
\mathcal{Y}$. We have that $\Psi^{-1}(0, y)=\xi_y$ is continuous in $y
\in \mathcal{Y}$, so we must have $$d_H(\Psi^{-1}(x_k,y_k) , \Psi^{-1}(0, y))
\not \to 0.$$ Hence, by compactness of $\Xi$, there is a further subsequence $u_l \in
\Psi^{-1}(x_l,y_l) $ with $u_l \to u$ in $\Xi$, such
that $u \neq \Psi^{-1}(0, y)=\xi_y$, and such that $x_l= \Psi(u_l,y_l)
\to 0$. But, by continuity $\Psi(u_{l}, y_{l}) \to
\Psi(u,y) \neq 0$ since $u \neq \Psi^{-1}(0, y)$, yielding a contradiction. \qed

\subsection{Proof of Theorem \ref{thm:fclt}}
We only consider the case where $\mathcal{Y}$ is a compact interval of $\mathbb{R}$. The case where $\mathcal{Y}$ is a finite set is simpler.  The proof follows the same steps as the proof of Theorem 5.2 of \citen{chernozhukov+13inference} for the DR-estimator without sample selection using Lemma \ref{cont inverse} in place of Lemma E.1 of \citen{chernozhukov+13inference}. Let $\Psi(\xi,y) = P[\varphi_{y,\xi}]$ and $\widehat \Psi(\xi,y) = P_n[\varphi_{y,\xi}]$, where $P_n$ is the empirical measure and $P$ is the corresponding probability measure. From the first order conditions, the two-step estimator obeys $\widehat \xi_y = \phi(\widehat \Psi(\cdot,y),0)$ for each $y \in \mathcal{Y}$, where $\phi$ is the $Z$-map defined in Appendix E.1 of \citen{chernozhukov+13inference}. The random vector  $\widehat \xi_y$ is the estimator of $\xi_y = \phi(\Psi(\cdot,y),0)$ in the notation of this framework. Then, by step \ref{step1} below,
$$
\sqrt{n} (\widehat \Psi - \Psi) \leadsto Z_{\Psi}  \text{ in } \ell^{\infty}(\mathcal{Y} \times \mathbb{R}^{d_{\xi}} )^{d_{\xi}},\ Z_{\Psi}(y,\xi) = \mathbb{G} \varphi_{y,\xi},
$$
where $d_{\xi} := \dim \xi_y$, $\mathbb{G} $ is a $P$-Brownian bridge, and $Z_{\Psi}$ has continuous paths a.s. Step \ref{step2} verifies the conditions of Lemma \ref{cont inverse} for $\dot{\Psi}(\xi_y,y) = J(y),$ the Hessian matrix defined in \eqref{eq:hess}, which also implies that $y \mapsto \xi_y$ is continuously differentiable in the interval $\mathcal{Y}$. Then, by Lemma E.2 of  \citen{chernozhukov+13inference}, the map $\phi$ is Hadamard differentiable with derivative map $(\psi,0) \mapsto -J^{-1} \psi$ at $(\Psi,0)$. Therefore, we can conclude by the functional delta method that 
\begin{equation}\label{eq:fclt}
\sqrt{n}(\widehat \xi_y - \xi_y) \leadsto Z_{\xi_y}:=- J^{-1}(y)Z_{\Psi}(y,\xi_y) \text{ in } \ell^{\infty}(\mathcal{Y})^{d_{\xi}},
\end{equation}
where $y \mapsto Z_{\xi_y}$ has continuous paths a.s.

\begin{step}[Donskerness]\label{step1} We verify that $\mathcal{G} = \{\varphi_{y,\xi}(W) : (y,\xi) \in \mathcal{Y} \times \mathbb{R}^{d_{\xi}} \}$ is $P$-Donsker with a square-integrable envelope. By inspection of the expression of  $\varphi_{y,\xi}(W) = [ S_{1,\xi}(W)', S_{2y,\xi}(W)'  ]'$ in Appendix \ref{app:scores},  $\varphi_{y,\xi}(W)$ is a Lipschitz transformation of VC functions with Lipschitz coefficient bounded by  $c \| Z \|$ for some constant $c$ and envelope function $c \| Z \|$, which is square-integrable. Hence $\mathcal{G}$ is $P$-Donsker by Example 19.9 in \citen{vdV}. 
\end{step}

\begin{step}[Verification of the Conditions of Lemma \ref{cont inverse}]\label{step2} Conditions (a) and (b) are immediate by Assumption \ref{ass:fclt}. To verify (c), note that for $(\tilde \xi,\tilde y)$ in the neighborhood of $(\xi_y,y)$,
$$
\frac{\partial \Psi(\tilde \xi,\tilde y)}{\partial (\tilde \xi', \tilde y)} = [J(\tilde \xi,\tilde y), R(\tilde \xi,\tilde y)],
$$
where 
$$
R(\tilde \xi,\tilde y) = - \Ep
\left\{
\begin{array}{ccc}
  0   \\
  f_{Y \mid Z, D}(\tilde y \mid Z,1) \Phi_{\pi}(Z) \Phi_{\tilde \pi}(Z)   \left[
\begin{array}{ccc}
  G_{2,\tilde \xi}(Z) \\
  G_{3,\tilde \xi}(Z)  
\end{array}
\right] \otimes X
\end{array}
\right\},
$$
for $\tilde \xi = (\tilde \pi', \tilde \beta', \tilde \rho')'$, and
$$
J(\tilde \xi,\tilde y) = 
\left[
\begin{array}{ccc}
  J_{11}(\tilde \xi,\tilde y)   &   J_{12}(\tilde \xi,\tilde y) \\
  J_{21}(\tilde \xi,\tilde y)   &   J_{22}(\tilde \xi,\tilde y)
\end{array}
\right],
$$
for
$$
 J_{11}(\tilde \xi,\tilde y) = \Ep\left[\{g_1(Z'\tilde \pi)(D - \Phi_{\tilde \pi}(Z)) - G_1(Z'\tilde \pi) \phi(Z'\tilde \pi) \} ZZ' \right],
$$
with $g_1(u) = d G_1(u)/ du$;
$
J_{12}(\tilde \xi,\tilde y) = 0;
$
\begin{multline*}
J_{21}(\tilde \xi,\tilde y)  =  \Ep \left\{ [\Phi_{\pi}(Z) \Phi_{2,\tilde \xi}^{\nu}(Z) - \phi(Z'\pi)\Phi_{2,\xi_{\tilde y}}(Z) ]  \left[
\begin{array}{ccc}
  G_{2,\tilde \xi}(Z)\\
  G_{3,\tilde \xi}(Z) 
\end{array}
\right] \otimes XZ'\right\} \\
+ \Ep \left\{ (\Phi_{\pi}(Z)\Phi_{2,\tilde \xi}(Z) - \Phi_{\tilde \pi}(Z)\Phi_{2, \xi_{\tilde y}}(Z))
\left[
\begin{array}{ccc}
   G_{2,\tilde \xi}^{\nu}(Z)  \\
  \rho'(X'\tilde \delta) G_{3,\tilde \xi}^{\nu}(Z) 
\end{array}
\right] \otimes XZ'\right\},
\end{multline*}
with $G_{j,\tilde \xi}^{\nu}(Z) := G_j^{\nu}\left(-X^\prime \tilde \beta,Z^\prime \tilde \pi; -\rho(X'\tilde \delta)\right)$ and $G_j^{\nu}(\mu,\nu;\rho) = \partial G_j(\mu,\nu;\rho)/\partial \nu$ for $j \in \{2,3\}$; and
\footnotesize \begin{multline*}
J_{22}(\tilde \xi,\tilde y) =  - \Ep \left\{ \Phi_{\pi}(Z)   \left[
\begin{array}{ccc}
  \Phi_{2,\tilde \xi}^{\mu}(Z) G_{2,\tilde \xi}(Z) &  \Phi_{2,\tilde \xi}^{\rho}(Z) G_{2,\tilde \xi}(Z)\\
  \Phi_{2,\tilde \xi}^{\mu}(Z) \rho'(X'\tilde \delta) G_{3,\tilde \xi}(Z) & \Phi_{2,\tilde \xi}^{\rho}(Z) \rho'(X'\tilde \delta) G_{3,\tilde \xi}(Z)
\end{array}
\right] \otimes XX' \right\} \\
+ \Ep \left\{ (\Phi_{\pi}(Z)\Phi_{2,\tilde \xi}(Z) - \Phi_{\tilde \pi}(Z)\Phi_{2, \xi_{\tilde y}}(Z))
\left[
\begin{array}{ccc}
  G_{2,\tilde \xi}^{\mu}(Z) &  G_{2,\tilde \xi}^{\rho}(Z) \\
  \rho'(X'\tilde \delta) G_{3,\tilde \xi}^{\mu}(Z)  & \rho'(X'\tilde \delta)^2 G_{3,\tilde \xi}^{\rho}(Z) +  \rho''(X'\tilde \delta) G_{3,\tilde \xi}(Z) 
\end{array}
\right] \otimes XX' \right\},
\end{multline*}
\normalsize
with  $ G_{j,\tilde \xi}^{a}(Z) := G_j^{a}\left(-X^\prime \tilde \beta,Z^\prime \tilde \pi; -\rho(X'\tilde \delta)\right) $ and  $G_j^{a}(\mu,\nu;\rho) = \partial G_j(\mu,\nu;\rho)/\partial a$ for $j \in \{2,3\}$ and $a \in \{\mu,\rho\}$. In the previous expressions we use some notation defined in Appendix \ref{app:scores}.

Both $(\tilde \xi,\tilde y) \mapsto R(\tilde \xi,\tilde y)$ and $(\tilde \xi,\tilde y) \mapsto J(\tilde \xi,\tilde y)$ are continuous at $(\xi_y,y)$ for each $y \in \mathcal{Y}$. The computation above as well as the verification of the continuity follow from using  the expressions of $\varphi_{y,\xi}$ in Appendix \ref{app:scores}, the dominated convergence theorem, and the following ingredients: (i) a.s. continuity of the map $(\tilde \xi,\tilde y) \mapsto \partial \varphi_{\tilde y,\tilde \xi}(W)/\partial \tilde \xi'$, (ii) domination of $\|\partial \varphi_{y,\xi}(W)/\partial \xi' \|$ by a square-integrable function $\| c Z\|$ for some constant $c$, (iii) a.s. continuity and uniform boundedness of the conditional density function $y \mapsto f_{Y \mid X,D}(y \mid X,1)$ by Assumption \ref{ass:fclt}, and (iv) $G_1(Z'\tilde \pi)$,  $G_{2,\tilde \xi}(Z)$ and $G_{3,\tilde \xi}(Z)$ being bounded uniformly on $\tilde \xi \in \mathbb{R}^{d_{\xi}}$, a.s. By assumption, $J(y) = J(\xi_y,y)$ is positive-definite uniformly in $y \in \mathcal{Y}$.
\end{step}
The expressions of the limit processes given in the theorem follow by partitioning $Z_{\xi_y} = (Z_{\pi}', Z_{\theta_y}')'$ and using the expressions of $J^{-1}(y)$ and $\Ep[\varphi_{y,\xi}(W) \varphi_{\tilde y,\xi}(W)'] $ given in \eqref{eq:invhess} and \eqref{eq:vscore}.
\qed

\subsection{Proof of Theorem \ref{thm:bfclt}} Let $\widehat \xi_y^b := (\widehat \pi^{b'}, \widehat \theta_y^{b'})'$.  By definition of the multiplier bootstrap draw of the estimator
$$
\sqrt{n}(\widehat \xi_y^b  - \widehat \xi_y) = \mathbb{G}_n \omega^b \varphi_{y,\widehat \xi} = \mathbb{G}_n \omega^b \varphi_{y, \xi}  + r_y,
$$
where $\omega^b \sim N(0,1)$ independently of the data and $r_y :=  \mathbb{G}_n \omega^b (\varphi_{y,\widehat \xi} - \varphi_{y, \xi})$. Then the result follows from $\mathbb{G}_n \omega^b \varphi_{y, \xi} \leadsto_{\Pr} Z_{\xi_y}$ in step \ref{step1b} and $r_y \leadsto_{\Pr} 0$ in step \ref{step2b}.

\begin{step}\label{step1b} Recall that $\varphi_{y,\xi}$ is $P$-Donsker by step \ref{step1}  of the proof of Theorem \ref{thm:fclt}. Then, by $\Ep \omega^b = 0$, $\Ep (\omega^b)^2 = 1$ and the Conditional Multiplier Functional Central Limit Theorem \cite[Theorem 2.9.6]{vdV-W},
$$
\mathbb{G}_n \omega^b \varphi_{y, \xi} \leadsto_{\Pr} Z_{\xi_y},
$$ 
where $Z_{\xi_y}$ is the same limit process as in \eqref{eq:fclt}. 
\end{step}

\begin{step}\label{step2b}  Note that $r_y \leadsto 0$ because $\varphi_{y,\xi}$ is $P$-Donsker and $\sqrt{n}(\widehat \xi_y - \xi_y) = O_{\Pr}(1)$ uniformly in $y \in \mathcal{Y}$ by  Theorem \ref{thm:fclt}.   To show that $r_y \leadsto_{\Pr} 0$, we use that this statement means that for any $\epsilon > 0$, $\Ep^b1(\| r_y\|_2 > \epsilon) = o_{\Pr}(1)$ uniformly in $y \in \mathcal{Y}$. Then, the result follows by the Markov inequality and
$$
\Ep \Ep^b1(\| r_y\|_2 > \epsilon) = \Pr(\| r_y\|_2 > \epsilon) = o(1),
$$
uniformly in $y \in \mathcal{Y}$, 
where the latter holds by the Law of Iterated Expectations and $r_y \leadsto 0$.
\end{step}\qed

\subsection{Expressions of the Score and Expected Hessian}\label{app:scores}
\subsection{Score} Let $\Phi_{\pi}(Z) := \Phi(Z'\pi)$ and $\Phi_{2,\xi_y}(Z) := \Phi_2\left(-X^\prime \beta(y),Z^\prime \pi; -\rho(X'\delta(y))\right)$. Note that by the properties of the standard bivariate normal distribution $\Phi_2\left(X^\prime \beta(y),Z^\prime \pi; \rho(X'\delta(y))\right) =  \Phi_{\pi}(Z) - \Phi_{2,\xi_y}(Z)$. Then, straighforward calculations yield
$$
S_{1,\xi}(W) =  \frac{\partial \ell_{1,\xi}(W)}{\partial \pi}  = G_1(Z'\pi) [D - \Phi_{\pi}(Z)] Z,
$$
where $G_1(u) = \phi(u)/[\Phi(u)\Phi(-u)]$, and
$$
 S_{2y,\xi}(W)  = \frac{\partial \ell_{2y,\xi}(W)}{\partial \theta_y} = D(\Phi_{2,\xi_y}(Z) - \Phi_{\pi}(Z) I_y)
\left[
\begin{array}{ccc}
  G_{2,\xi_y}(Z)  \\
  \rho'(X'\delta(y)) G_{3,\xi_y}(Z)    
\end{array}
\right] \otimes X,
$$
where $$G_{2,\xi_y}(Z) := G_2\left(-X^\prime \beta(y),Z^\prime \pi; -\rho(X'\delta(y))\right), G_{3,\xi_y}(Z) := G_3\left(-X^\prime \beta(y),Z^\prime \pi; -\rho(X'\delta(y))\right)$$ with
$$G_2(\mu,\nu; \rho) = \frac{\Phi_2^{\mu}(\mu,\nu;\rho)}{\Phi_2(\mu,\nu;\rho)[\Phi(\nu) - \Phi_2(\mu,\nu;\rho)]}, \ \ 
G_3(\mu,\nu; \rho) = \frac{\Phi_2^{\rho}(\mu,\nu;\rho)}{\Phi_2(\mu,\nu;\rho)[\Phi(\nu) - \Phi_2(\mu,\nu;\rho)]},$$ for
\begin{equation}\label{eq:phi2mu}
\Phi_2^{\mu}(\mu,\nu;\rho) = \frac{\partial \Phi_2(\mu,\nu;\rho)}{\partial \mu} = \Phi\left(\frac{\nu - \rho \mu}{\sqrt{1 - \rho^2}} \right) \phi(\mu),
\end{equation}
and
\begin{equation}\label{eq:phi2rho}
\Phi_2^{\rho}(\mu,\nu;\rho) = \frac{\partial \Phi_2(\mu,\nu;\rho)}{\partial \rho} = \phi_2(\mu,\nu;\rho).
\end{equation}

To show \eqref{eq:phi2mu} and \eqref{eq:phi2rho}, start from the factorization
$$
\Phi_2(\mu,\nu;\rho) = \int_{-\infty}^{\mu} \Phi\left(\frac{\nu-\rho v}{\sqrt{1-\rho^2}}\right)\phi(v) dv.
$$
Then,  \eqref{eq:phi2mu} follows from taking the partial derivative with respect to $\mu$ using the Leibniz integral rule.  Taking the partial derivative with respect to $\rho$ yields
    \begin{align*}
        \frac{\partial \Phi_2(\mu,\nu;\rho)}{\partial \rho} &= \int_{-\infty}^{\mu} \phi\left(\frac{\nu-\rho v}{\sqrt{1-\rho^2}}\right)\frac{\rho \nu - v}{(1-\rho^2)^\frac{3}{2}}\phi(v)dv \\
        &= \int_{-\infty}^\mu \frac{1}{\sqrt{2\pi}}\exp{\left[-\frac{(\nu-\rho v)^2}{2(1-\rho^2)}\right]} \frac{1}{\sqrt{2\pi}}\exp{\left[-\frac{v^2}{2}\right]} \frac{\rho \nu - v}{(1-\rho^2)^\frac{3}{2}} dv \\
        &= \int_{-\infty}^\mu \frac{\rho \nu -v}{2\pi(1-\rho^2)^\frac{3}{2}} \exp{\left[-\frac{\nu^2-2\rho v\nu +v^2}{2(1-\rho^2)}\right]}dv \\
        &= \frac{1}{2\pi\sqrt{1-\rho^2}}\exp{\left[-\frac{\nu^2-2\rho \mu\nu +\mu^2}{2(1-\rho^2)}\right]} = \phi_2(\mu,\nu;\rho)
    \end{align*}

\subsection{Expected Hessian} Straighforward calculations yield
$$
H_1 = \Ep \left[ \frac{\partial \ell_{1,\xi}(W)}{\partial \pi \partial \pi'} \right] = - \Ep\left[ G_1(Z'\pi) \phi(Z'\pi) ZZ'\right], \ \ \Ep \left[ \frac{\partial \ell_{1,\xi}(W)}{\partial \pi \partial \theta_y'} \right] = 0,
$$
$$
J_{21y} = \frac{\partial \ell_{2y,\xi}(W)}{\partial \theta_y \partial \pi'} =  \Ep \left\{ [\Phi_{\pi}(Z) \Phi_{2,\xi_y}^{\nu}(Z) - \phi(Z'\pi)\Phi_{2,\xi_y}(Z) ]  \left[
\begin{array}{ccc}
  G_{2,\xi_y}(Z) \\
  \rho'(X'\delta(y)) G_{3,\xi_y}(Z) 
\end{array}
\right] \otimes XZ'\right\},
$$
where $\Phi_{2,\xi_y}^{\nu}(Z) = \Phi_2^{\nu}\left(-X^\prime \beta(y),Z^\prime \pi; -\rho(X'\delta(y))\right)$ with 
$$
\Phi_2^{\nu}(\mu,\nu;\rho) = \frac{\partial \Phi_2(\mu,\nu;\rho)}{\partial \nu} = \Phi\left(\frac{\mu - \rho \nu}{\sqrt{1 - \rho^2}} \right) \phi(\nu),
$$ 
by a symmetric argument to \eqref{eq:phi2mu}, 
and
$$
H_{2y} = \frac{\partial \ell_{2y,\xi}(W)}{\partial \theta_y \partial \theta_y'} = - \Ep \left\{ \Phi_{\pi}(Z)   \left[
\footnotesize \begin{array}{cc}
  \Phi_{2,\xi_y}^{\mu}(Z) G_{2,\xi_y}(Z) & \Phi_{2,\xi_y}^{\rho}(Z) G_{2,\xi_y}(Z)  \\
  \Phi_{2,\xi_y}^{\mu}(Z) \rho'(X'\delta(y)) G_{3,\xi_y}(Z) & \Phi_{2,\xi_y}^{\rho}(Z) \rho'(X'\delta(y)) G_{3,\xi_y}(Z)   
\end{array}
\right] \otimes XX' \right\},
$$
\normalsize
where $$\Phi_{2,\xi_y}^{\mu}(Z) := \Phi_{2}^{\mu}\left(-X^\prime \beta(y),Z^\prime \pi; -\rho(X'\delta(y))\right)$$  $$\Phi_{2,\xi_y}^{\rho}(Z) := \Phi_{2}^{\rho}\left(-X^\prime \beta(y),Z^\prime \pi; -\rho(X'\delta(y))\right).$$

\section{Monte Carlo Simulation}\label{app:mc}

We conduct a Monte Carlo simulation calibrated to the empirical application to study the properties of the estimation and inference methods in small samples. The data generating process is the HSM of Example \ref{ex:hsm} with the values of the covariates and parameters calibrated to the data for women in the last ten years of the sample (2004--2013). We do not use the entire dataset to speed up computation. We generate 500 artificial datasets and estimate the DR-model with the same specifications for the selection and outcome equations as in the empirical application and specification 1 for the selection sorting function, i.e. $\rho(x'\delta(y)) = \rho(y)$. 

\begin{figure}[htbp]
    \includegraphics[width=0.32\textwidth,page=1]{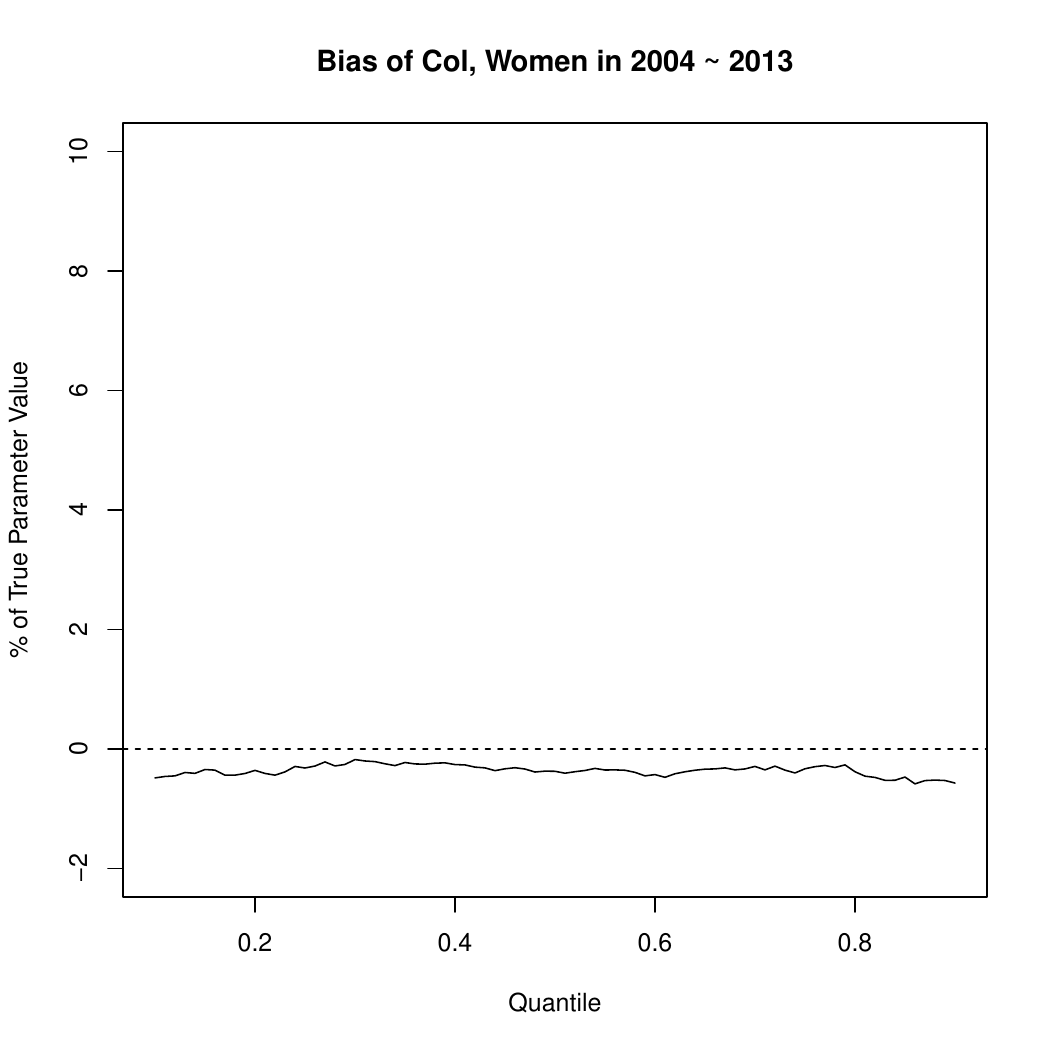}
    \includegraphics[width=0.32\textwidth,page=2]{Figures/Sim_f_113104_coef.pdf}
    \includegraphics[width=0.32\textwidth,page=3]{Figures/Sim_f_113104_coef.pdf}
    \caption{Bias, SD and RMSE for the coefficient of the college indicator in the outcome equation}
    \label{figure:Sim_col}
\end{figure}
\begin{figure}[htbp]
    \includegraphics[width=0.32\textwidth,page=5]{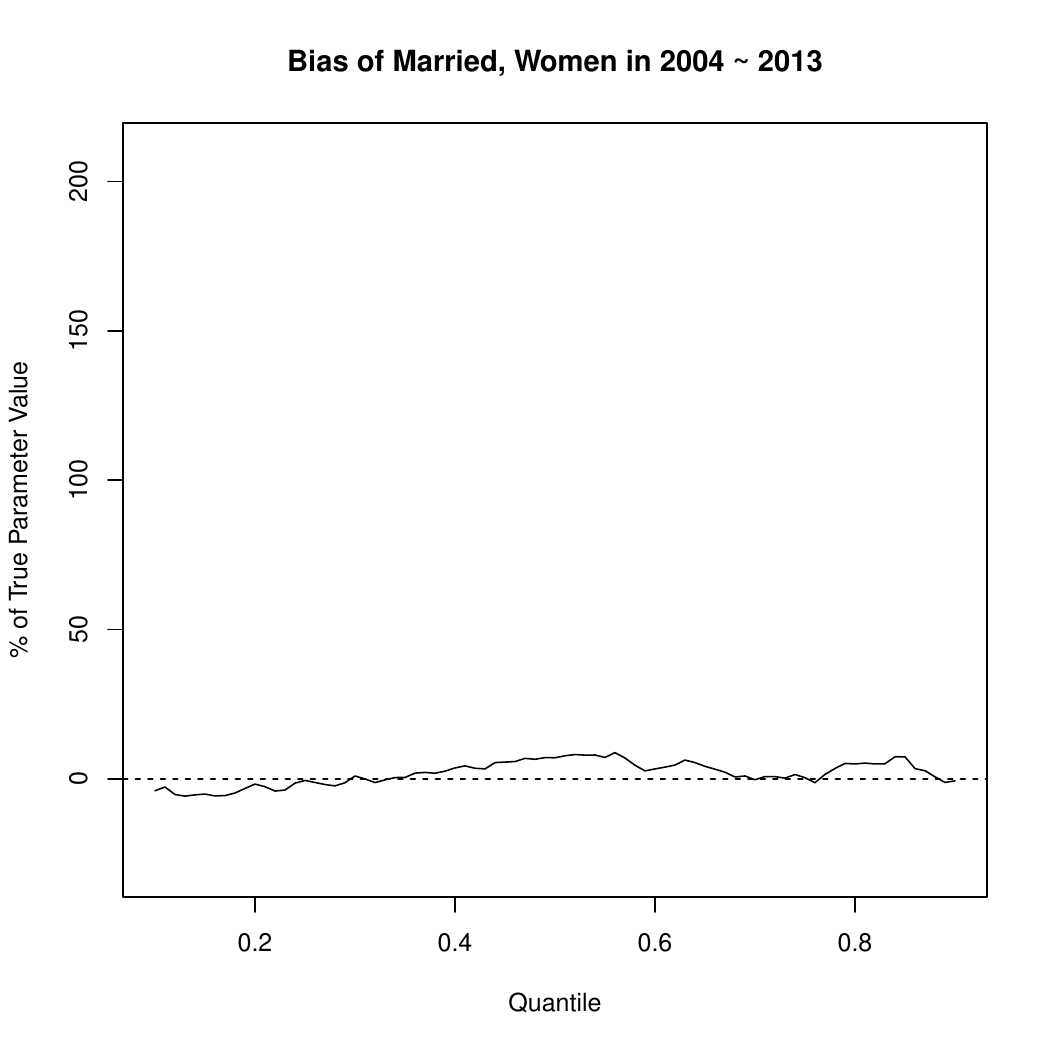}
    \includegraphics[width=0.32\textwidth,page=6]{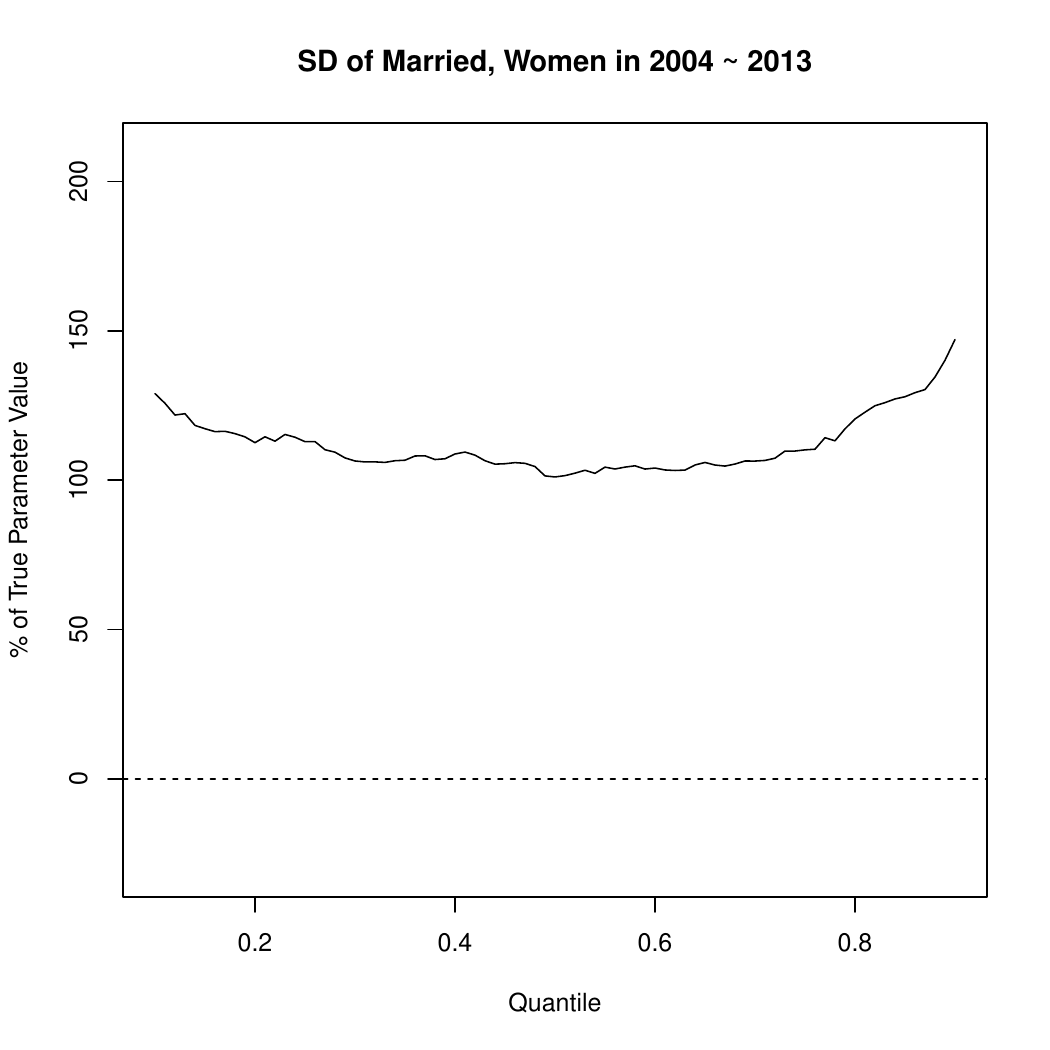}
    \includegraphics[width=0.32\textwidth,page=7]{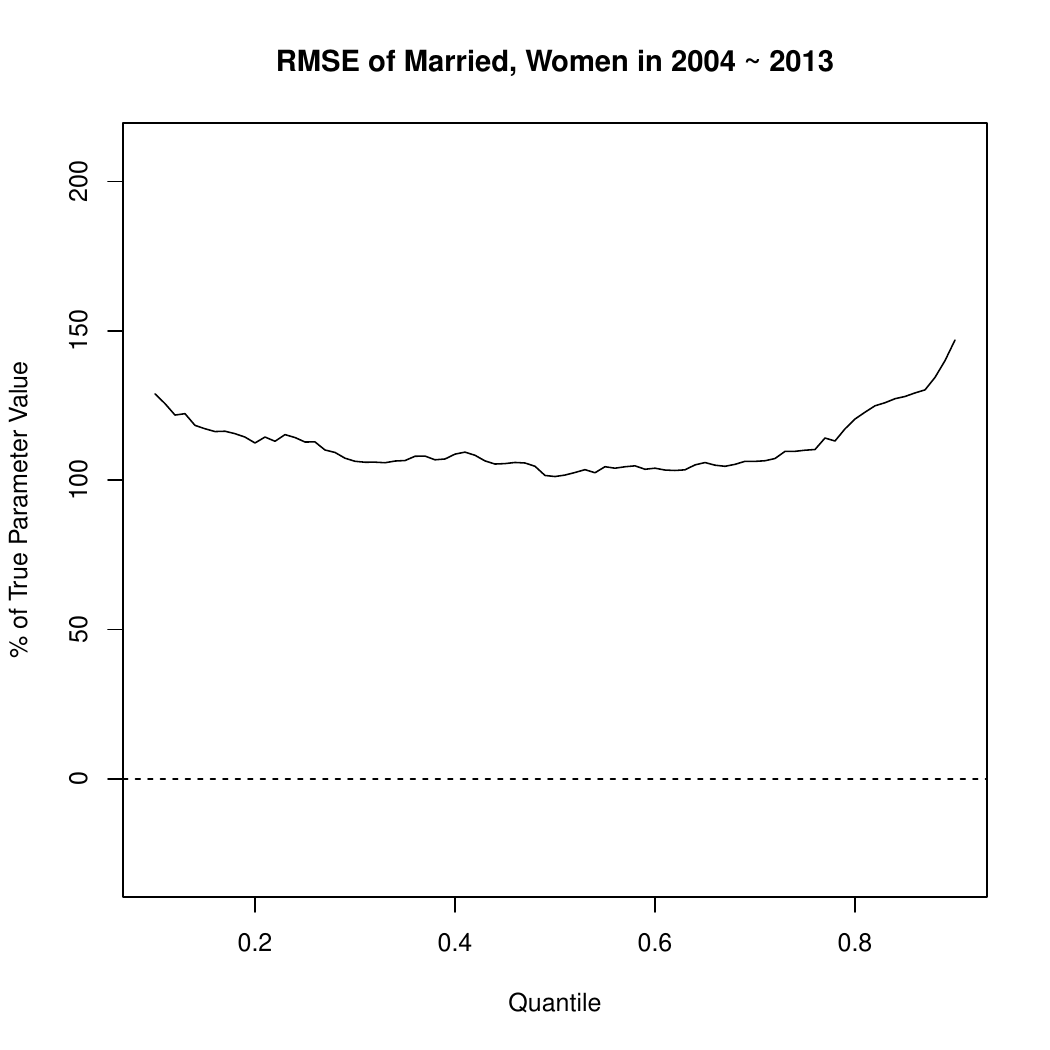}
    \caption{Bias, SD and RMSE for the coefficient of the marital status indicator in the outcome equation}
        \label{figure:Sim_married}
\end{figure}
\begin{figure}[htbp]
    \includegraphics[width=0.32\textwidth,page=9]{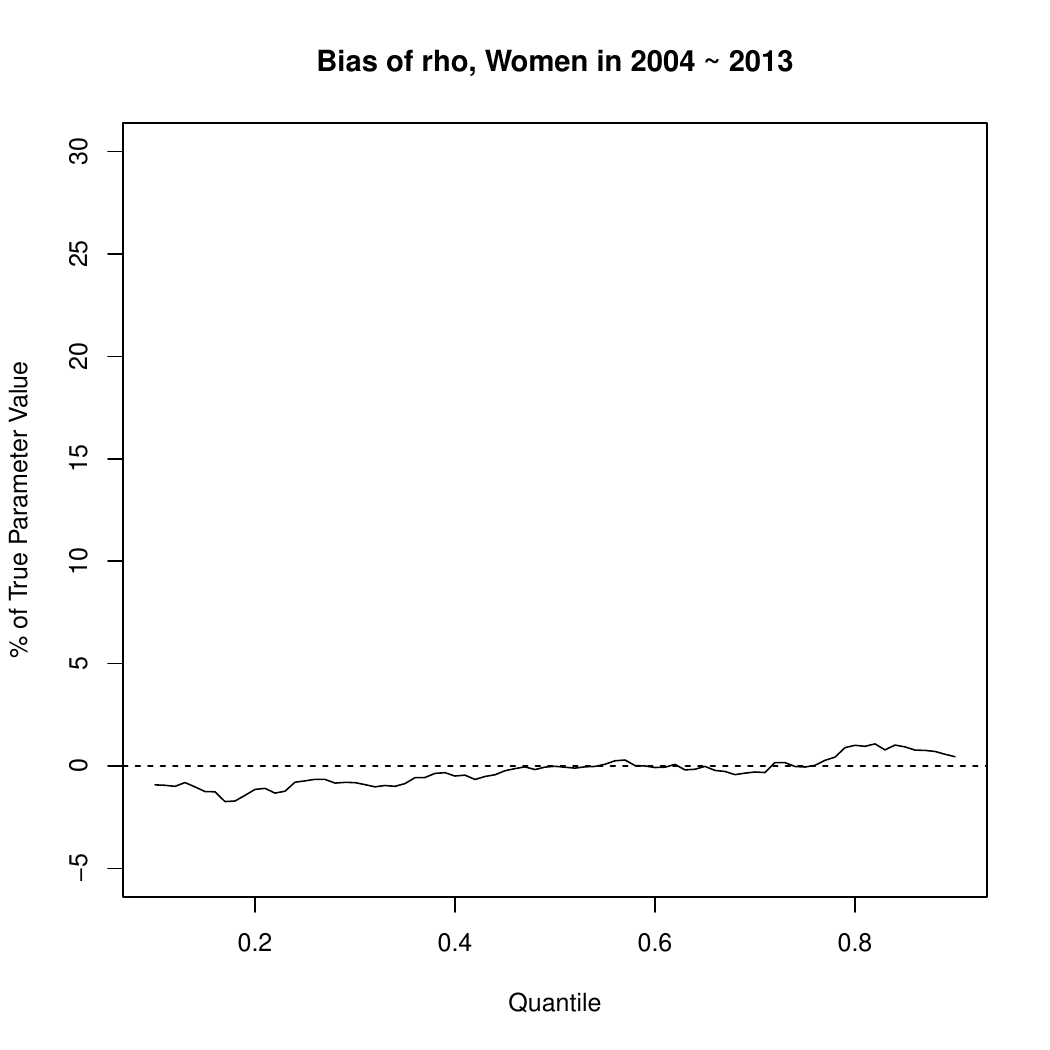}
    \includegraphics[width=0.32\textwidth,page=10]{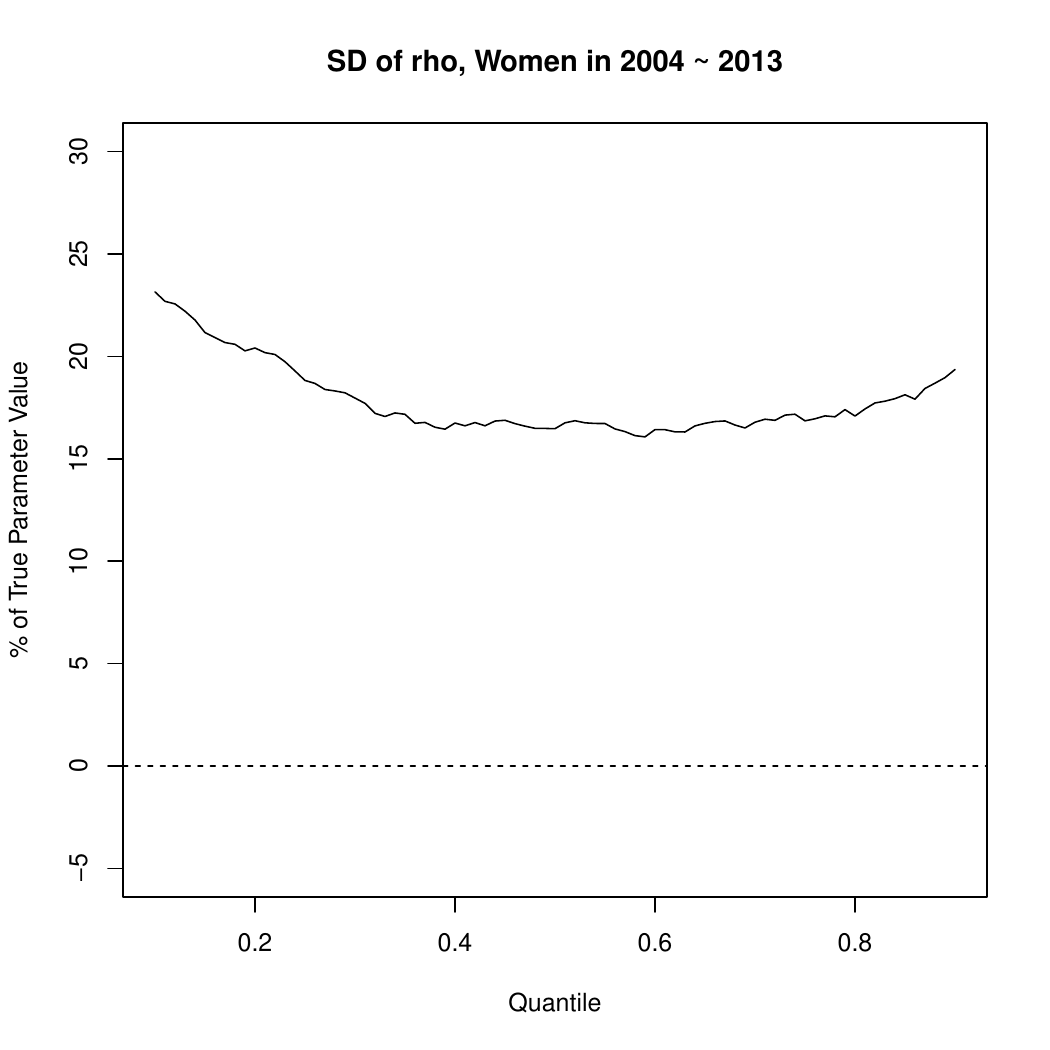}
    \includegraphics[width=0.32\textwidth,page=11]{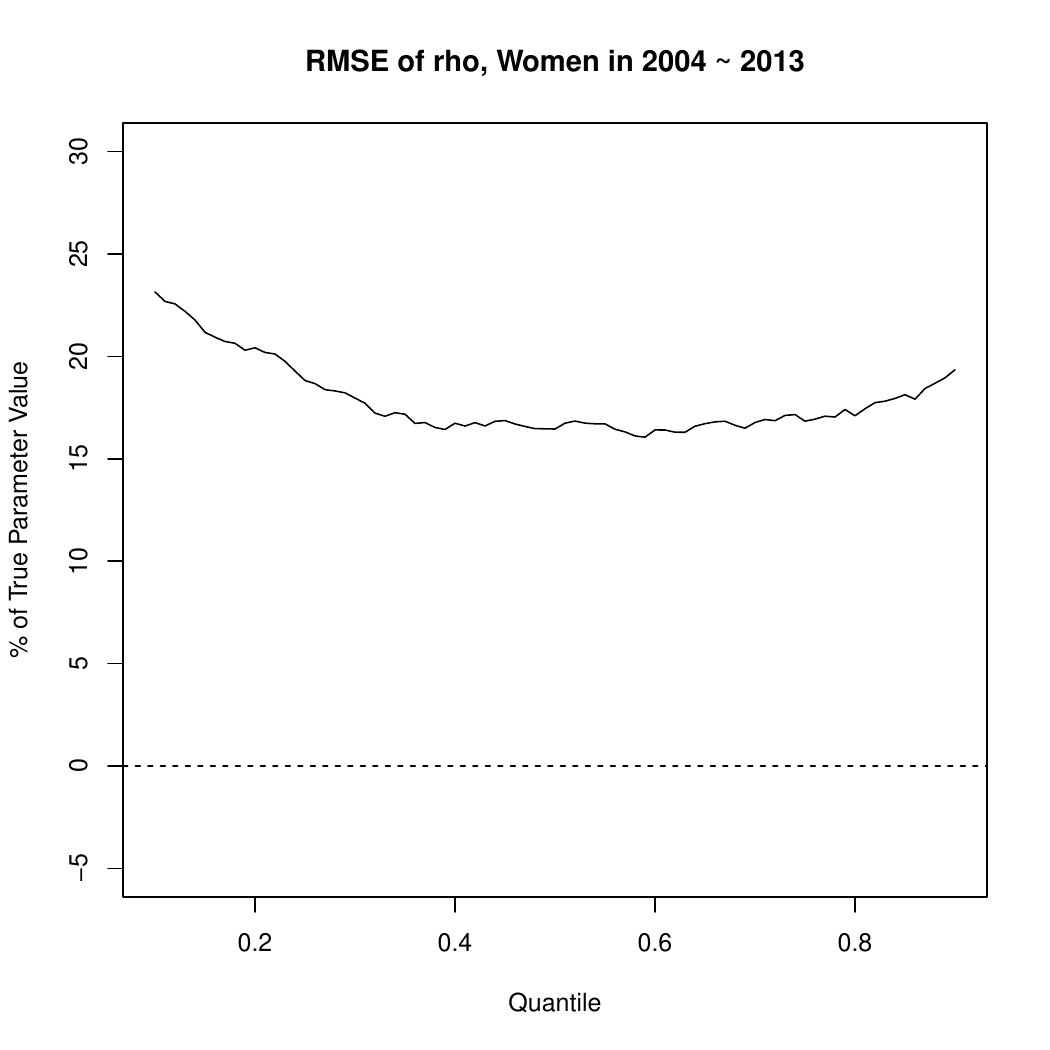}
    \caption{Bias, SD and RMSE for coefficient $\rho(y)$ in the selection sorting equation}
    \label{figure:Sim_rho}
\end{figure}

Figures \ref{figure:Sim_col}, \ref{figure:Sim_married} and \ref{figure:Sim_rho} report the biases, standard deviations and root mean square errors for the estimators of the coefficients of the college (age when ceasing school 21--22) and marital status indicators in the outcome equation, and  $\rho(y)$ in the selection sorting function, as a function of the quantile indexes of the values of log real hourly wage in the data used in the calibration.\footnote{We find similar results for the other coefficients of the outcome equation. We do not report these reports for the sake of brevity.} Although these coefficients are constant in the HSM, we do not impose this condition in the estimation. The estimates are obtained with Algorithm \ref{alg:tsdr} replacing $\mathcal{Y}$ by a finite grid containing the sample quantiles of log real hourly wage with indexes $\{0.10, 0.11, \ldots, 0.90 \}$ in the original subsample of women in the last ten years of the sample.  All the results are in percentage of the true value of the parameter. As predicted by the asymptotic theory, the biases are all small relative to the standard deviations and root mean squared errors. The estimation error increases for all the coefficients as we move away from the median towards tail values of the outcome.

Table \ref{table:simcb} shows results on the finite sample properties of 95\% confidence bands for the coefficients of the indicators of college and marital status in the outcome equation and  $\rho(y)$ of the selection sorting function. The confidence bands are constructed by Algorithm \ref{alg:mb} with $B=200$ bootstrap repetitions and the same grid of values $\bar{\mathcal{Y}}$ as for the estimators.  We report the average length of the confidence bands integrated over threshold values, the average value of the estimated critical values, and the empirical coverages of the confidence bands. For comparison, we also report the coverage of pointwise confidence bands using the normal distribution, i.e. with critical value equal to 1.96. The last row computes the ratio of the standard error averaged across simulations to the simulation standard deviation, integrated over threshold values. We find that the bands have coverages close to the nominal level. As expected, pointwise bands severely undercover the entire functions. The standard errors based on the asymptotic distribution provide a fair approximation to the sampling variability of the estimator.

\begin{table}[h]
    \caption{Properties of 95\% Confidence Bands}
    \centering
    \label{table:simcb}
    \begin{tabular}{lccc}
        \hline\hline
        & College & Married & $\rho(y)$ \\
        \hline
        Average Length & 0.38 & 0.16 & 0.35 \\
        Average Critical Value & 2.91 & 2.89 & 2.88 \\
        Coverage uniform band (\%) & 96 & 98 & 96 \\
        Coverage pointwise band (\%) & 68 & 64 & 67 \\
        Average SE/SD & 1.04 & 1.05 & 1.07 \\
        \hline\hline
        \multicolumn{4}{l}{\begin{minipage}{9cm}\footnotesize Notes: Nominal level of critical values is 95\%. 500 simulations with 200 bootstrap draws.\end{minipage}}
    \end{tabular}
\end{table}

%

\section{Wage Decompositions in the UK: Additional Results}\label{app:empirics}

\subsection{Descriptive statistics and background on U.K. labor market} Table \ref{table:summstat} reports means and standard deviations of all the variables used in the analysis. We report these statistics for the entire sample, and by employment status and gender. The overall employment rate is 74\%. Women are 17\% less likely to be employed than men, and the unconditional gender wage gap is 33\%. Overall, women and men are similar in terms of covariates. Both working men and women are relatively more highly educated, younger, and more likely to be married than their non-working counterparts. Having young children and high out-of-work benefits is negatively associated with employment for women but not for men.  


\begin{footnotesize}
\begin{table}
\caption{Summary Statistics}\label{table:summstat}
\def\sym#1{\ifmmode^{#1}\else\(^{#1}\)\fi}
\begin{tabular}{l*{6}{c}}
\hline\hline
                &\multicolumn{2}{c}{Full}         &\multicolumn{2}{c}{Male}         &\multicolumn{2}{c}{Female}    \\
                &     All        &  Employed         &     All        &   Employed         &   All         & Employed \\
\hline
Log Hourly Wage &                  &    2.38         &                  &    2.54         &                  &    2.21         \\
                &                  &  (0.54)         &                  &  (0.51)         &                  &  (0.52)         \\
Employed        &    0.74         &                  &    0.83         &                  &    0.66        &                  \\
                &  (0.44)         &                  &  (0.38)         &                  &  (0.47)         &                \\
\multicolumn{7}{l}{Ceased School at} \\
\hspace{3mm} $\leq$ 15       &    0.33         &    0.30         &    0.33         &    0.31         &    0.33         &    0.29         \\
                &  (0.47)         &  (0.46)         &  (0.47)         &  (0.46)         &  (0.47)         &  (0.45)         \\

\hspace{3mm}16       &    0.31         &    0.30         &    0.32         &    0.32         &    0.30         &    0.29         \\
                &  (0.46)         &  (0.46)         &  (0.47)         &  (0.47)         &  (0.46)         &  (0.45)         \\

\hspace{3mm}17-18     &    0.18         &    0.19         &    0.16         &    0.17         &    0.20         &    0.22         \\
                &  (0.38)         &  (0.39)         &  (0.37)         &  (0.37)         &  (0.40)         &  (0.41)         \\

\hspace{3mm}19-20      &   0.04         &   0.05         &   0.04         &   0.04         &   0.04         &   0.05         \\
                &  (0.20)         &  (0.21)         &  (0.20)         &  (0.20)         &  (0.21)         &  (0.22)         \\

\hspace{3mm}21-22      &   0.09         &    0.11         &   0.09         &    0.10         &   0.09         &    0.12         \\
                &  (0.29)         &  (0.31)         &  (0.29)         &  (0.30)         &  (0.29)         &  (0.32)         \\

\hspace{3mm}$\geq$23     &   0.05         &   0.05         &   0.06         &   0.06         &   0.04         &   0.04         \\
                &  (0.21)         &  (0.22)         &  (0.23)         &  (0.24)         &  (0.19)         &  (0.20)         \\
Age            &    40.13         &    39.84         &    40.22         &    39.76         &    40.06         &    39.92         \\
                &  (10.43)         &  (10.10)         &  (10.40)         &  (10.11)         &  (10.45)         &  (10.08)         \\

Married        &    0.76         &    0.79         &    0.78         &    0.81         &    0.75         &    0.76         \\
                &  (0.43)         &  (0.41)         &  (0.42)         &  (0.39)         &  (0.43)         &  (0.43)         \\
\multicolumn{7}{l}{Number of children with age} \\
\hspace{3mm} 0--1          &   0.06         &   0.05         &   0.06         &   0.06         &   0.06         &   0.03         \\
                &  (0.24)         &  (0.22)         &  (0.24)         &  (0.25)         &  (0.24)         &  (0.18)         \\

\hspace{3mm} 2          &   0.05         &   0.04         &   0.05         &   0.05         &   0.05         &   0.03         \\
                &  (0.23)         &  (0.21)         &  (0.23)         &  (0.23)         &  (0.23)         &  (0.18)         \\

\hspace{3mm} 3--4         &    0.10         &   0.09         &   0.10         &    0.10         &    0.11         &   0.07         \\
                &  (0.32)         &  (0.29)         &  (0.31)         &  (0.32)         &  (0.32)         &  (0.27)         \\

\hspace{3mm} 5--10        &    0.32         &    0.29         &    0.30         &    0.30         &    0.33         &    0.28         \\
                &  (0.64)         &  (0.61)         &  (0.62)         &  (0.62)         &  (0.65)         &  (0.59)         \\

\hspace{3mm} 11--16       &    0.30         &    0.30         &    0.28         &    0.29         &    0.32         &    0.32         \\
                &  (0.63)         &  (0.62)         &  (0.61)         &  (0.61)         &  (0.64)         &  (0.63)         \\

\hspace{3mm} 17--18       &   0.03         &   0.04         &   0.03         &   0.03         &   0.04         &   0.04         \\
                &  (0.19)         &  (0.19)         &  (0.18)         &  (0.18)         &  (0.19)         &  (0.20)         \\

Benefit Income  &    5.44         &    5.50         &    5.25         &    5.29         &    5.60         &    5.73         \\
                &  (0.74)         &  (0.78)         &  (0.70)         &  (0.72)         &  (0.73)         &  (0.78)         \\
\hline
Observations    &   258,900         &   190,765         &   119,396         &    98,764         &   139,504         &    92,001         \\
\hline\hline
\multicolumn{7}{l}{\footnotesize \emph{Notes}: all the entries are means with standard deviations in parentheses.}\\
\multicolumn{7}{l}{\footnotesize \emph{Source:} FES/EFS/LCFS Data.}\\
\end{tabular}

\end{table}

\end{footnotesize}

\begin{figure}    \includegraphics[width=0.32\textwidth,page=1]{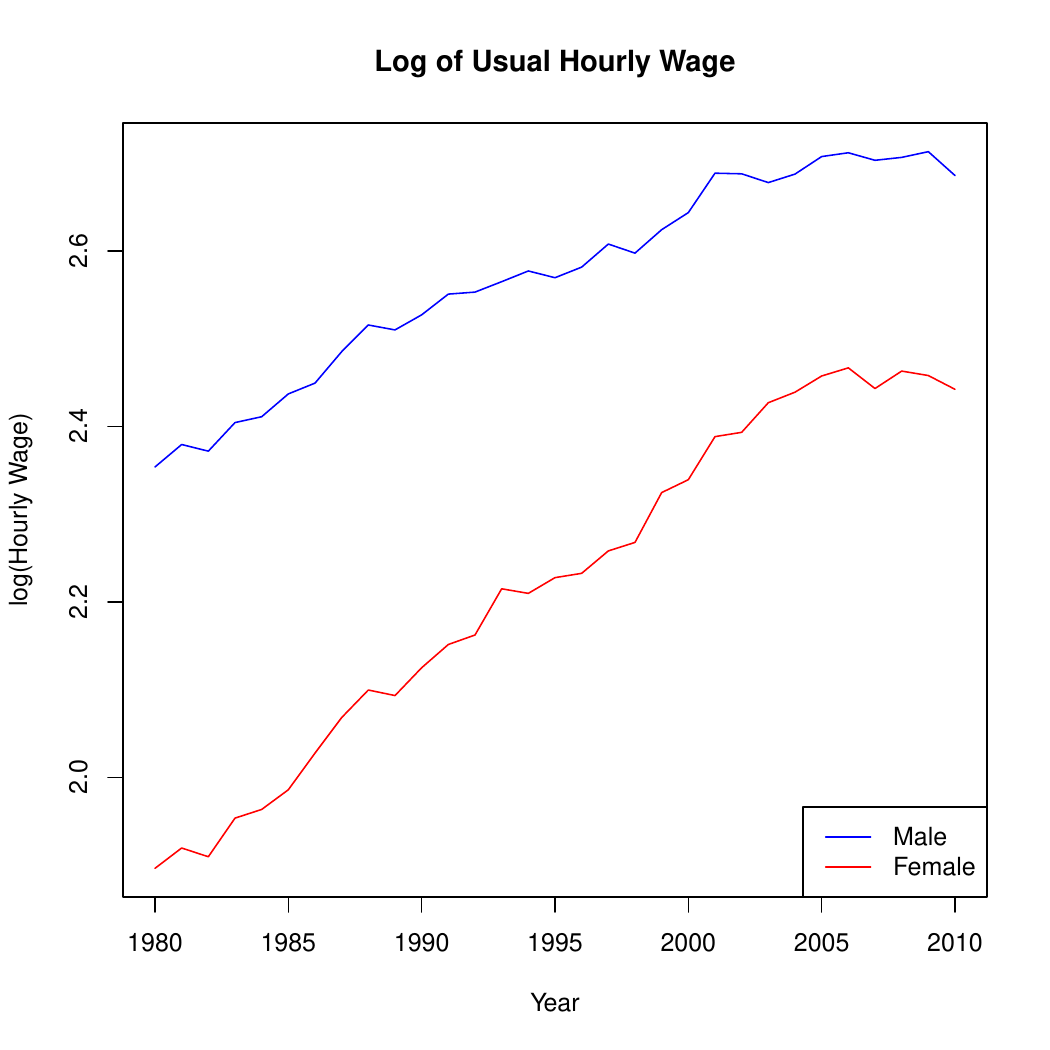}
    \includegraphics[width=0.32\textwidth,page=2]{Figures/Inequality.pdf}
    \includegraphics[width=0.32\textwidth,page=3]{Figures/Inequality.pdf}
\caption{Trends in U.K. labor market 1978-2013 by gender: left panel reports the average of the log wage rate, the middle panel reports the 90-10 percentile  spread of the log wage rate, and the right panel reports the employment rate}\label{fig:trends}
\end{figure}

Figure \ref{fig:trends} provides some background on the U.K. labor market using our data. The left panel shows that  over 36 years the average wages of working men and women have continuously grown and the unconditional gender wage gap  has progressively narrowed from $46\%$ to $24\%$. The middle panel indicates that the growth of average wage has come together with an increase in wage inequality for both working men and women until 2000. The positive trend in wage inequality has continued for men after 2000, but not for women. The right panel shows opposite trends in the employment rate for men and women, where the gender employment gap has steadily and sharply reduced from 34\% to 8\%. 

\subsection{Additional Empirical Results}
\begin{figure}
    \includegraphics[height=0.75\textheight, width=0.85\textwidth,page=1]{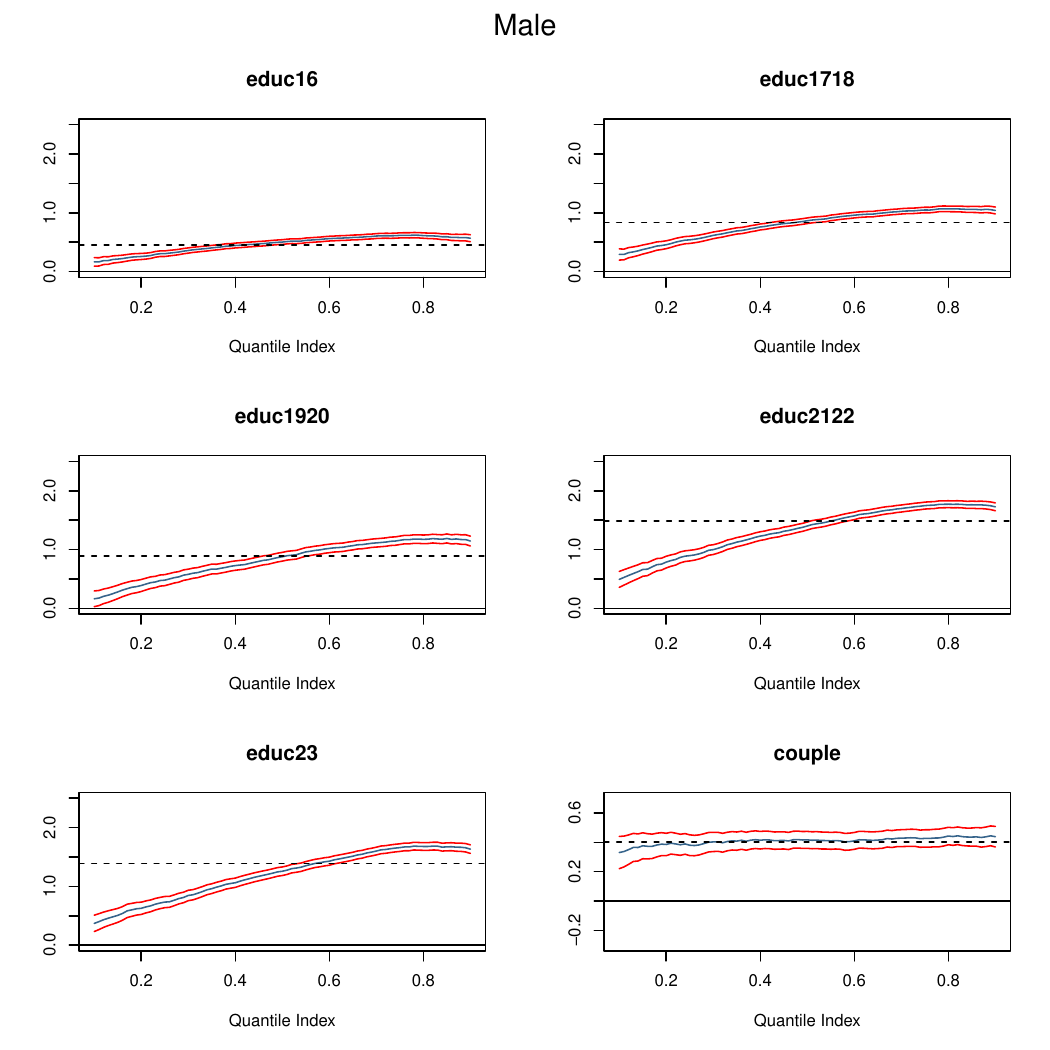}
    \caption{Estimates and 95\% confidence bands for coefficients of education and marital status in the outcome equation: specification 1 for men}\label{fig:coef_men_spec1}
\end{figure}

\begin{figure}
    \includegraphics[height=0.75\textheight, width=0.85\textwidth,page=3]{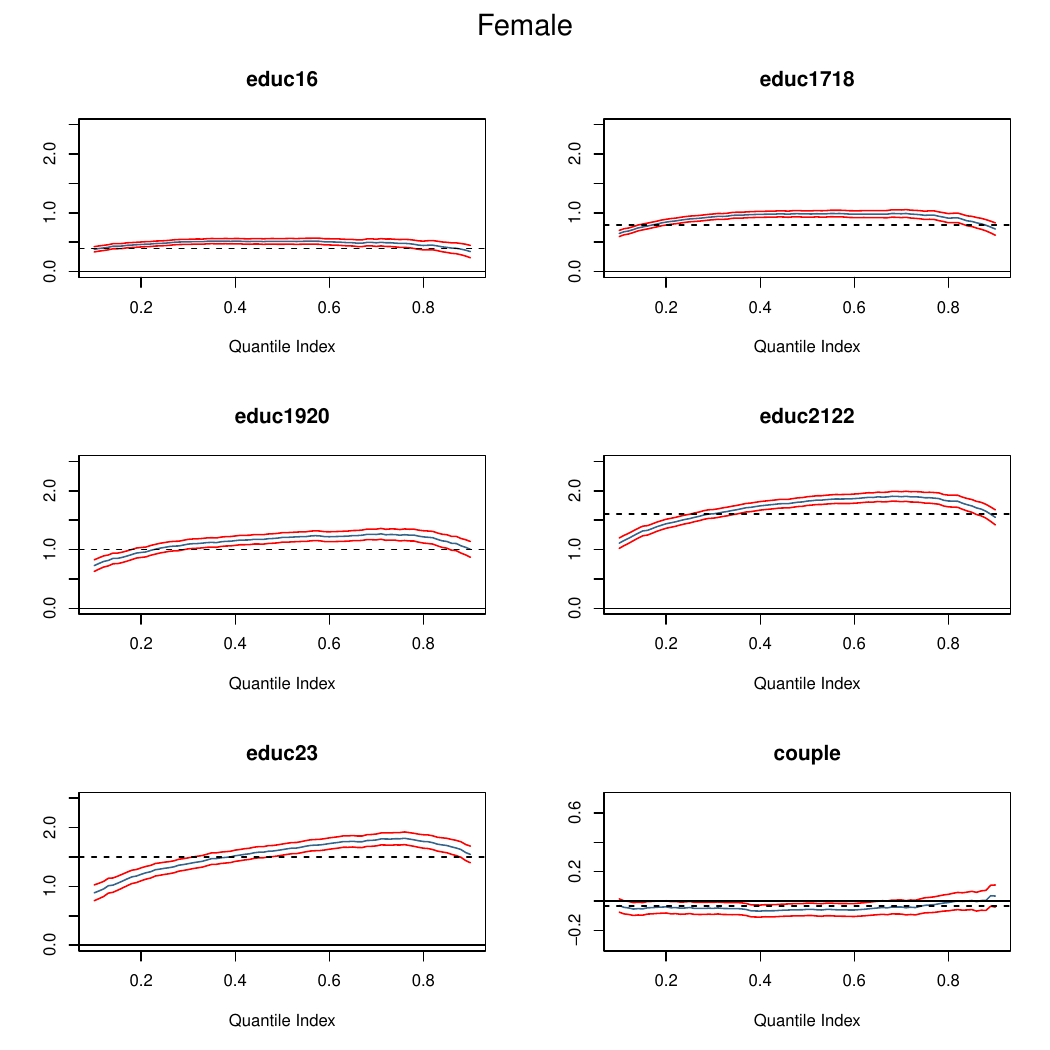}
    \caption{Estimates and 95\% confidence bands for coefficients of education and marital status in the outcome equation:  specification 1 for women}\label{fig:coef_women_spec1}
\end{figure}

\begin{figure}
    \includegraphics[height=0.75\textheight, width=0.85\textwidth,page=1]{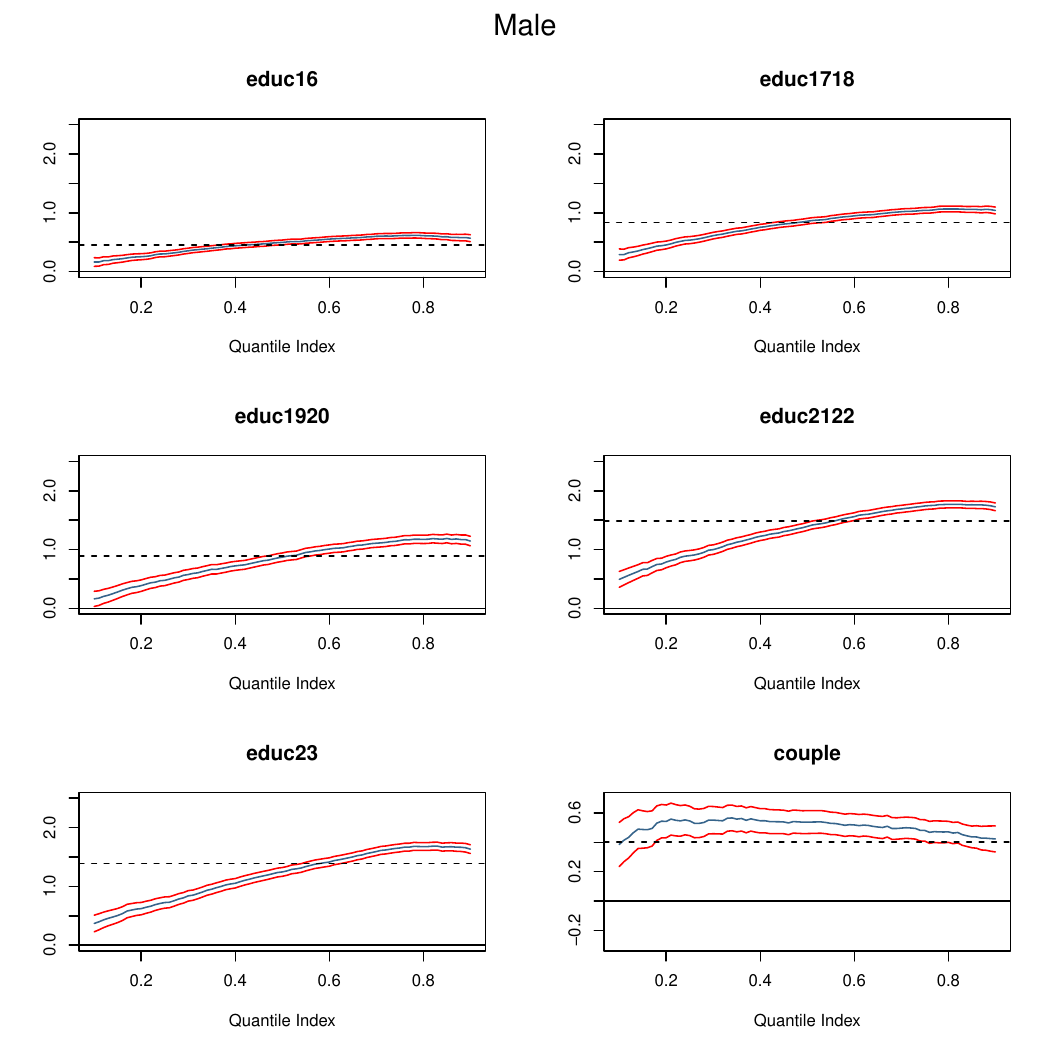}
    \caption{Estimates and 95\% confidence bands for coefficients of education and marital status in the outcome equation: specification 2 for men}\label{fig:coef_men_spec2}
\end{figure}

\begin{figure}
    \includegraphics[height=0.75\textheight, width=0.85\textwidth,page=3]{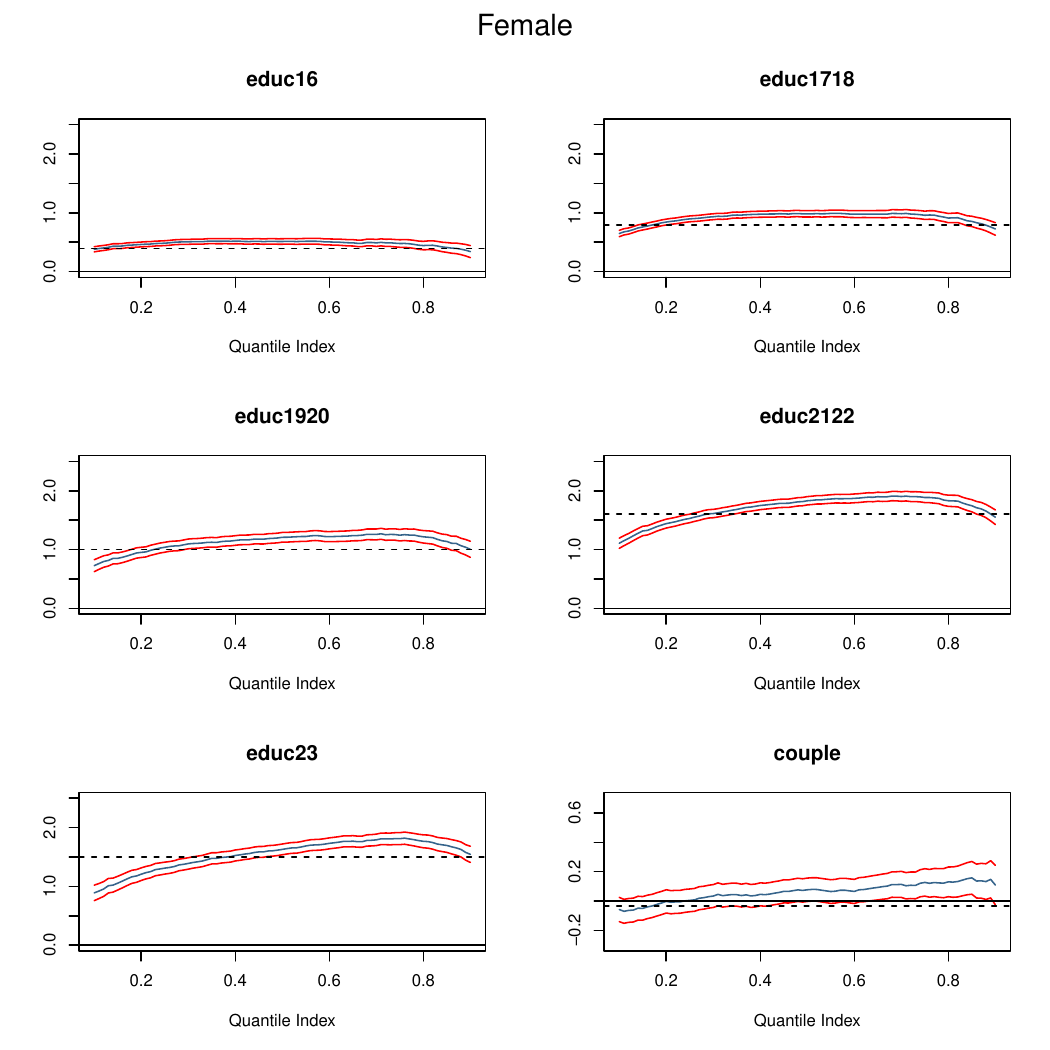}
    \caption{Estimates and 95\% confidence bands for coefficients of education and marital status in the outcome equation:  specification 2 for women}\label{fig:coef_women_spec2}
\end{figure}

\begin{figure}
    \includegraphics[height=0.75\textheight, width=0.85\textwidth,page=1]{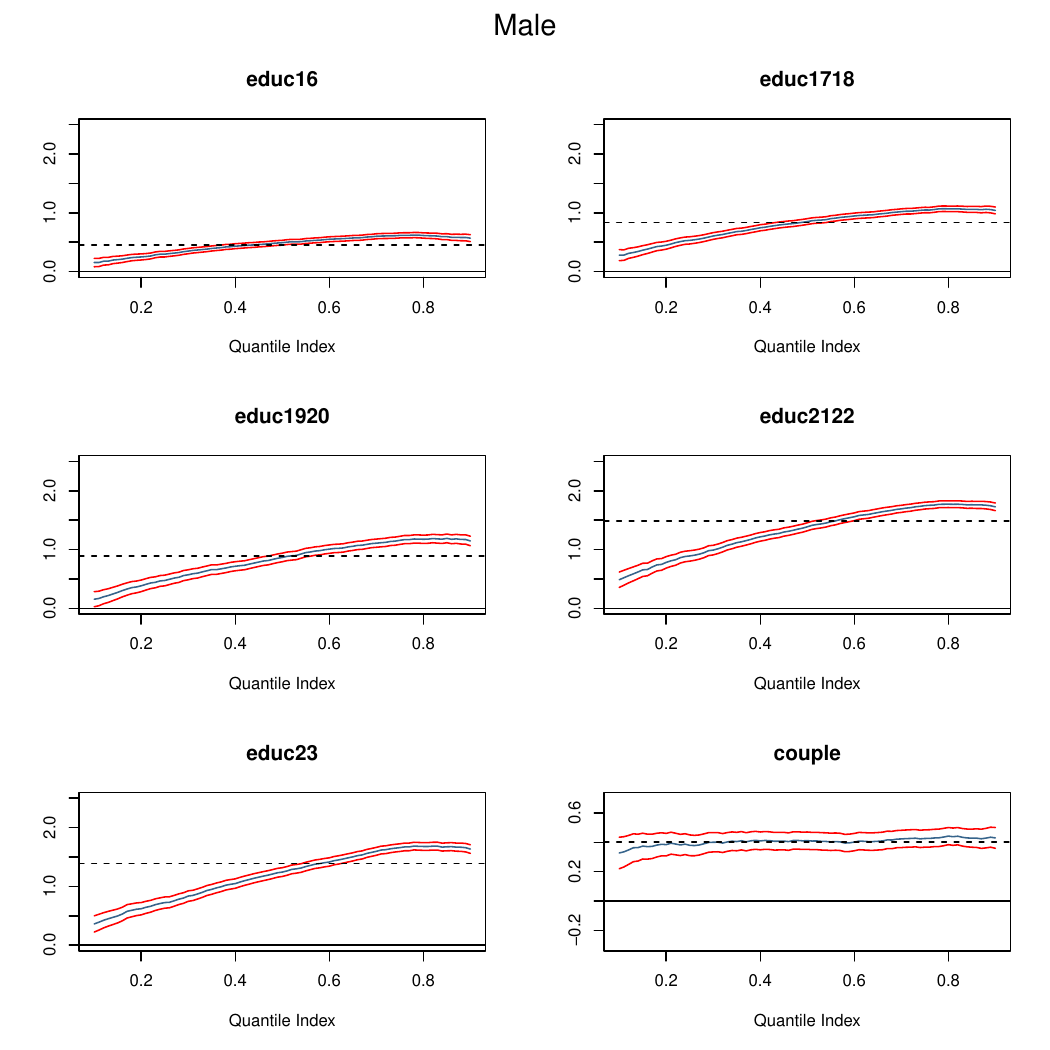}
    \caption{Estimates and 95\% confidence bands for coefficients of education and marital status in the outcome equation: specification 3 for men}\label{fig:coef_men_spec3}
\end{figure}

\begin{figure}
    \includegraphics[height=0.75\textheight, width=0.85\textwidth,page=4]{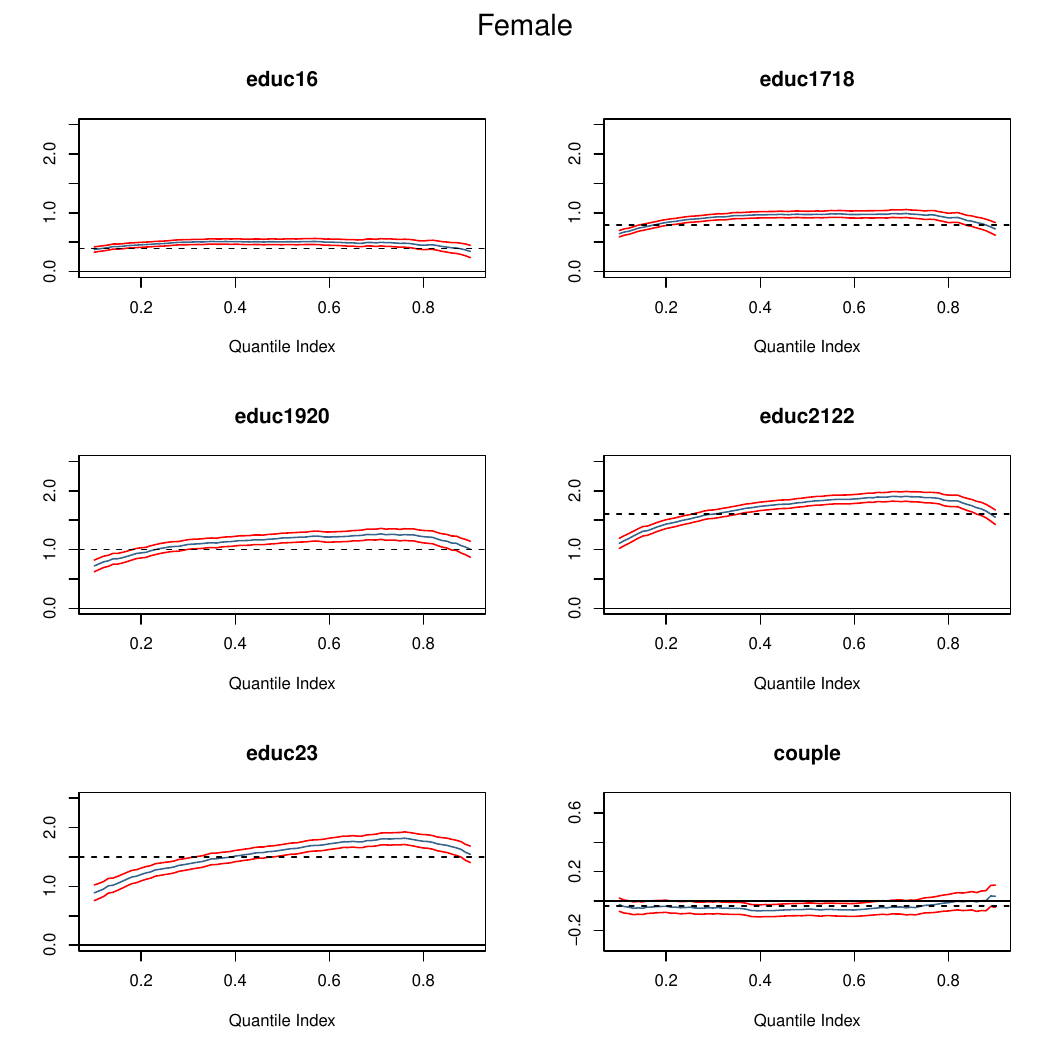}
    \caption{Estimates and 95\% confidence bands for coefficients of education and marital status in the outcome equation:  specification 3 for women}\label{fig:coef_women_spec3}
\end{figure}

\begin{figure}
    \includegraphics[height=0.75\textheight, width=0.85\textwidth,page=1]{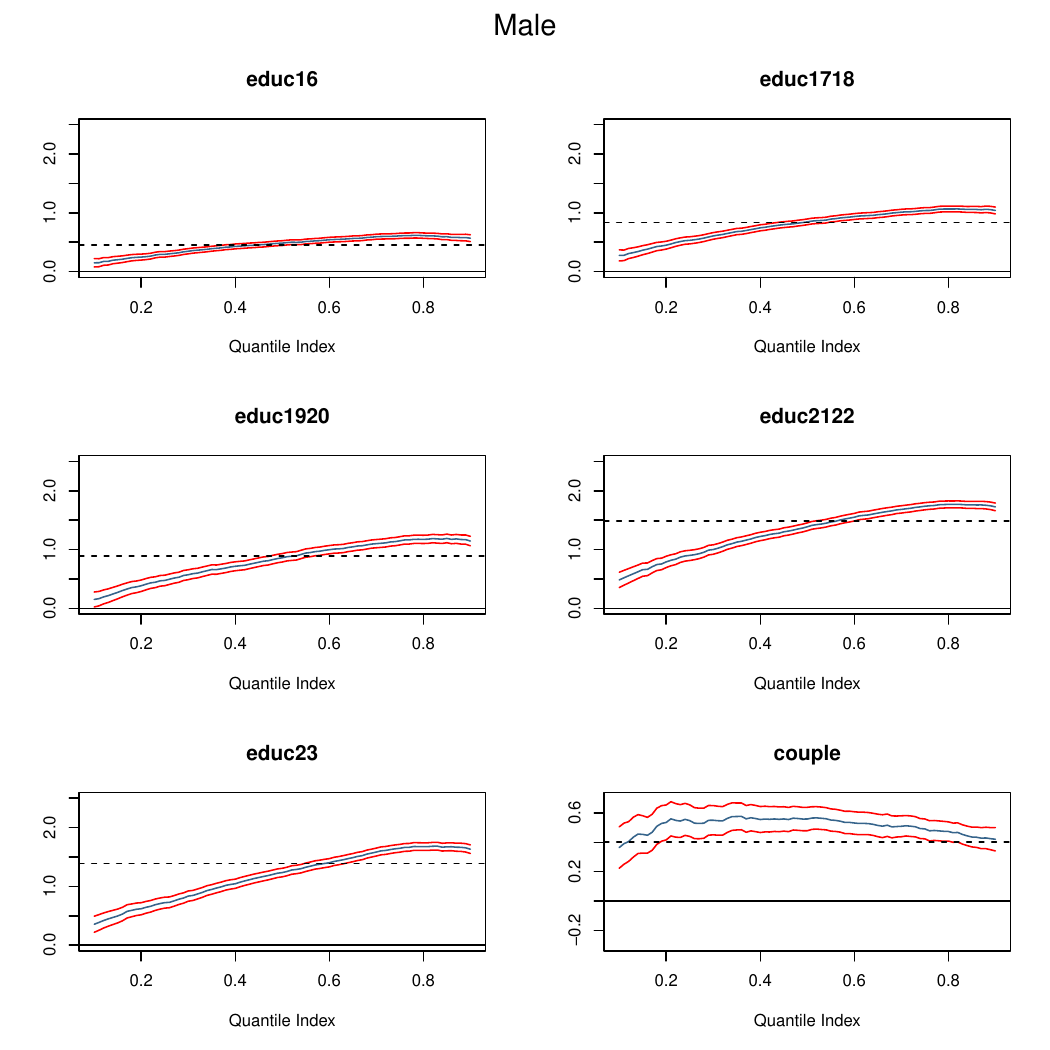}
    \caption{Estimates and 95\% confidence bands for coefficients of education and marital status in the outcome equation: specification 4 for men}\label{fig:coef_men_spec4}
\end{figure}

\begin{figure}
    \includegraphics[height=0.75\textheight, width=0.85\textwidth,page=4]{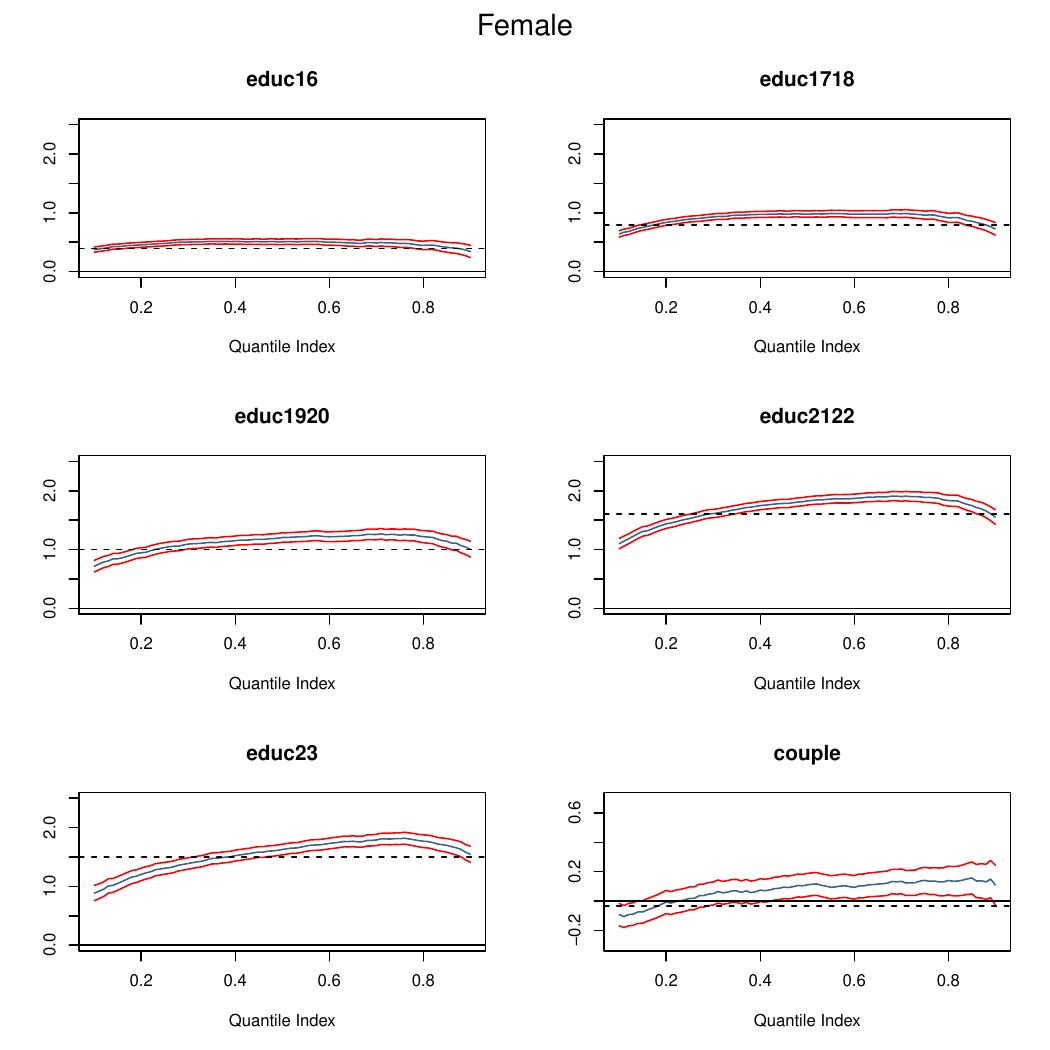}
    \caption{Estimates and 95\% confidence bands for coefficients of education and marital status in the outcome equation:  specification 4 for women}\label{fig:coef_women_spec4}
\end{figure}

\begin{figure}
    \includegraphics[height=0.75\textheight, width=0.85\textwidth,page=1]{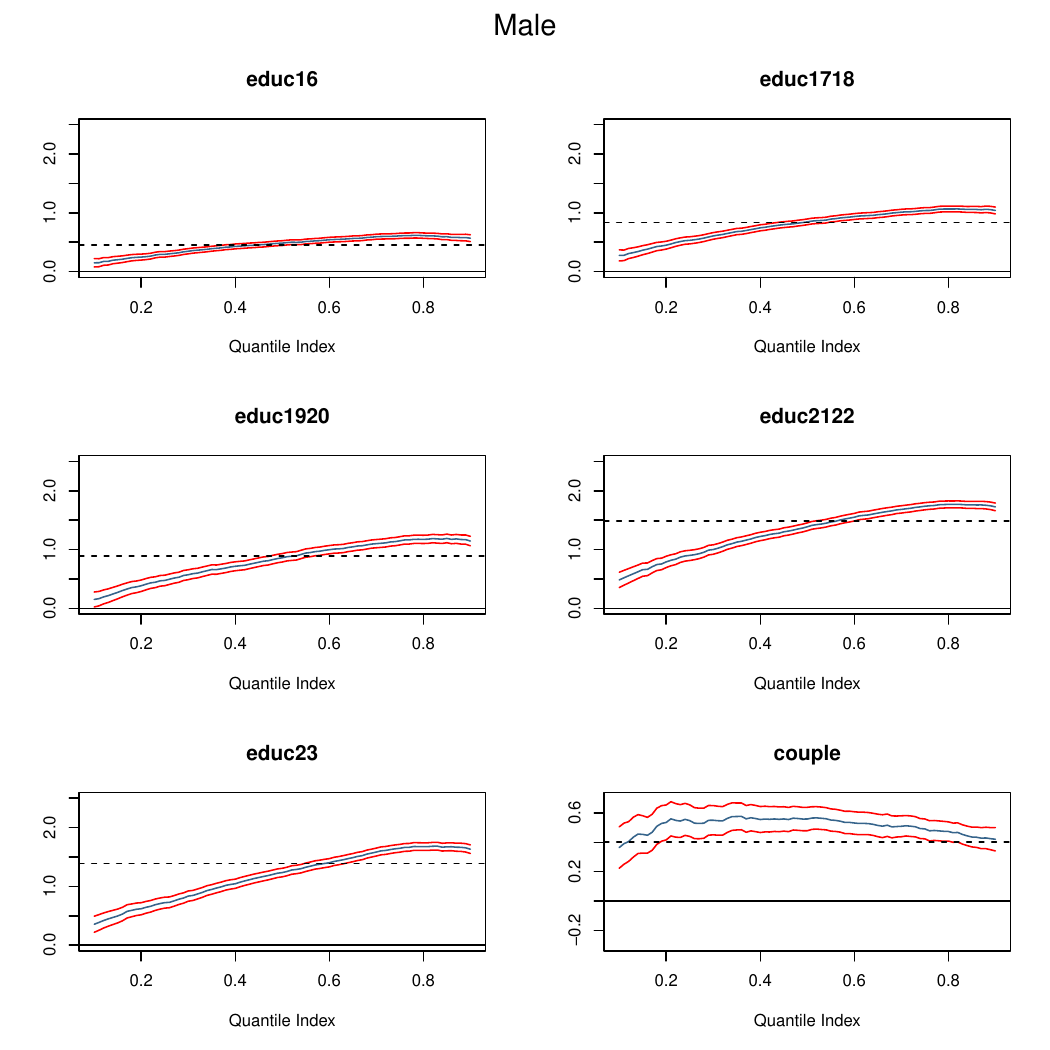}
    \caption{Estimates and 95\% confidence bands for coefficients of fertility in the outcome equation: specification 4 for men}\label{fig:coef2_men_spec1}
\end{figure}

\begin{figure}
    \includegraphics[height=0.75\textheight, width=0.85\textwidth,page=3]{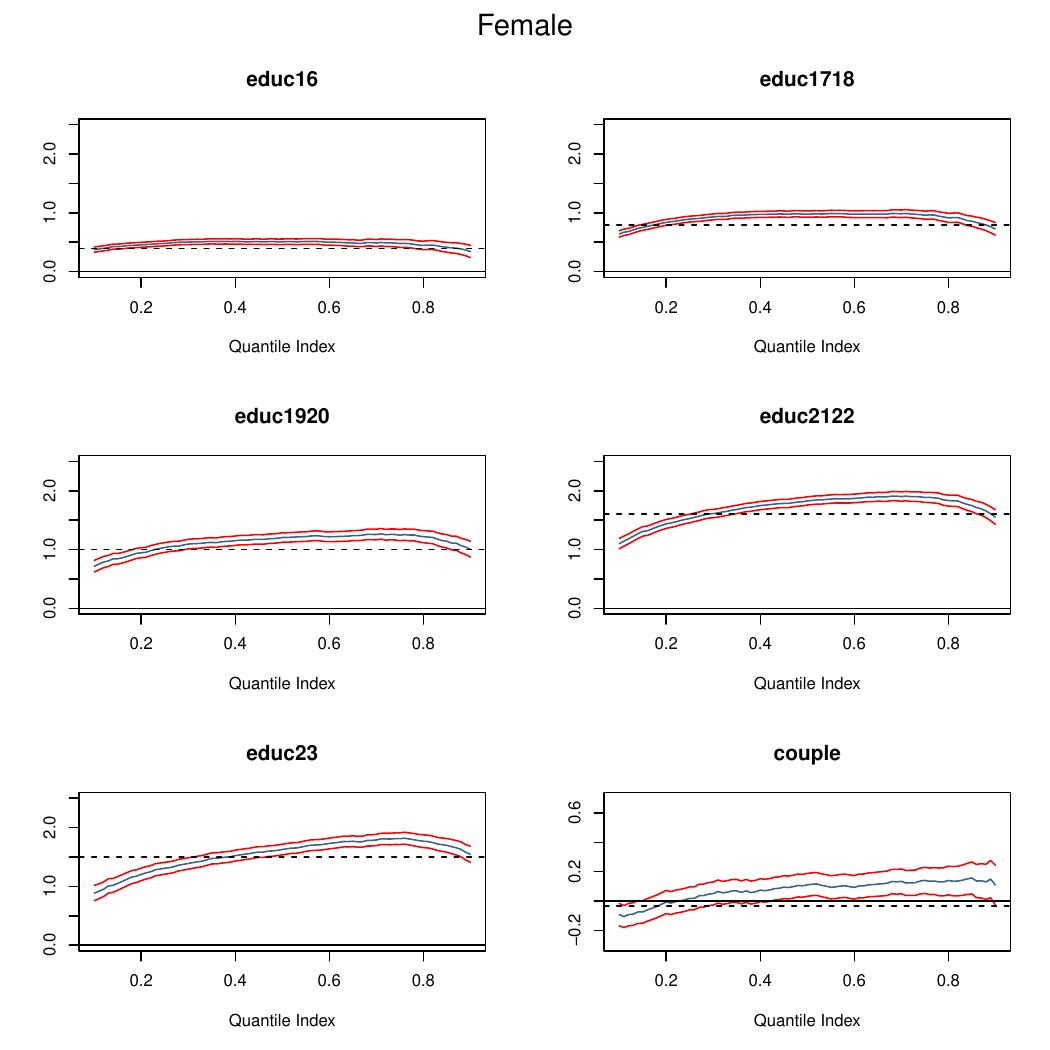}
    \caption{Estimates and 95\% confidence bands for coefficients of fertility in the outcome equation:  specification 4 for women}\label{fig:coef2_women_spec1}
\end{figure}

\begin{figure}
    \includegraphics[height=.8\textwidth, width=0.4\textwidth,page=2]{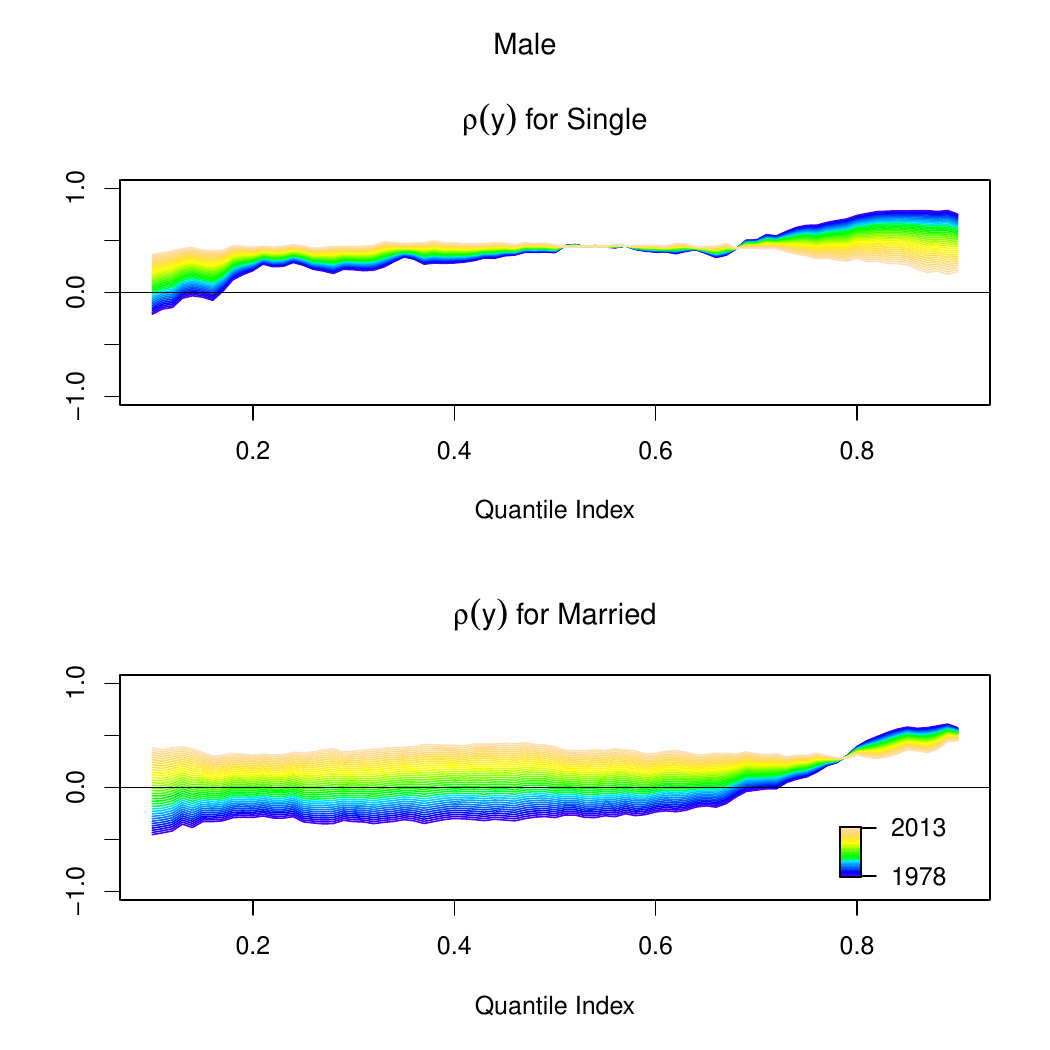}
    \includegraphics[height=.8\textwidth, width=0.4\textwidth,page=5]{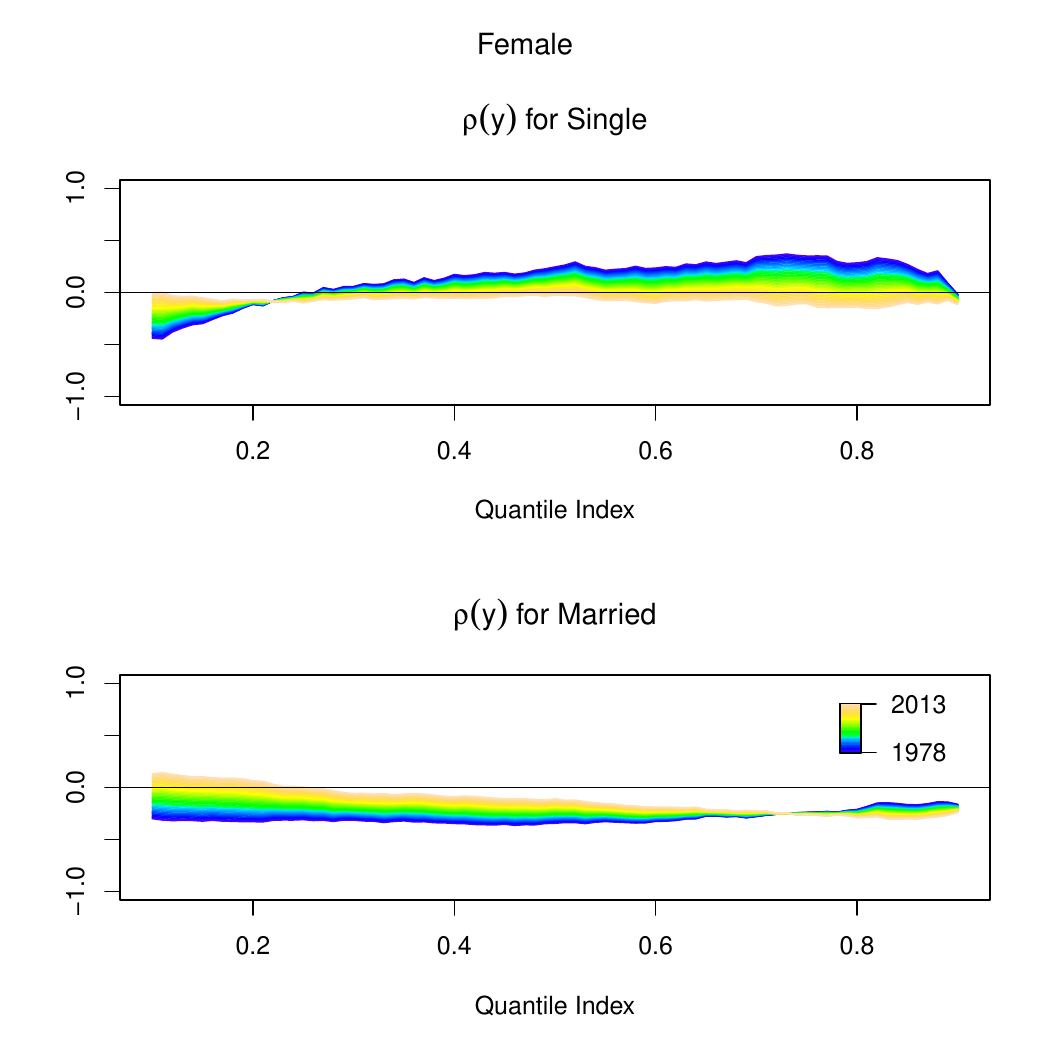}
    \caption{Estimates and 95\% confidence bands for the selection sorting function: specification 4}\label{fig:rho_men_spec4}
\end{figure}

\begin{figure}
    \includegraphics[height=.85\textwidth, width=0.49\textwidth,page=3]{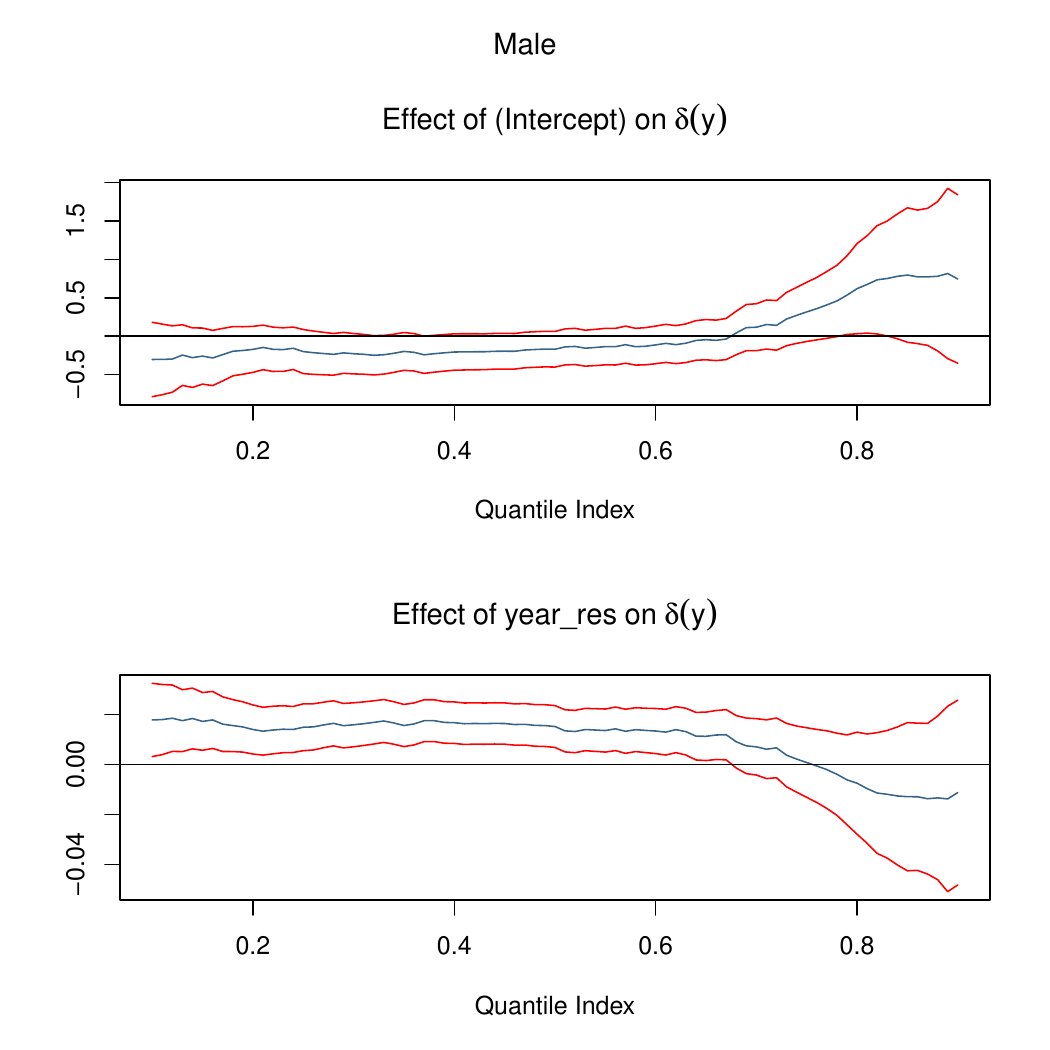}
    \includegraphics[height=.85\textwidth, width=0.49\textwidth,page=6]{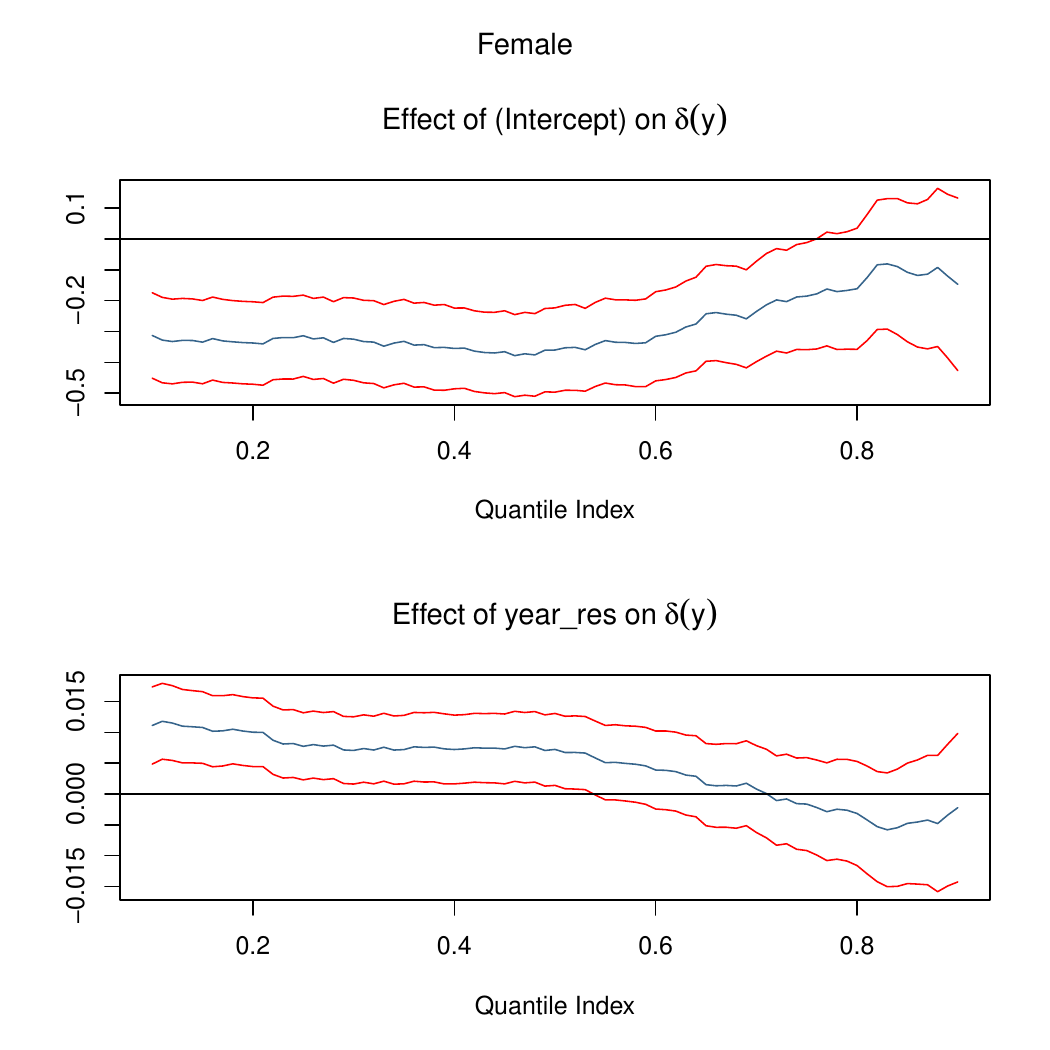}
    \caption{Estimates and 95\% confidence bands for coefficients of the selection sorting function: specification 3}\label{fig:coef_sorting_spec3}
\end{figure}

\begin{figure}
    \includegraphics[height=.75\textwidth, width=0.49\textwidth,page=3]{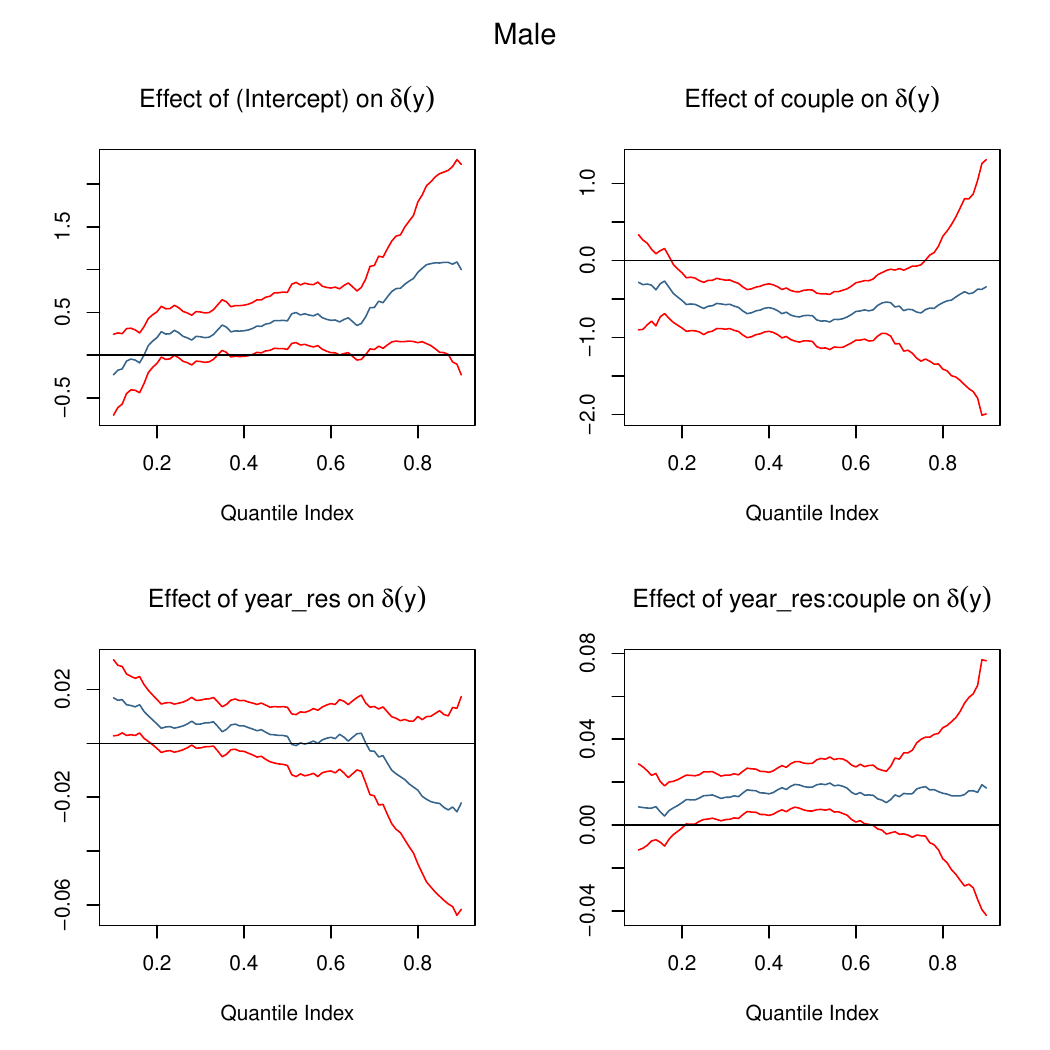}
    \includegraphics[height=.75\textwidth, width=0.49\textwidth,page=6]{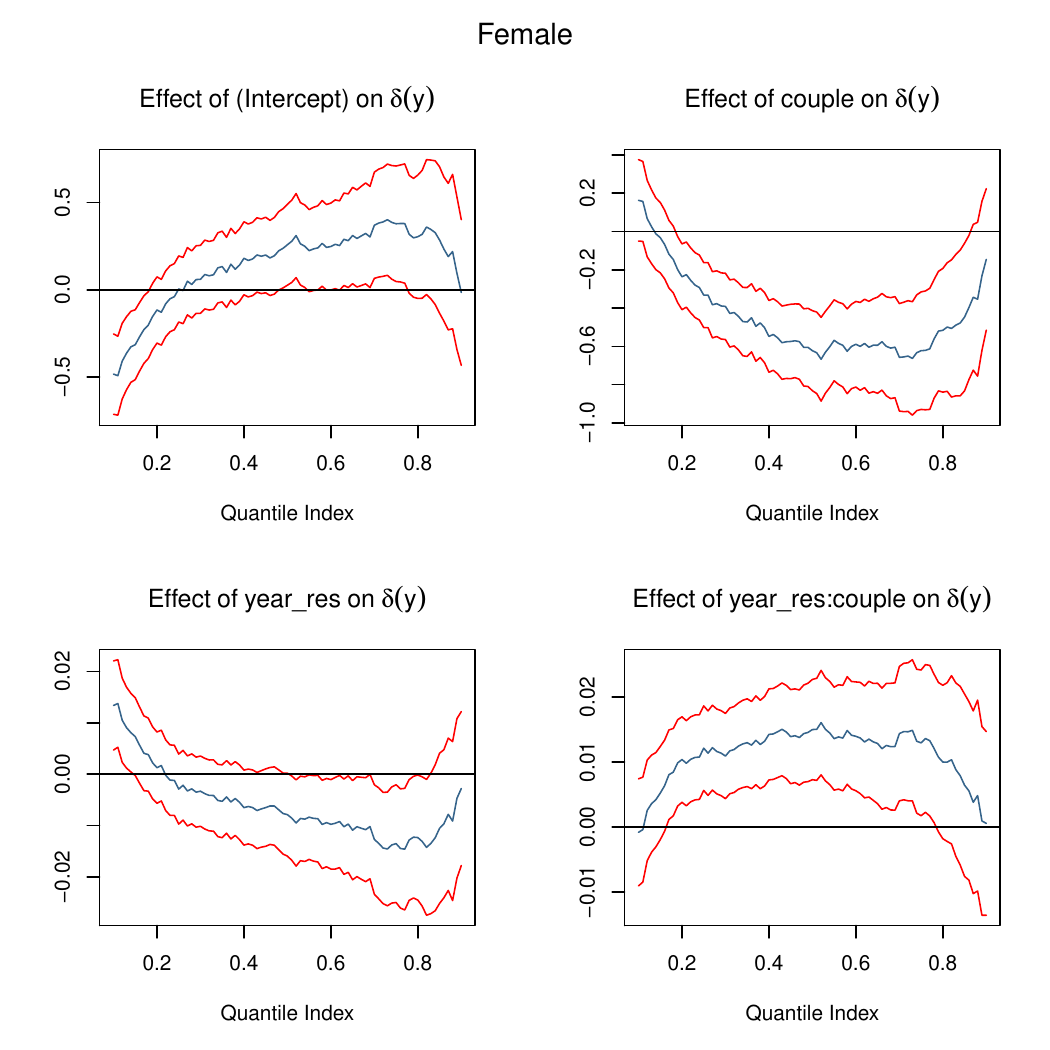}
    \caption{Estimates and 95\% confidence bands for coefficients of the selection sorting function: specification 4}\label{fig:coef_sorting_spec4}
\end{figure}

\begin{figure}
    \includegraphics[height=0.49\textwidth, width=0.49\textwidth,page=13]{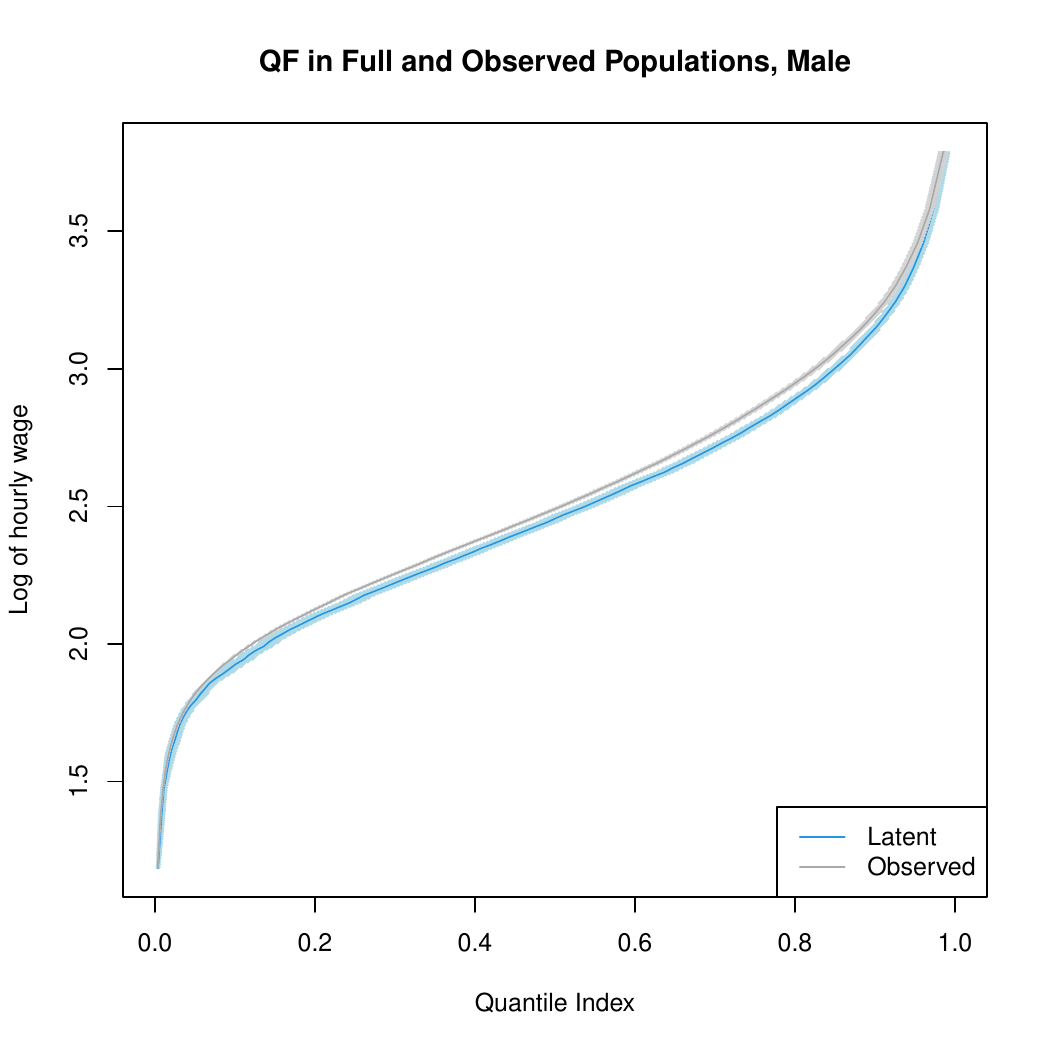}
    \includegraphics[height=0.49\textwidth, width=0.49\textwidth,page=15]{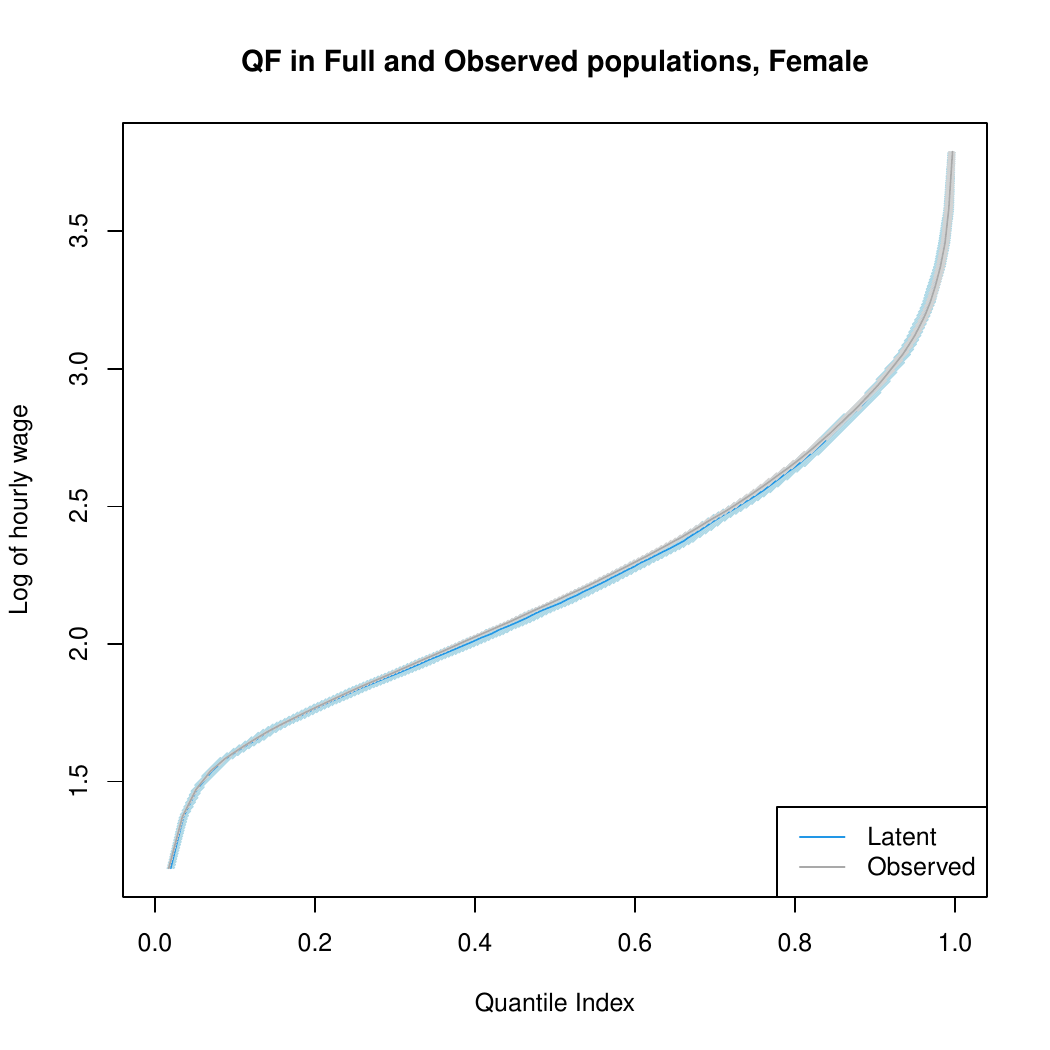}
    \caption{Estimates and 95\% confidence bands for the quantiles of observed and offered (latent) wages: specification 4}\label{fig:dist_latent_spec1}
\end{figure}

\begin{figure}
    \includegraphics[height=0.33\textwidth, width=0.32\textwidth,page=13]{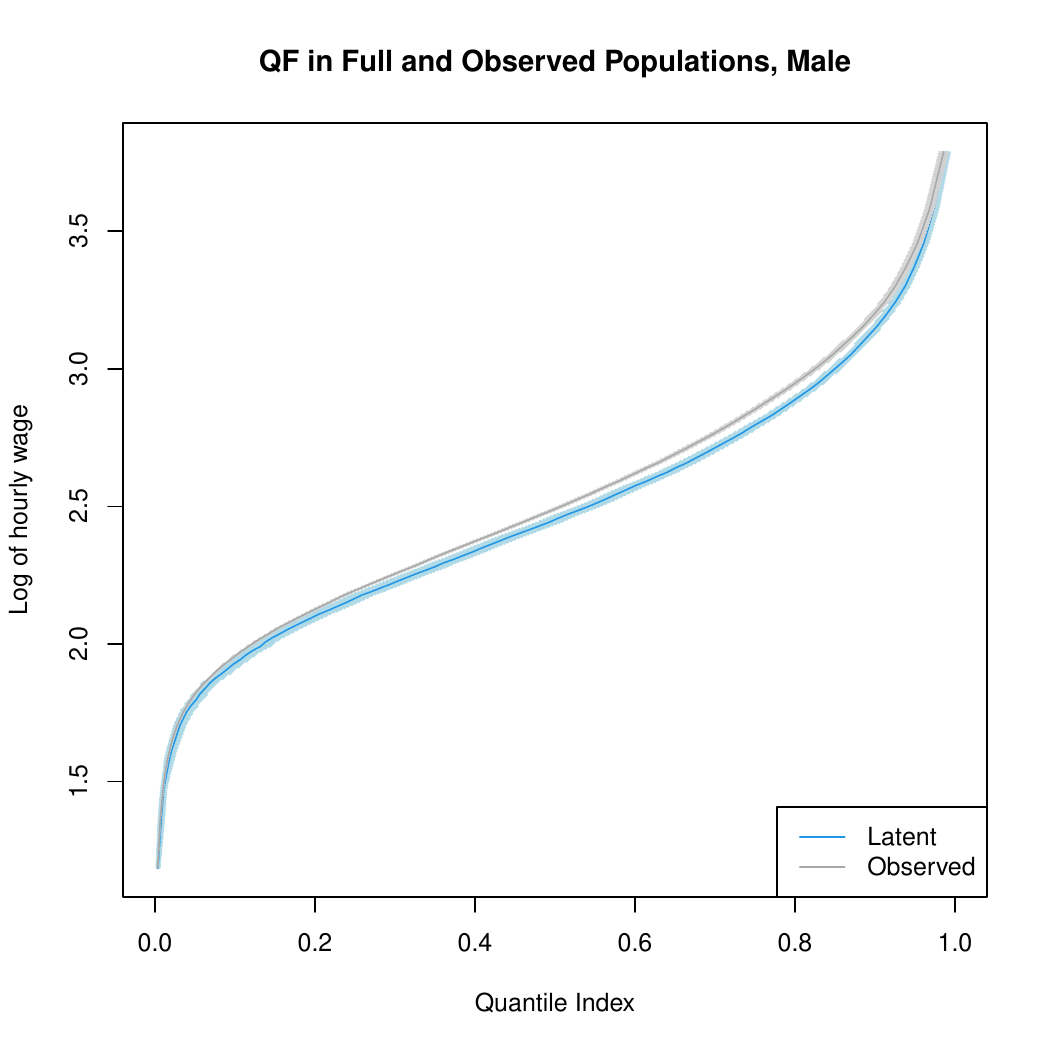}
    \includegraphics[height=0.33\textwidth, width=0.32\textwidth,page=15]{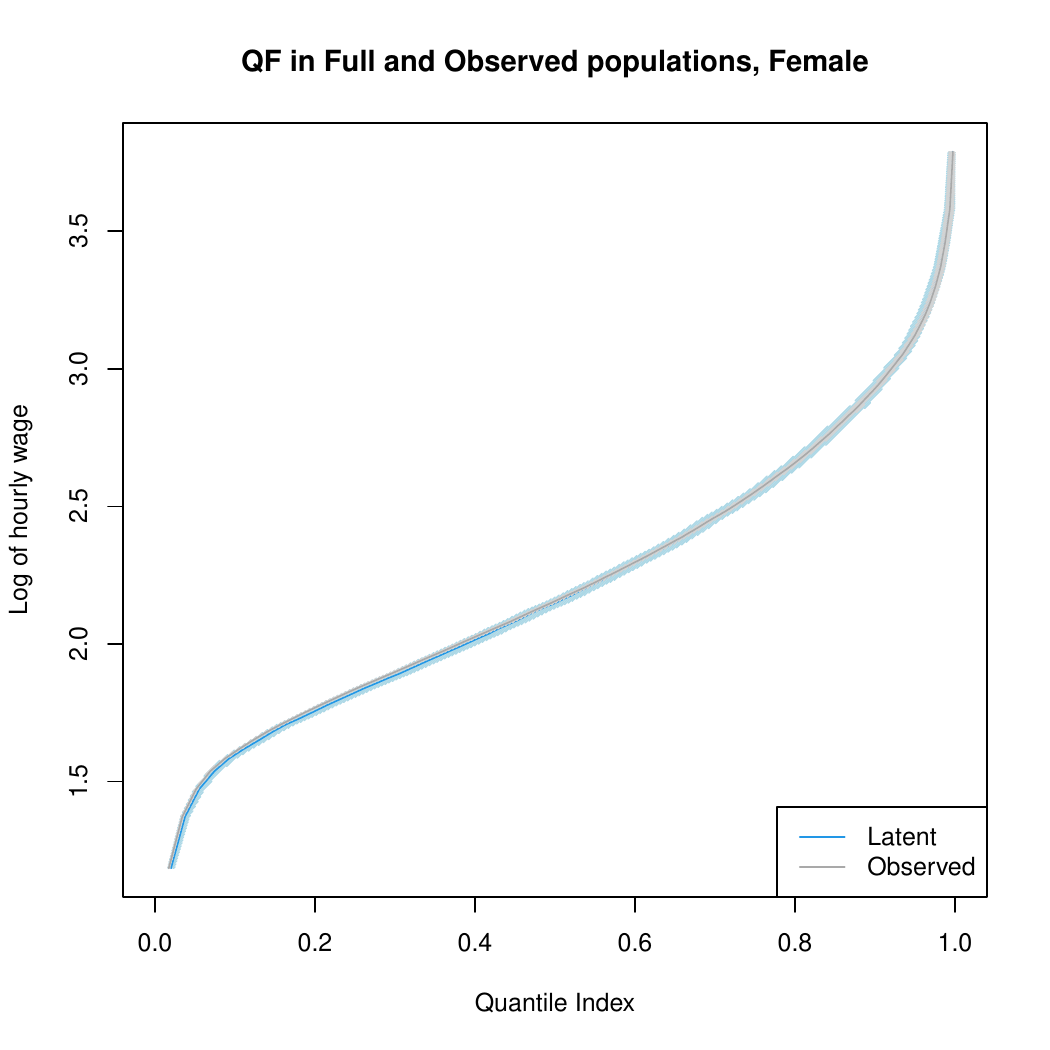}
    \includegraphics[height=0.33\textwidth, width=0.32\textwidth,page=21]{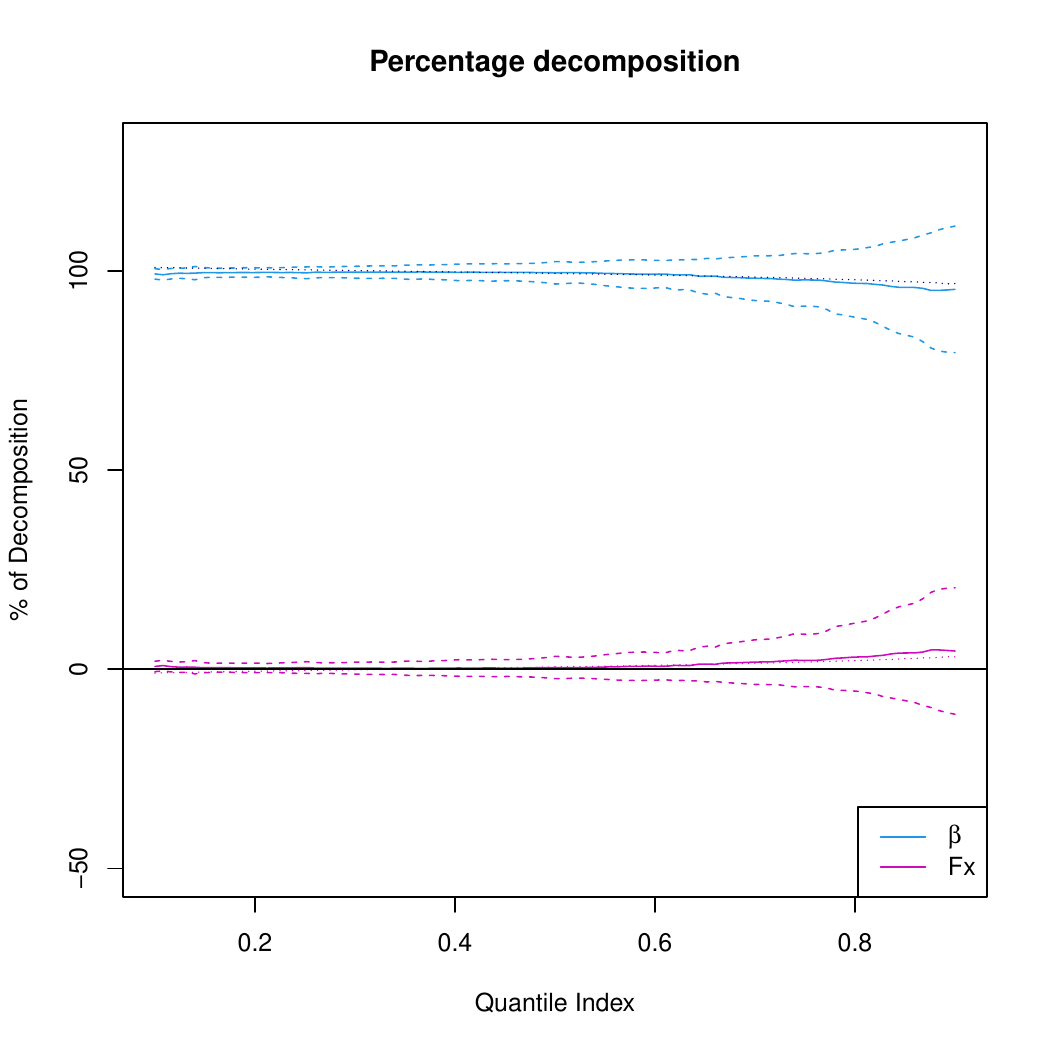}
    \caption{Estimates and 95\% confidence bands for the quantiles of observed and offered (latent) wages and decomposition of offered wages between women and men: specification 1}\label{fig:decomp_spec4}
\end{figure}

\begin{figure}
    \includegraphics[height=0.33\textwidth, width=0.32\textwidth,page=13]{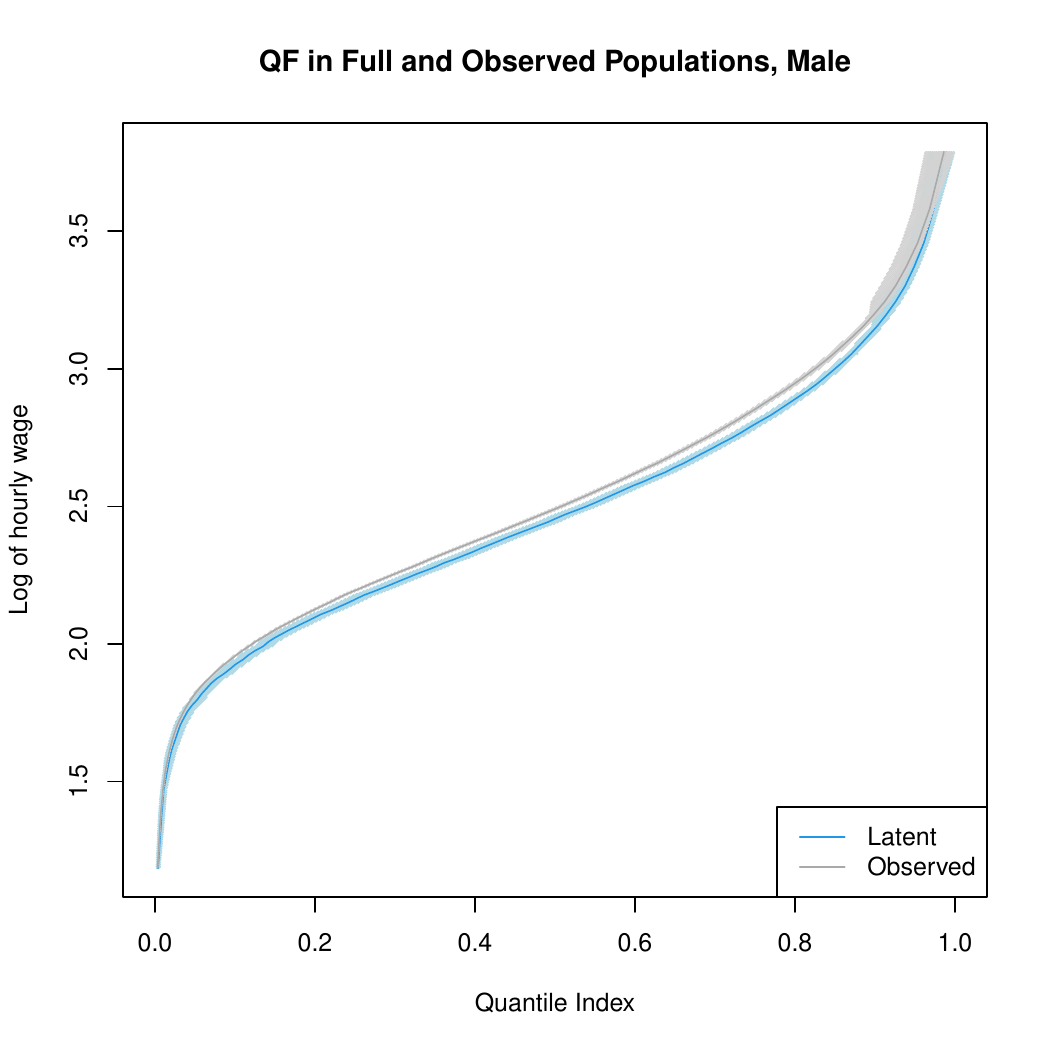}
    \includegraphics[height=0.33\textwidth, width=0.32\textwidth,page=15]{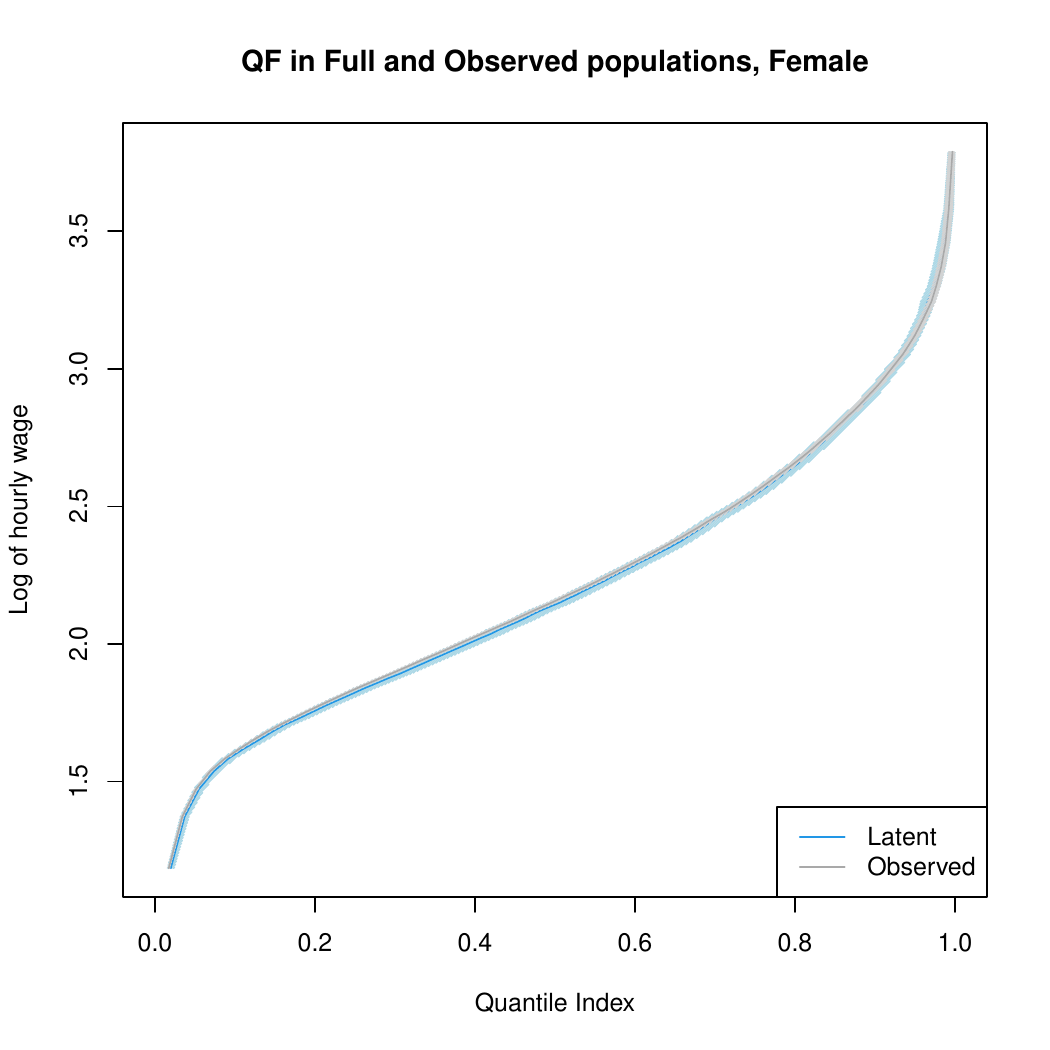}
    \includegraphics[height=0.33\textwidth, width=0.32\textwidth,page=21]{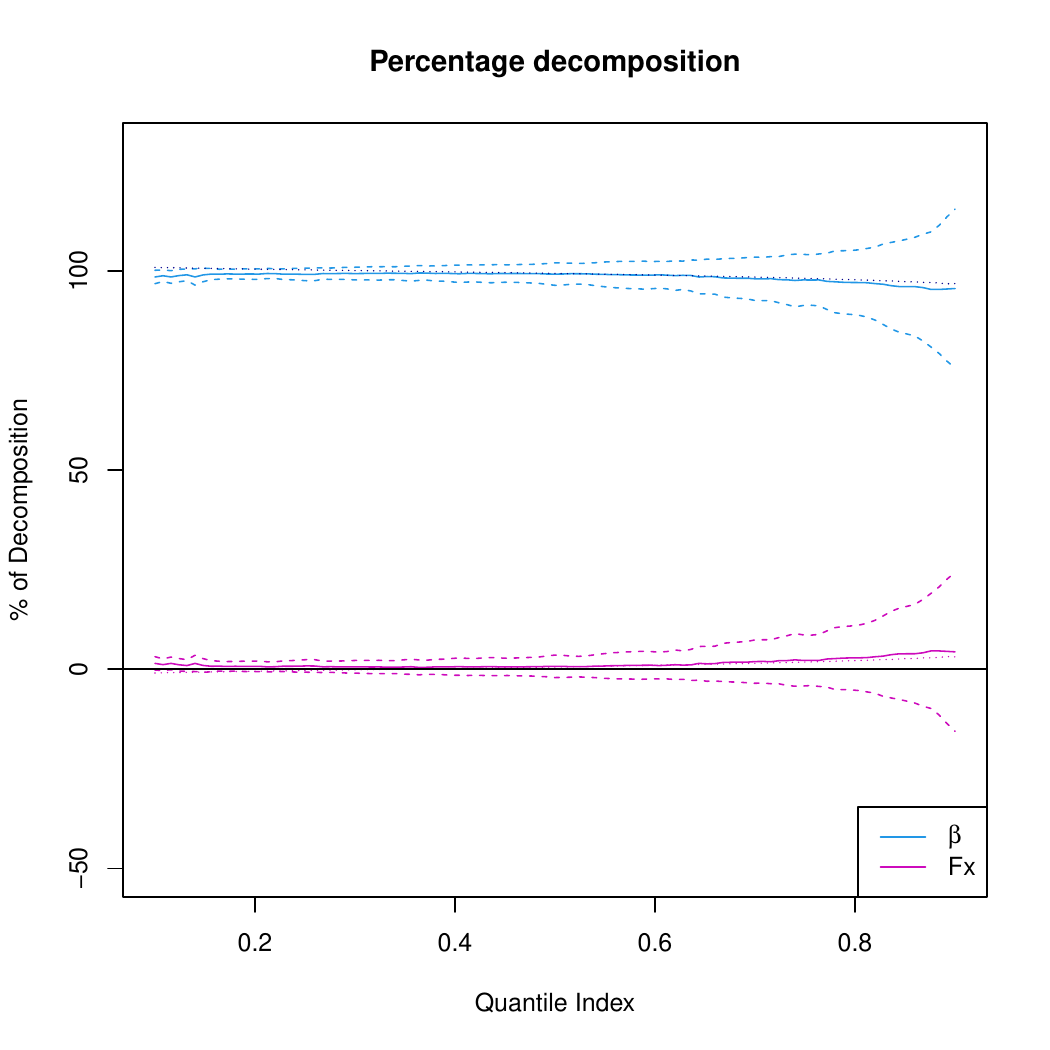}
    \caption{Estimates and 95\% confidence bands for the quantiles of observed and offered (latent) wages and decomposition of offered wages between women and men: specification 2}\label{fig:decomp_spec2}
\end{figure}

%
This section includes the following empirical results omitted from the main text:
\begin{itemize}
\item Table \ref{table:eq}: estimates and standard errors of the coefficients of the employment equation. 
\item Table \ref{table:dec_emp}: estimates and 95\% confidence intervals for components of employment rate decomposition between women and men. 
\item Figures \ref{fig:coef_men_spec1}--\ref{fig:coef_women_spec4}: estimates and 95\% confidence bands of the coefficients of education and marital status indicators in the wage equation for men and women in all the specifications.\footnote{The dashed lines report estimates of $\beta/\sigma$ in the outcome equation of the HSM.} 
\item Figures \ref{fig:coef2_men_spec1} and \ref{fig:coef2_women_spec1}: estimates and 95\% confidence bands of the coefficients of fertility in the wage equation for men and women in specification 4. 
\item Figures \ref{fig:rho_men_spec4}-\ref{fig:coef_sorting_spec4}: estimates and 95\% confidence bands of the sorting equation.  
\item Figure \ref{fig:dist_latent_spec1}: estimates and 95\% confidence bands for the quantiles of the observed and latent wages in specification 4. 
\item Figure \ref{fig:decomp_spec2}-\ref{fig:decomp_spec4}: estimates and 95\% confidence bands for the decomposition of offered wages between women and men for specifications 1--3. 
\item Figure  \ref{fig:adec}: estimates and 95\% confidence bands for decomposition between men and women with aggregated selection effects. 
\item Figure \ref{fig:dec_spec2to4}: estimates and 95\% confidence bands for the quantiles of observed wages and decomposition between men and women in specifications 1--3. 
\item Figures \ref{fig:cdec_spec1}--\ref{fig:cdec_spec4}: estimates and 95\% confidence bands for components of wage decomposition between women and men. 
\item Figures \ref{fig:cdec_men_spec1} and \ref{fig:cdec_women_spec1}:  estimates and 95\% confidence bands for components of wage decomposition between first and second half of sample period for men and women in specification 2.
\end{itemize}

\begin{table}\caption{Estimates of Coefficients of the Employment Equation}\label{table:eq}
    \begin{tabular}{lcc|lcc}
        \hline\hline
       Variable & Male & Female & Variable & Male & Female \\
      \hline
      educ16 &  0.25 &  0.06 &    numch34 & -0.18 & -0.63 \\
       & (0.01) & (0.01) & & (0.02) & (0.01) \\
      educ1718 &  0.46 &  0.20 &  numch510 & -0.18 & -0.33 \\
       & (0.02) & (0.01) &  & (0.01) & (0.01) \\
      educ1920 &  0.42 &  0.16 & numch1116 & -0.16 & -0.15\\
       & (0.03) & (0.02) & & (0.01) & (0.01) \\
      educ2122 &  0.74 &  0.28 & numch1718 & -0.02 & -0.11 \\
       & (0.02) & (0.02) & & (0.03) & (0.02) \\
      educ23 &  0.51 &  0.15 & benefit & -0.35 & -0.42   \\
       & (0.02) & (0.02) &  & (0.01) & (0.01) \\
      couple & -4.02 & -8.14 & benefit$\times$couple &  0.87 &  1.40 \\
       & (0.09) & (0.08) & & (0.02) & (0.02)\\
       numch1 & -0.16 & -0.90 & constant &   2.50 &  2.75 \\
       & (0.02) & (0.02) & & (0.08) & (0.07) \\
      numch2 & -0.18 & -0.77 &  \\
       & (0.02) & (0.02) &  \\
       \hline\hline
               \multicolumn{6}{l}{\begin{minipage}{9cm}\footnotesize Notes: standard errors in parentheses. The specification includes a quartic polynomial in age, region indicators and survey year indicators. \end{minipage}}
    \end{tabular}
\end{table} 

\begin{table}\caption{Employment rate decomposition between men and women}\label{table:dec_emp}
   \begin{tabular} {c c | c c} \\
        \hline\hline
        
        \multirow{2}{*}{Employment (\%)} & &  \multicolumn{2}{c}{Structure ($\pi$)} \\
        & & Male & Female \\
        \hline
        \multirow{4}{*}{Composition ($F_Z$)} & \multirow{2}{*}{Male} & 83 & 59 \\
                                                                                                       & & (82, 83) & (59, 59) \\
                                                                  & \multirow{2}{*}{Female} & 83 & 66 \\
                                                                                                       & & (83, 83) & (66, 66) \\
                                                                  \hline\hline
          \multicolumn{4}{l}{\footnotesize{95\% bootstrap confidence intervals in parentheses}}                                                         
    \end{tabular}
\end{table}

\begin{figure}
    \includegraphics[height=0.33\textwidth, width=0.32\textwidth,page=13]{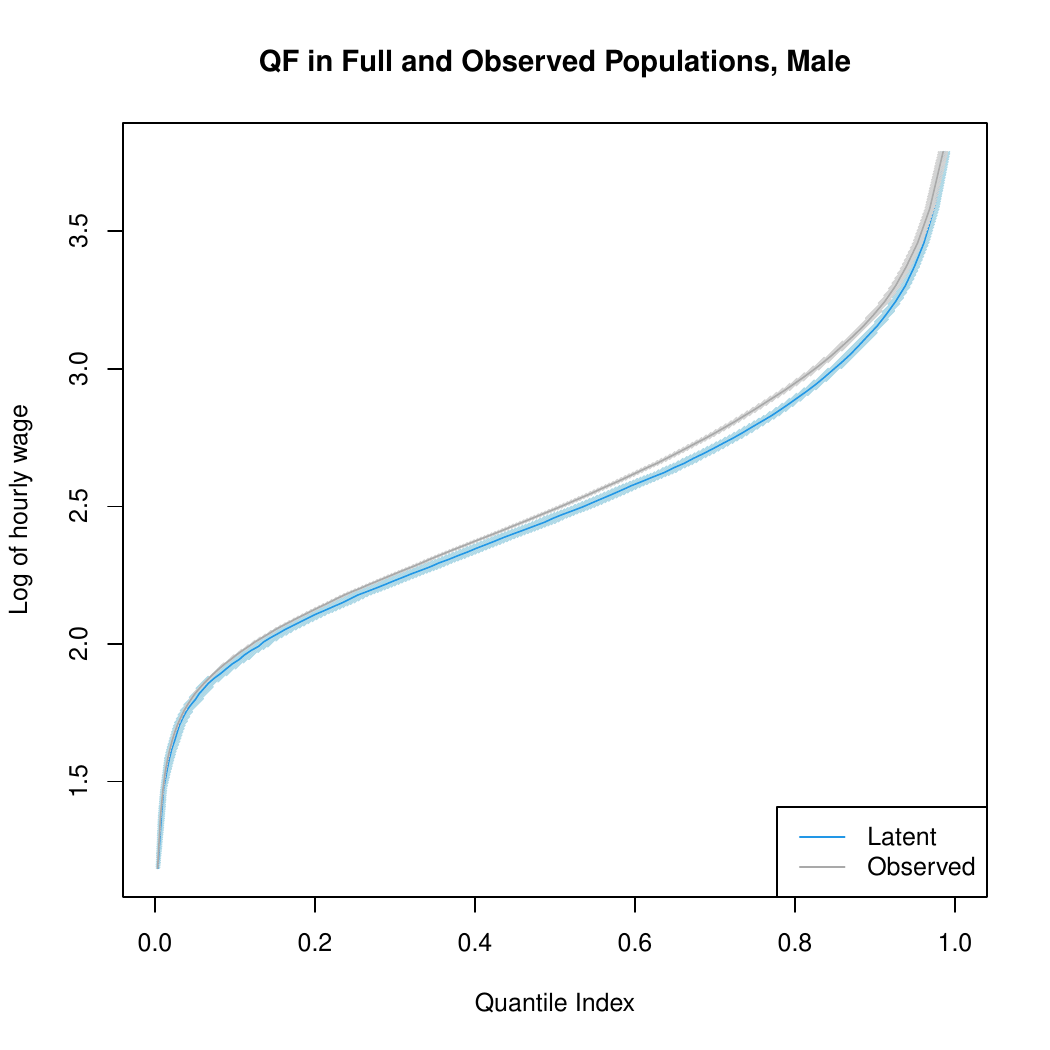}
    \includegraphics[height=0.33\textwidth, width=0.32\textwidth,page=15]{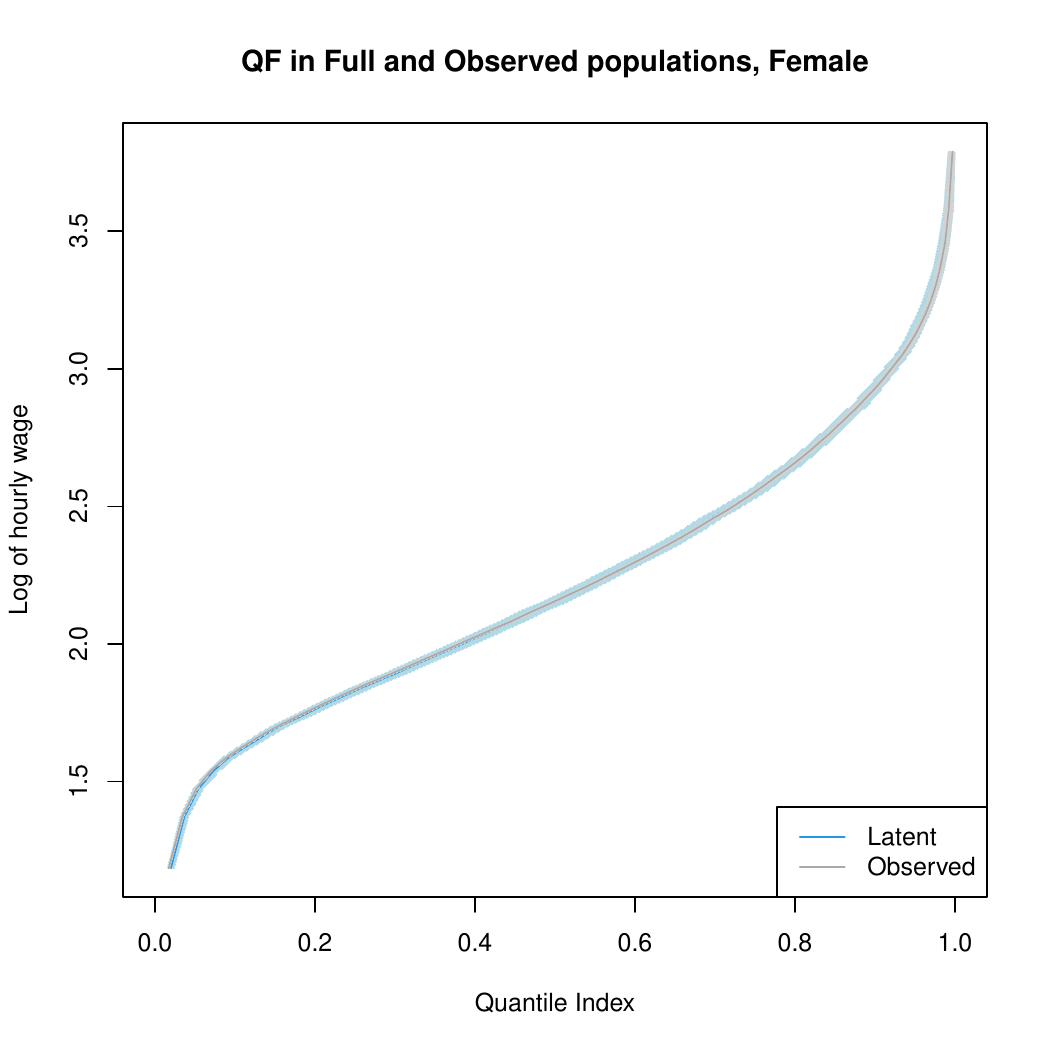}
    \includegraphics[height=0.33\textwidth, width=0.32\textwidth,page=21]{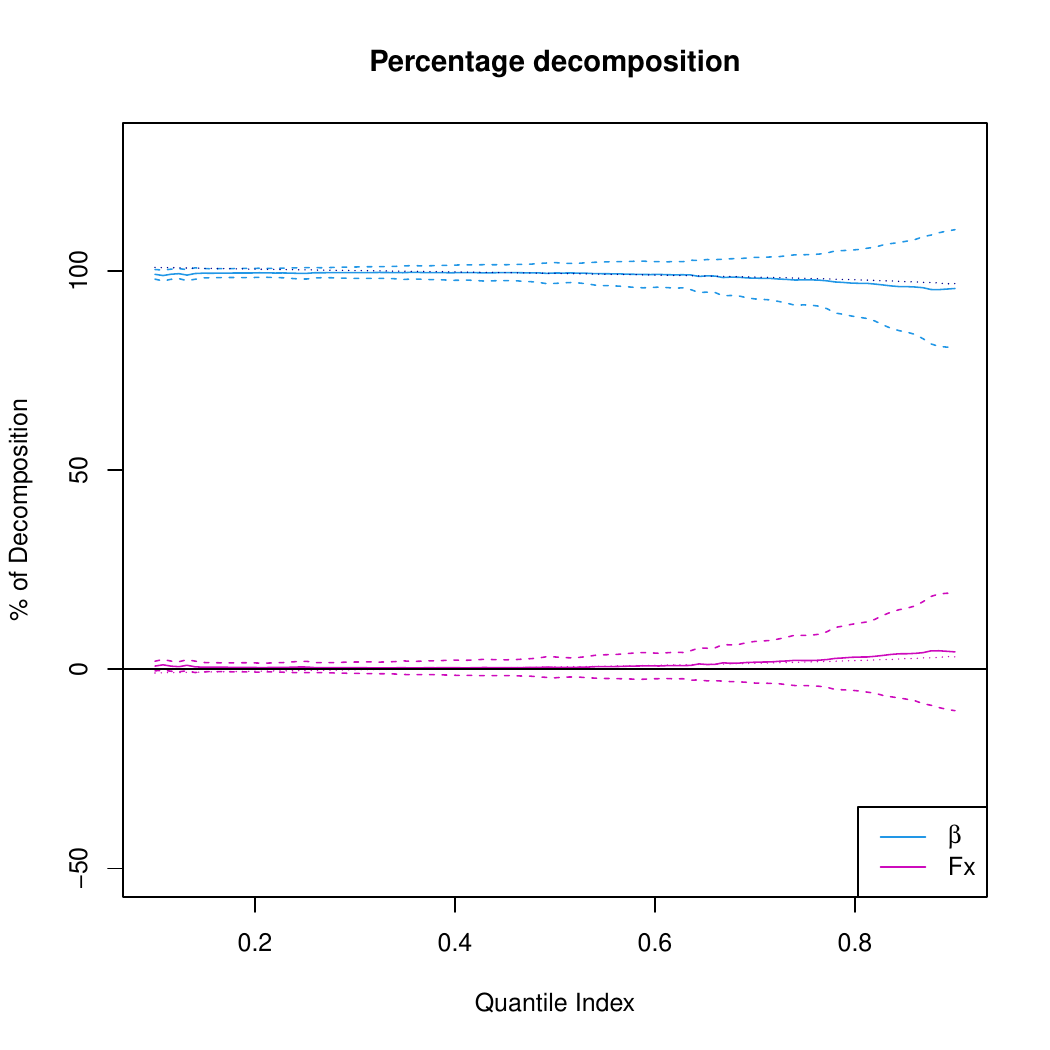}
    \caption{Estimates and 95\% confidence bands for the quantiles of observed and offered (latent) wages and decomposition of offered wages between women and men: specification 3}
\end{figure}


\begin{figure}
        \includegraphics[height=.49\textwidth, width=.49\textwidth,page=47]{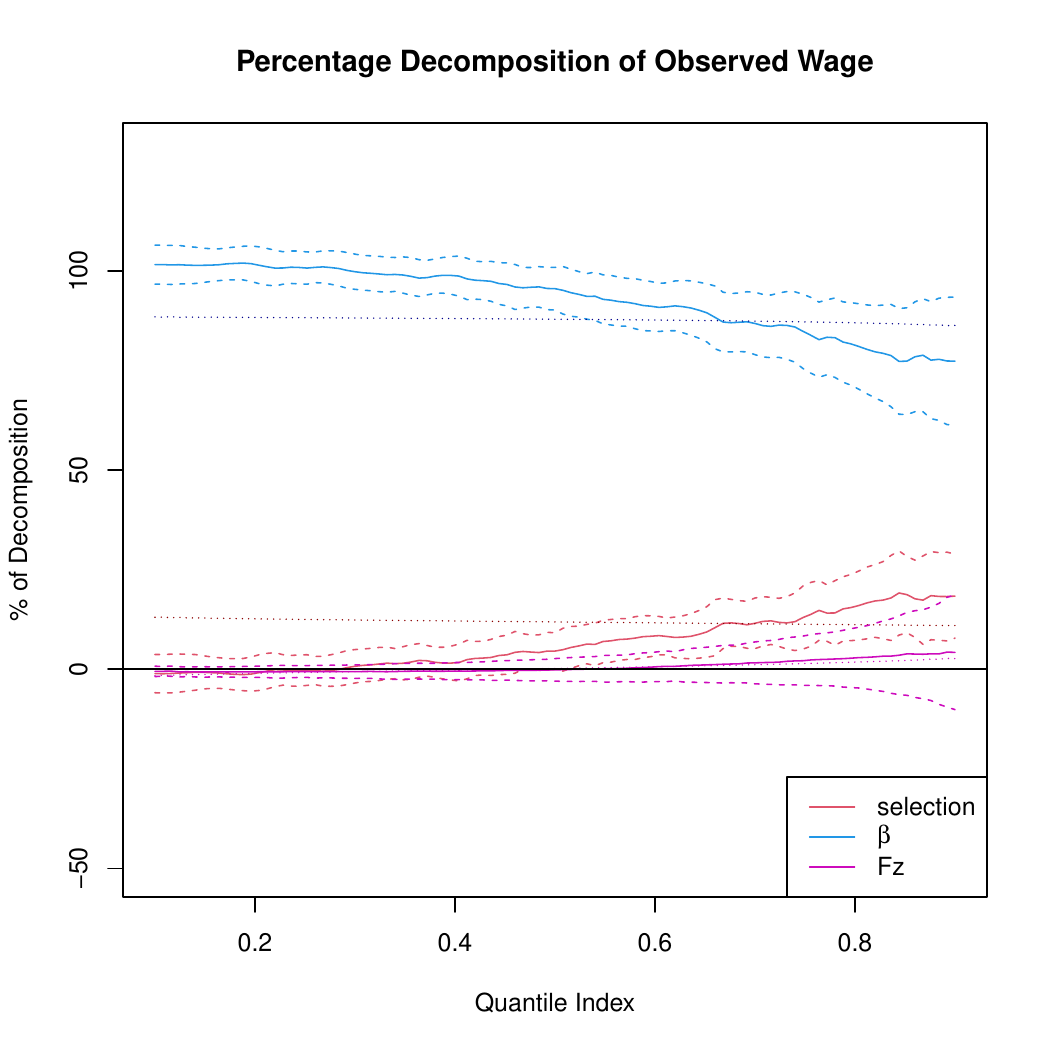}
        \includegraphics[height=.49\textwidth, width=.49\textwidth,page=47]{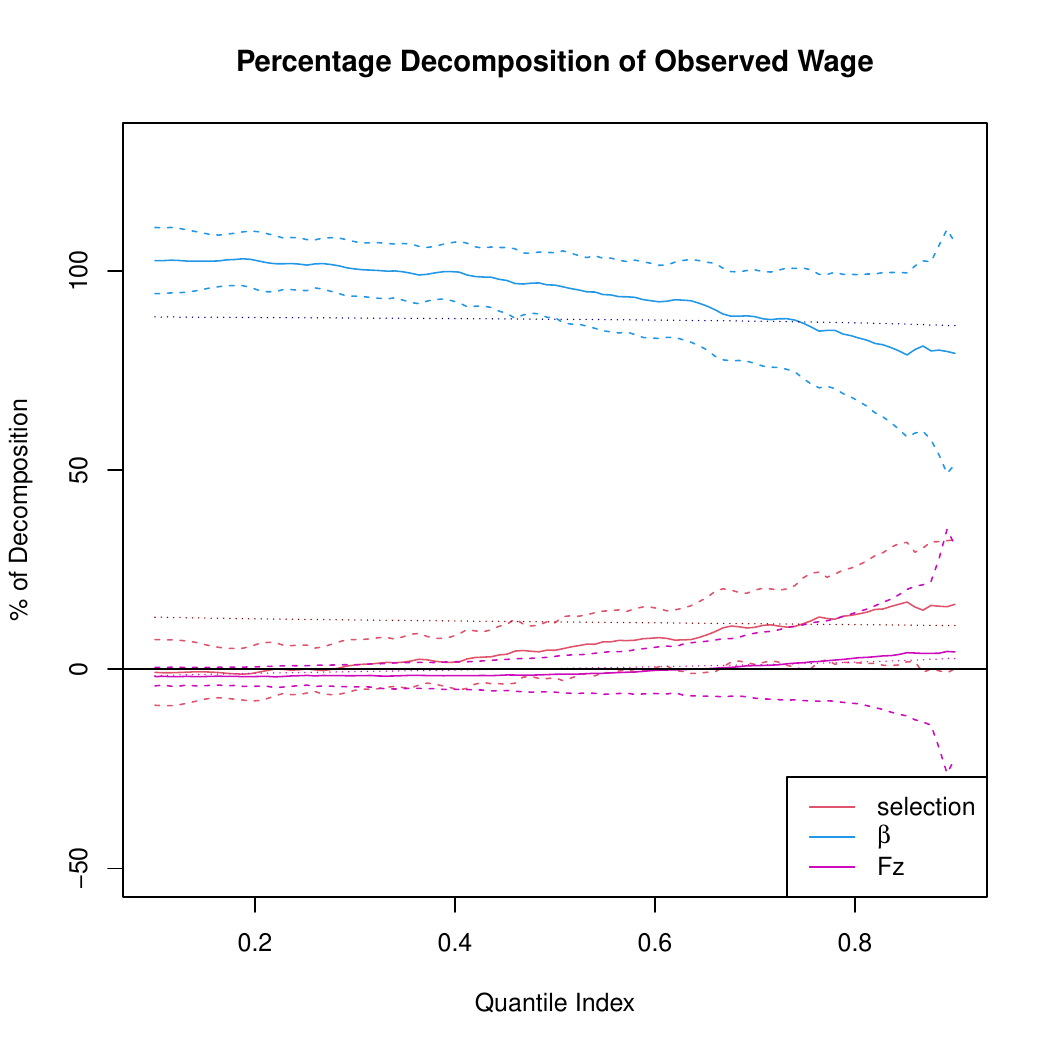}
        \includegraphics[height=.49\textwidth, width=.49\textwidth,page=47]{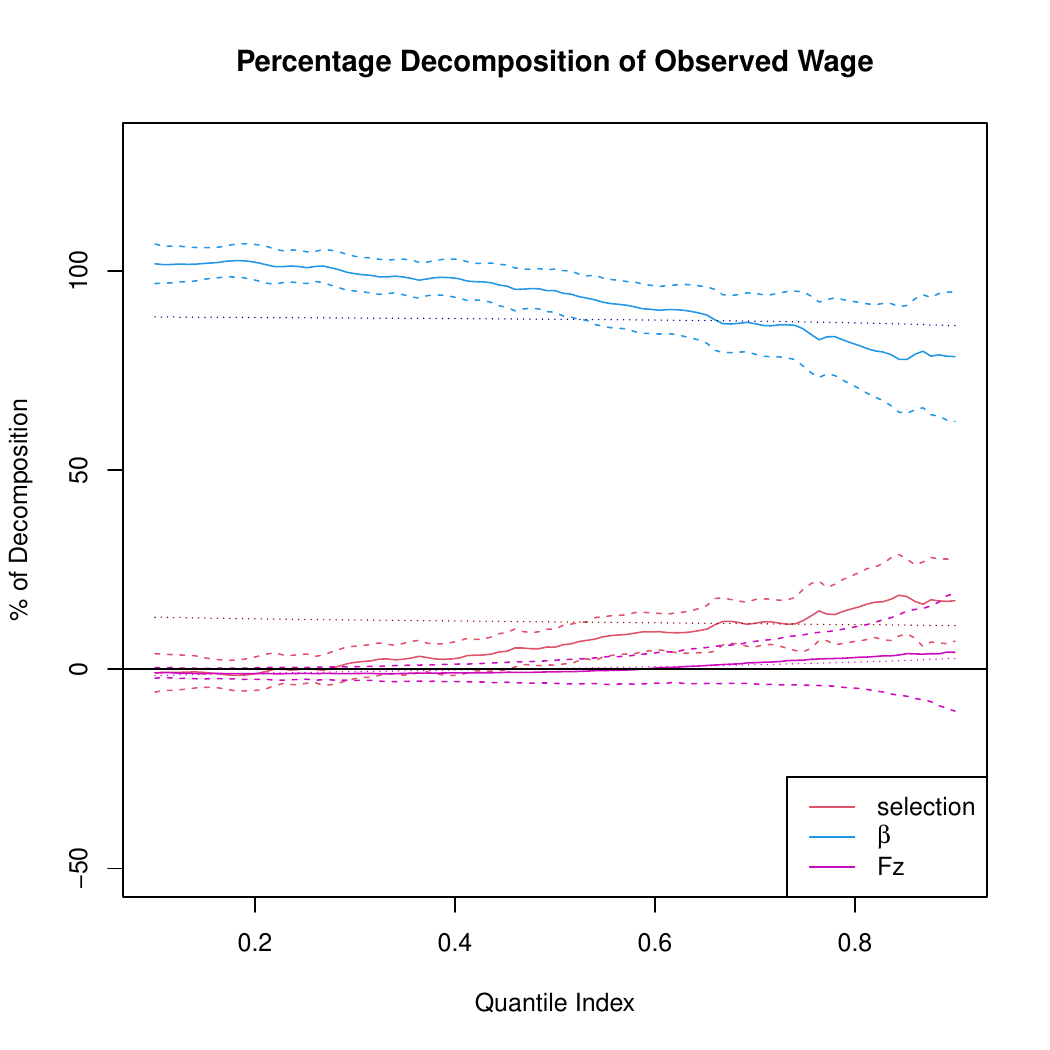}
        \includegraphics[height=.49\textwidth, width=.49\textwidth,page=47]{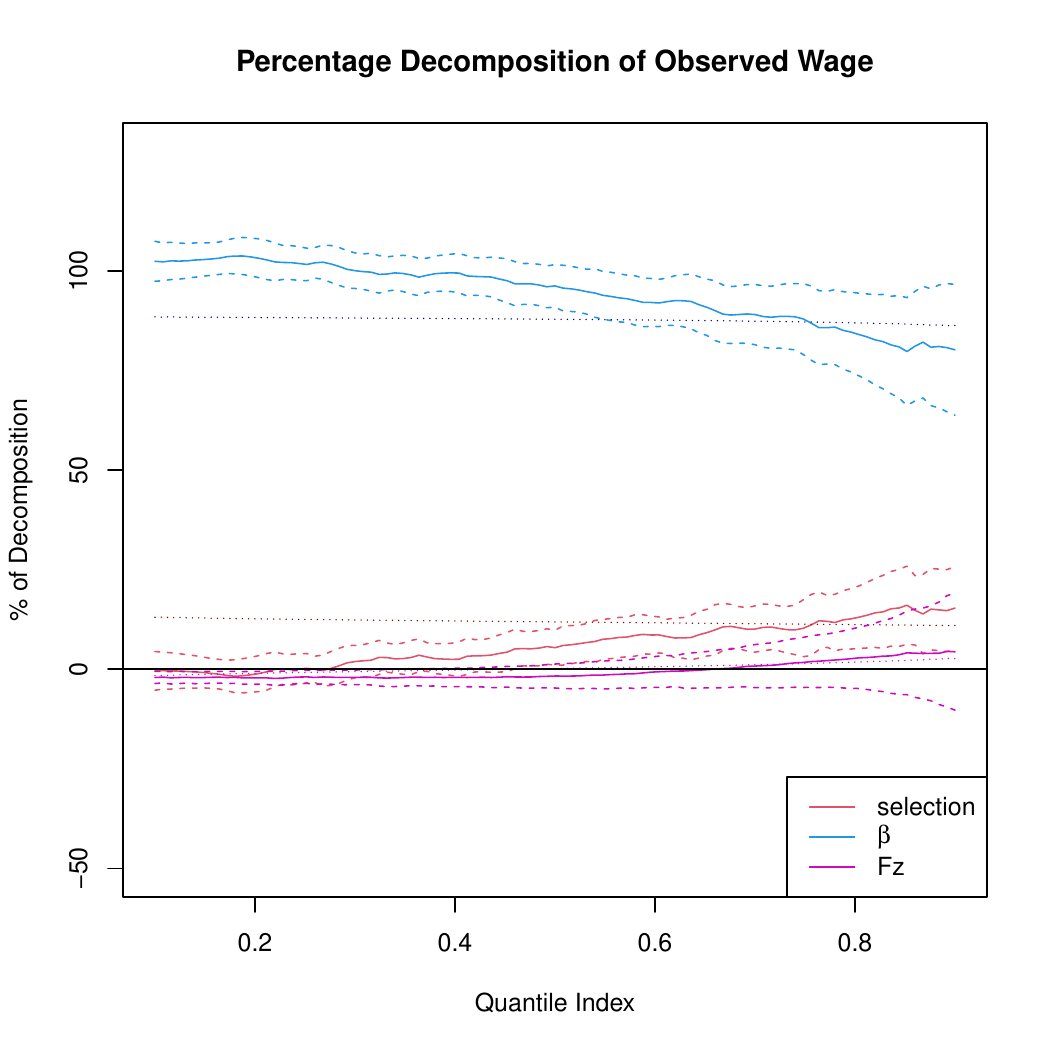}
    \caption{Estimates and 95\% confidence bands for the quantiles of observed wages and decomposition between men and women with aggregated selection effect: (upper left) specification 1, (upper right) specification 2, (bottom left) specification 3, and (bottom right) specification 4}\label{fig:adec}
\end{figure}

\begin{figure}
        \includegraphics[height=.32\textwidth, width=.32\textwidth,page=43]{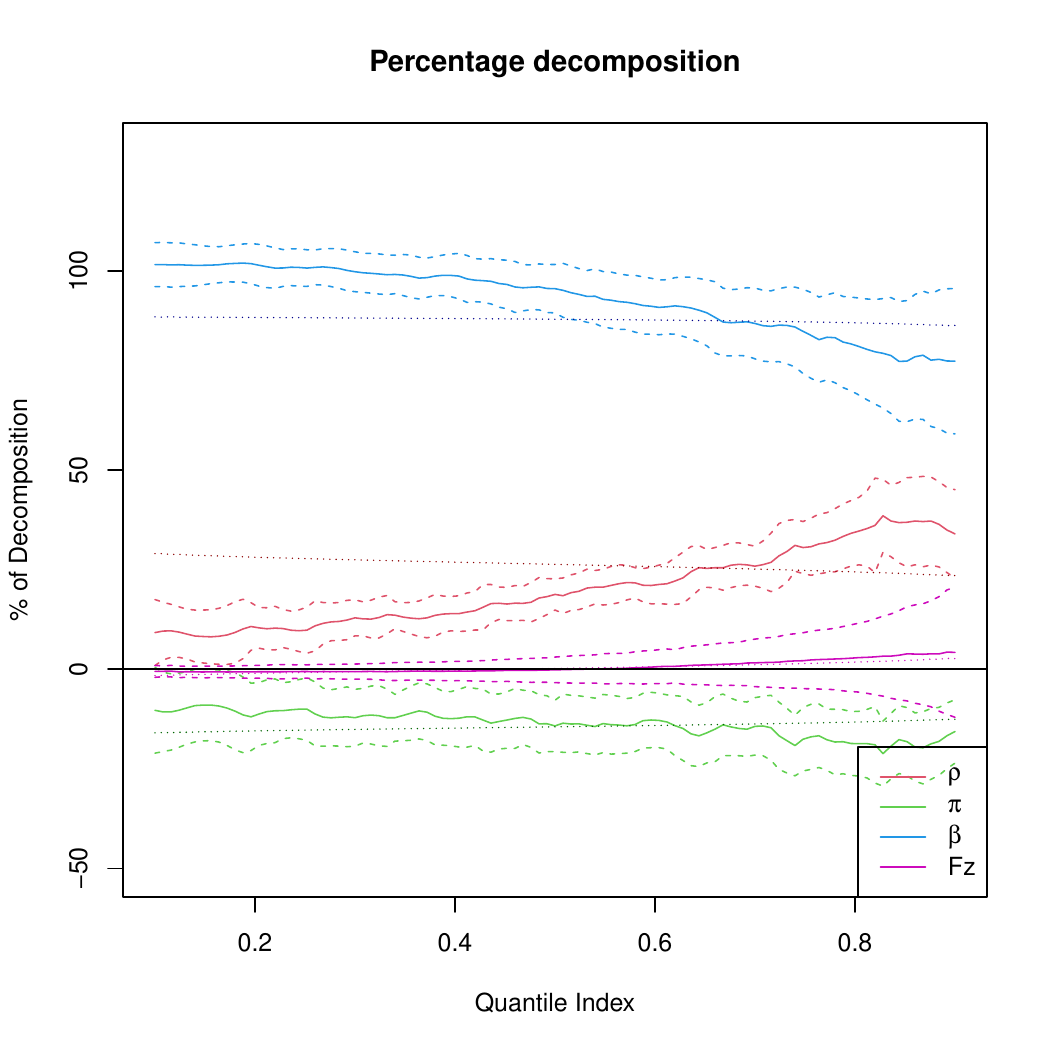}
        \includegraphics[height=.32\textwidth, width=.32\textwidth,page=43]{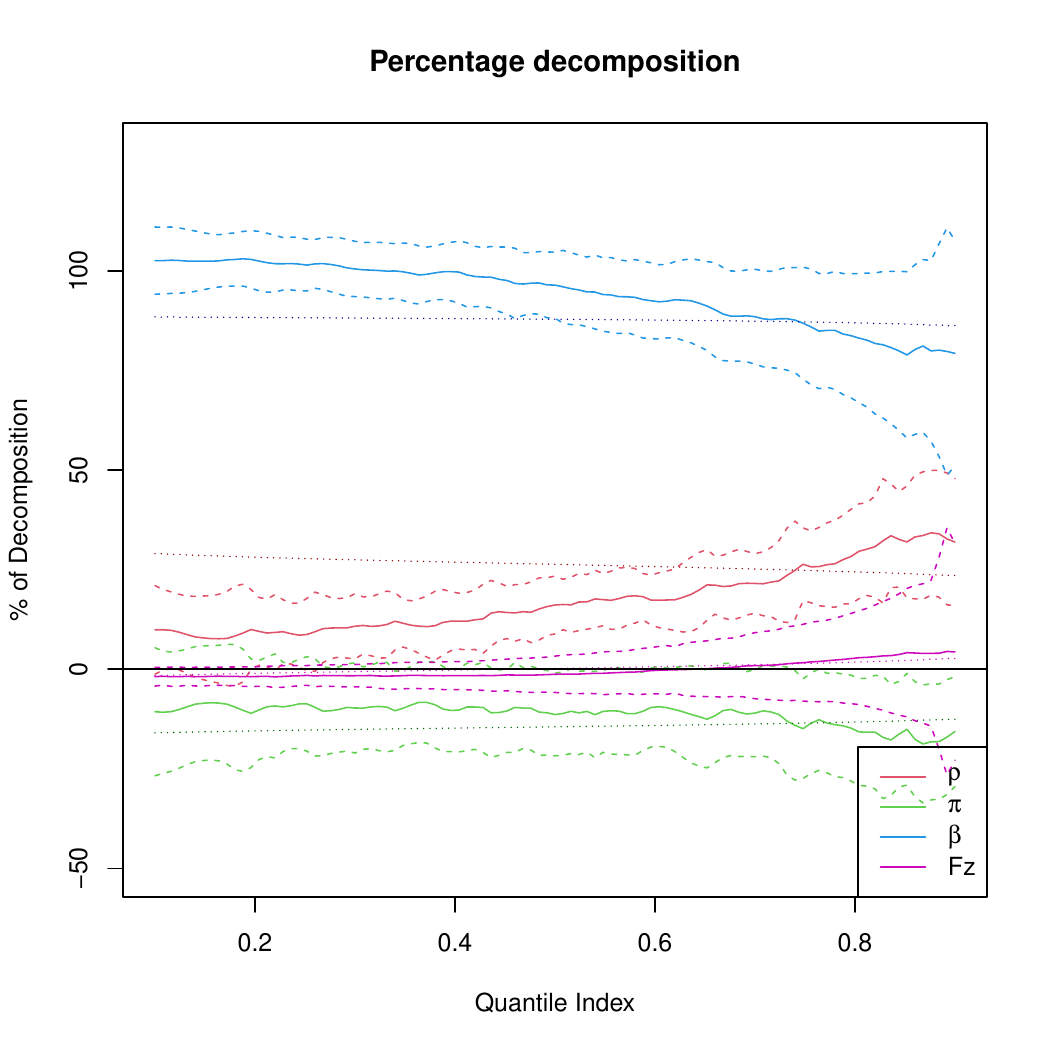}
        \includegraphics[height=.32\textwidth, width=.32\textwidth,page=43]{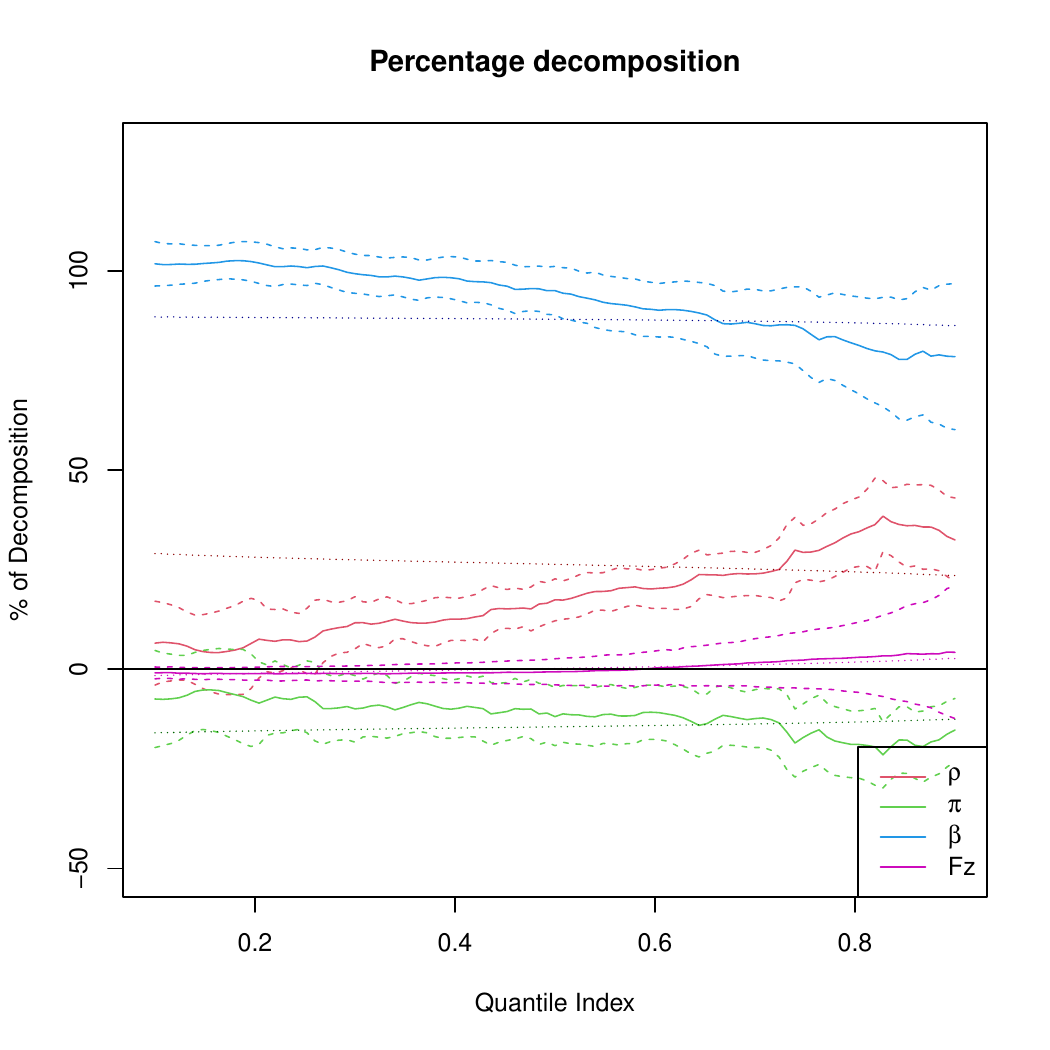}
    \caption{Estimates and 95\% confidence bands for the quantiles of observed wages and decomposition between men and women: (left) specification 1, (middle) specification 2, and (right) specification 3}\label{fig:dec_spec2to4}
\end{figure}

\begin{figure}
    \includegraphics[height=0.45\textwidth, width=0.45\textwidth,page=29]{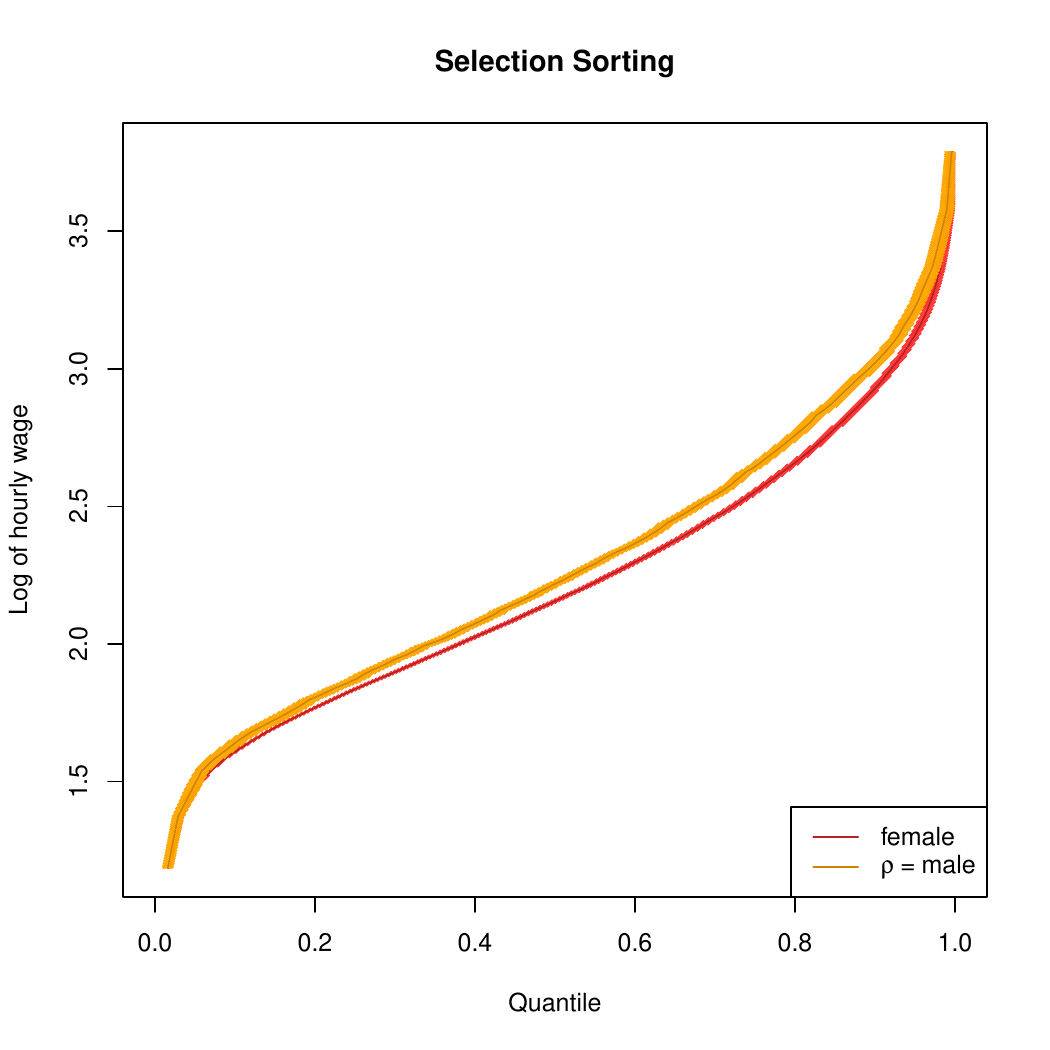}
    \includegraphics[height=0.45\textwidth, width=0.45\textwidth,page=30]{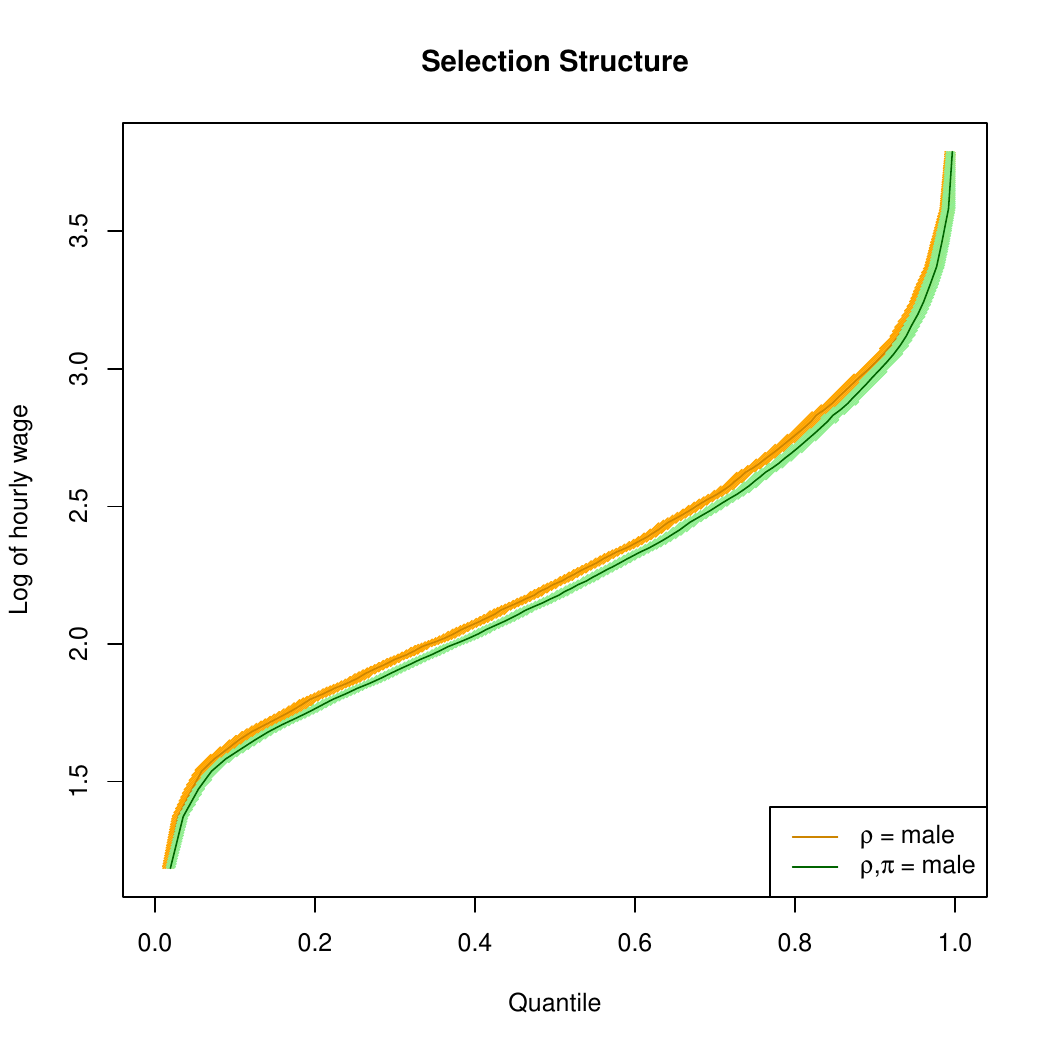}\\
    \includegraphics[height=0.45\textwidth, width=0.45\textwidth,page=31]{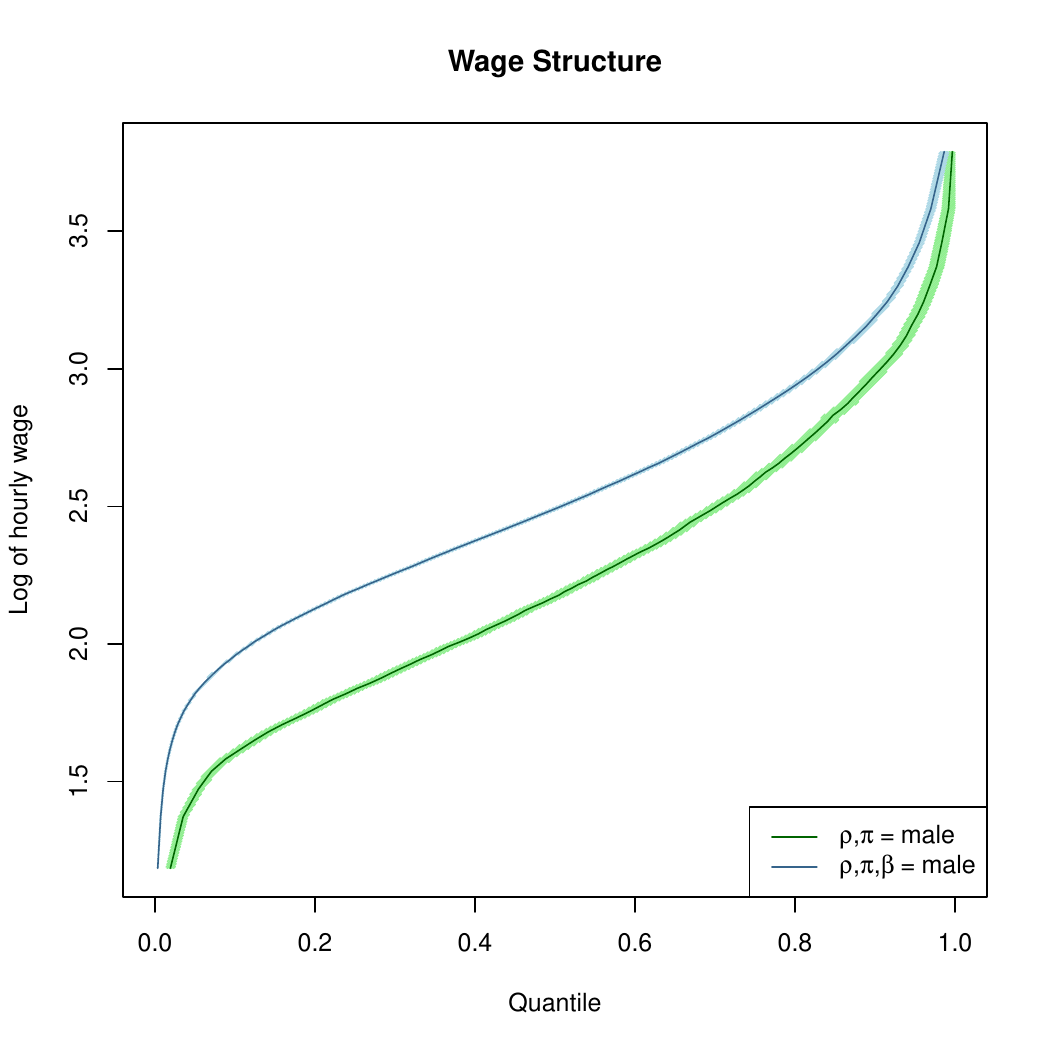}
    \includegraphics[height=0.45\textwidth, width=0.45\textwidth,page=32]{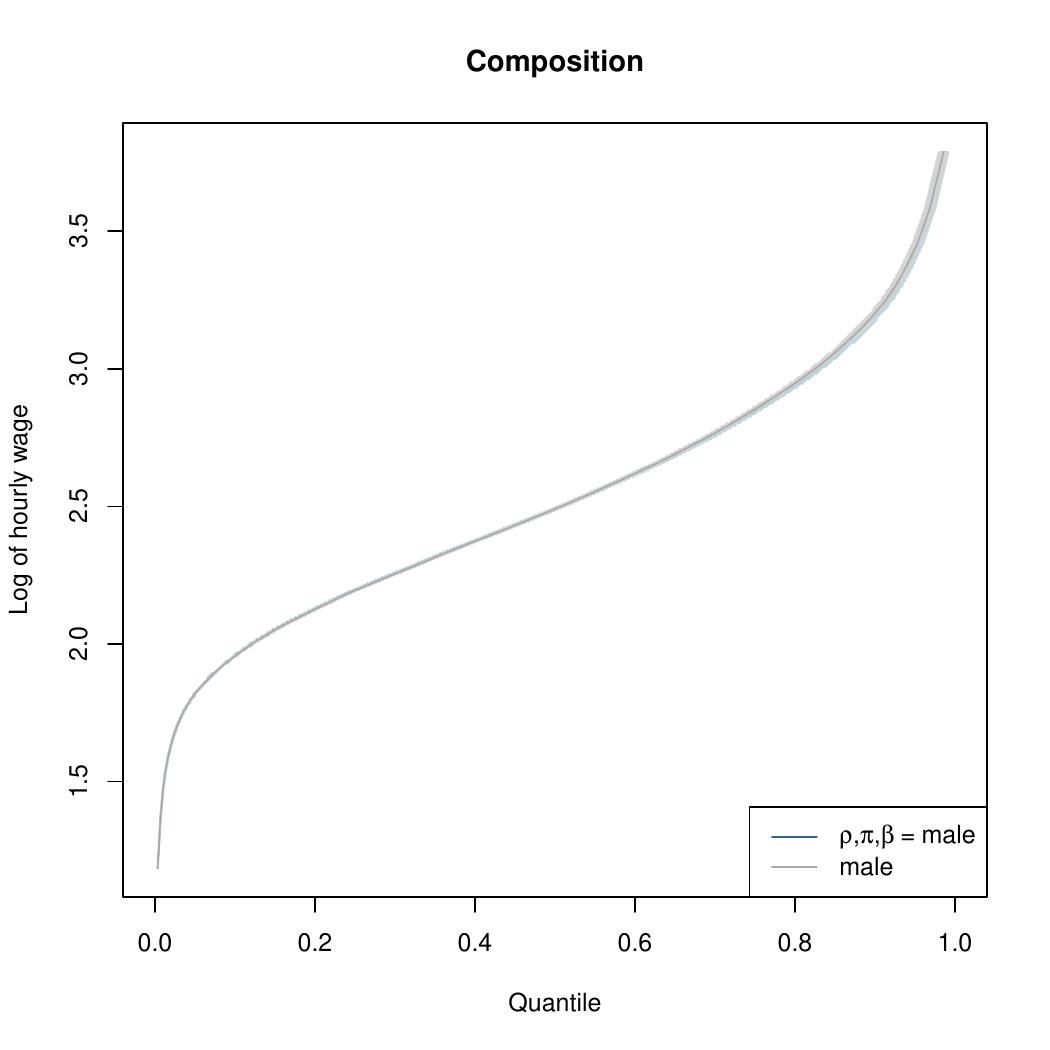}
    \caption{Estimates and 95\% confidence bands for components of wage decomposition between women and men in specification 1}\label{fig:cdec_spec1}
\end{figure}

\begin{figure}
    \includegraphics[height=0.45\textwidth, width=0.45\textwidth,page=29]{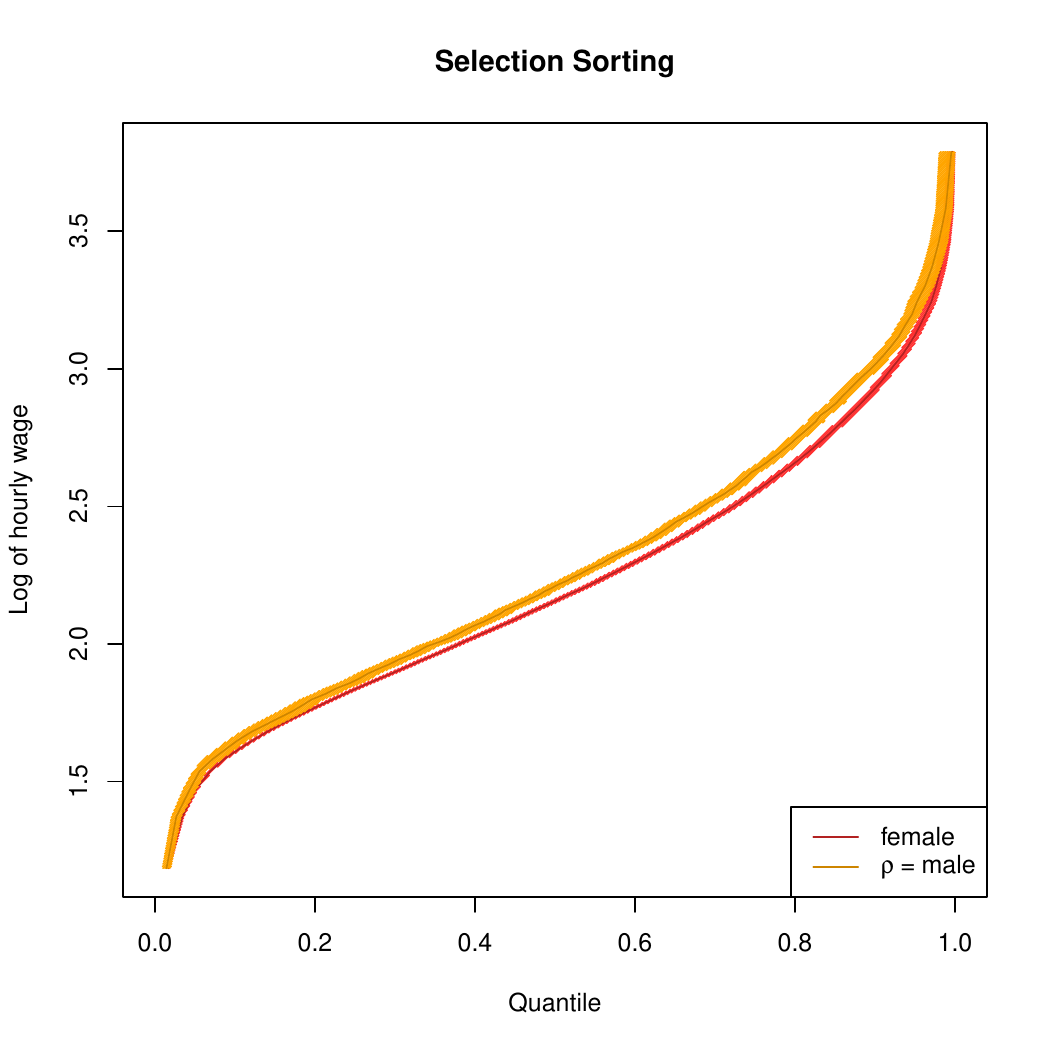}
    \includegraphics[height=0.45\textwidth, width=0.45\textwidth,page=30]{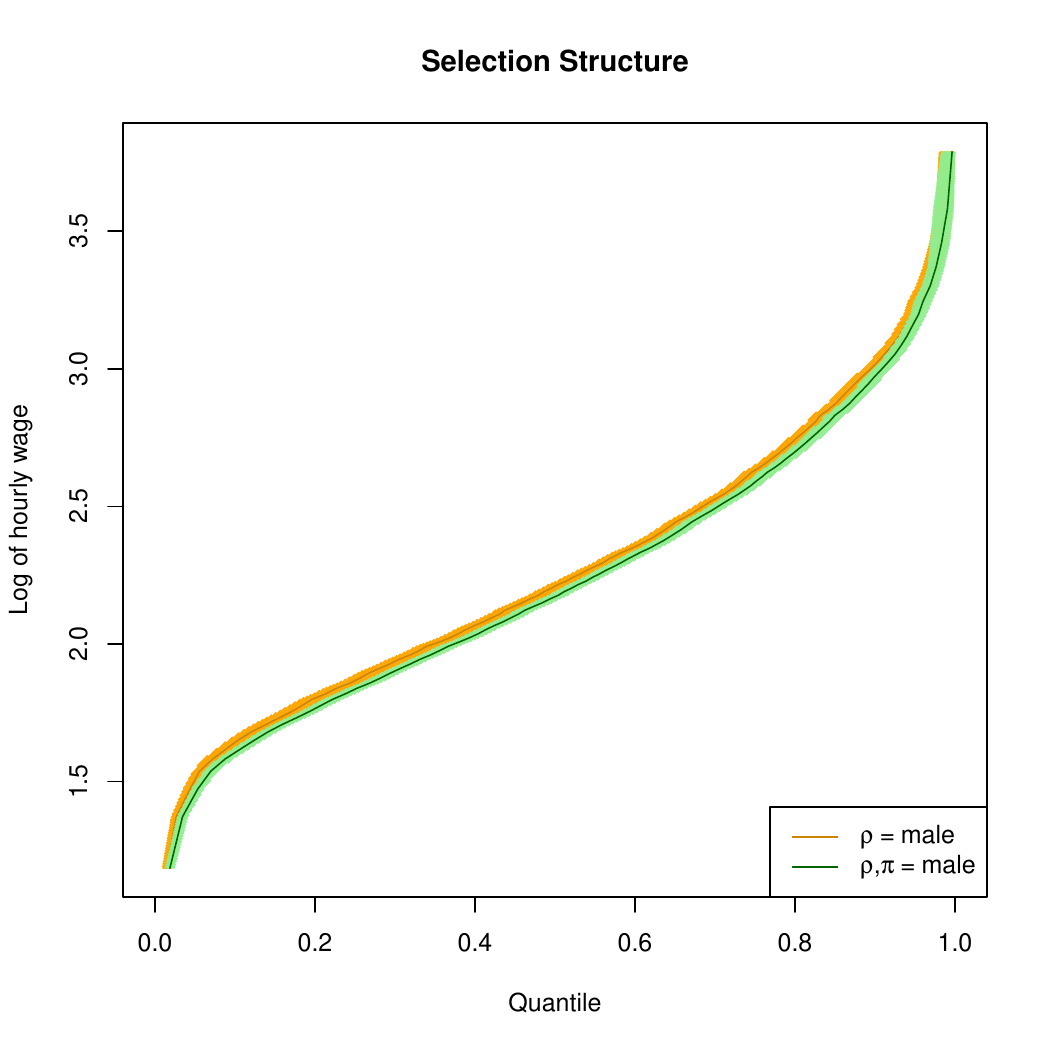}\\
    \includegraphics[height=0.45\textwidth, width=0.45\textwidth,page=31]{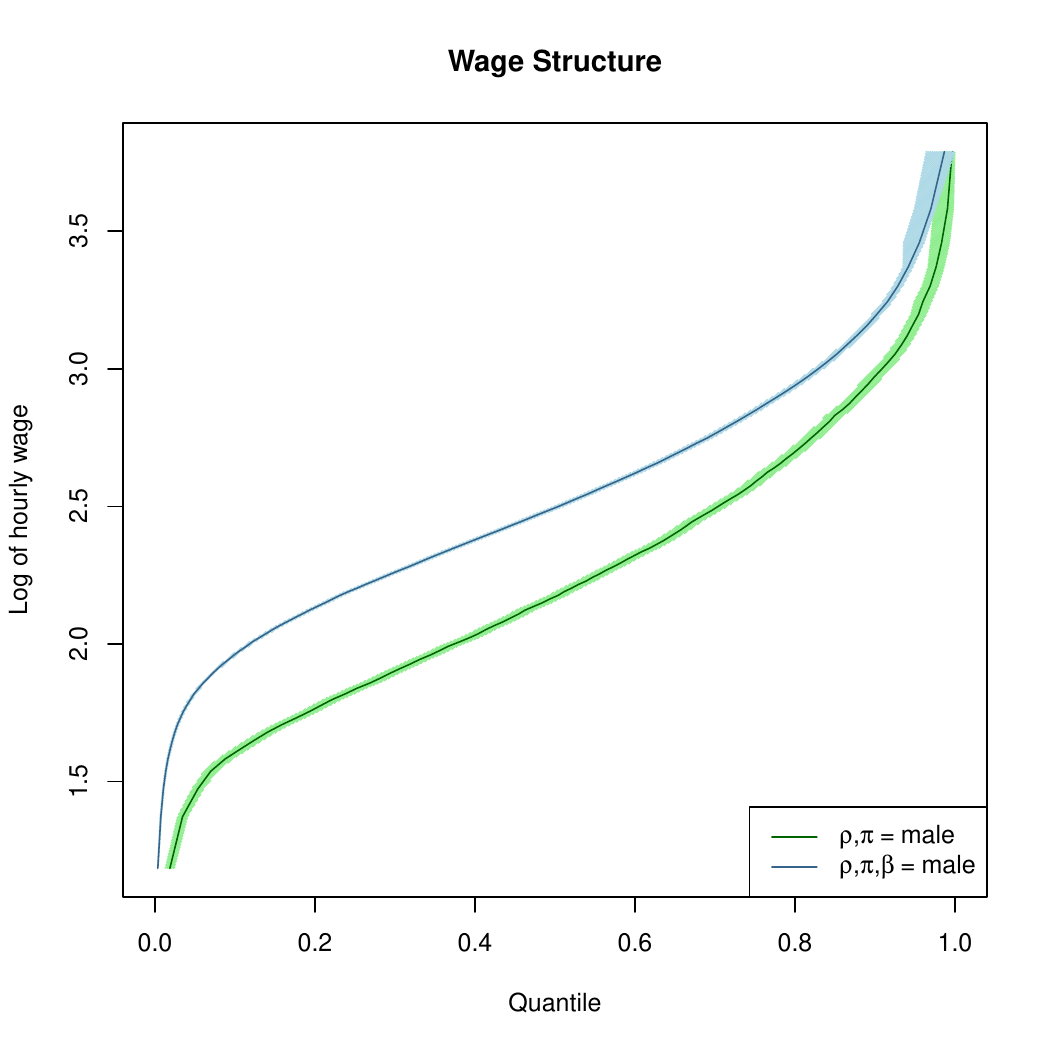}
    \includegraphics[height=0.45\textwidth, width=0.45\textwidth,page=32]{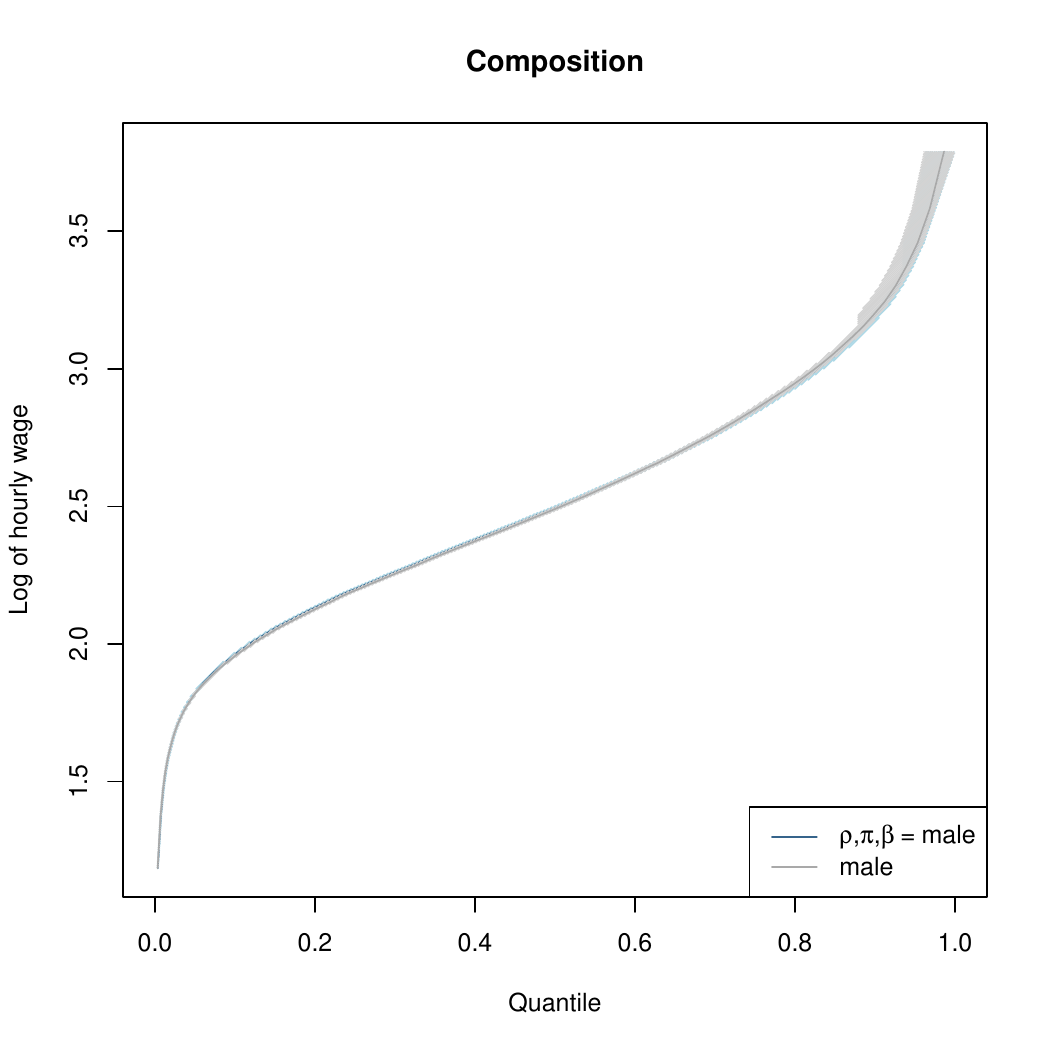}
    \caption{Estimates and 95\% confidence bands for components of wage decomposition between women and men in specification 2}
\end{figure}

\begin{figure}
    \includegraphics[height=0.45\textwidth, width=0.45\textwidth,page=29]{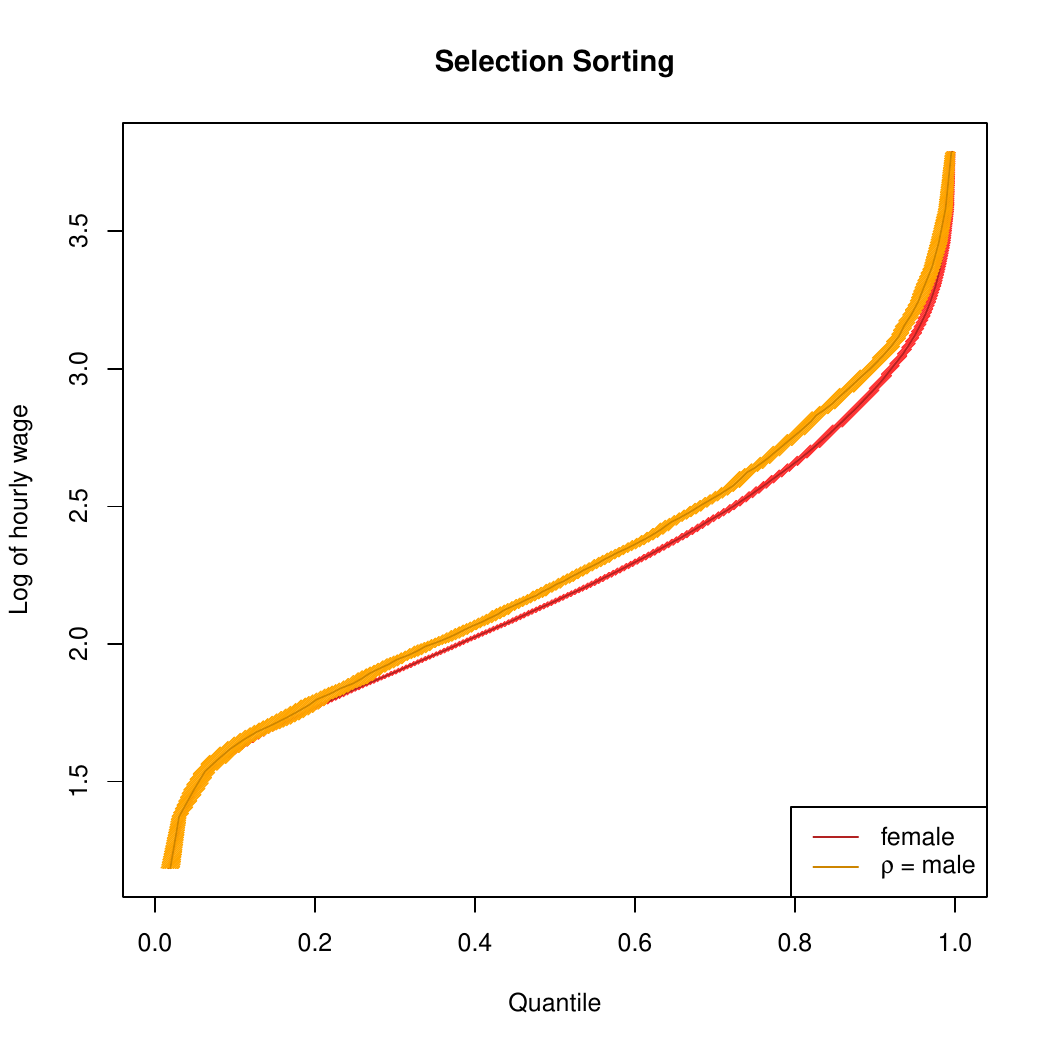}
    \includegraphics[height=0.45\textwidth, width=0.45\textwidth,page=30]{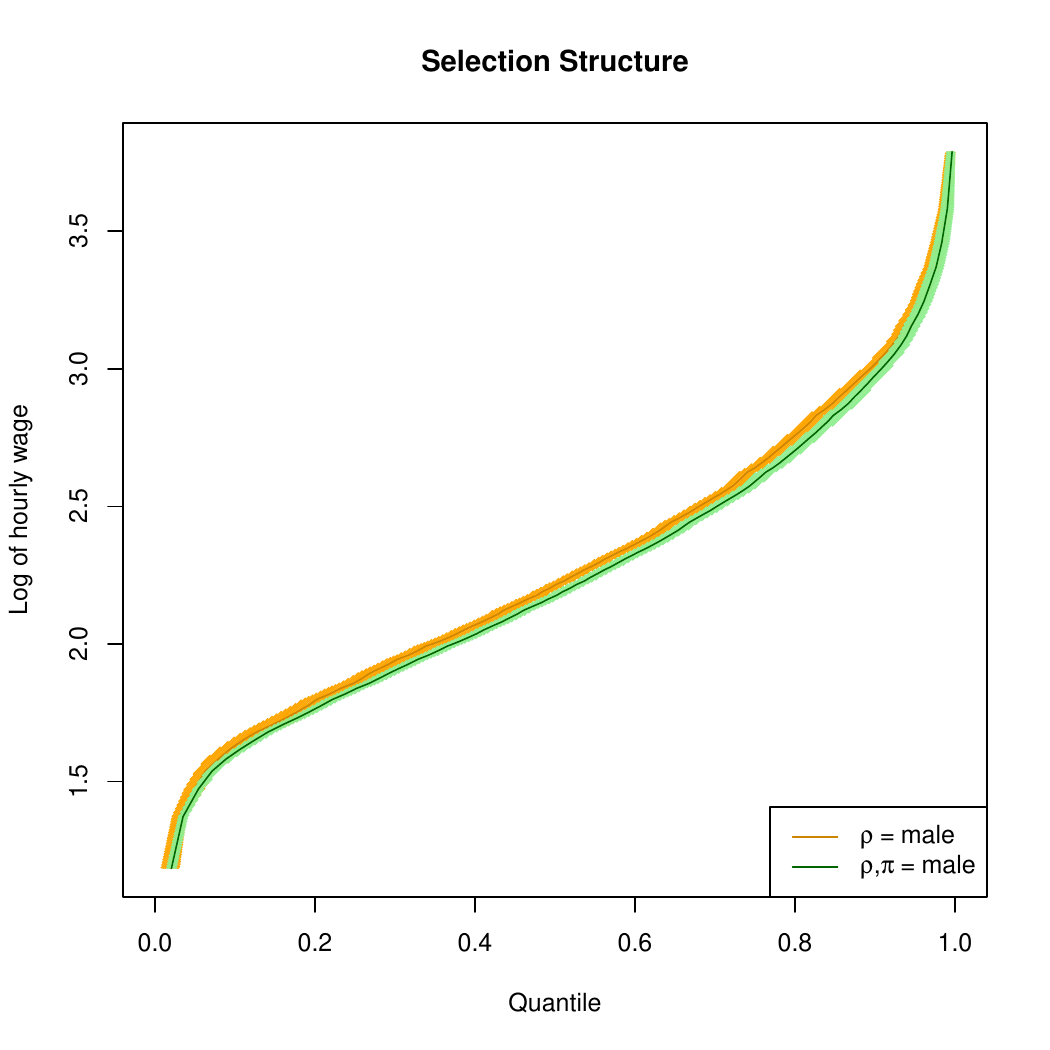}\\
    \includegraphics[height=0.45\textwidth, width=0.45\textwidth,page=31]{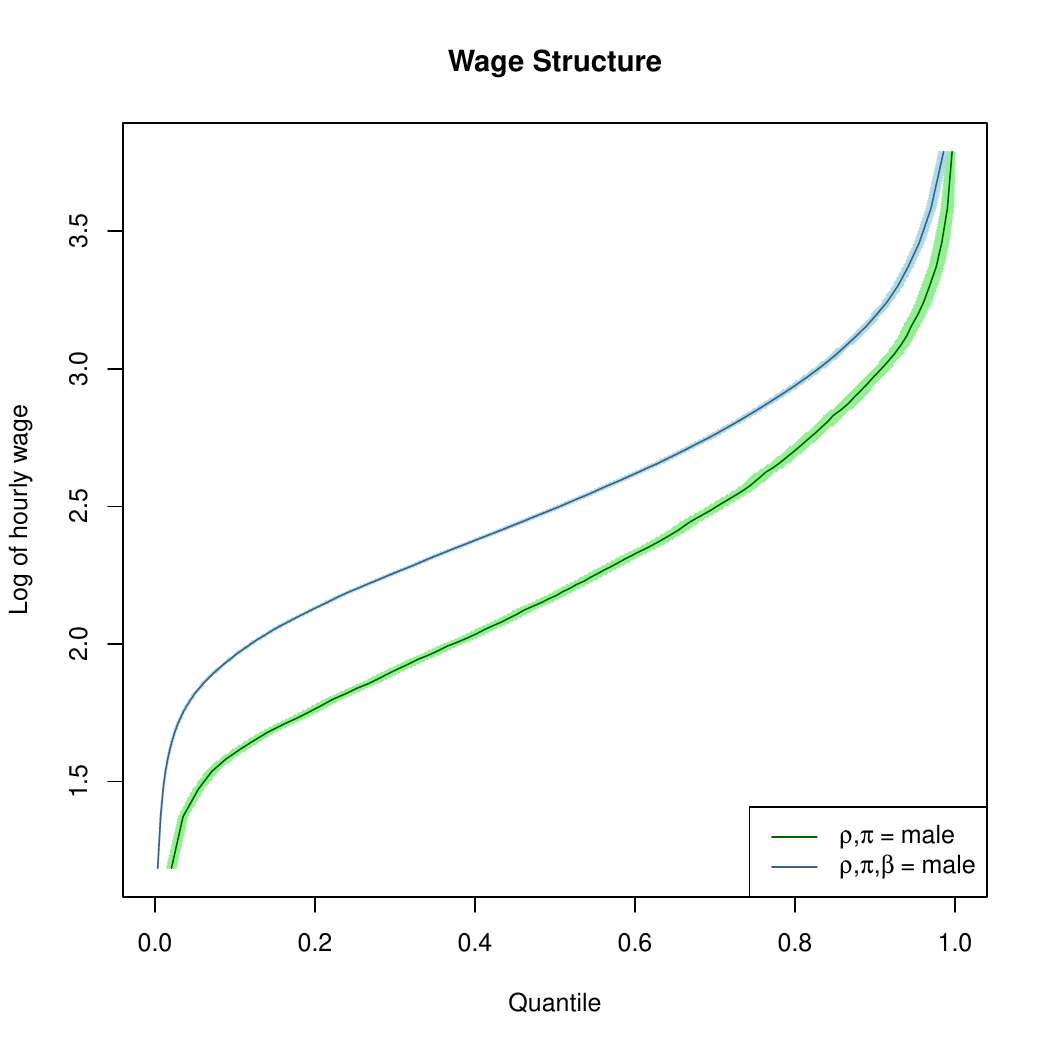}
    \includegraphics[height=0.45\textwidth, width=0.45\textwidth,page=32]{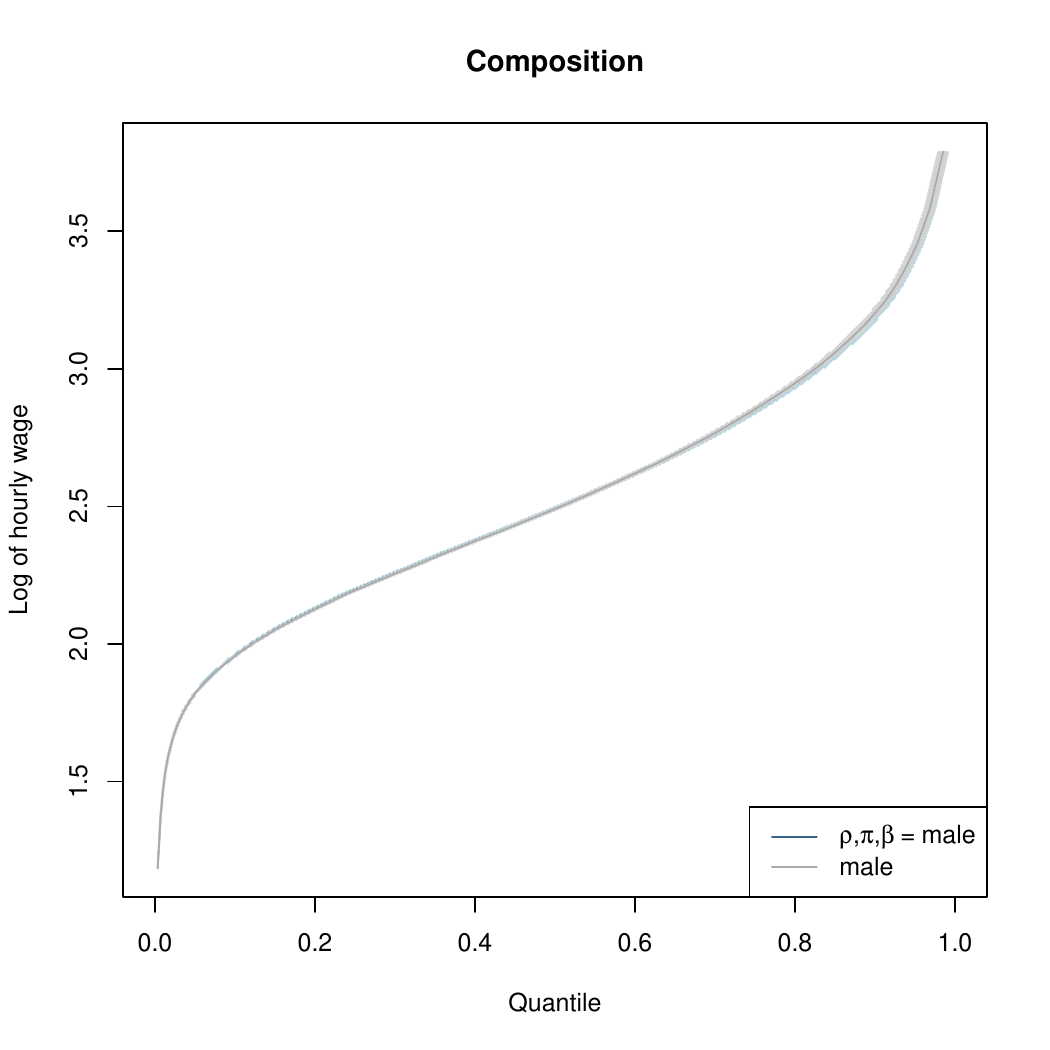}
    \caption{Estimates and 95\% confidence bands for components of wage decomposition between women and men in specification 3}
\end{figure}

\begin{figure}
    \includegraphics[height=0.45\textwidth, width=0.45\textwidth,page=29]{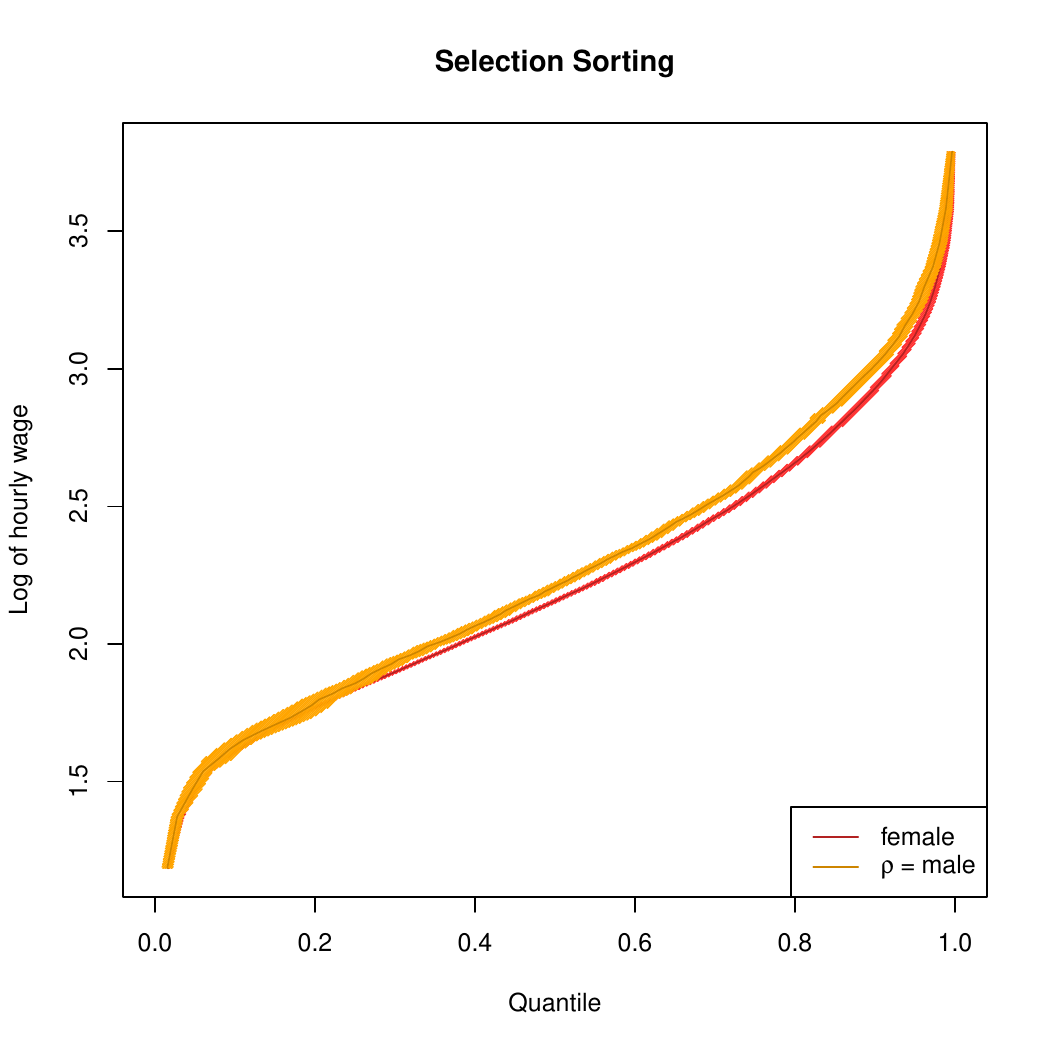}
    \includegraphics[height=0.45\textwidth, width=0.45\textwidth,page=30]{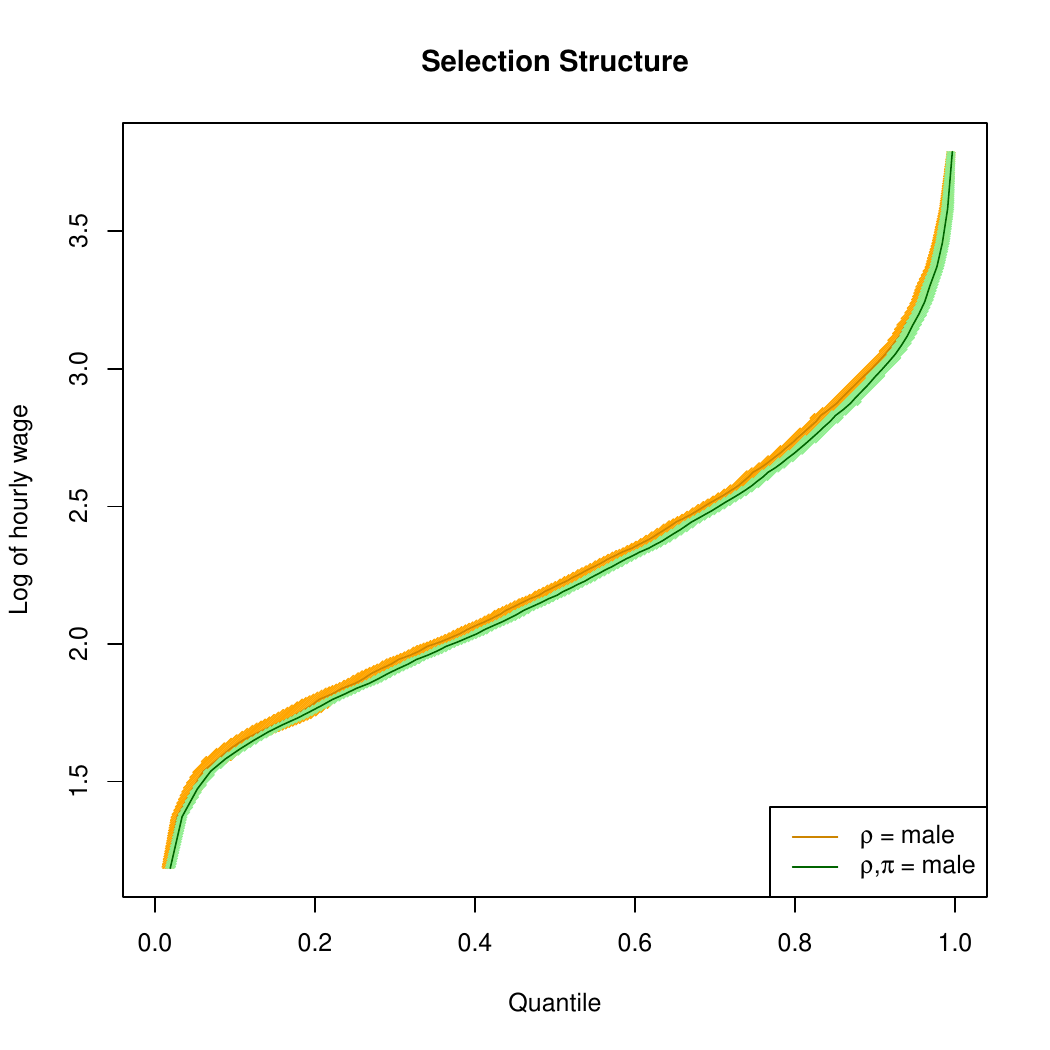}\\
    \includegraphics[height=0.45\textwidth, width=0.45\textwidth,page=31]{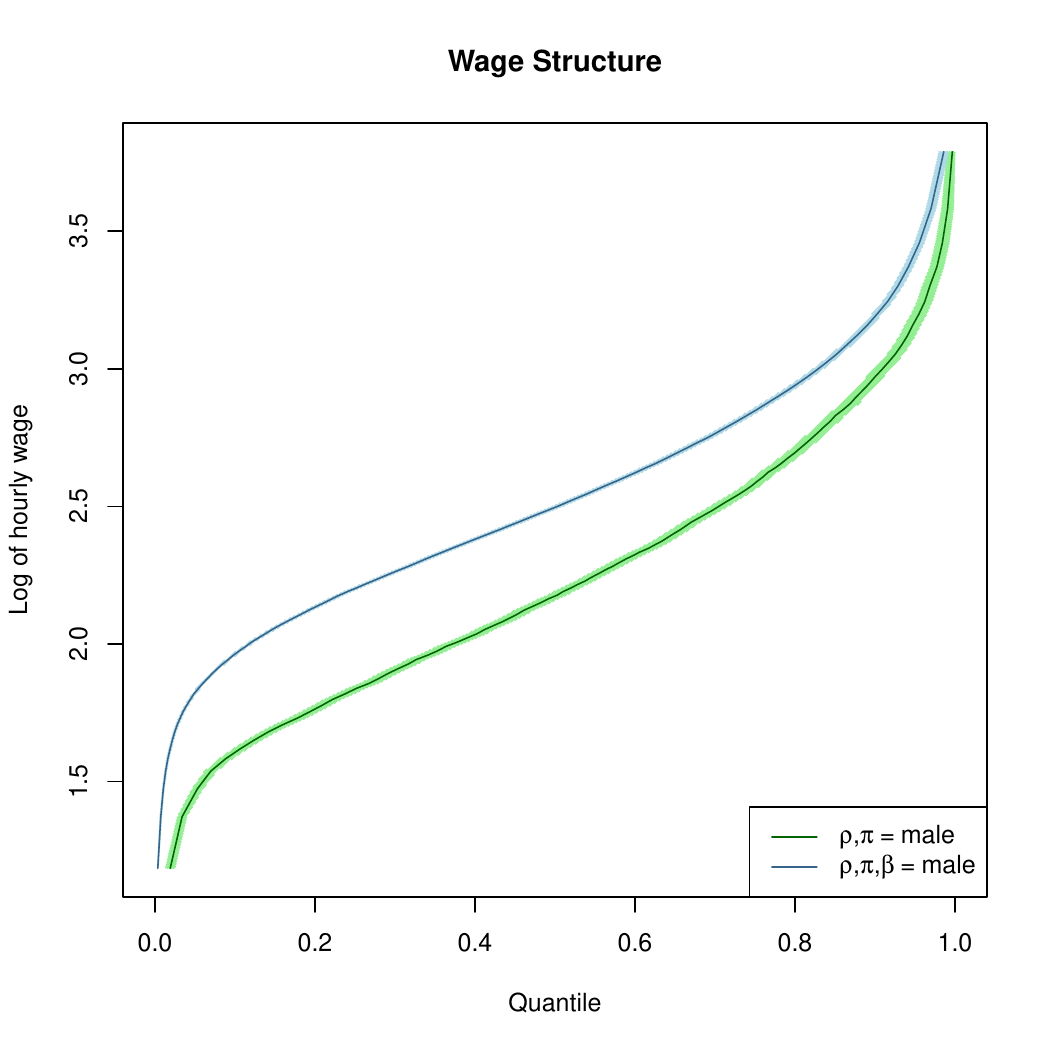}
    \includegraphics[height=0.45\textwidth, width=0.45\textwidth,page=32]{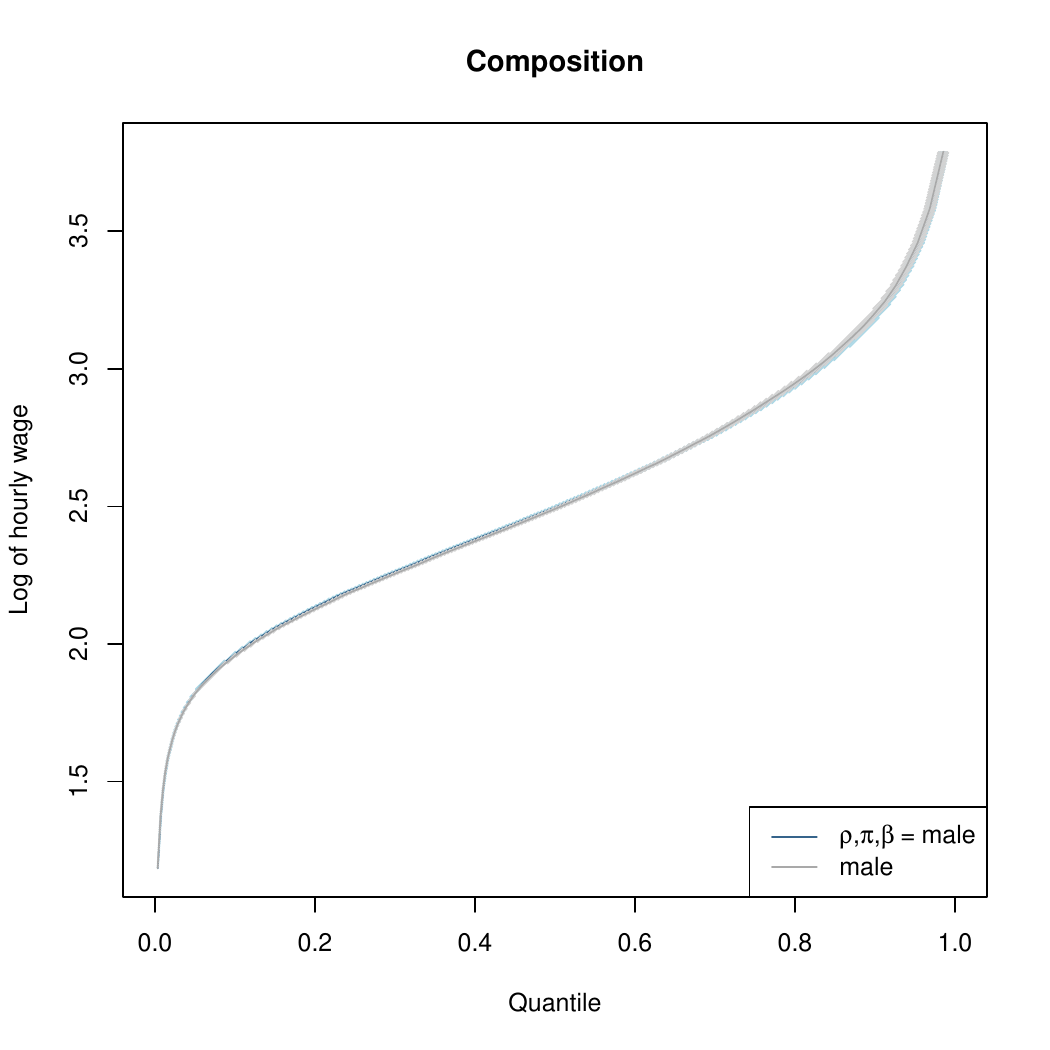}
    \caption{Estimates and 95\% confidence bands for components of wage decomposition between women and men in specification 4}\label{fig:cdec_spec4}
\end{figure}

\begin{figure}
    \includegraphics[height=0.45\textwidth, width=0.45\textwidth,page=29]{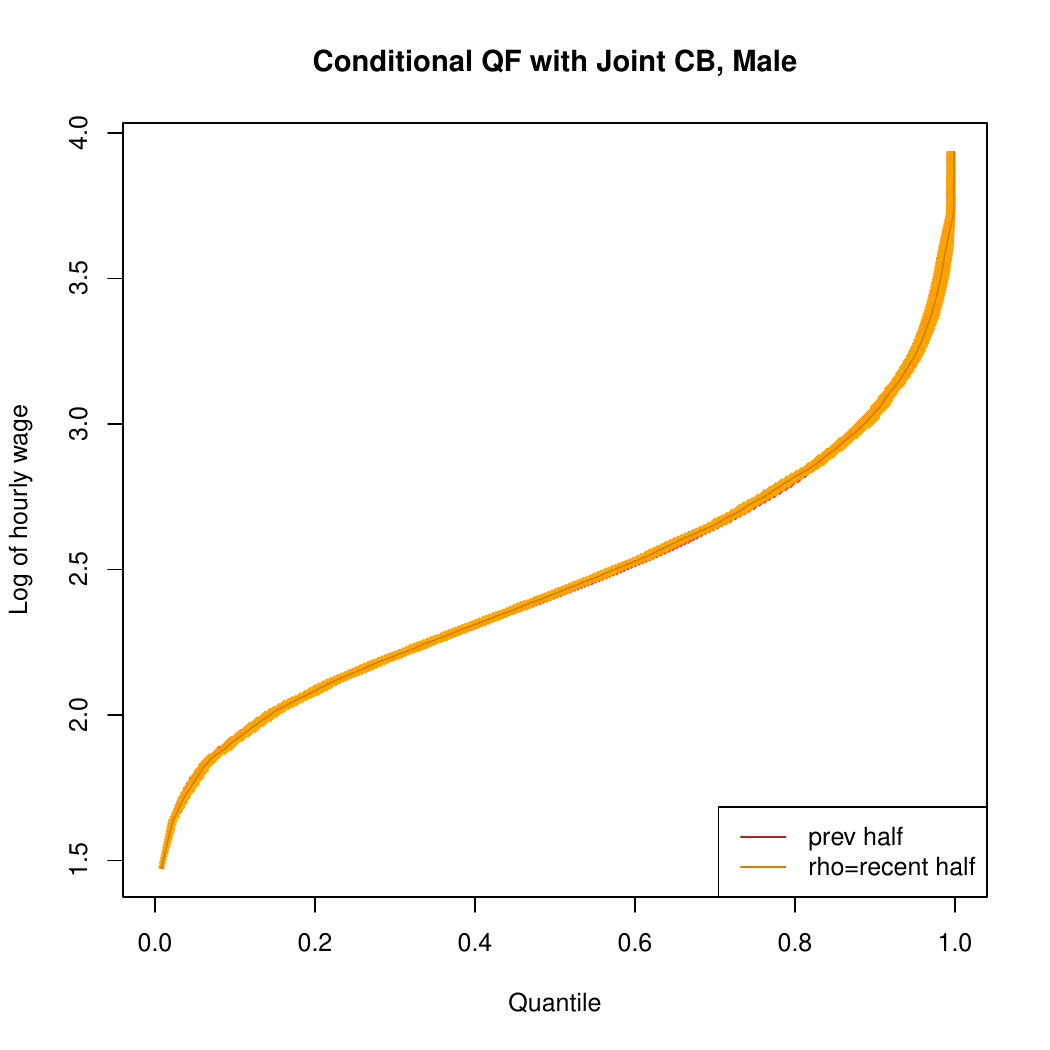}
    \includegraphics[height=0.45\textwidth, width=0.45\textwidth,page=30]{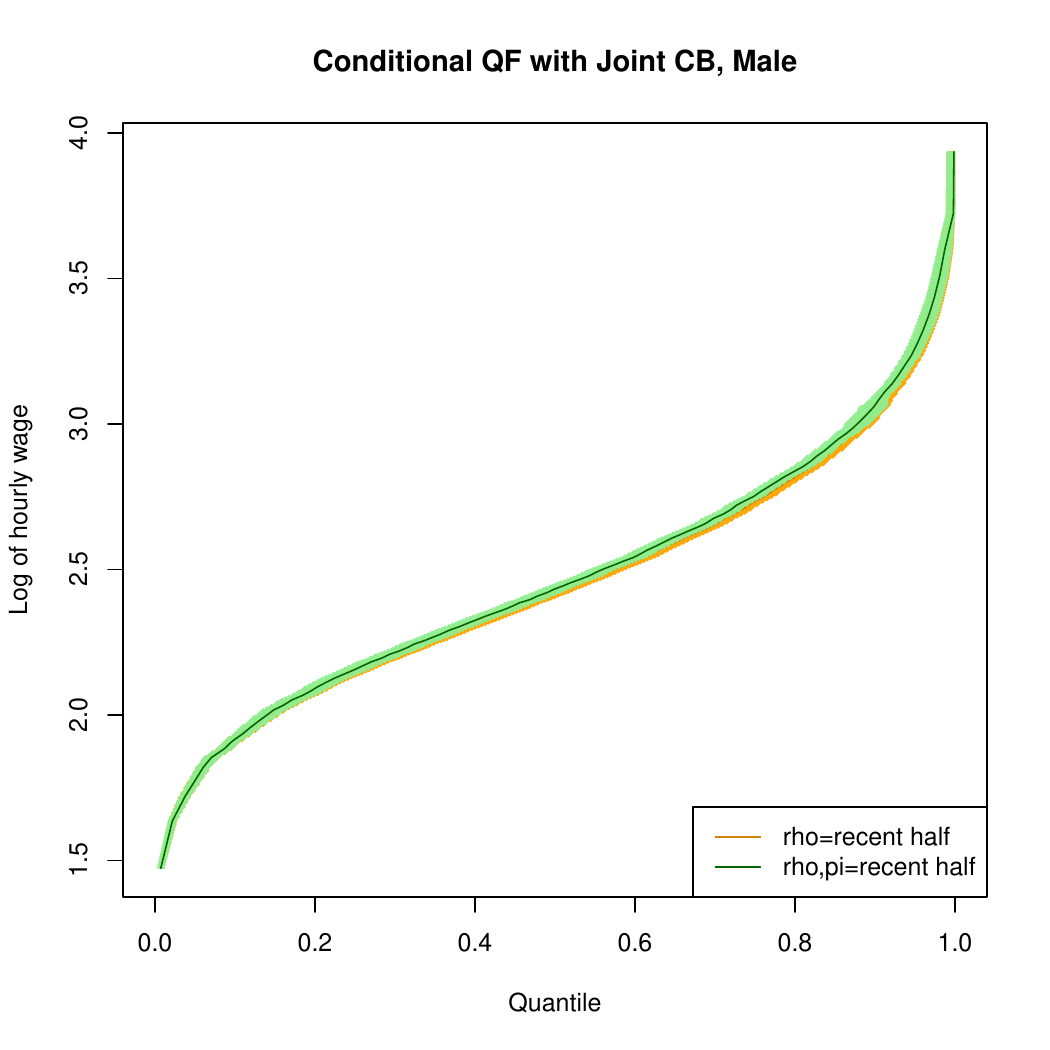}\\
    \includegraphics[height=0.45\textwidth, width=0.45\textwidth,page=31]{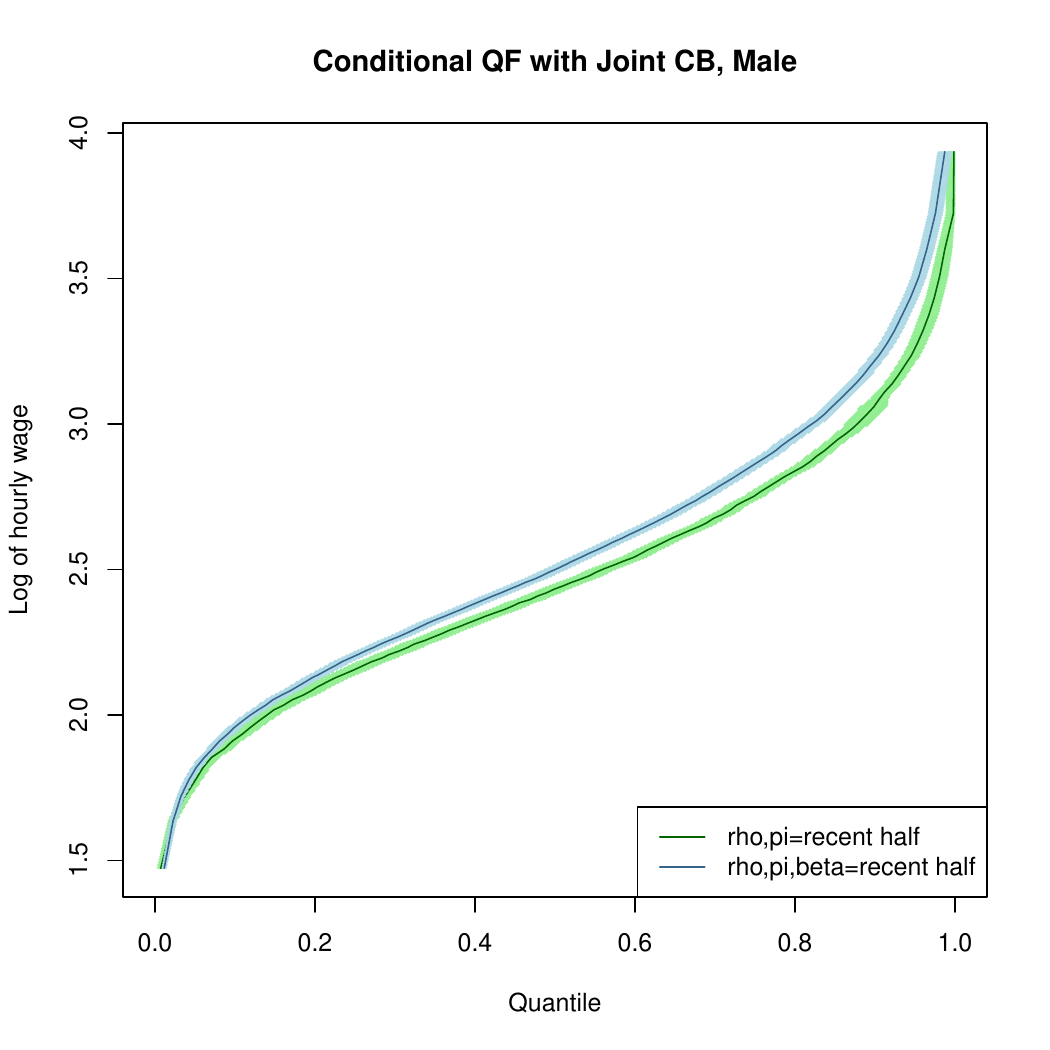}
    \includegraphics[height=0.45\textwidth, width=0.45\textwidth,page=32]{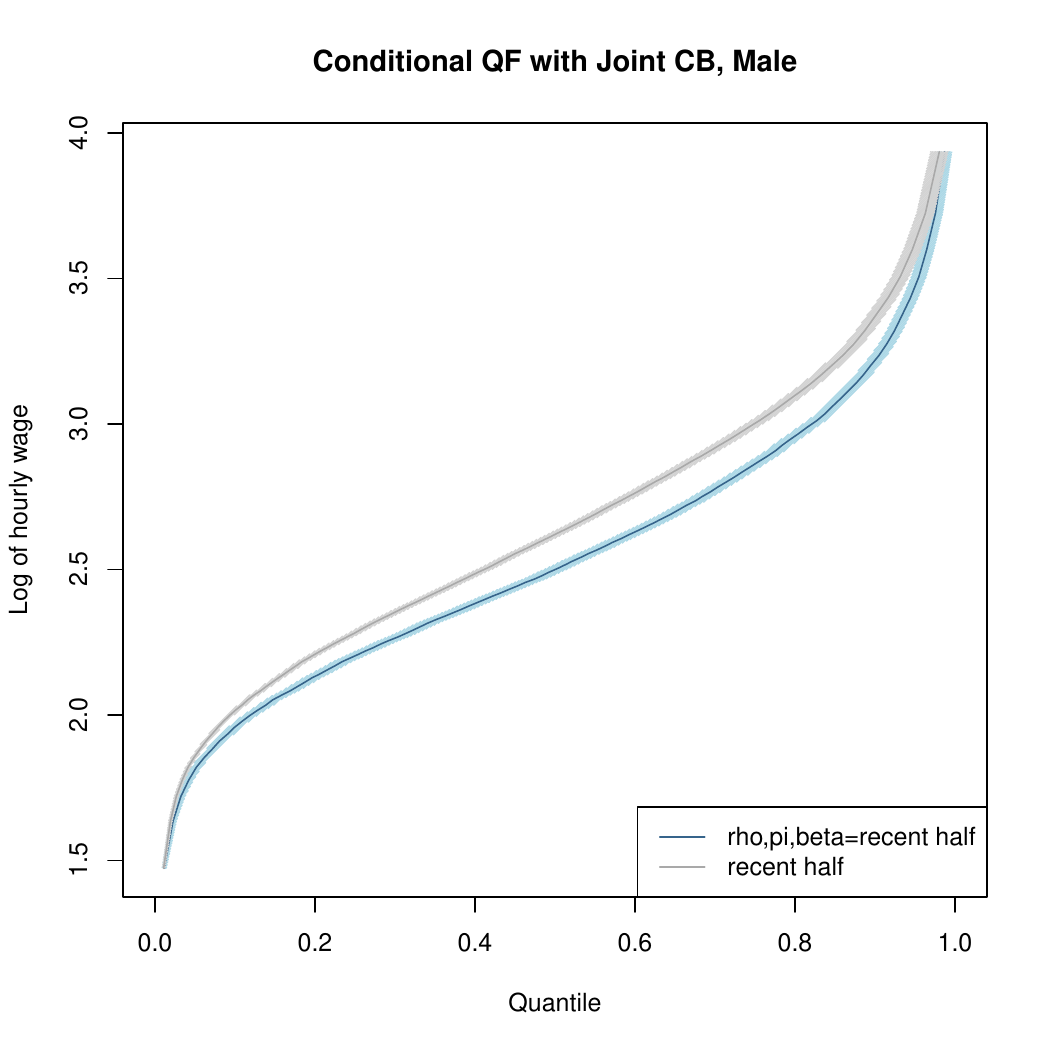}
    \caption{Estimates and 95\% confidence bands for components of wage decomposition between first and second half of sample period for men in specification 2}\label{fig:cdec_men_spec1}
\end{figure}

\begin{figure}
    \includegraphics[height=0.45\textwidth, width=0.45\textwidth,page=29]{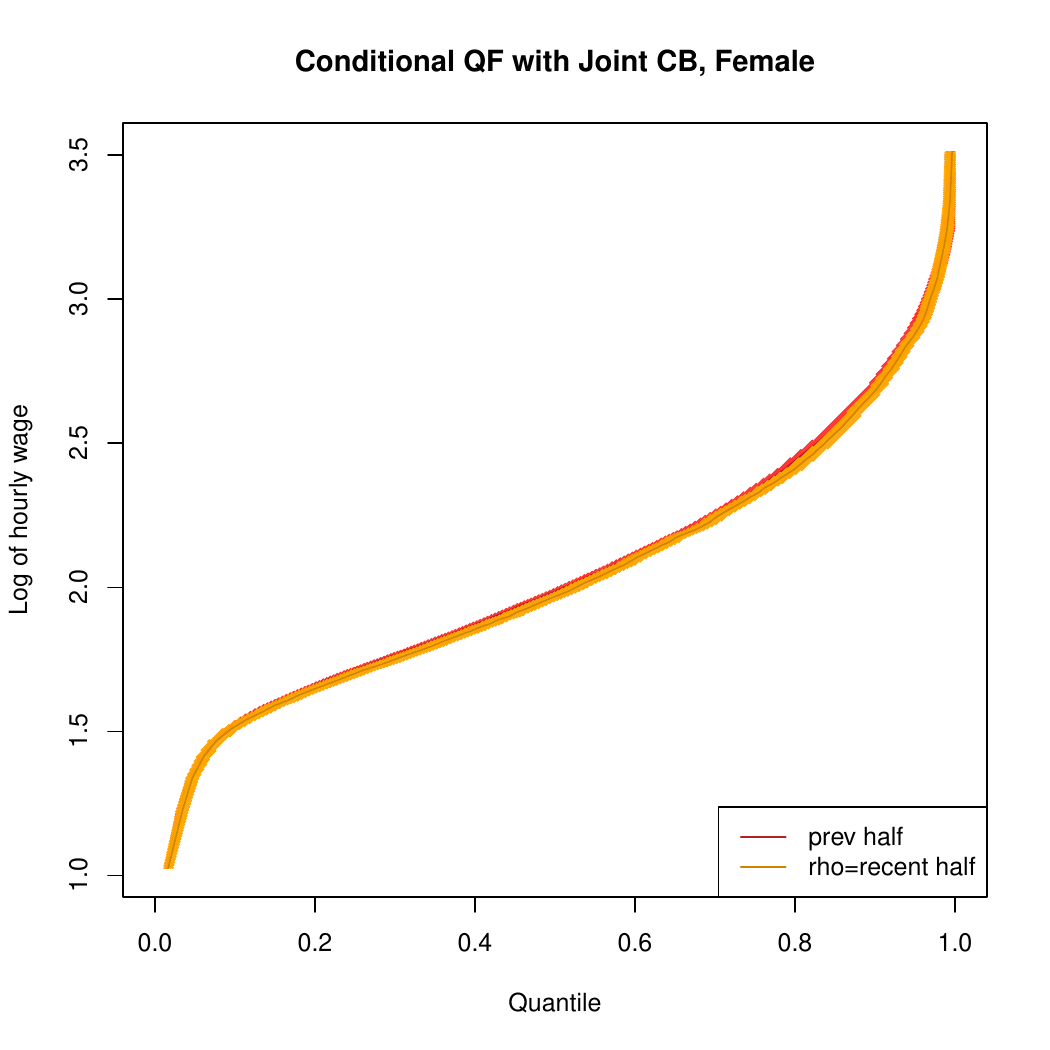}
    \includegraphics[height=0.45\textwidth, width=0.45\textwidth,page=30]{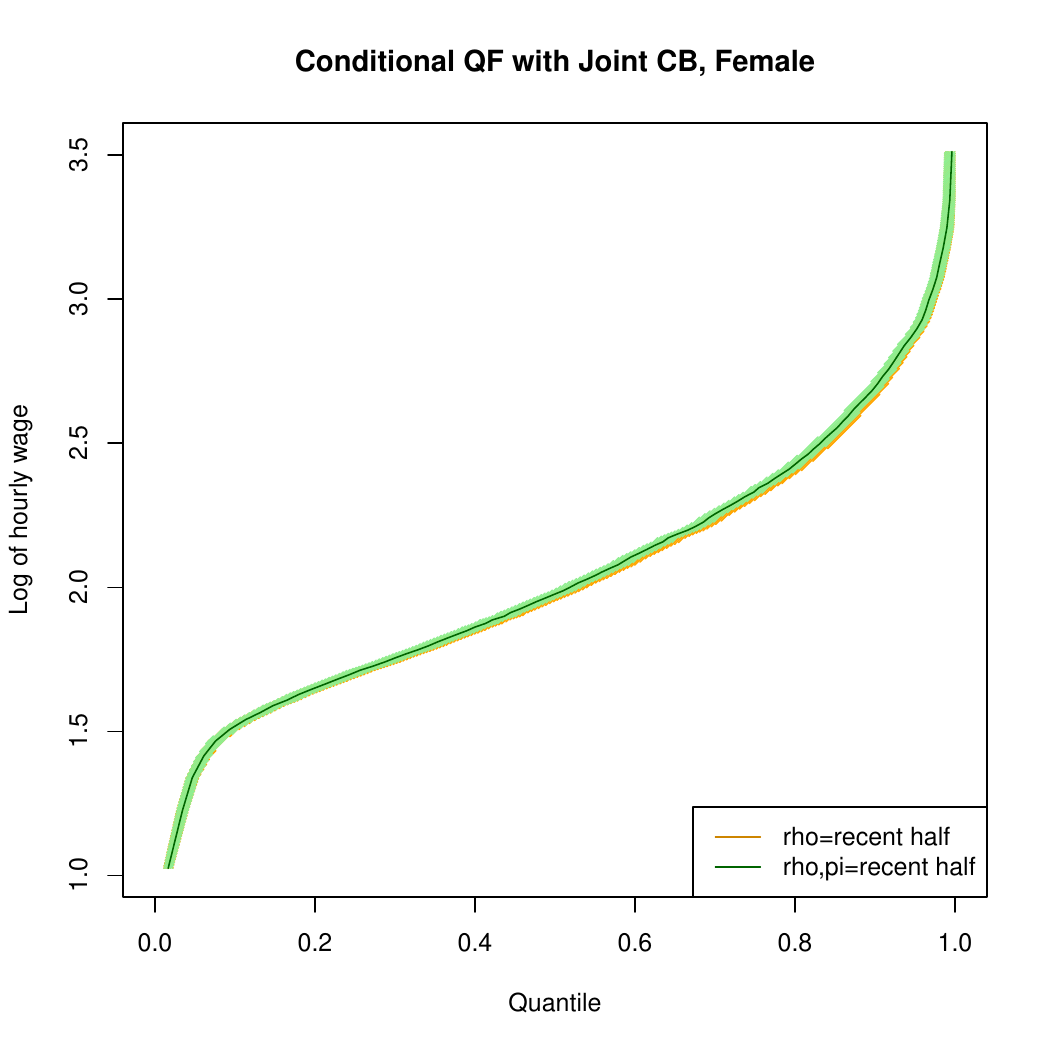}\\
    \includegraphics[height=0.45\textwidth, width=0.45\textwidth,page=31]{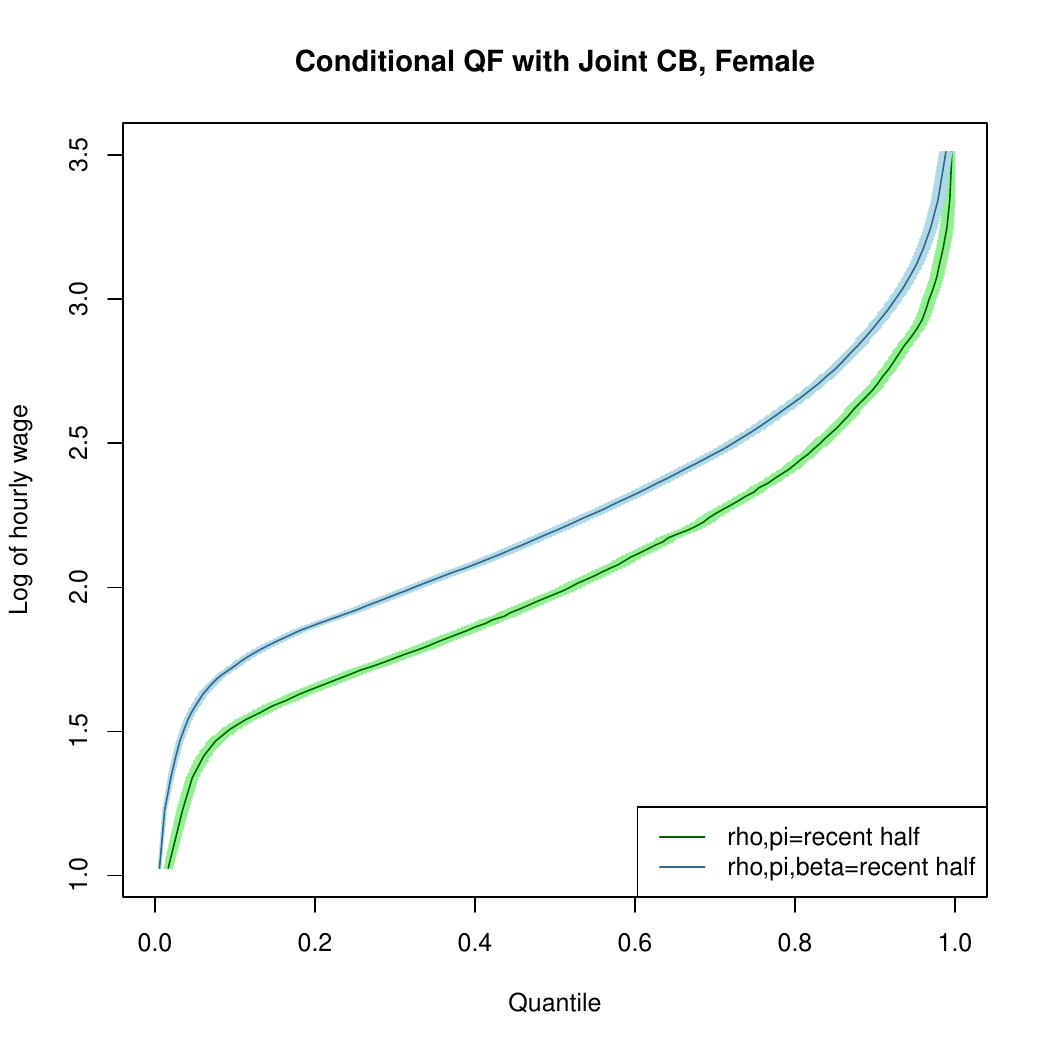}
    \includegraphics[height=0.45\textwidth, width=0.45\textwidth,page=32]{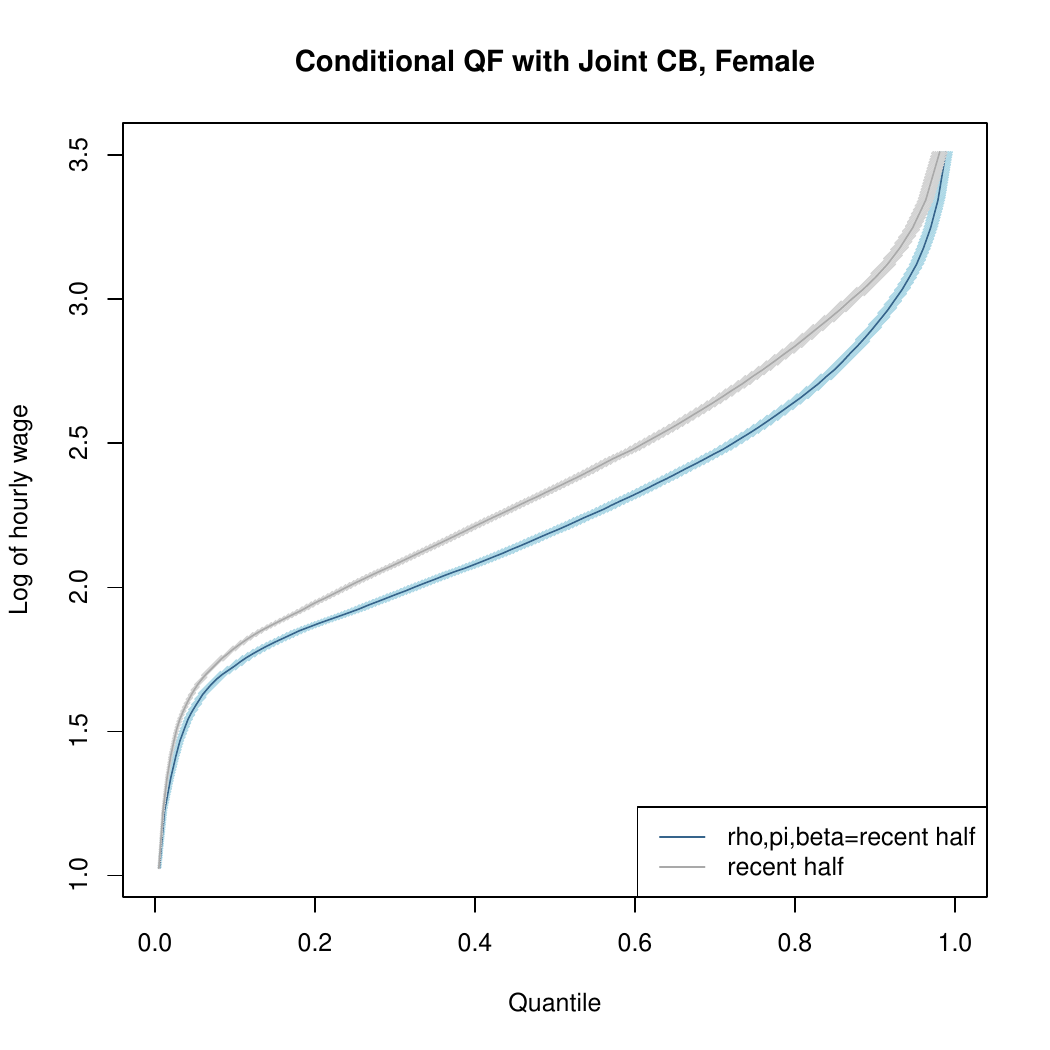}
    \caption{Estimates and 95\% confidence bands for components of wage decomposition between first and second half of sample period for women in specification 2}\label{fig:cdec_women_spec1}
\end{figure}

\end{document}